\documentclass[11pt]{article}

\usepackage{amsmath, amssymb, amsfonts}
\usepackage{graphicx}
\usepackage{float} 
\usepackage{multirow}

\usepackage{authblk}

\usepackage[usenames,dvipsnames,svgnames,table]{xcolor}
\definecolor{dullpurple}{rgb}{0.431,0.188,0.534}
\definecolor{darkgreen}{rgb}{0.075,0.302,0.047}
\definecolor{darkergreen}{rgb}{0,0.196,0.125}
\definecolor{darkergreen2}{rgb}{0,0.294,0.188}
\definecolor{dullred}{rgb}{0.706,0.208,0.192}
\definecolor{darkred}{rgb}{0.545,0,0}
\definecolor{dullblue}{rgb}{0,0.298,0.49}
\definecolor{blue3}{RGB}{31, 119, 180}

\usepackage{csquotes}

\usepackage[margin=1in]{geometry}   

\usepackage{multicol}
\usepackage{subfig} 

\newcommand\snowmass{
\begin{center}
  \rule[-0.2in]{\hsize}{0.01in}\\
  \rule{\hsize}{0.01in}\\
  \vskip 0.1in
  Submitted to the Proceedings of the US Community Study\\ 
  on the Future of Particle Physics (Snowmass 2021)\\
  \rule{\hsize}{0.01in}\\
  \rule[+0.2in]{\hsize}{0.01in}\\[-2em]
\end{center}
}

\usepackage[firstpage=true]{background}
\backgroundsetup{contents={\parbox{6.5in}{\snowmass}}, scale=1,placement=top,opacity=1,color=black,position={3.25in,1.2in}}

\usepackage[
colorlinks=true,
urlcolor=dullblue,
anchorcolor=blue3,
citecolor=dullblue,
filecolor=blue3,
linkcolor=darkred,
menucolor=blue,
linktocpage=true
]{hyperref}

\title{\Huge{Snowmass 2021 Community Survey Report}

\normalsize{from the Snowmass Early Career Survey Core Initiative Team}}

\author[1]{Garvita~Agarwal}
\author[2,3]{Joshua~L.~Barrow}
\author[4]{Mateus~F.~Carneiro}
\author[5]{Erin~Conley} 
\author[6]{Maria~Elidaiana~da~Silva~Pereira}
\author[5]{Sam~Hedges}
\author[7]{Samuel~Homiller}
\author[8]{Ivan~Lepetic}
\author[9]{Tianhuan~Luo}
\affil[1]{State University of New York at Buffalo, Buffalo, NY, USA}
\affil[2]{Massachusetts Institute of Technology, Cambridge, MA, USA}
\affil[3]{Tel-Aviv University, Tel-Aviv, Israel}
\affil[4]{Brookhaven National Laboratory, Upton, NY, USA}
\affil[5]{Duke University, Durham, NC, USA}
\affil[6]{Hamburger Sternwarte, Universit\"{a}t Hamburg, HH, Germany}
\affil[7]{Harvard University, Cambridge, MA, USA}
\affil[8]{Rutgers University, Piscataway, NJ, USA}
\affil[9]{Lawrence Berkeley National Laboratory, Berkeley, CA, USA}

\date{\today}

\begin{document}

\maketitle
\thispagestyle{empty}

\begin{abstract}
The Snowmass Community Survey was designed by the Snowmass Early Career (SEC) Survey Core Initiative team between April 2020 and June 2021, and released to the community on June 28, 2021. It aims to be a comprehensive assessment of the state of the high-energy particle and astrophysics (HEPA) community, if not the field, though the Snowmass process is largely based within the United States. Among other topics, some of the central foci of the Survey were to gather demographic, career, physics outlook, and workplace culture data on a large segment of the Snowmass community. With nearly $1500$ total interactions with the Survey, the SEC Survey team hopes the findings and discussions within this report will be of service to the community over the next decade. Some conclusions should reinforce the aspects of HEPA which are already functional and productive, while others should strengthen arguments for cultural and policy changes within the field.
\end{abstract}


\clearpage
\tableofcontents
\vspace{0.5cm}


\section{Introduction}
\label{sec:intro}

The first Snowmass Process began on June 28, 1982, in Snowmass, Colorado \cite{Donaldson:1983mq}. The series took the form of a several weeks-long summer study deciding the top scientific priorities for particle physics. Today, the Snowmass 2021 Process \cite{Snowmass2021} consists of ten active Frontiers of research made up of topical and subtopical working groups. These groups have come together over the past two years to discern the most important future research directions for their collective work. The process will culminate in the Community Summer Study, expected to be held at the University of Washington in Summer 2022, collecting reports from these Frontiers in the forms of individual white papers and Frontier-level summarized reviews. From these white papers and summaries, the final Snowmass Report will be authored, which in turn will inform the Particle Physics Project Prioritization Panel (P5) \cite{P5Website}. The P5 report, distilled from these inputs to Snowmass, finally points funding agencies such as the United States Department of Energy and the National Science Foundation toward future physics initiatives, reorienting the field for the next decade or more.

As a largely volunteer-based organization, Snowmass 2021 has generally welcomed the involvement of early career (``young'') scientists across its Frontiers and Topical working groups. A key facilitator in this regard has been the semi-separate Snowmass Early Career (SEC) group \cite{SECWebsite}, a parallel conglomerate of young scientists with $\lesssim 10$ years of professional experience. This group was initially nominated by their American Physical Society (APS) Division of Particles and Fields (DPF) \cite{APS_DPF} colleagues to join SEC in recognition of their past and current work, as well as their future professional promise.

Through the SEC, these future leaders in particle physics took on several initiatives in service to themselves and the broader community:
\begin{enumerate}
    \item Snowmass Coordination
    \item Inreach
    \item Diversity, Equity, and Inclusion (DEI)
    \item Long-Term Organization
    \item Snowmass Community Survey
\end{enumerate}
Snowmass Coordination aimed to integrate early career voices within the various Frontiers, establishing liaison positions throughout the Snowmass consortia. Inreach aimed to broadly educate early career physicists on the current and future physics theory and experiment while encouraging collaboration across the Frontiers. DEI aimed to facilitate a welcoming and respectful environment with the SEC and serve as a leading example to the Snowmass community in such efforts. Long-Term organization aimed to establish an independent arm of advocacy and networking for early career physicists throughout the fields of particle physics. Finally, the survey aimed to inform the community on the people and opinions that make up the whole of the field. This document will review the techniques, results, and some interpretations of the 2021 Snowmass Community Survey.

Previous versions of the Snowmass Survey took place in 2001~\cite{Fleming:2001zk} and 2013~\cite{Anderson:2013fca} and were similarly led by early career physicists, many of them continuing to work both in the particle physics community and within the 2021 Snowmass process. The major motivations of those surveys was to collect demographic and career-oriented information, as well as to gauge interest on potential future particle physics projects. The SEC Survey team studied these past survey in some detail, evaluating their content and style in order to best construct the Snowmass 2021 Survey.

From discussions among the SEC Survey group members and several past survey authors, it was decided that the Snowmass 2021 Survey would focus more on human factors, experiences, trajectories, and broad opinions within the field. In this way, the 2021 Survey could potentially serve the community in a more informative manner, exploring and focusing on the time-independent human experiences which make up the field, rather than focusing on shorter time scales and specific experimental concerns which could be interpreted as effectively voting on projects' priorities. Given that the main tenets of the Snowmass process are already to address these concerns to inform the Snowmass Report and thus the P5, we targeted a different path. Given the scope of the 2021 Snowmass process, along with the nearly impossible task of adequately exploring all possible projects and future directions HEPA might take, our primary focus was to understand human experiences while working in the field.

With these goals in mind, the SEC Survey team established a manifold approach in designing the survey's questions. Over the course of several months, the questions were developed and agreed upon. The questions broadly fall into the following seven categories, here presented in the order that they will be discussed in this document:
\begin{enumerate}
    \item Demographics
    \item Physics Outlook
    \item Careers
    \item Workplace Culture
    \item Diversity and Racism
    \item Caregiving Responsibilities 
    \item Impacts of COVID-19
\end{enumerate}
The remainder of this report is structured as follows. In Section~\ref{sec:methods} we detail the methodologies behind the construction, distribution, and analysis of the 2021 Snowmass Survey. Section~\ref{sec:results} is devoted to discussing the responses to the survey. We discuss the results in Section~\ref{sec:discussions}, and Section~\ref{sec:conclusions} contains our conclusions. Two appendices contain a list of all questions asked in the 2021 Snowmass Survey and additional results. The final appendix details ideas for effectively conducting the survey in future Snowmass processes. Questions not covered in the Results Section will appear in Appendix \ref{appendix_surveyQs} and will be discussed in an updated version of this white paper.

\section{Methodology}
\label{sec:methods}
\subsection{Survey Development}

The construction of the Snowmass 2021 Community Survey began with an evaluation of questions which had appeared in previous Snowmass Surveys. The evaluation consisted of an ``efficacy survey'' using Qualtrics software~\cite{Qualtrics} which contained of all questions appearing in the 2001~\cite{Fleming:2001zk} and 2013~\cite{Anderson:2013fca} surveys. In the efficacy survey, respondents provided their opinions on whether to keep or remove a question similar to one that appeared in 2001 or 2013. Neutral responses were accepted in the efficacy survey, and respondents could provide additional comments or suggestions via a text box. All members of the SEC Survey Initiative as well as SEC DEI members \cite{SEC_DEI} completed the efficacy survey between July and August 2020. 
 
The Survey initiative organized weekly discussions centered around the 2001 and 2013 survey questions based on the responses from all recorded interactions in the efficacy survey. We grouped questions into sets based upon two primary factors: (1) content of the original question and (2) average response based on the efficacy survey results. Over a period of several weeks, the weekly discussions drove the development of the seven main categories of questions we presented in the 2021 survey. This process also organically led to conversations about how to best combine and arrange the different categories. Some sets of questions contained a strong consensus to keep some version of those questions; the discussions consisted of changing syntax or answer options. For other groups, the questions received a mixed response, so we discussed whether to include those questions in the final survey. The overall process compelled us to develop questions that would hopefully stand the test of time and could potentially be reused in a future survey. Given these discussions and the enormity of the Snowmass process, it was decided to give lower priority toward rankings of particular experimental concepts, and we instead focused on gathering opinions about the overall Snowmass process.

Once an initial version of the 2021 survey was agreed upon, the SEC Survey Initiative sent the survey to a small group of people for feedback and to estimate the survey's duration. Each SEC Survey Initiative member chose one or two people with the attempt to cover a wide variety of career stages and life experiences. Many of the suggestions and feedback from this exercise were incorporated into the final version of the survey.

\subsection{Survey Methods}

The Snowmass 2021 Community Survey was implemented using online Qualtrics~\cite{Qualtrics} software hosted by Duke University. We released the survey to the community on June 28, 2021 with the following preface:

\begin{displayquote}
{\small
The Snowmass Survey represents a core initiative within the Snowmass Early Career organization. A team of early career scientists has prepared an in-depth set of questions for the entire High Energy Physics \& Astrophysics (HEPA) community (including students, faculty, scientists, engineers, technicians, and people who have recently left the field). Please make your voice heard by participating in this survey; your input will inform both the HEPA field and its supporting agencies about demographics, career opportunities, and experiences within HEPA.

Topics of this survey include but are not limited to careers, physics outlook, workplace culture, harassment, racism, visa policies, the impacts of COVID-19, and demographics. Some of the questions may make you uncomfortable due to the subject matter. These sections are optional, and anything you share with us will be kept confidential. 

Your answering of these questions in full is greatly appreciated. The survey will take 20-60 minutes to complete. The survey will save your progress as you complete it, allowing you to leave and come back anytime within the survey’s data-taking time window (provided you use the same machine and browser). The survey will be accessible from June 28 to August 15, 2021 at 11:59PM CST. You may pause taking the survey at any point during this time window. 

The answers you provide for this survey will be used for investigative purposes only and will not reveal who you are. The information you share will be kept private from those outside the SEC Survey/Career \& Professional Development (CPD) teams. Any personal information that could identify you will be removed or changed before results are made public, published, or used for future research. Raw data will not be made publicly available. This survey does not require Institutional Review Board approval because it falls under “generalized knowledge” (i.e., is not a “systematic investigation”) and therefore does not count as “regulated research”. By continuing, you are agreeing to participate in this survey.

We want to foster more involvement and maximize the impact of this survey. Feel free to distribute this survey to others involved in HEPA or who have recently left the field. If you have suggestions for specific groups, organizations, or collaborations to distribute this survey to, you can additionally reach out to the Survey Team via email: \url{SNOWMASS21-SURVEY@listserv.fnal.gov}. We hope to make this process as inclusive as possible, so please make your voice heard.

Thank you for your participation in this survey!
}
\end{displayquote}

\begin{figure}[H]
    \includegraphics[scale=0.56]{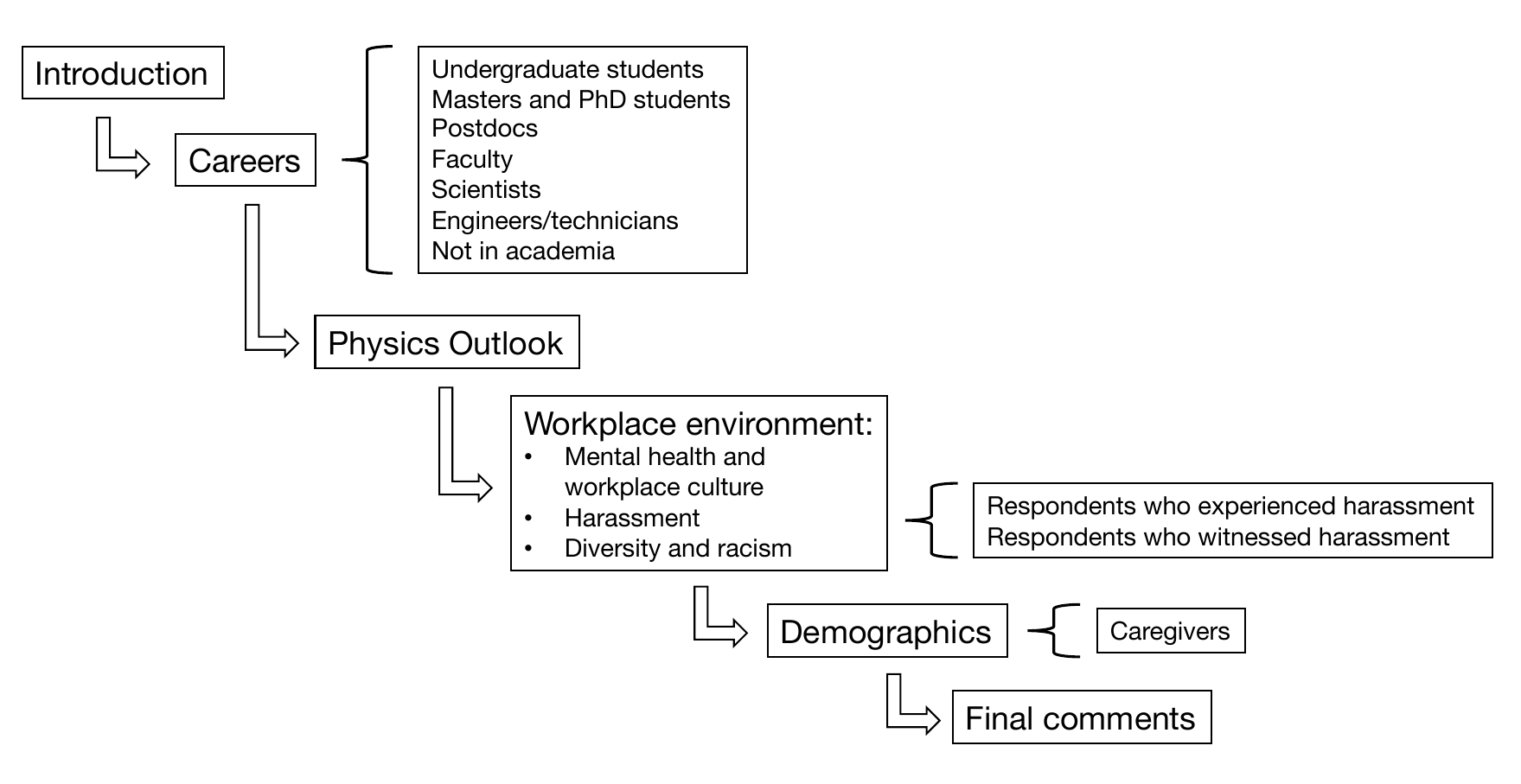}
    \caption{General overview of the 2021 Snowmass Community Survey. The arrows represent the overall flow of the survey ordered by the questions respondents viewed earliest, and the brackets present some of the display logic utilized in the survey.}
    \label{snowmass_survey_flowchart}
\end{figure}

Figure \ref{snowmass_survey_flowchart} shows the outline of the 2021 Snowmass Community Survey, and a list of all questions asked can be found in Appendix \ref{appendix_surveyQs}. The only required question in the survey was ``Are you currently in academia?'' which was presented immediately following the preface. All other questions were optional; combined with the fact that we used display logic to ask unique sets of questions to different groups of respondents, all questions save ``Are you currently in academia?'' do not have a fixed number of respondents. We primarily used display logic for questions about careers, harassment, and caregiving responsibilities; the display logic is represented by the brackets in Fig. \ref{snowmass_survey_flowchart}. Also using display logic, a majority of the physics outlook questions were only shown to respondents who answered ``Yes'' to ``Are you currently in academia?''.

The survey was accessible using either a URL link or via QR code. We created two separate but identical copies of the survey;  one copy was for ``safer'' virtual areas such as the Snowmass Slack workspace \cite{SnowmassSlack} or various institutional mailing lists where we expected little to no spam or bots. The other copy was for ``less safe'' virtual areas such as social media (only accessible via a URL link). We initially distributed the survey to the Snowmass community via Slack and email, and then expanded to the Fermilab community via an article in Fermilab At Work \cite{FNAL_SurveyAdvertisement} and an email to the Fermilab users mailing list. We also requested that collaboration spokespeople and representatives forward the survey to their collaboration members. The survey advertisement also appeared in the July 2021 DPF newsletter \cite{APS_DPF_2021JulyNewsletter}. Finally, we posted survey advertisements featuring the ``social media'' link on forums for software such as Geant4 and ROOT, along with various APS forums, and finally on social media sites such as Reddit \cite{Reddit}. 

The survey used Qualtrics fraud detection features that detect duplicate submissions and to detect bots \cite{Qualtrics_Fraud}. No submissions were tagged as duplicate. There were, however, a handful of survey interactions flagged as potential bots: One of these interactions was empty (i.e., no questions were answered), and we examined the other interactions to check for comments. We determined that the flagged interactions contained logical responses and comments, so we did not remove these from our pool of interactions. 

\subsection{Analysis Methods}

The primary focus for the 2021 survey analysis was to protect the respondents' privacy. For many of the plots shown in Section \ref{sec:results}, raw numbers were removed and distributions were re-normalized. We present the raw number of responses for the Demographics results in Section \ref{demographics}. In the figures, we only show the breakdown of categories with $\geq 5$ responses to preserve anonymity \cite{ipc2015, tufts2022}. For the case of categories with $<5$ responses, we discuss in the text relevant information contained in those answers.

One common analysis strategy was to separate responses based on provided demographic information such as gender, race, age, Snowmass frontier, and primary workplace. Some questions received only a handful of responses from respondents in a given demographic group; when finalizing plots based on this analysis method, we opted for groupings with at least 30 responses to keep results private and anonymous. On many occasions, respondents were not represented by any demographic group (e.g., they did not provide demographic information). To account for this, final plots always show  the combined responses from all respondents in order for complete representation.

To compare responses by race, we used racial splittings of ``White'' and ``other racial groups''. With the primary motivation being privacy, these splittings allowed for adequate groupings to show plots or report hard numbers. Because scientists with different racial backgrounds do not share a homogeneous experience in HEPA, we are limited in the conclusions we can draw with this analysis strategy. At the same time, by breaking down responses using these generic splittings, we could search for key differences in the way these groups responded to certain questions without exposing individuals. 

When comparing responses by gender, we found that we received a handful of responses from scientists who identify as nonbinary, agender, genderqueer, self-identify, or multi-gendered (shortened to NB-GQ-AG). Every question analyzed in this method yielded too small of groups to produce plots or report hard numbers for NB-GQ-AG scientists. To account for this, we discuss general trends from these responses when applicable. 

Some questions in the survey allowed for selection of multiple answer choices. To gauge whether there were any commonly selected pairs of answers, we calculated correlation matrices between all pairs of answers. Correlations above 0.5 were considered strong enough to make comments on.

Some questions presented in the 2021 survey were not analyzed for this white paper and will thus not be discussed beyond this section. The majority of these questions were added to the 2021 survey as part of an agreement with groups within the Snowmass Community Engagement Frontier (CEF) \cite{LOI_CommF2}, and they will handle the analysis and results of those questions in other white papers. Some questions are currently undergoing analysis and will appear in a future version of this paper.

\section{Results of the Survey}
\label{sec:results}

\subsection{\label{metadata}Metadata}

The 2021 survey was originally intended to run for 49 days (from June 28 until August 15, 2021). The survey was officially live for a total of 59 days, including a deadline extension to August 26. The survey accumulated 1462 total interactions over its lifetime -- these interactions ranged from opening the first page of the survey to fully completing the survey. 127 of these interactions came from the social media copy of the survey. The vast majority of interactions occurred via the URL link, and only a handful of interactions originated from the QR code.

Each interaction contains a ``progress'' variable that stores the percentage of survey seen by the respondent. We cut down the interactions into our final set of respondents using a cut of $\textrm{progress}=44$. This corresponds to a survey interaction which reached the end of (or continued beyond) the section containing career-related questions. With this selection cut, the survey received a total of 1014 respondents. 

\begin{figure}[H]
    \includegraphics[scale=0.65]{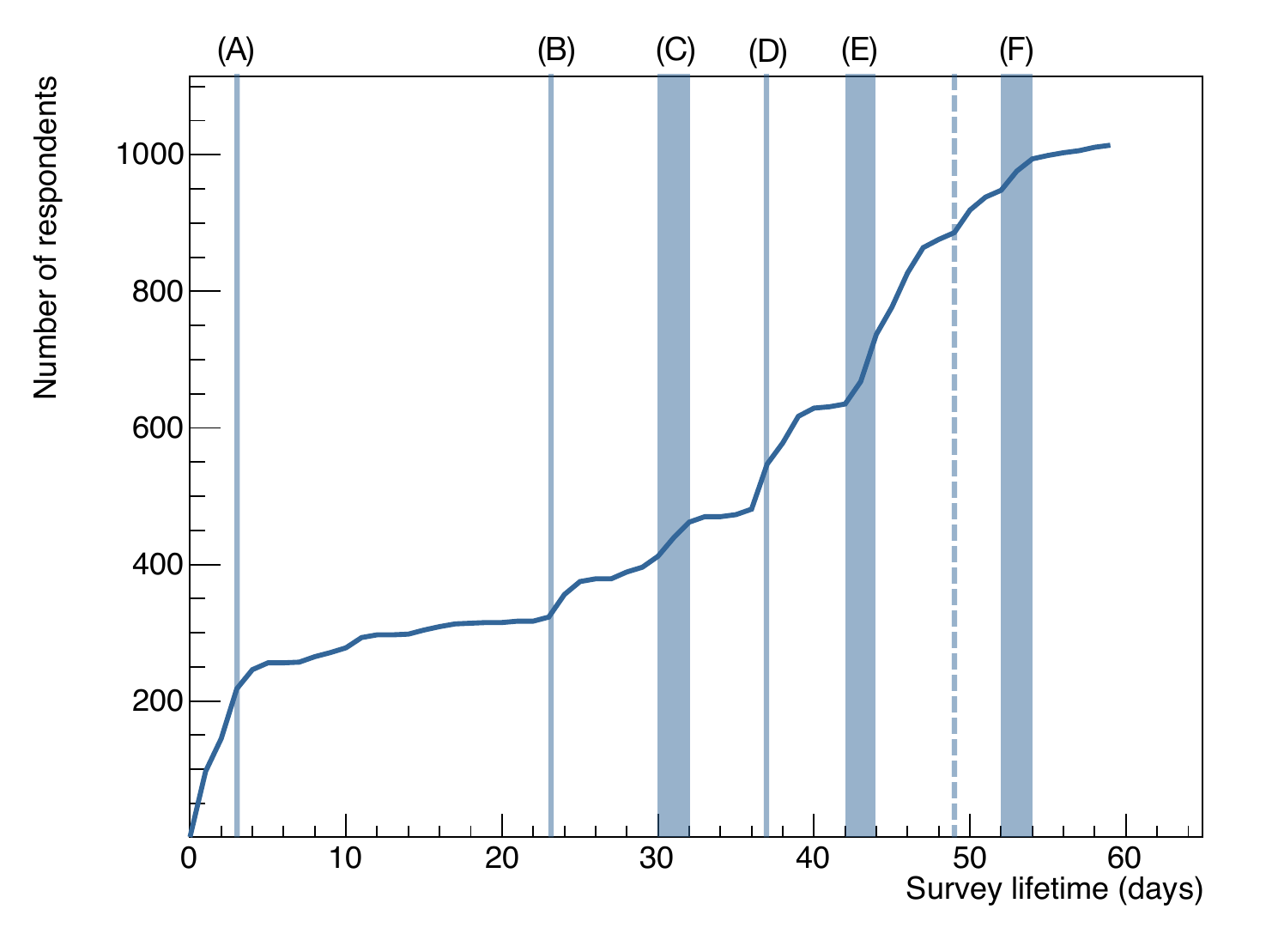}
    \caption{Number of respondents over the survey lifetime. The original deadline of August 15 is indicated by the dashed line. Various advertising campaigns are noted by the solid lines or bands, including (A) an article in Fermilab Today \cite{FNAL_SurveyAdvertisement} and email to the Fermilab users mailing list; (B) an advertisement to the Deep Underground Neutrino Experiment collaboration; (C) targeted advertising to various neutrino and astrophysics collaborations; (D) an appearance in the DPF Newsletter along with an APS+DAP email; (E) a reminder email sent to the Fermilab and Snowmass mailing lists; and (F) emails notifying the Fermilab and Snowmass communities about the deadline extension.}
    \label{overall_num_respondents_dates}
\end{figure}

\subsection{Demographics}

\label{demographics}

In this section, we present the following demographic information for the 2021 survey: gender, race/ethnicity, employment, age, citizenship and visa status. 

\subsubsection{Careers, Identity, and Origins}

\begin{figure}[H]
  \centering
  	\includegraphics[width=120mm]{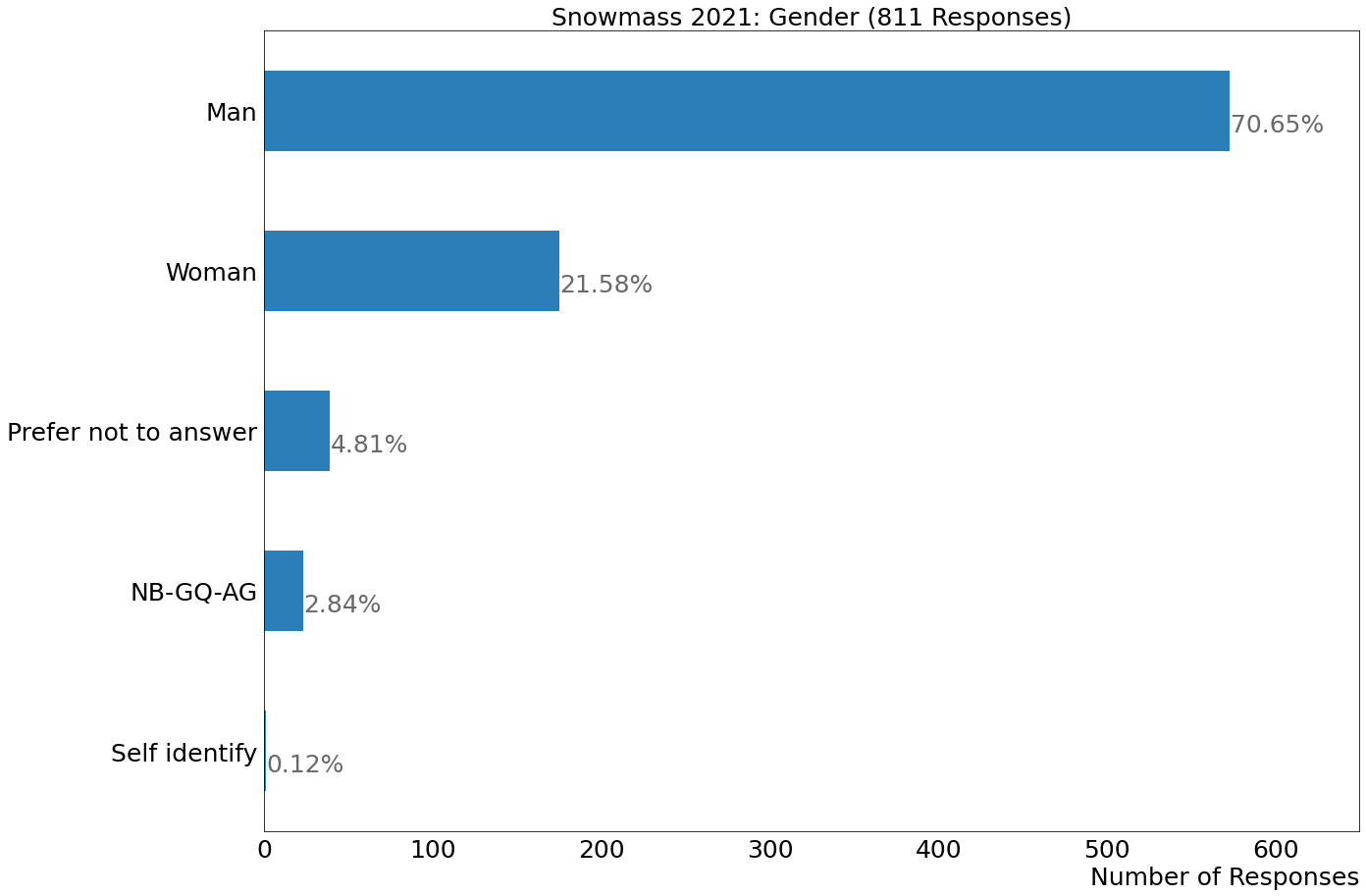}
  \caption{Gender of the respondents. The legend ``NB-GQ-AG'' stands for non-binary/gender queer/a-gender. Respondents could have selected multiple options.}
  \label{fig:demogender}  
\end{figure}

\begin{figure}[H]
  \centering
  	\includegraphics[width=120mm]{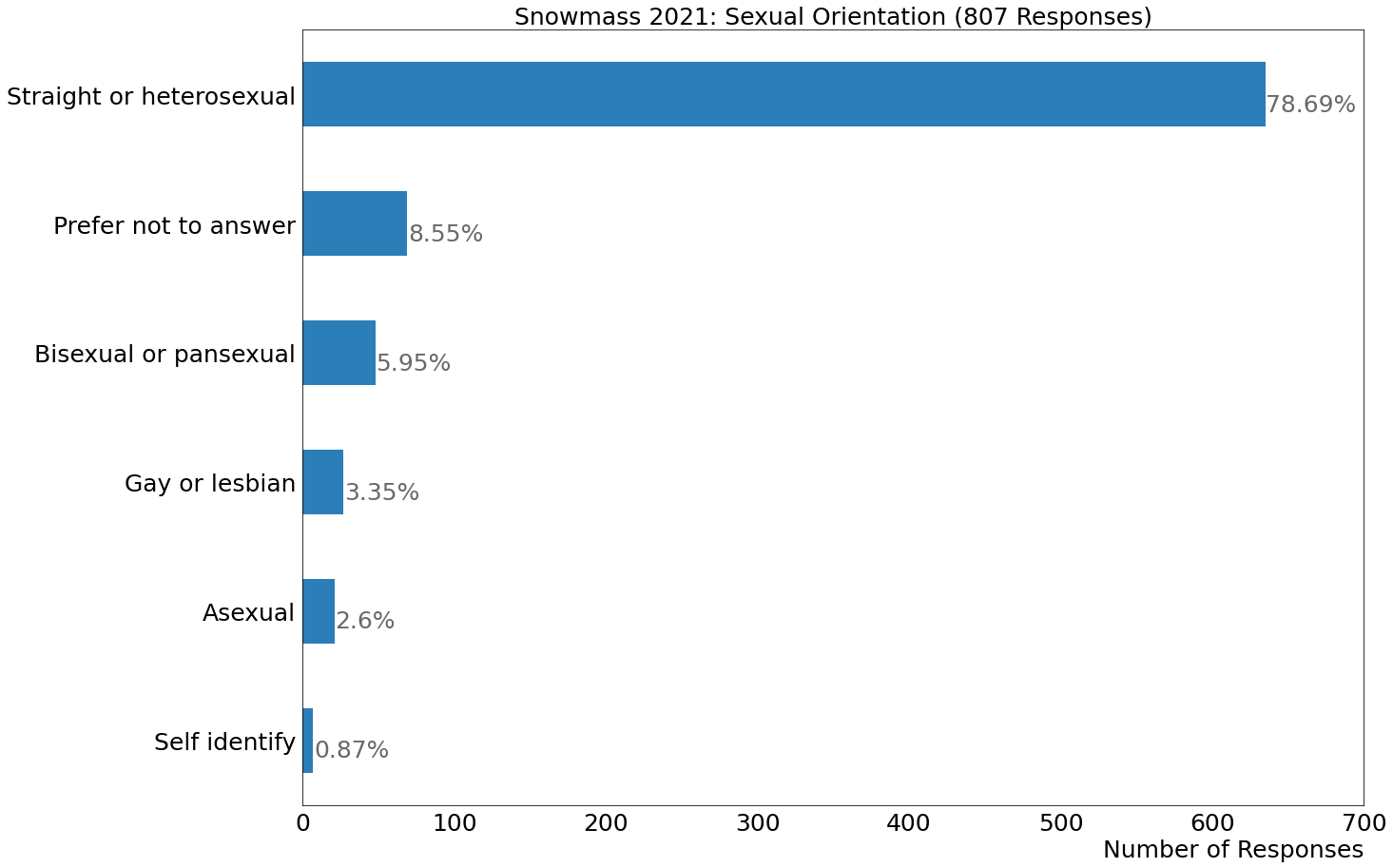}
  \caption{Sexual orientation of the respondents. Respondents could have selected multiple options.}
  \label{fig:demosex}  
\end{figure}

From \autoref{fig:demogender}, we can see that in around 70\% of the total responses, the respondents identify themselves as men; nearly 22\% of respondents identify as women, and around 3\% identify as non-binary, gender queer, or a-gender (NB-GQ-AG). We also asked respondents if they identify as transgender: 95\% of survey respondents do not identify themselves as transgender, around 4\% preferred not to answer, and 1\% of respondents identify as transgender. We also asked about the sexual orientation of the respondents, and \autoref{fig:demosex} shows the results.The majority of the respondents (78.7\%) identify themselves as straight or heterosexual.

\begin{figure}[H]
  \centering
  	\includegraphics[width=120mm]{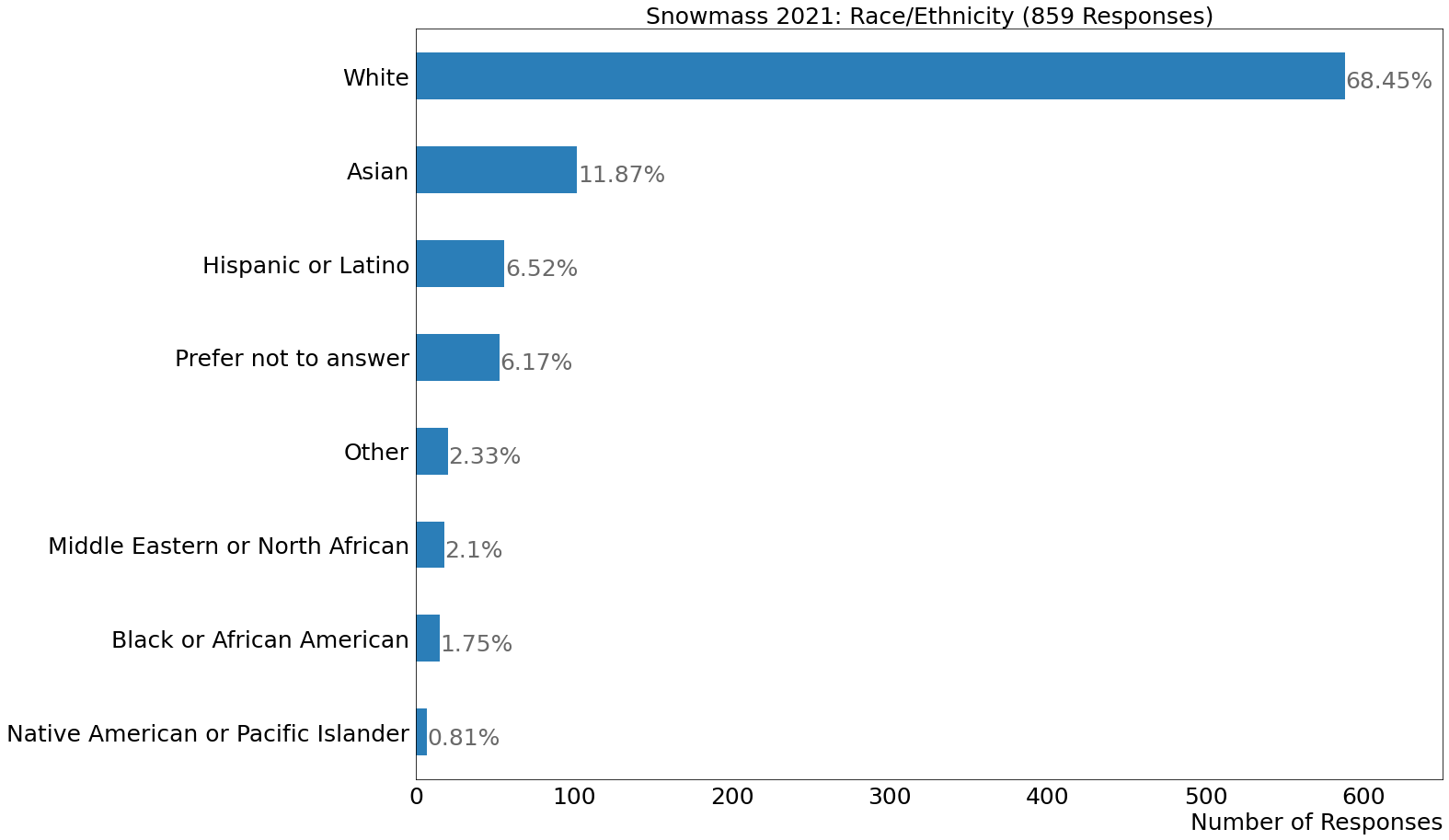}
  \caption{Race/ethnicity of the respondents. Respondents could have selected multiple options.}
  \label{fig:demopyrrace}  
\end{figure}

\autoref{fig:demopyrrace} shows that around 68\% of respondents identify themselves as White. The second-largest racial/ethnic group by percentage are Asians with nearly 12\% of respondents, followed by Hispanic and Latinos with 6.5\%. Groups with fewer respondents include Middle Eastern or North African (2.1\%), Black or African American (1.8\%), and Native American or Pacific Islander (0.8\%).            
\begin{figure}[H]
  \centering
  	\includegraphics[width=120mm]{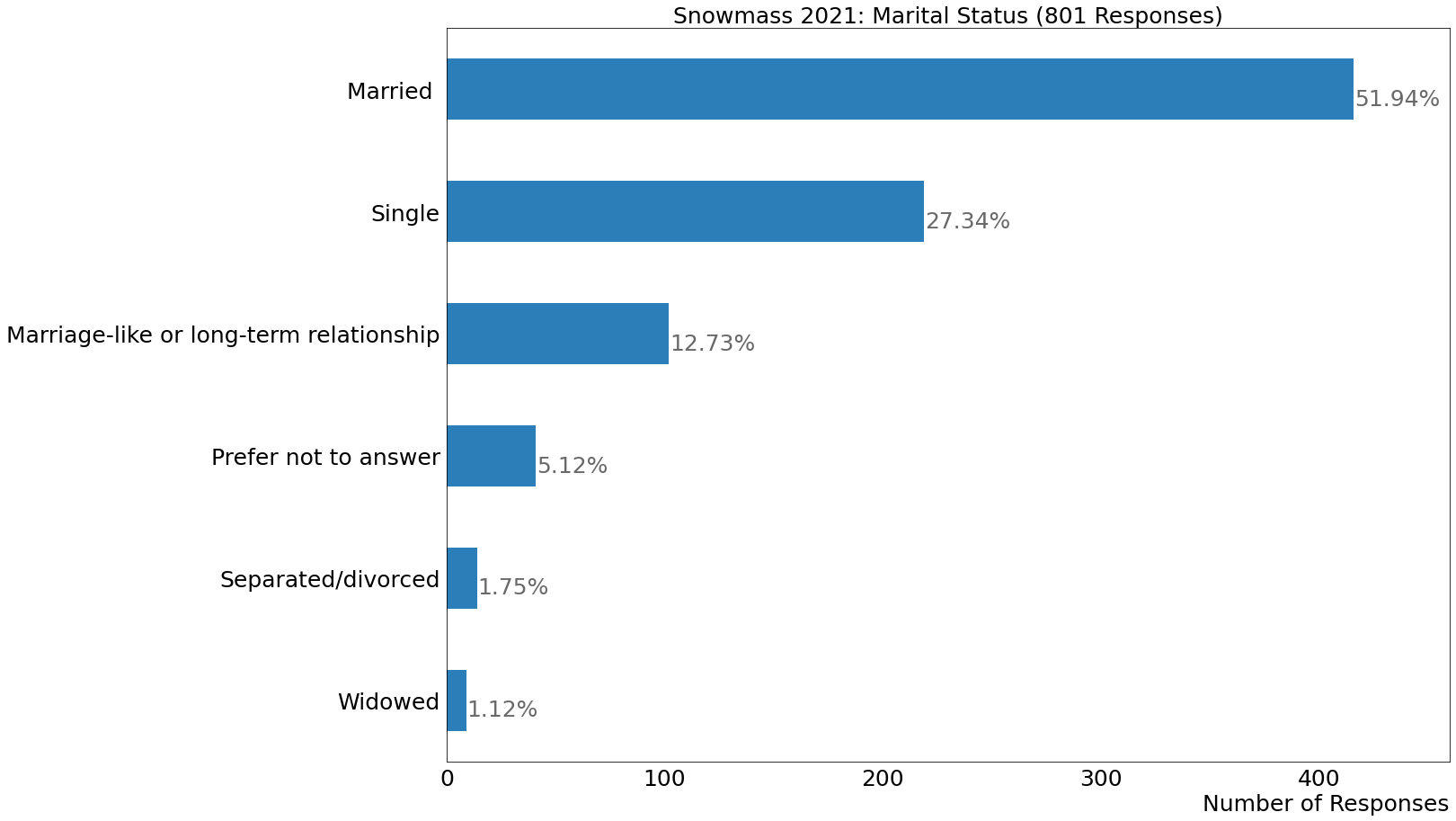}
  \caption{Marital status of the respondents. Respondents could have selected multiple options.}
  \label{fig:demomarital}  
\end{figure}

\autoref{fig:demomarital} shows that majority of the respondents are married (51.9\%) or in a long-term relationship (12.7\%). Another group of respondents (27.3\%) identified themselves as single. Note that respondents could have selected multiple options for gender, sexual orientation, race/ethnicity and marital status. Thus, the total number of responses are different in each of the figures.  

For respondents in a relationship, we asked if they lived together with their partners. Around 91\% of respondents reported that they live with their partner(s), 8\% do not live with their partner(s), and around 1\% answered with ``other''. For those who selected ``other'', respondents wrote that the main reason for not living with their partner(s) is the ``2-body problem''. We also asked these respondents if their partner(s) work in academia: Around two-thirds of respondents answered ``no'' while one-third answered ``yes''. 

\begin{figure}[H]
  \centering
  	\includegraphics[width=120mm]{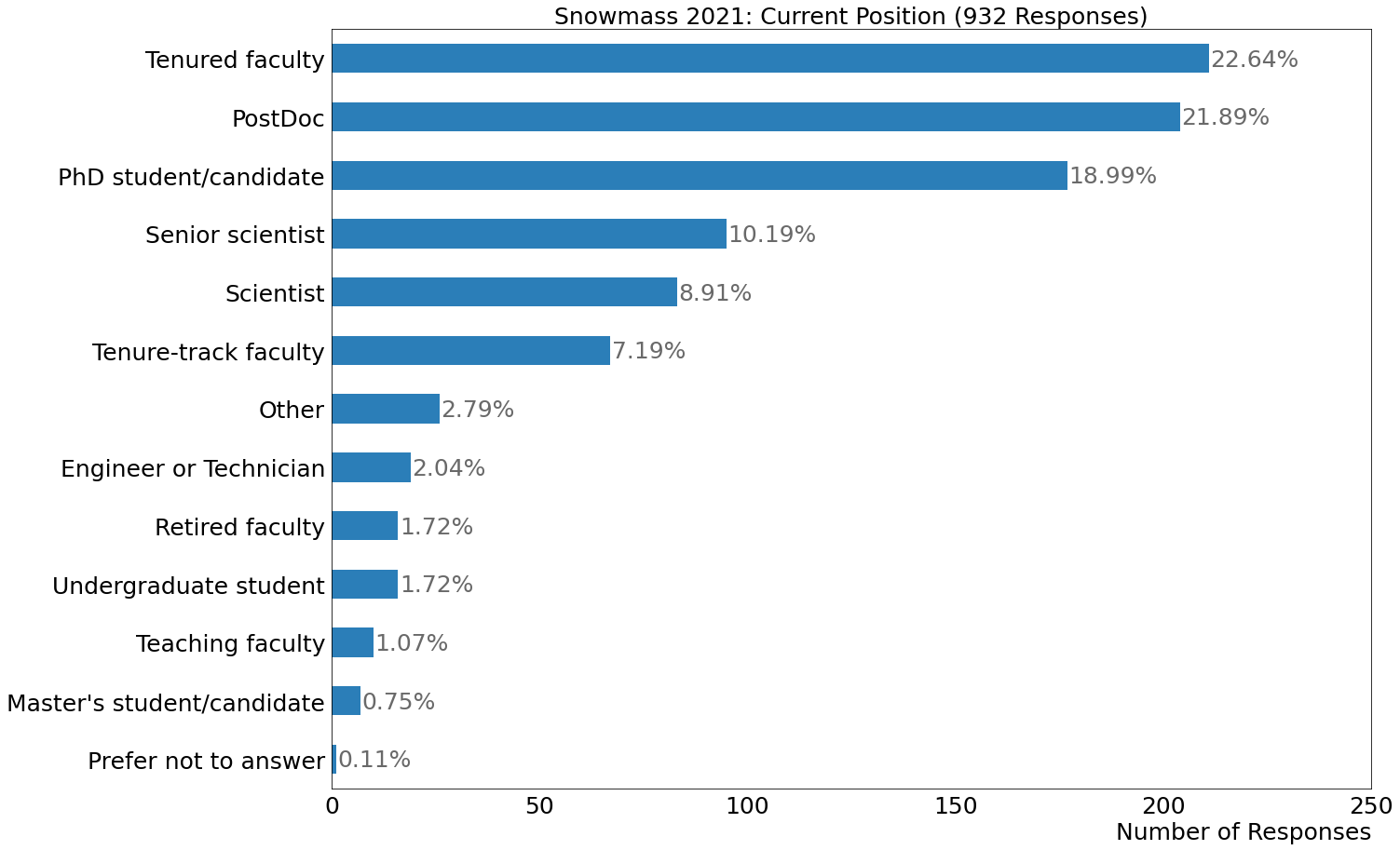}
  \caption{Current position of the respondents.}
  \label{fig:demoposition}  
\end{figure}

\autoref{fig:demoposition} shows that majority of the survey respondents are tenured faculty members (22.6\%), followed by postdocs (21.9\%) and PhD students or candidates (19\%). 

\begin{figure}[H]
  \centering
  	\includegraphics[width=165mm]{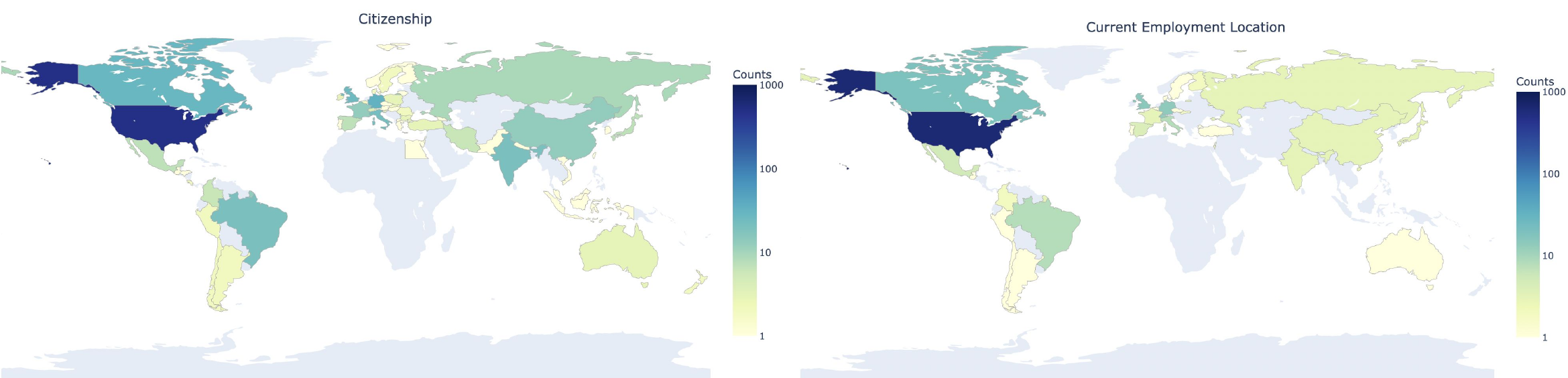}
  \caption{Citizenship and employment location of the respondents.}
  \label{fig:demomaps}  
\end{figure}

As expected, the majority of survey respondents are U.S. citizens and currently work in the United States (see \autoref{fig:demomaps}). However, the 2021 survey also received respondents from South America, Europe and Asia.     

\subsubsection{Respondents with U.S. Visas}

For the non-American researchers, we asked a few questions about their visa status. For those who work in the United States, around 35\% of respondents have a green card or permanent residence, while the majority (around 65\%) said they do not. When asked about the visa type of the non-residents or those who are not green card holders, the most common U.S. Visas were the J1 visa (37\%), F1 visa (35\%), and the H1-B visas (25\%).

\begin{figure}[H]
  \centering
  	\includegraphics[width=160mm]{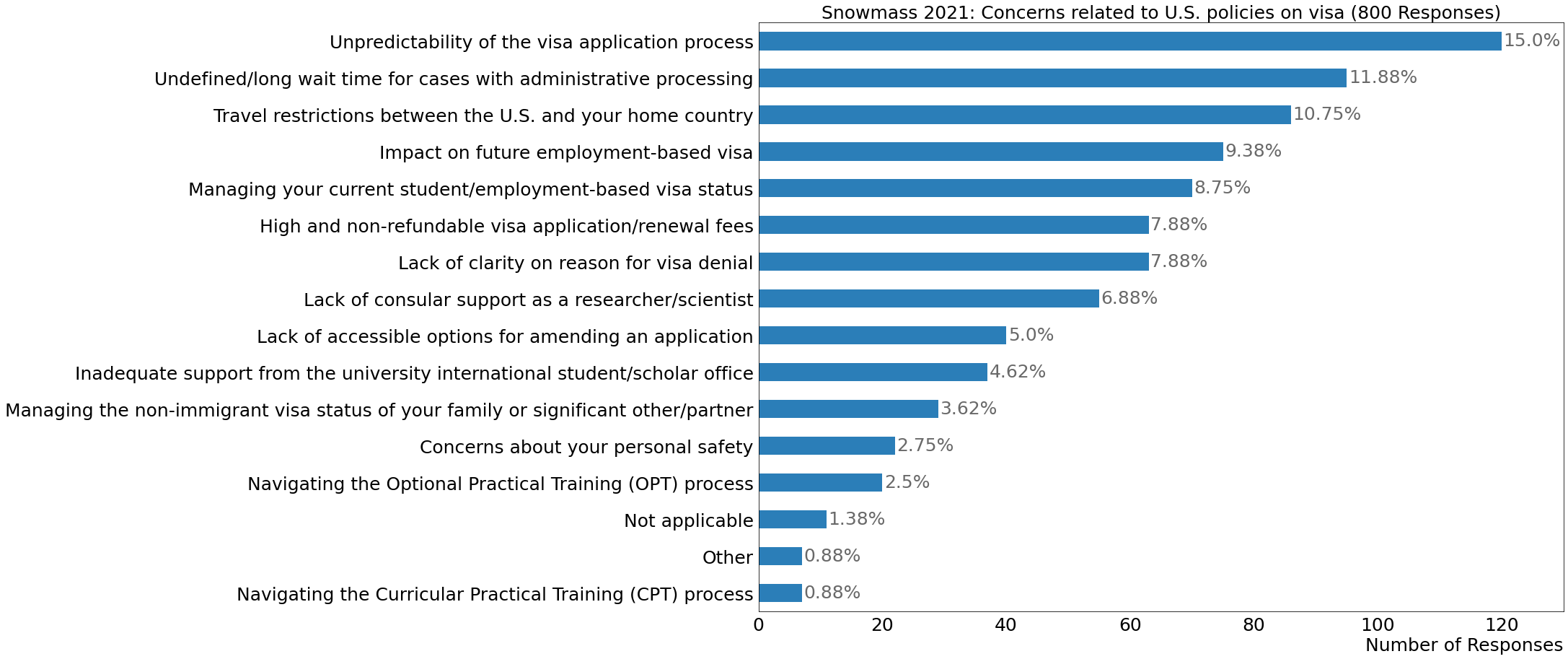}
  \caption{Concerns that visa holder respondents have related to the U.S. Visa policies. Respondents could have selected multiple options.}
  \label{fig:visaconcerns}  
\end{figure}

\begin{figure}[H]
  \centering
  	\includegraphics[width=150mm]{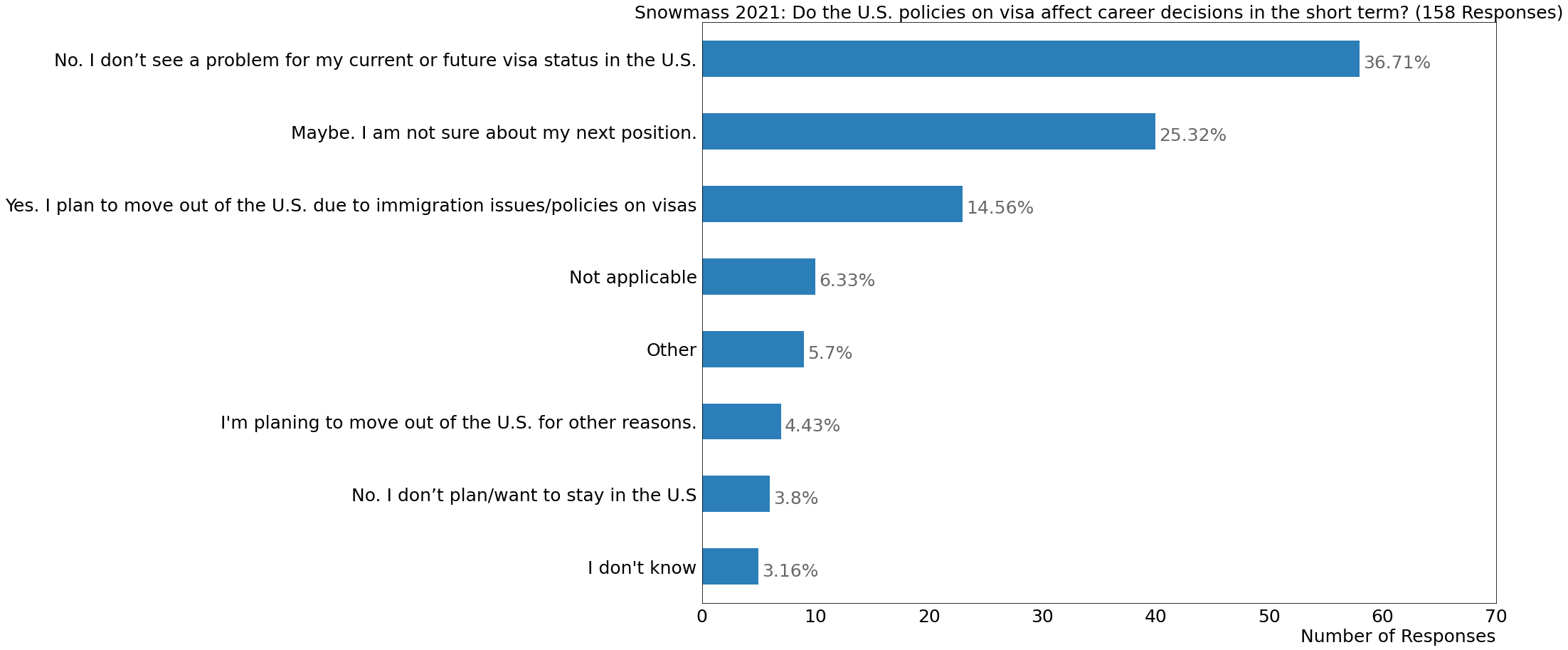}
  \caption{We asked respondents with U.S. Visas whether their career decisions are affected by Visa policies in the short term.}
  \label{fig:visaeffects}  
\end{figure}

When asked about their concerns related to the visa policies in the United States, respondents mainly mentioned the unpredictability, the long time of the visa process, and the travel restrictions between the United States and their home countries. More concerns chosen by the respondents can be found in \autoref{fig:visaconcerns}. When asked if the United States visa policies would affect their career decisions in the next two years (see \autoref{fig:visaeffects}), around 37\% of the respondents answered ``no'', but the majority of this group are green card holders or residents. A quarter of respondents were not sure because they did not know what their next position would be. For around 15\% of respondents, the visa policies would affect their career, and they see themselves moving out from the United States in the short term.           

\begin{figure}[H]
  \centering
  	\includegraphics[width=120mm]{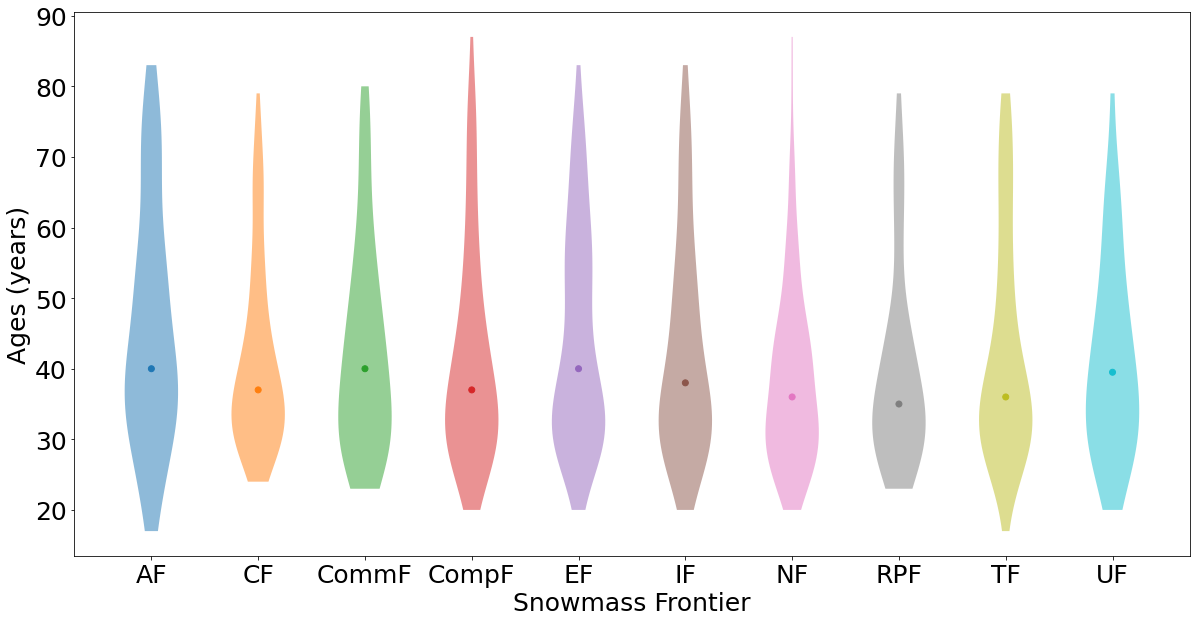}
  \caption{Age distribution of the respondents by their main frontier. The dots are the median age for each frontier.}
  \label{fig:agefront}  
\end{figure}

In \autoref{fig:agefront}, we show the age distributions split by Snowmass Frontiers. The dots represent the median age in each Frontier, and the median age is around 40 for all Frontiers. The Rare Processes and Precision Frontier (RTF) has the youngest median age (35), and the Accelerator (AF), Community Engagement (CommF), and Energy (EF) Frontiers have the oldest median age (40).        

From the demographic information collected, we can see that the typical profile of the 2021 survey respondents is a North American, white, straight man, between 30 and 40 years old, married, with a tenured faculty or Postdoc position. This observation corroborates with several other reports \cite{aipw2019, ncses2021} in the literature that demonstrates the low number of women and underrepresented minorities in Physics, which is reflected in the HEPA community.

\subsection{Physics Outlook}


Past Snowmass Surveys had focused on particular projects, but the breadth of the current Snowmass process made this effectively untenable. Thus, the Physics Outlook subteam took it upon themselves to design questions related to broad categories of interests across the field, including their involvement with particular Frontiers, thoughts on the current and future research directions within HEPA, and general experimental collaboration dynamics.

\begin{figure}
    \centering
    \includegraphics[width=1.0\linewidth]{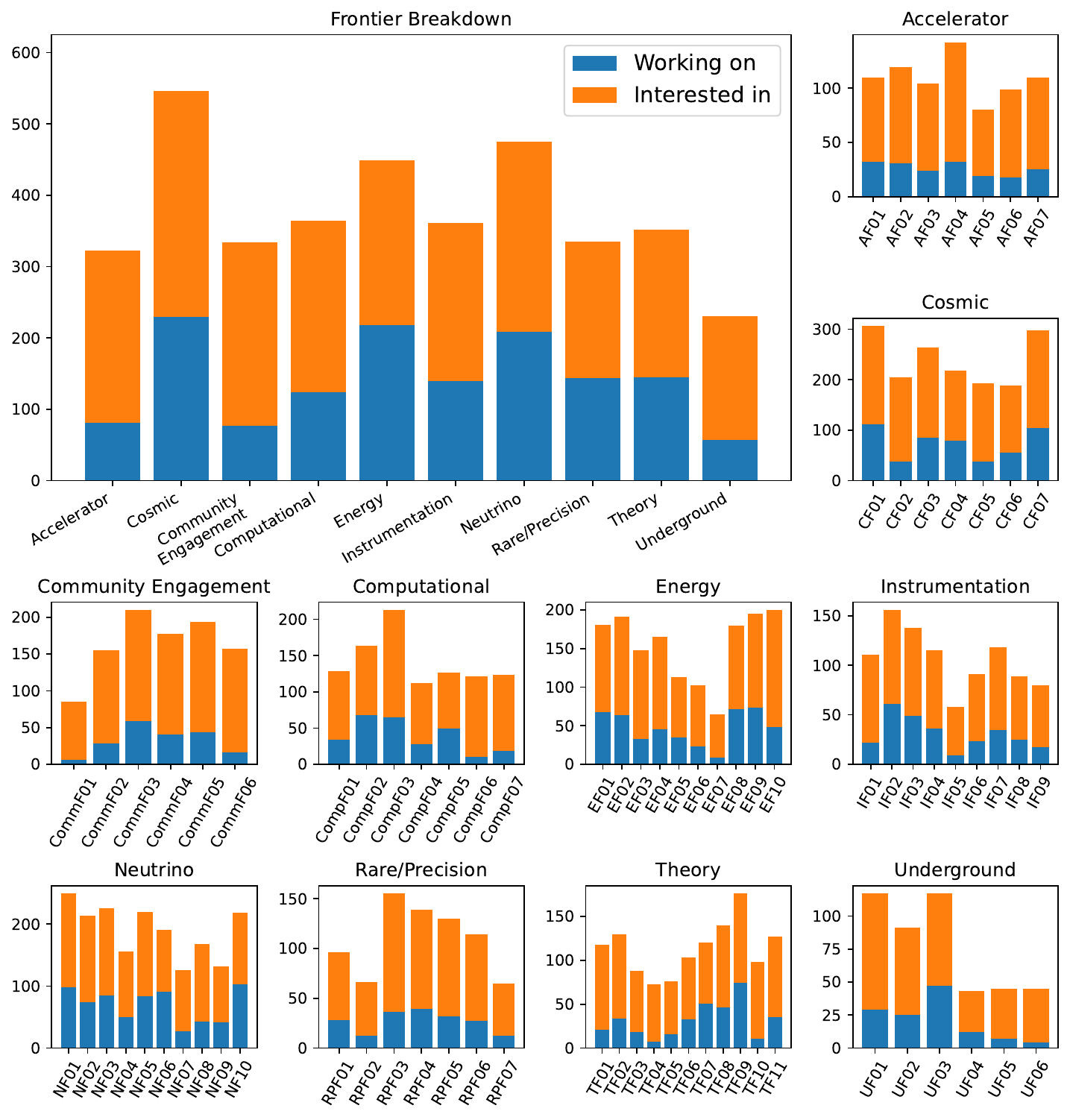}
    \caption{We asked respondents to indicate which Frontiers and Topical Groups they were working in or had an interest in, even if they were not directly working on such physics. A list of the full Topical Group names can be found in Appendix \ref{appendix_surveyQs} or at \href{https://snowmass21.org}{snowmass21.org}.}
    \label{fig:frontier_breakdown}
\end{figure}

The large scope of the fields within HEPA are reflected across the many different Frontiers and Topical Groups which have been established as part of the Snowmass process. Respondents were asked to indicate which Frontiers and Topical Groups they were actively working in, and also to note which other Frontiers or Topical Groups they maintained an interest in, even if they were not currently working in that particular field. The results are shown in \autoref{fig:frontier_breakdown}. The results demonstrate broad interest and participation across all of the different Frontiers, with a great deal of respondents indicating that they had interests outside their particular line of research. Note that respondents were allowed to multi-select which Frontier(s) they worked in, and over $40\%$ of respondents took advantage of this, reflecting that a great deal of research in HEPA is cross-disciplinary. 

\begin{figure}
    \centering
    \hskip 1cm
    \includegraphics[width=0.9\linewidth]{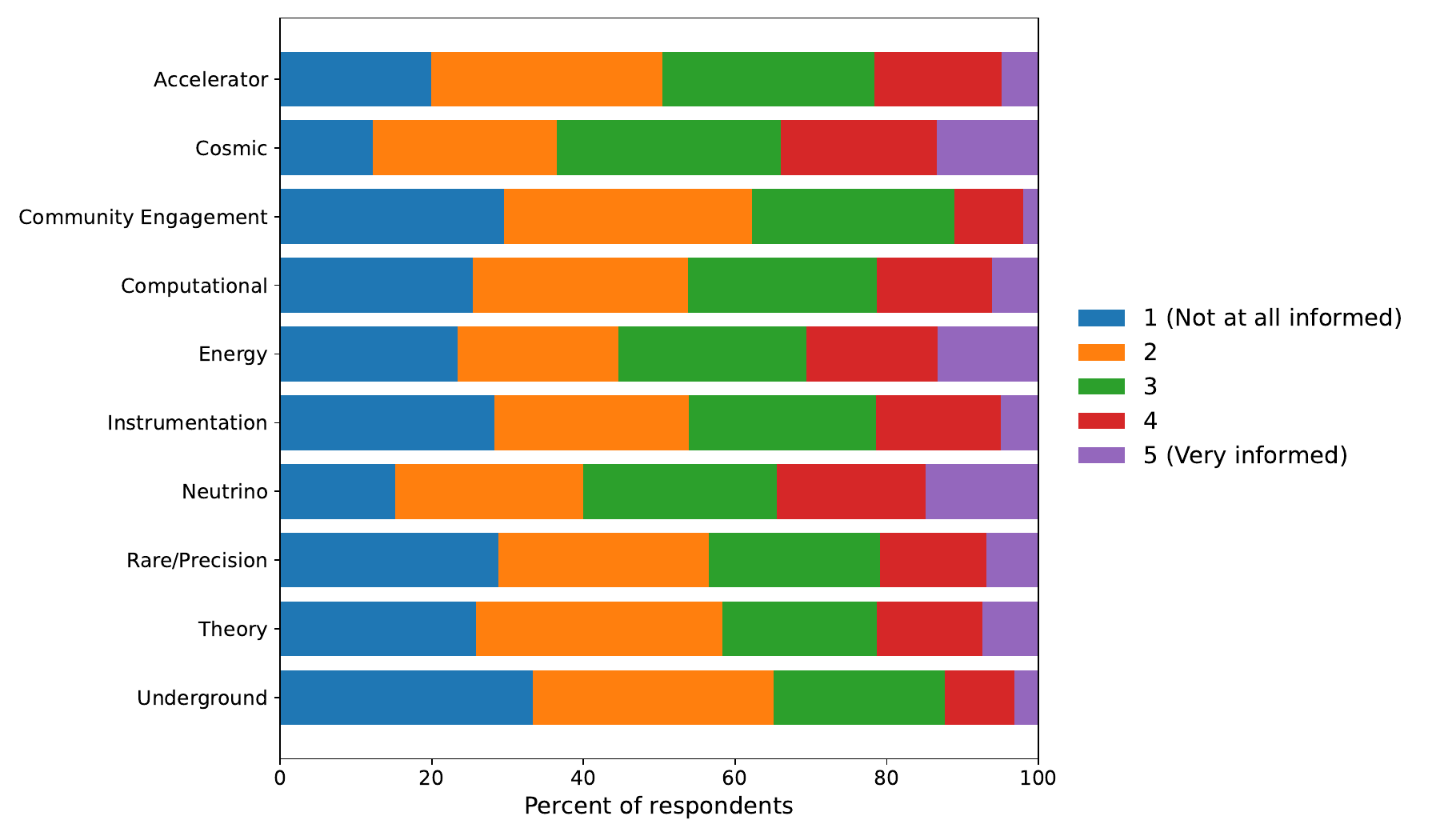}
    \caption{Respondents were asked how well informed they felt about future scientific directions within different Frontiers.}
    \label{fig:physics_informed}
\end{figure}

Given the broad scope of topics included in the Snowmass process, respondents were also asked to indicate how informed they felt about future scientific directions in the different Frontiers. The results are summarized for each Frontier in \autoref{fig:physics_informed}. In general, most respondents rated how informed they felt at a 2 or 3, with 1 indicating ``Not at all informed'' and 5 indicating ``Very informed''. Respondents felt the least informed about the Community Engagement and Underground Frontiers, but most felt they were informed about the Cosmic, Energy and Neutrino Frontiers (the three Frontiers that also had the largest number of respondents, as indicated by \autoref{fig:frontier_breakdown}).

\begin{figure}
    \centering
    \includegraphics[width=\linewidth]{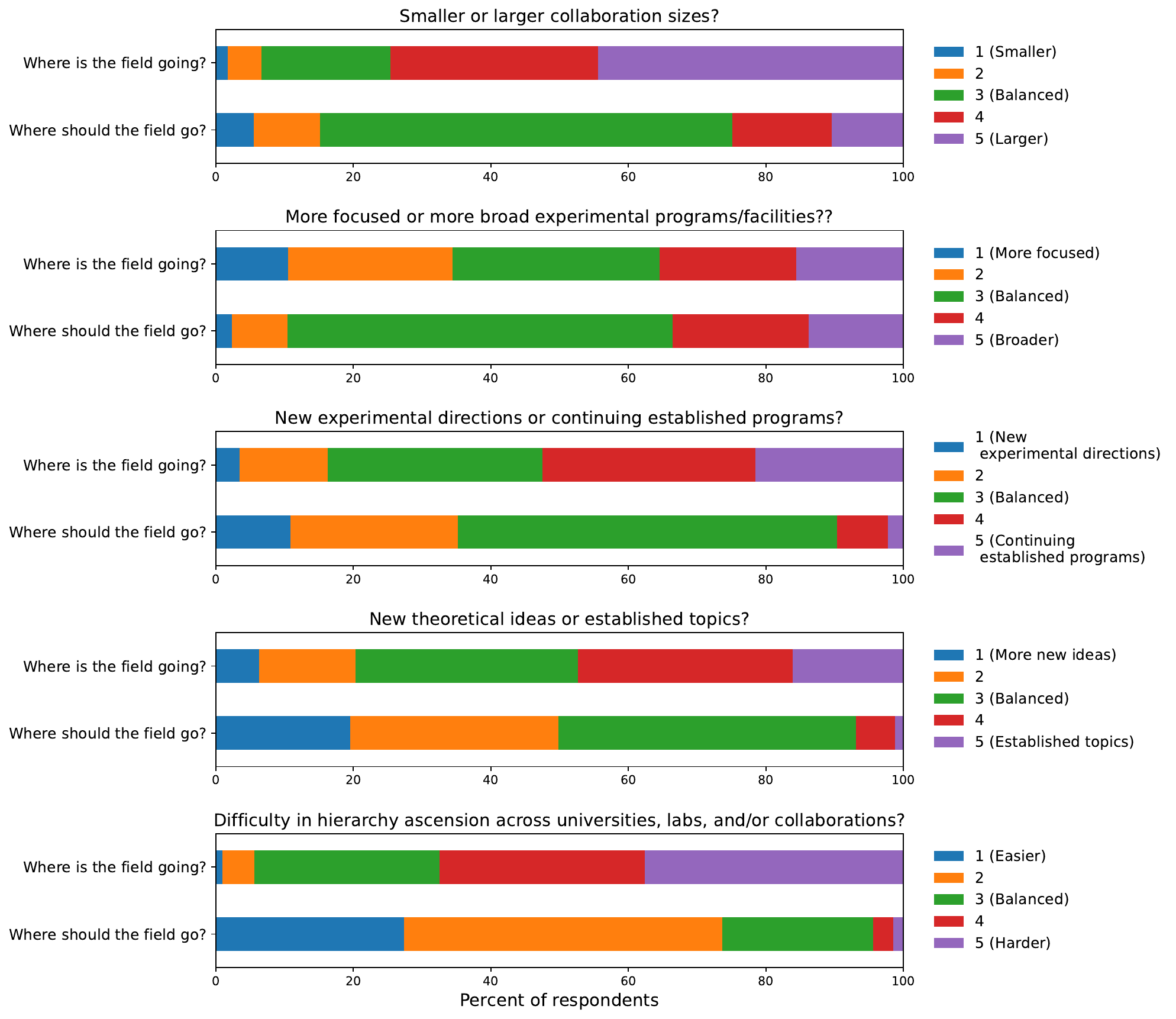}
    \caption{For a range of topics, respondents were asked to rate (on a 1-5 scale) which direction they thought the field was currently going and where they thought the field should go.}
    \label{fig:physics_directions}
\end{figure}

In consideration of the previous discussion on difficulties in establishing general consensus on a given experimental idea given the breadth of Snowmass 2021, another strategy was implemented as a part of this Survey. Instead, respondents were asked to rate their feelings across a range of topics including what direction they felt the field \textit{was moving} in certain aspects, and then to gauge which direction they felt the field \textit{should be moving}. The results of these questions are shown in \autoref{fig:physics_directions}. Respondents were asked whether they thought the field was trending towards smaller or larger collaboration sizes; towards more focused or broad experimental programs or facilities; towards new experimental directions or established programs; towards new theoretical ideas or established topics; and whether hierarchy ascension across universities, labs or collaborations was becoming easier or more difficult.

For the first four questions, a plurality of respondents indicated they felt the field ``should be going'' for the most balanced response. However, there were still several notable asymmetries in the results which are worth remarking upon. Overwhelmingly, the biggest imbalance between where respondents felt the field was going compared to where it should be going was in the difficulty in hierarchy ascension. Nearly $70\%$ of respondents said the field was heading towards a 4 or 5 (with 5 implying that it was becoming still hardest to ascend the job ladder), though there was an overwhelming preference (with $> 70\%$) among respondents for this to become easier.

In contrast, as to whether the field was heading towards more focused or more broad experimental programs and facilities, respondents were roughly evenly divided on what direction the field was heading, with $\gtrsim 10\%$ of respondents selecting each answer choice, and no more than $30\%$ selecting any. Interestingly, despite this disagreement about the current direction of the field, this question had the most alignment between the feelings about where the field was going and where respondents felt it should go. A slightly larger share of respondents indicated that they thought the field should be heading towards broader experimental programs and facilities, compared to those who said it should be going towards narrower ones.

In terms of smaller or larger collaboration sizes, there was a significant majority of respondents indicating they thought the field was heading towards larger (4 or 5) collaboration sizes. As for where the field should be going, a significant majority indicated they thought there should be a balanced number of smaller and larger collaborations.

As for whether the community was interested exploring new experimental directions or continuing established programs within the field, a slight majority of respondents indicated that they thought the field was continuing established programs (either a 4 or 5). As for where the field should be going, most respondents indicated they preferred a balanced approach, but there was some preference for exploring new experimental directions.

Finally, as to whether the field was heading towards new theoretical ideas or established topics, just under $50\%$ of respondents indicated that they felt the field was going towards established topics, while $< 20\%$ felt as though it was going towards new theoretical ideas. In contrast, nearly $50\%$ of respondents said they thought the field should be exploring new theoretical ideas, with $< 10\%$ indicating they thought it should continue studying established topics. For both questions, though, a plurality of respondents preferred a balanced position.

\begin{figure}
    \centering
    \hskip 2cm 
    \includegraphics[width=0.8\linewidth]{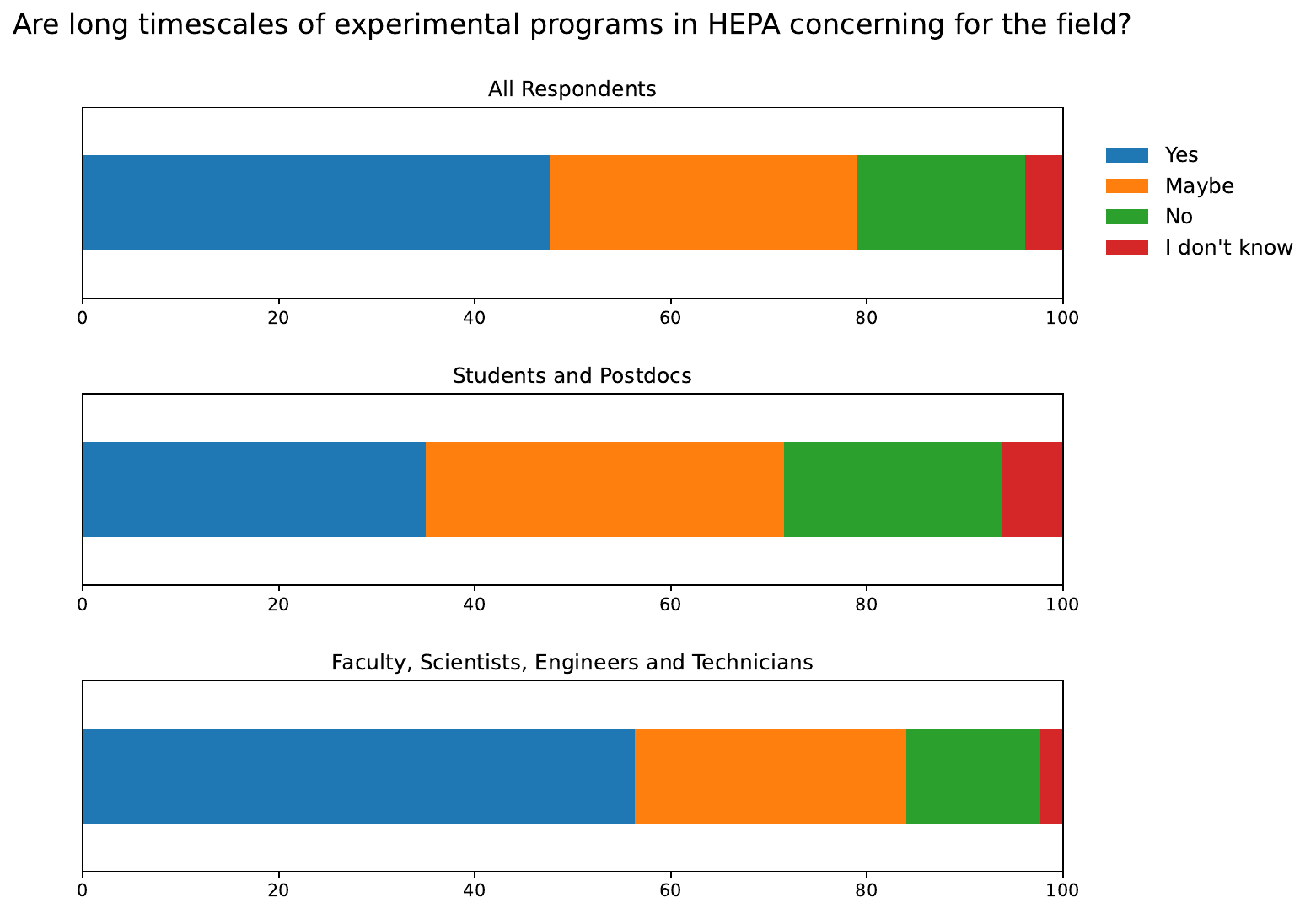}
    \caption{Respondents were asked whether they believed long timescales of experimental programs in HEPA were a concern for the field.}
    \label{fig:physics_timescales}
\end{figure}

A matter-of-need synthesis across a number of scientific disciplines, HEPA is unique in the relative importance of large-scale experimental projects, which can often span decades and even funding agencies. These long timescales are not without some challenges, particularly the impact of these timescales on the careers of scientists in the field, as discussed in Section~\ref{sec:careers}. Given the prominence of long timescales, and their wide impact, we asked respondents whether they believed long timescales of experimental programs in HEPA were concerning for the field. The results are shown in \autoref{fig:physics_timescales}. A plurality of respondents answered ``Yes'', with another $\sim 30\%$ of respondents indicating ``Maybe''. Only about $20\%$ of respondents answered a definitive ``No''.

The responses regarding timescales were also broken down between ``junior'' (students and postdocs) and ``senior'' (faculty, scientists, engineers and technicians) members of the field. The differences were relatively small, but notably, junior members were substantially less concerned about the long timescales of projects, with under $35\%$ responding ``Yes'' compared to $56\%$ of ``senior'' respondents.

\begin{figure}
    \centering
    \includegraphics[width=0.9\linewidth]{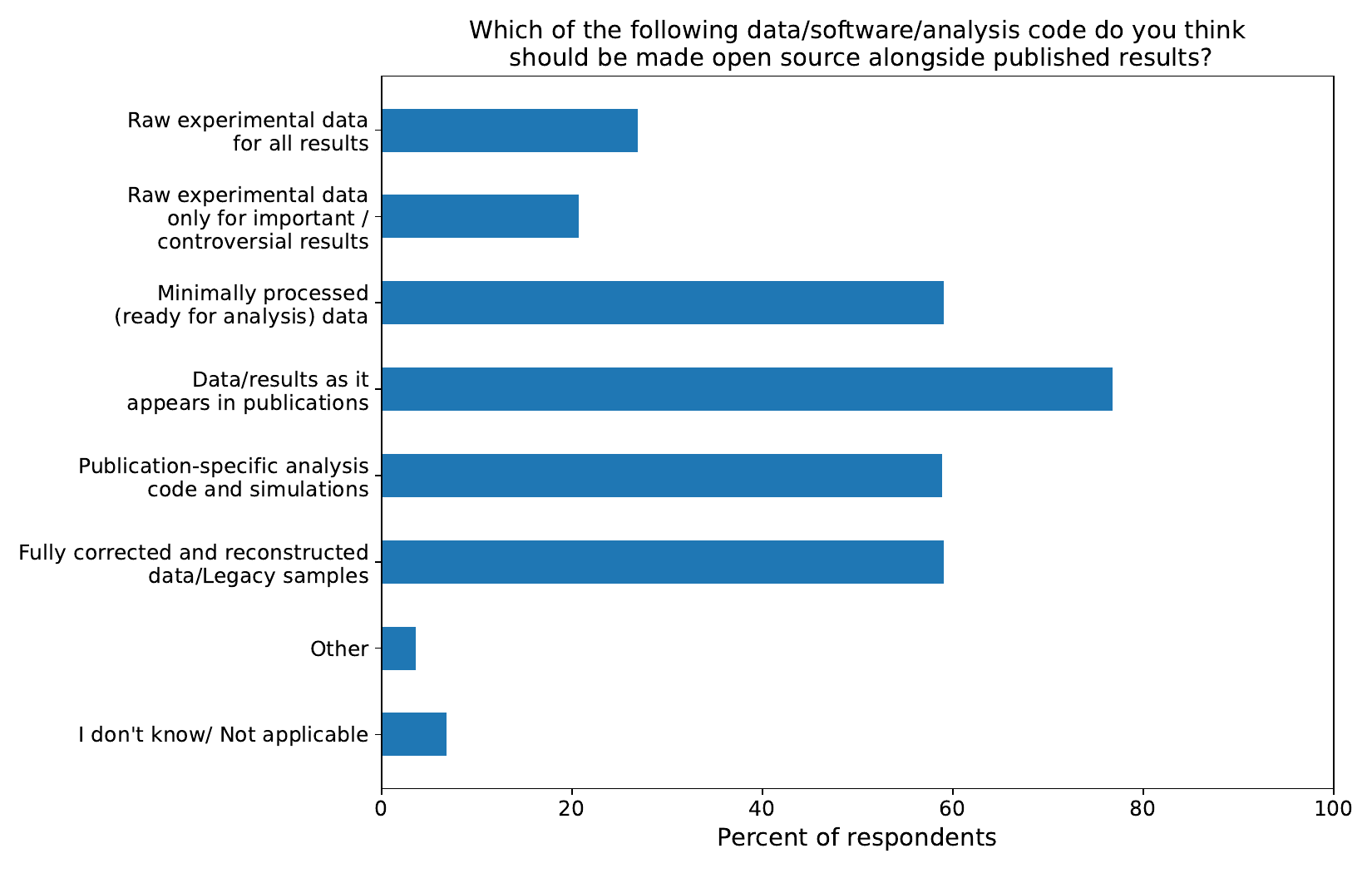}\hskip 0.5cm
    \caption{Respondents were asked which data/software/analysis code they believed should be made open source alongside published results.}
    \label{fig:outlook_6}
\end{figure}

The survey also asked respondents to select what types of data, software or analysis codes they believed should be made open source alongside published results. These results are tallied in \autoref{fig:outlook_6}, where the results are normalized to all those responding to this question. 
The most support was in favor of publishing data/results as they appear in publications, but a significant majority of respondents also said they supported ``Minimally processed (ready for analysis) data'', '   ``Publication-specific analysis code and simulations'', and ``Fully corrected and reconstructed data / legacy samples'' being made open source. The other categories all received less than $30\%$ of respondents support. 

\begin{figure}
    \centering
    \includegraphics[width=0.9\linewidth]{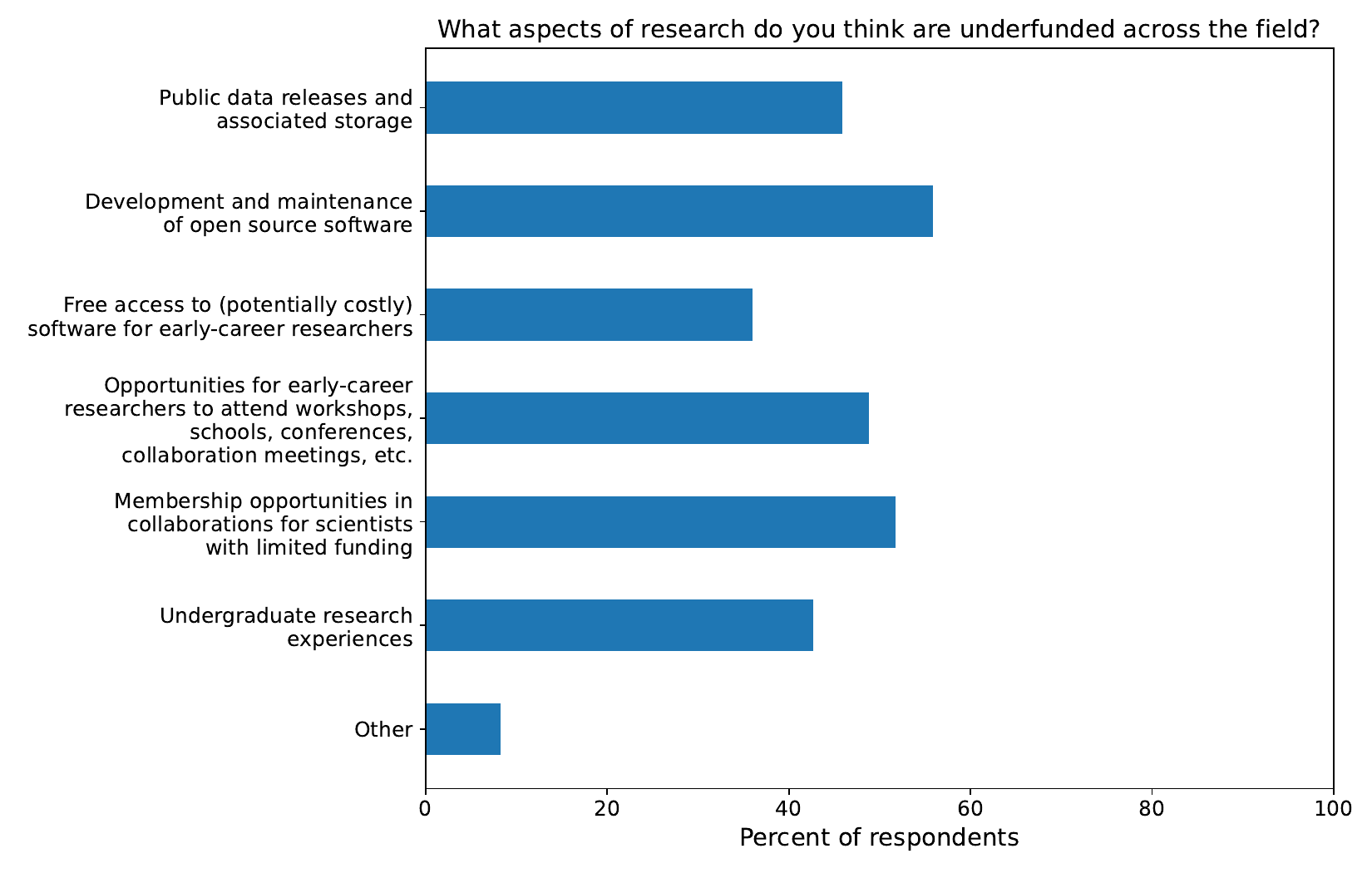}\hskip 0.5cm 
    \caption{Respondents were asked what aspects of research in HEPA they believed were underfunded.}
    \label{fig:physics_underfunded}
\end{figure}

We also asked respondents to select which aspects of research in HEPA they believed were underfunded. The results are shown in \autoref{fig:physics_underfunded}. The most frequently selected option was ``development and maintenance of open source software'', which over half of respondents selected.  Over $40\%$ of respondents also selected ``public data releases and associated storage'', ``Opportunities for early-career researchers to attend workshops, schools, conferences and meetings'', ''Membership opportunities in collaborations for scientists with limited funding'', and ``undergraduate research experiences''. 

\begin{figure}
    \centering
    \includegraphics[width=0.95\linewidth]{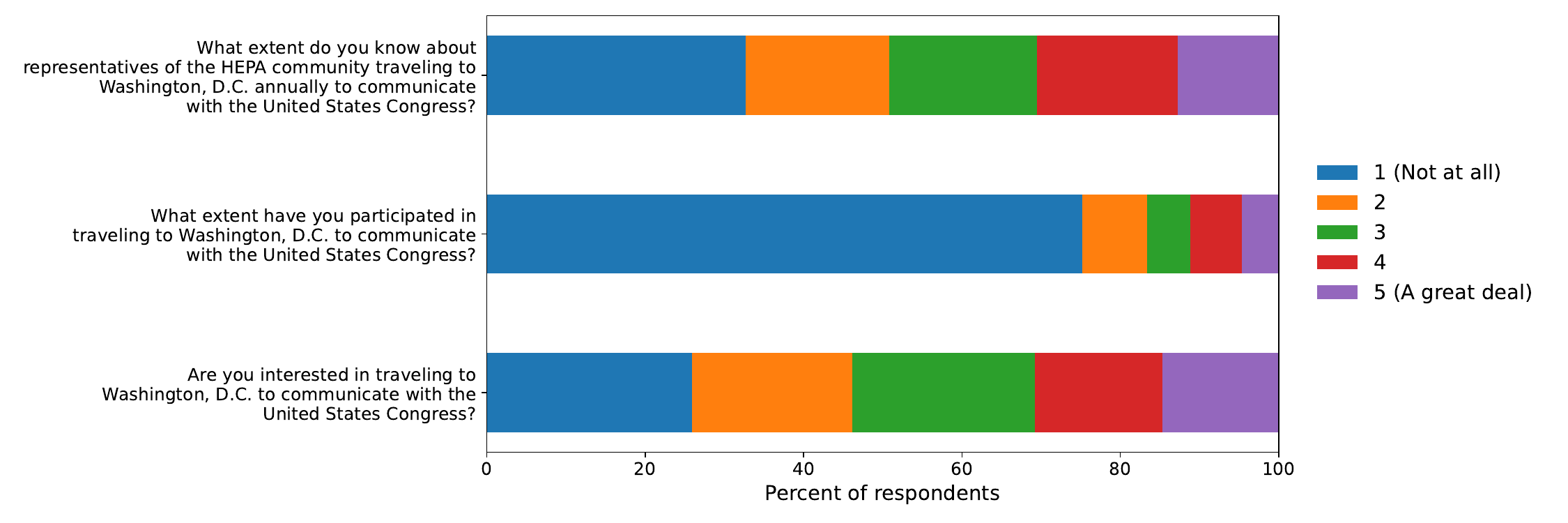} \hskip 0.5cm
    \caption{Respondents were asked to what extent they knew about, participated in, or were interested in participating in advocacy efforts in HEPA.}
    \label{fig:physics_advocacy}
\end{figure}

Respondents were also asked to what extent they were aware of, had participated in, or were interested in participating in HEPA advocacy---in particular in travelling to Washington D.C.~\cite{fnalnewdctrip} to meet with members of congress. The responses to this question are shown in Figure \ref{fig:physics_advocacy}. Nearly half of respondents said they were ``Not at all'' or ``only slightly'' aware of advocacy efforts, and $\sim 75\%$ of respondents said they had ``Not at all'' participated in them. However, a majority of respondents rated their level of interest in travelling to D.C. at a 3, 4 or 5.

\begin{figure}
    \centering
    \hskip 0.5cm
    \includegraphics[width=0.95\linewidth]{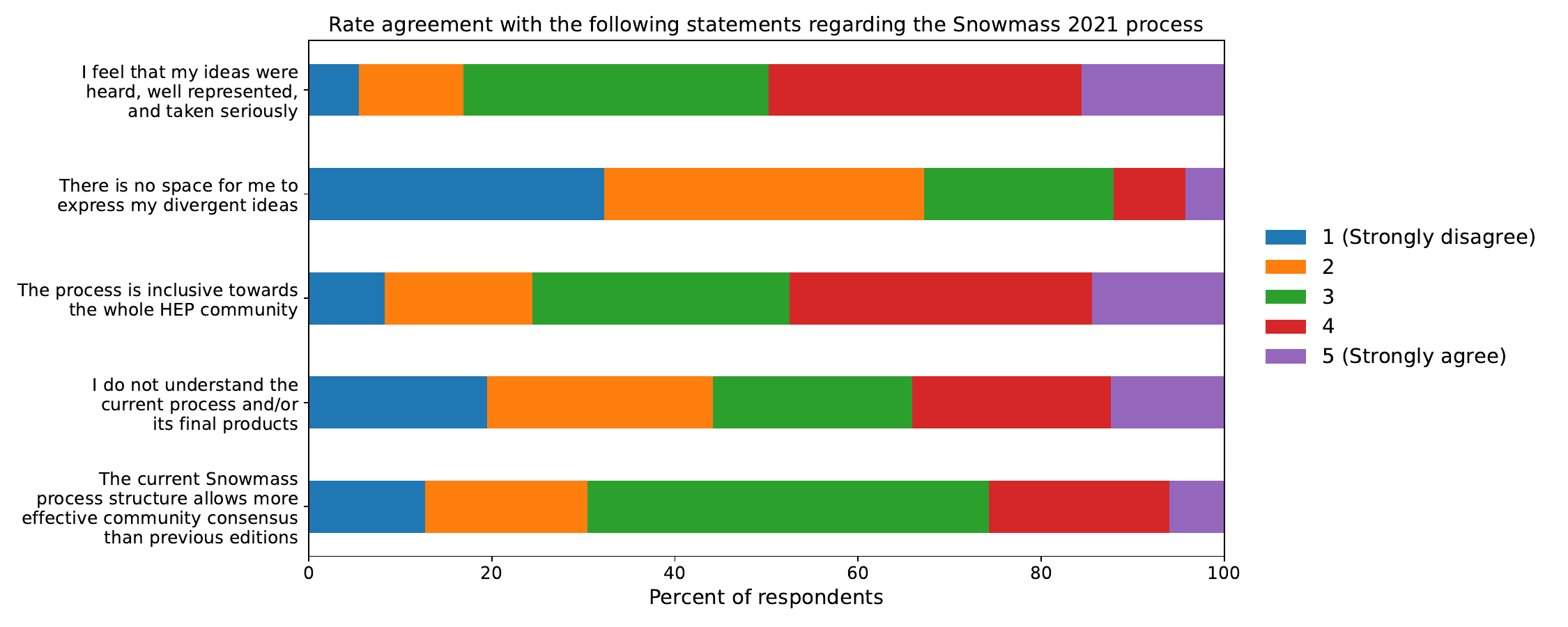}
    \caption{Respondents were asked to rate their agreement with several statements regarding the Snowmass 2021 process. 
    Respondents who answered ``I don't know'' were not included in the plot.}
    \label{fig:physics_snowmass_feelings}
\end{figure}

Finally, respondents were asked to rate their agreement with several statements regarding the Snowmass 21 process, in an attempt to gauge the utility of the process as a forum for ideas in the field.
Overall, the results shown in \autoref{fig:physics_snowmass_feelings} show favorable opinions about expressing their ideas in the Snowmass process. A majority of respondents rated agreement or strong agreement with the statement ``I feel that my ideas were heard, well represented, and taken seriously.'', with fewer than $20\%$ disagreeing. A solid majority also responded that they disagreed or strongly disagreed with the statement that there was no space for them to express their divergent ideas. A plurality of respondents also said they agreed that the process was inclusive towards the whole HEP community.

Responses were somewhat more mixed as to whether respondents understood the current process or its final products, with a roughly even distribution of answers across all five answer choices. Respondents were also somewhat mixed on whether the current Snowmass process structure allowed more effective community consensus than previous editions---the most popular response to this question by far, was neutral.

\subsubsection{\label{env}Climate emergency and HEPA}

\begin{figure}[H]
  \centering
  	\includegraphics[width=1.0\linewidth]{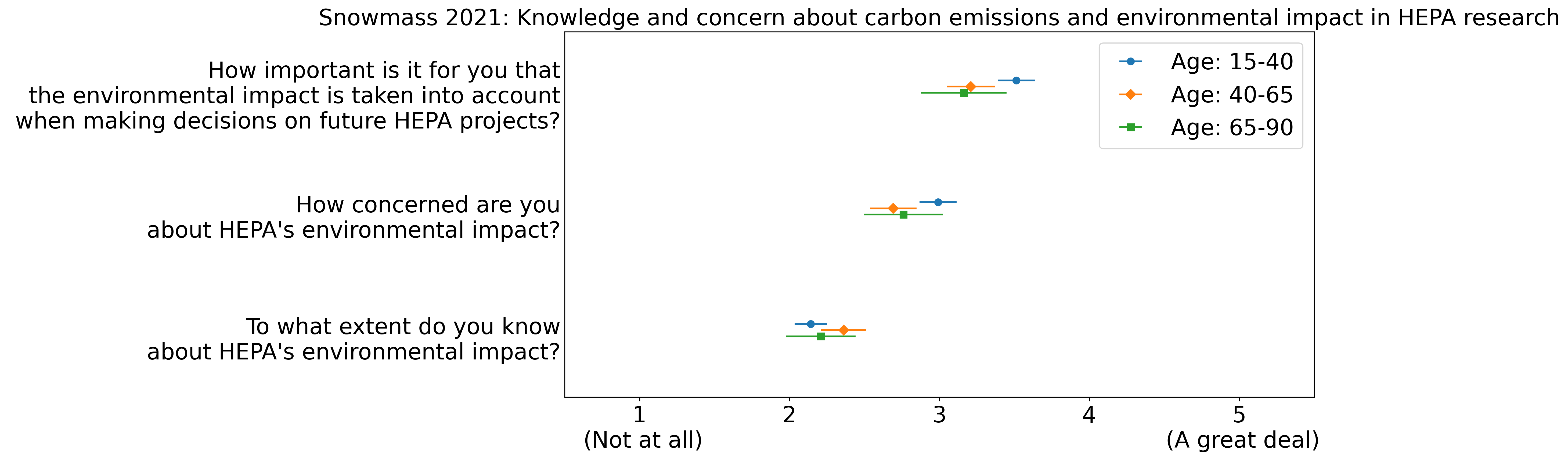}
  \caption{Perception of the community on the HEPA’s environmental impact.}
  \label{fig:env}  
\end{figure}

There has been a growing awareness about the climate emergency in our society. In particular, several members of the HEPA community are developing efforts to estimate the environmental impacts related to experiments and the way we advance research in our field \cite{2020NatAs...4..843S, 2020NatAs...4..812J, 2021APh...13102587A, 2022arXiv220107895A, 2022arXiv220108748K}. In the Snowmass 2021 survey, we have asked three questions related to this topic (see \autoref{fig:env}). The objective was to understand whether members of our community are well-informed and concerned and believe that HEPA endeavours must assume a share in the global effort to reduce greenhouse gas by adopting more sustainable practices in the future. We have separated the answers by age to access how these concerns change between different career stages. For example, ages 15-40 would be a good proxy for the responses of early-career scientists; from 40-65 years, we are likely to probe the responses from tenure-track and tenured faculty; and finally, 65-90 years from senior and retired researchers.     

When asked to what extent they know about the HEPA’s environmental impact, on average, respondents said they do not know much about it. Early career scientists report knowing slightly less than senior researchers. We also asked how concerned the respondents were about the HEPA’s environmental impact. The responses are, on average, more neutral. Nevertheless, early career members seem to be more concerned than senior researchers. Finally, the respondents think, on average, that it is important to take into account the environmental impact when making decisions on future HEPA projects. Particularly the earlier career researchers, who will most likely cope with more severe effects of the climate emergency.

\subsection{Careers}
\label{sec:careers}



The only question which received an answer from all survey respondents happened to be the only required question: ``Are you currently in academia?'', with 92.3\% of respondents indicating they are in academia, and 6.9\% of respondents indicating that they are not currently in academia. Of all survey respondents, 38.4\% were early career (graduate students or postdoctoral researchers), 31.3\% were faculty (either teaching, tenure-track, tenured, or retired), and 18\% were scientists. We also received responses from undergraduate students, engineers, and technicians, but we did not receive large enough samples in which to perform an in-depth analysis. General comments about their responses can be found in Sections \ref{undergraduate_section} and \ref{engineers_technicians_section} respectively.

\subsubsection{Undergraduate students}
\label{undergraduate_section}
Only 1.7\% of all respondents indicated they were undergraduate students. The majority of undergraduate respondents plan to go to graduate school and study physics or astrophysics. All undergraduate respondents indicated they currently participate in research, and the majority of respondents participate throughout the entire year (summer and academic year). The majority of undergraduate respondents indicated they contacted a professor to get into their research group. Only a select few indicated that a department or university-led effort got them into research. Undergraduate respondents reported committing the equivalent of a part-time job to their research. A majority also indicated they are compensated, either through payment or course credit, but a nonzero number of undergraduate respondents indicated they are not compensated for their research responsibilities. Nevertheless, undergraduate respondents reported that they enjoy the research they conduct.

\subsubsection{Graduate students and Postdocs}

Early career scientists, i.e. Master's students, PhD students, or postdoctoral researchers (Postdocs) formed 38.4\% of all the survey respondents. Of those indicating they are in academia, 19.8\% were graduate students (of which PhD students comprise 96\% of the total graduate student pool), while 21.8\% were Postdocs.

Despite making up less than half of the survey respondents, about half of all the women who responded to the survey were early career scientists. Early career scientists also make up about one third of all the male respondents. The vast majority of nonbinary, agender, or genderqueer (NB-GQ-AG) respondents were early career scientists. Over half of the early career scientists are located at universities, with 40.6\% at U.S. based universities. A smaller group of early career scientists (22.1\%) reported being based at national labs (both in and out of the U.S.). A large chunk (44.2\%) of early career scientists are White, and 12.9\% are Asian. Generally speaking, early career scientists comprised about half of all Asian and Hispanic/Latino respondents. The majority of the Black and Middle Eastern respondents who took the survey are early career scientists.  

Around 59\% of PhD students reported working 40 or more hours per week; significantly more Postdocs reported the same (+17.9\%). This gap lessens when we consider early career scientists who reported working even longer hours: Compared to PhD students, only 2.4\% more Postdocs reported working 60 or more hours per week. 

\begin{figure}[H]
    \includegraphics[scale=0.5]{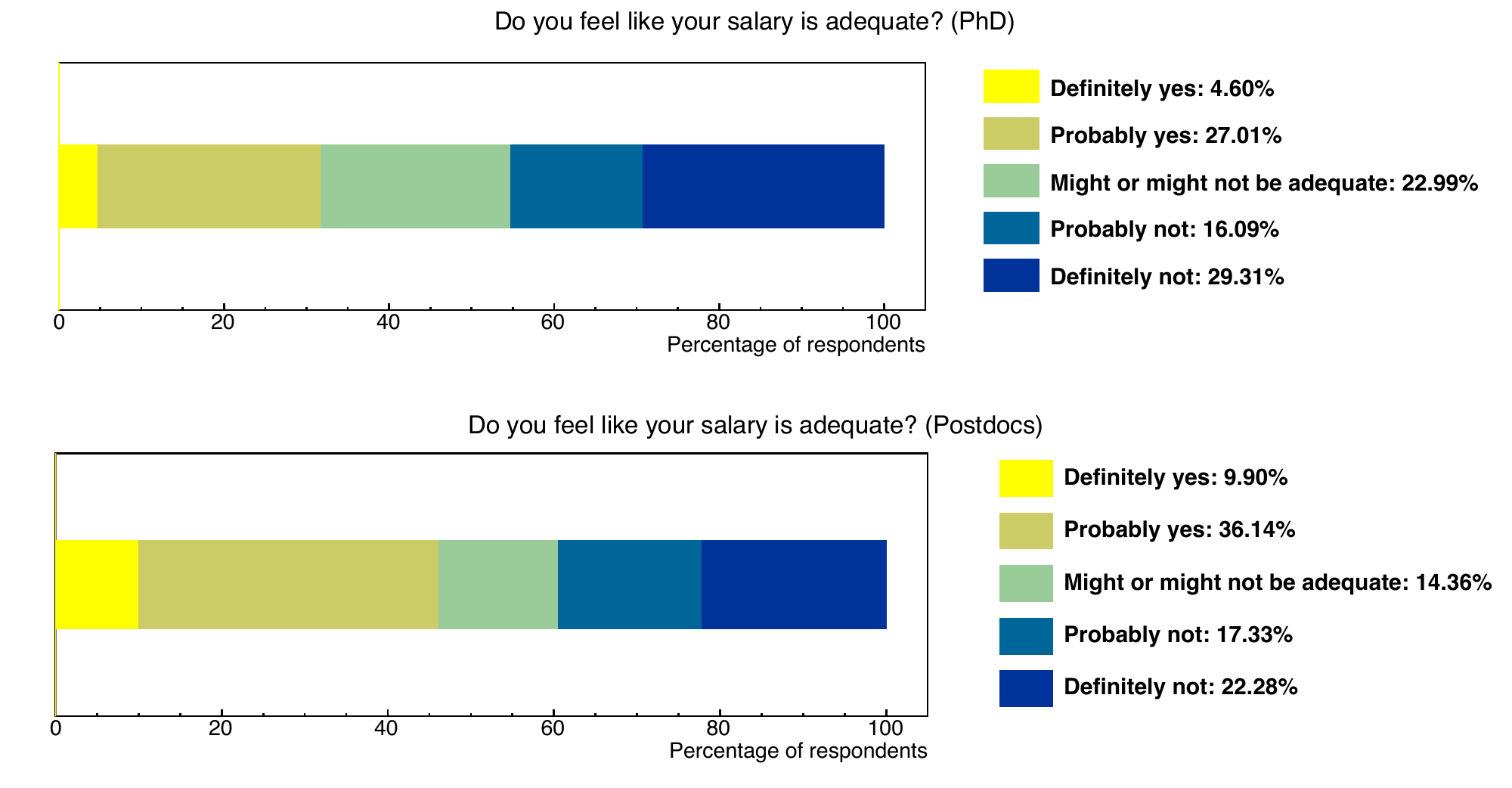}
    \caption{We asked PhD students and Postdocs how they felt about their salary. The majority of PhD students (97.8\%) and Postdocs (99\%) answered this question.}
    \label{Q48_phd_postdocs_combined}
\end{figure}

When asked about their salary, the majority of PhD students reported feeling like their salary isn't adequate -- of those who responded, 45.4\% of PhD students feel as if their salary probably or definitely isn't adequate. A smaller group of PhD students (31.6\%) reported that their salary probably or definitely is adequate, but only 4.6\% fell into the ``definitely'' end of the spectrum. 

On the other hand, Postdocs were quite mixed about their salaries; 46\% of Postdocs who responded think their salary probably or definitely is adequate, although the definitely subgroup only makes up 9.9\% of the Postdocs. On the other hand, 39.6\% of Postdocs who responded say their salary probably or definitely isn't adequate, with more of an even split between ``probably not'' and ``definitely not.'' 

\begin{figure}[H]
    \includegraphics[scale=0.8]{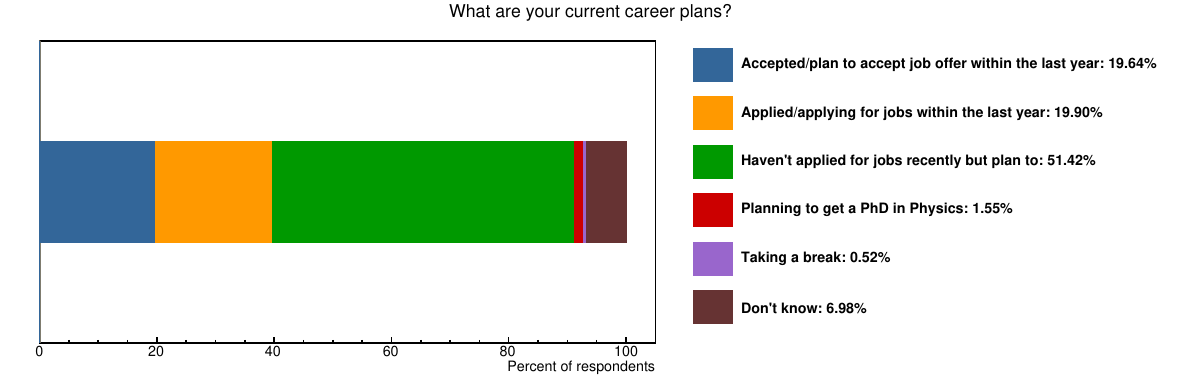}
    \caption{We asked graduate students and Postdocs about their career plans; around 99\% of early career scientists responded to this question.}
    \label{Q21_breakdownBar}
\end{figure}

When asked about their current career plans, the majority of Master's students are planning on getting a PhD in physics, while the majority of PhD students (77\%) and Postdocs (86\%) reported that they are currently applying or plan to apply for jobs. Some PhD students (9.6\%) and Postdocs (28.4\%) indicated that they accepted or plan to accept a job offer within the last year. Another set of early career scientists -- 12.9\% of PhD students and 3\% of Postdocs -- indicated that they didn't know their future plans, or they plan to take a break. Finally, around 3\% of early career scientists provided additional information in the ``other'' text box in which a majority indicated the question was not applicable to their current situation (e.g., they were not ready to graduate at the time of the survey).

\begin{figure}[H]
    \includegraphics[scale=0.82]{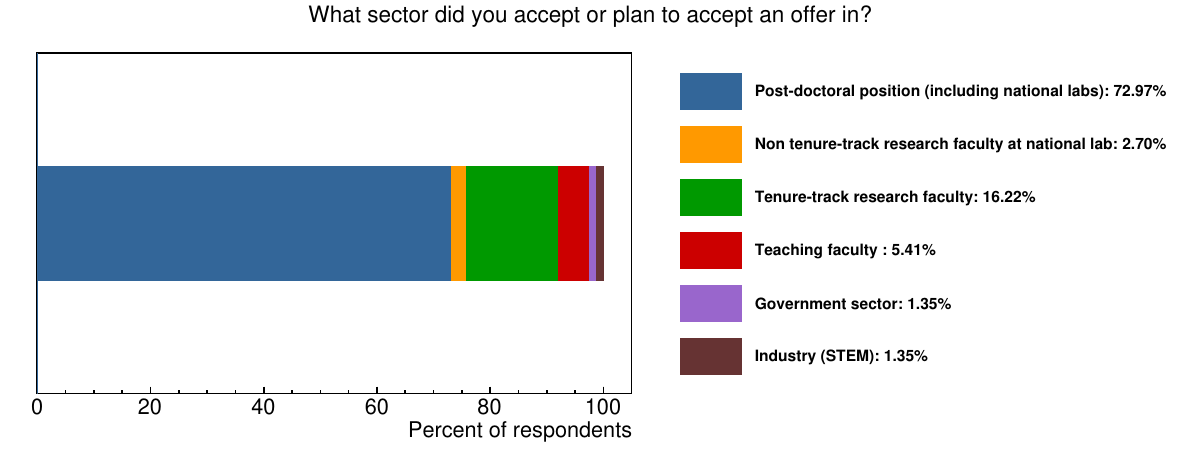}
    \caption{We asked early career scientists about the job sector in which they accepted or plan to accept a job. The majority (97.4\%) of early career scientists who saw this question provided a response.}
    \label{Q23_breakdownBar}
\end{figure}

For the 20\% of early career scientists who accepted or plan to accept a job, we asked about the type of job they accepted (or plan to accept). The vast majority of early career scientists accepted an academic position: Nearly 73\% of early career scientists accepted a postdoc position (including national lab positions), and around 16\% of early career scientists accepted a tenure-track research faculty position. For the respondents who accepted postdoc positions, three-quarters are themselves Postdocs, comprising about 20\% of all the Postdocs who took the survey. Half of these Postdocs had accepted their current position within the last year (although they could still be on their second or third postdoc position). About 7\% of PhD students who took the survey indicated that they accepted a postdoc position.

We also asked why this group of early career scientists accepted their job offer by multi-selecting from the following answer options: 
\begin{multicols}{2}
\begin{itemize}
        \itemsep0em
        \item Compensation 
        \item Personal or family-related reasons 
        \item Intellectually challenging/found it\\ interesting 
        \item Job availability  
        \item Freedom in research topics  
        \item Possibility for career advancement  
        \item Opportunity to work with a specific person  
        \item Switch to different sector  
        \item Necessary step to get future position  
        \item Visa restrictions limited my options  
        \item More inclusive working environment  
        \item Other reason (included a text box for\\ respondent to elaborate)
    \end{itemize}
  \end{multicols}  
On average, early career scientists chose between three and four answer choices. All early career scientists were likely to select challenging/interesting, ``necessary step'', and ``career advancement'' in some combination of two or all three. Less than 1\% of early career scientists provided an additional reason beyond the list presented in the survey, but further reasons include the desire for a permanent position. We examined pairs of answer options using a correlation matrix, and when graduate students indicated that they were switching to a different job sector, they were also more likely to select compensation and job availability as additional reasons. Graduate students did not indicate visa restrictions on the whole, although a nonzero number did select this option. 

Postdocs who selected ``freedom in research'' as a reason were also likely to select challenging/interesting and career advancement. Compared to graduate students, more Postdocs indicated they were switching to a different job sector; this group was also likely to select career advancement, freedom in research, and challenging/interesting as additional reasons. More Postdocs also selected ``visa restrictions'' as a reason compared to the graduate students who responded. 

Finally, we asked whether these early career scientists applied to jobs in other sectors. The majority of graduate students did not apply to other sectors, although the PhD students who accepted non-Postdoc positions also applied for Postdocs. Postdocs who had started within the last year were more likely to apply to both academic and industry positions on fairly equal levels. About one third of Postdocs who started a year ago or longer indicated that they only applied for postdoc positions. 

\begin{figure}[H]
    \includegraphics[scale=0.85]{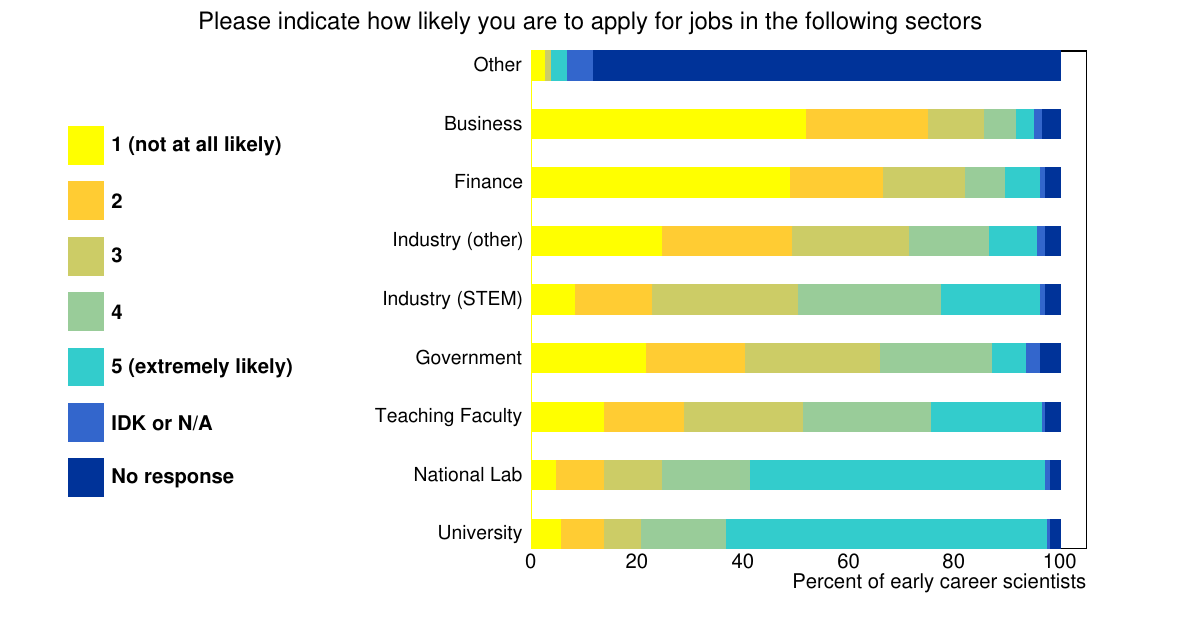}
    \caption{We asked early career scientists who are planning to apply or are currently applying for jobs to indicate in which sectors they were interested in applying. A little over half of early career scientists surveyed saw this question, and 96-98\% of those early career scientists responded to each sector (with the exception of the ``Other'' category).}
    \label{Q27_overall}
\end{figure}

For the early career scientists who are (thinking about) applying for jobs, we asked those respondents to indicate how likely they were to apply for jobs in different sectors. The highest rated sectors were academic positions at universities and national laboratories, followed by teaching faculty and industry positions in STEM. Early career scientists applying for teaching faculty positions tended to be more open to applying for finance and industry positions. Respondents were the least likely to apply for finance, entrepreneurship, or non-STEM industry positions. 

\begin{figure}[H]
    \includegraphics[scale=0.55]{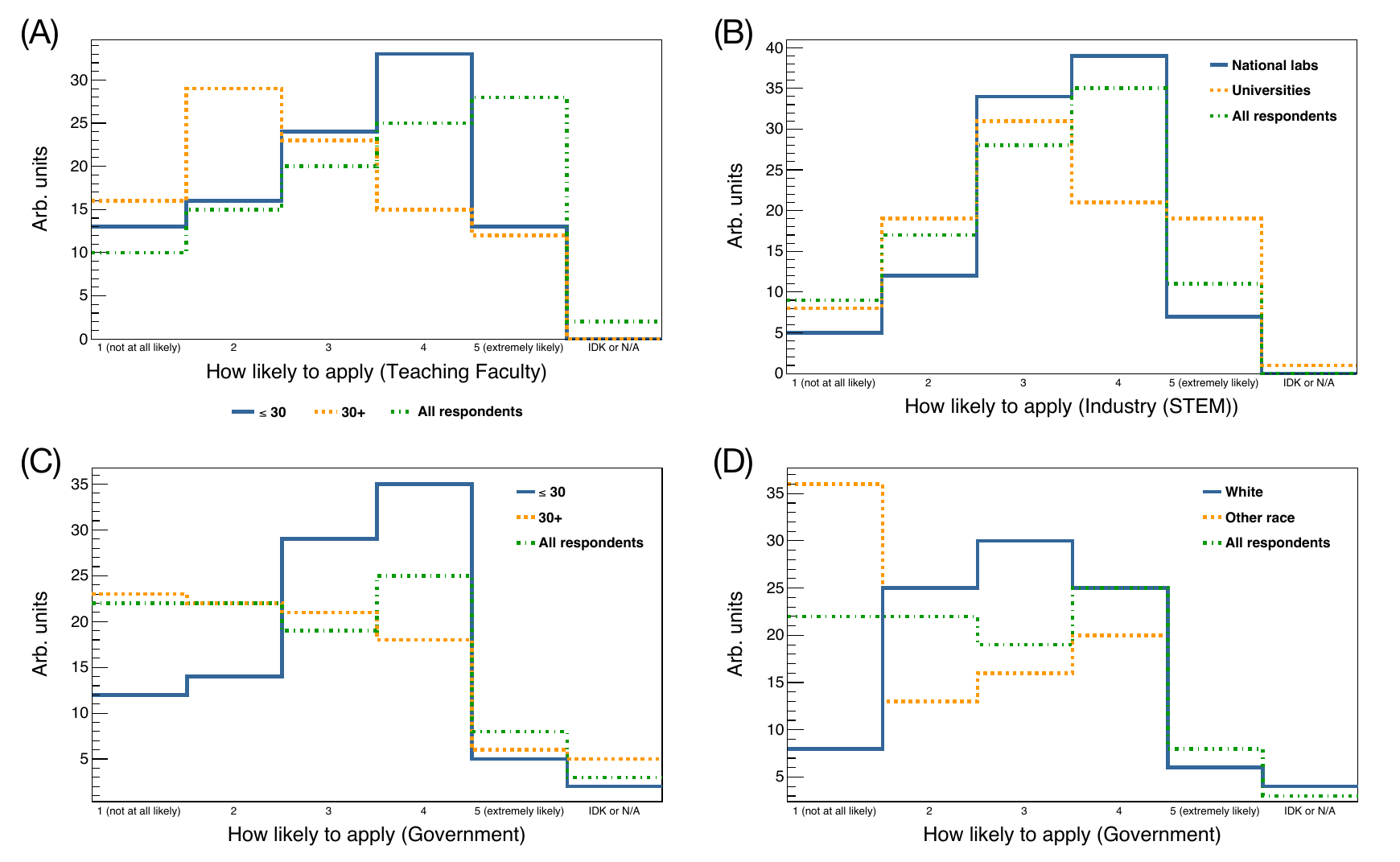}
    \caption{We asked early career scientists who are planning to apply or are currently applying for jobs to indicate in which sectors they were interested in applying. Using demographic information, we broke down (A) the teaching faculty responses by age; (B) the industry (STEM) responses by primary workplace; and the government responses by (C) age and (D) race.}
    \label{Q27_breakdown_combined}
\end{figure}

Broken down by primary workplace (Figure \ref{Q27_breakdown_combined}B), early career scientists located outside of universities were more likely to apply for STEM industry positions, while respondents at universities were more mixed in their opinions. Broken down by race (Figure \ref{Q27_breakdown_combined}D), white early career scientists were more likely to apply to government positions compared to early career scientists in other racial groups. Broken down by gender, men indicated they are more likely to apply for STEM industry positions than women. Men were also more likely to apply for finance positions. Younger male respondents (less than 30 years old) are more likely to apply for teaching faculty positions compared to other demographic groups. Early career scientists falling into other gender groups (non-binary, agender, genderqueer, or self-identify) indicated that they are more likely to apply for academic positions and less likely to apply to industry and finance positions. When broken down by Snowmass frontiers, respondents in the Energy and Instrumentation frontiers were less likely to apply for university, national laboratory, and teaching faculty positions compared to respondents in other frontiers. Respondents working in the Computational and Rare Processes frontiers indicated they were more likely to apply for national laboratory positions compared to university positions; respondents in all other frontiers reported the opposite scenario. 

\begin{figure}[H]
    \includegraphics[scale=0.52]{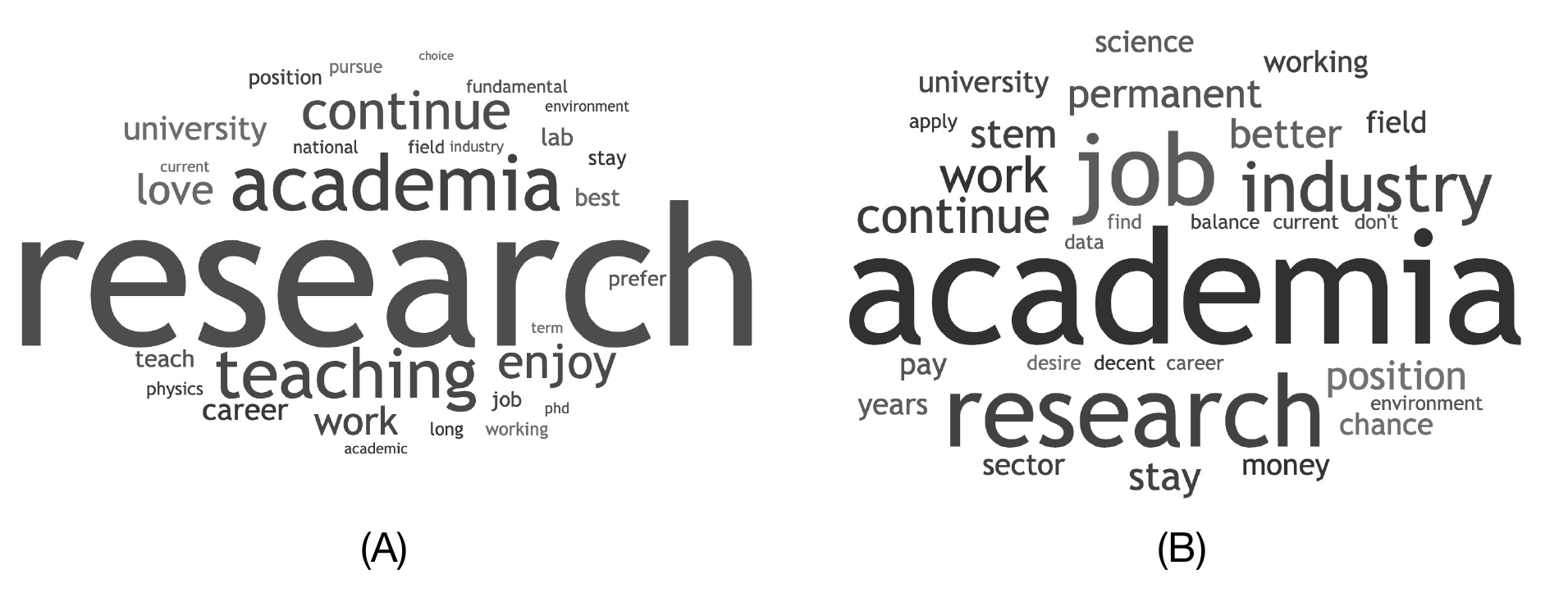}
    \caption{The top 30 words most commonly used by early career scientists planning to apply or currently applying for jobs when asked to comment on why they chose their top sector. Words such as ``I'd'' and ``I'm'' were removed from the top 30 list. (A) This group of early career scientists indicated they were extremely likely to apply for academic research positions and less likely to apply for industry positions. (B) This group of early career scientists indicated they were extremely likely to apply for industry positions and less likely to apply for academic positions.}
    \label{Q28_wordcloud_top30}
\end{figure}

We asked for early career respondents to comment on why they chose their top job sector. A word cloud with the 30 most commonly used words is shown in Figure \ref{Q28_wordcloud_top30} for different groups of early career scientists. Because we did not ask the respondents to rank the sectors, we defined the ``top sector'' as the sector in which the respondent was extremely likely to apply. Additionally, we removed respondents who were also likely to apply to other job sectors. The group in Fig. \ref{Q28_wordcloud_top30}A are early career scientists who reported they were extremely likely to apply to research academic positions (either at national labs or universities), but indicated less than a 4 for industry positions, STEM or otherwise. Meanwhile, the group in Fig. \ref{Q28_wordcloud_top30}B are early career scientists who stated they were extremely likely to apply to industry positions but are less likely to apply to any academic positions (including teaching). 

Words that appeared in both groups include research, academia, and continue. For the group interested in academic positions, the majority of comments indicated that early career scientists love research or teaching and want to stay in academia. There is an additional group who are likely to apply to a variety of positions in the academic, industry, and government sectors. There were several comments which expressed a desire to continue working in academia while also pointing out obstacles preventing them from achieving that goal (e.g., not enough permanent positions, the expectation to complete multiple Postdocs, etc.). Themes of a toxic culture in academia appeared more than once in these comments. There were also early career scientists expressing their dissatisfaction with academia and a desire to switch to a different sector -- this group almost always indicated they are extremely likely to apply to non-academic sectors. Finally, early career scientists who selected finance as their top sector tended to reference money and job opportunities in their comments; however, the same dissatisfaction about academia appeared here as well. 

\begin{figure}[H]
    \includegraphics[scale=0.83]{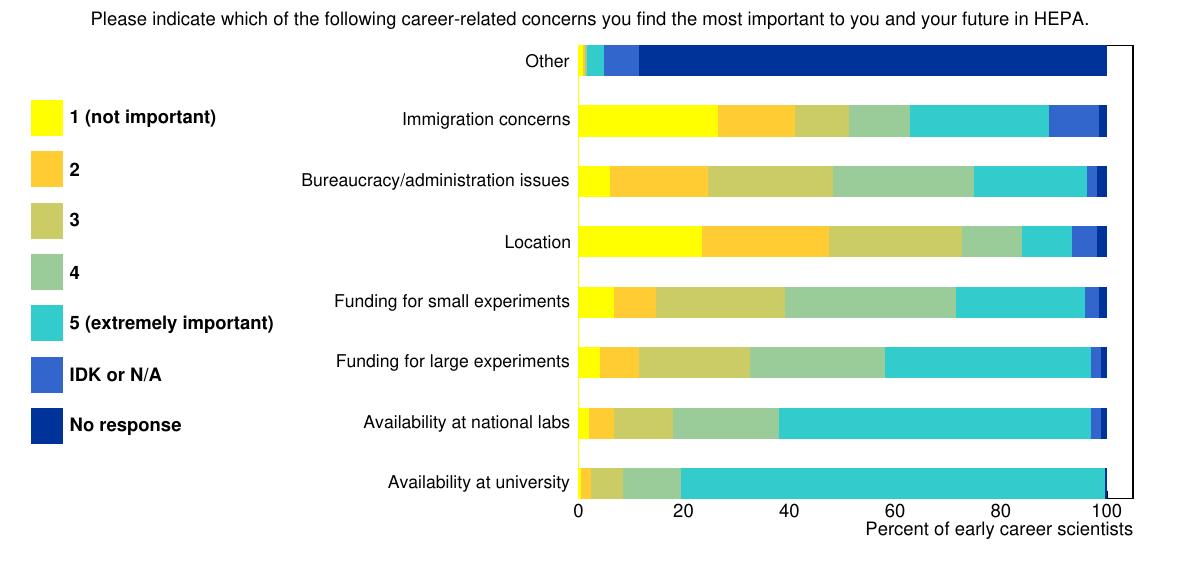}
    \caption{We asked early career scientists who were likely to apply to academic positions about their career-related concerns.}
    \label{Q29_overall}
\end{figure}

\begin{figure}[H]
    \includegraphics[scale=0.5]{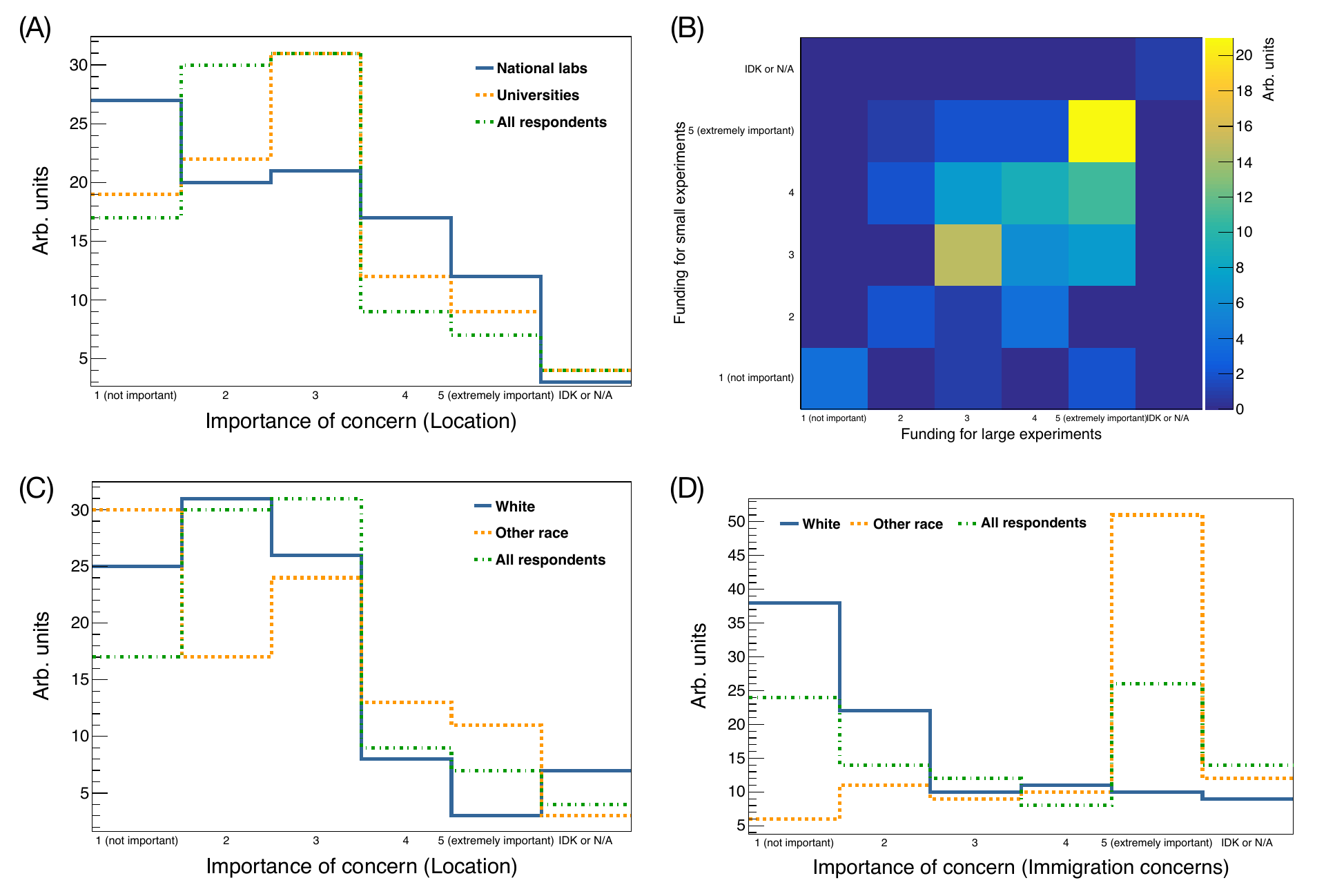}
    \caption{We asked early career scientists who were likely to apply to academic positions about their career-related concerns. Using demographic information, we broke down location concerns by (A) primary workplace and (C) race; we also broke down immigration concerns by race (D). We also compared responses for funding concerns for large and small experiments (B).}
    \label{Q29_breakdown_combined}
\end{figure}

For the 65.3\% of early career respondents who indicated that they were somewhat likely (or more) to apply to academic positions, we asked respondents to indicate which of the following career-related concerns they find the most important to their future in HEPA. Based on whether respondents indicated a concern as extremely important, by and far the most important concerns were availability at universities (80.6\%) and national labs (59.8\%). While the majority of early career scientists equally rated their concerns for funding for large and small experiments, there was slightly more concern about funding for larger experiments compared to smaller experiments (Fig. \ref{Q29_breakdown_combined}B). When broken down by Snowmass frontier, we found that respondents in some frontiers were (on average) more concerned about funding for smaller experiments. The respondents in Snowmass frontiers with the largest discrepancies, i.e., showed the most concern for funding for smaller experiments, were respondents in the Instrumentation, Neutrino, Computational, and Rare Processes frontiers. 

Bureaucracy, immigration concerns, and location were all rated lower concerns compared to the funding and availability concerns, with location being rated as the least important concern for respondents' future in HEPA -- only 9.6\% of respondents rated it as extremely important. Broken down by race, white early career scientists tended to rate location as less important compared to scientists in other racial groups (Fig. \ref{Q29_breakdown_combined}C). Early career scientists at universities tended to rate location as more important compared to respondents at other institutions (Fig. \ref{Q29_breakdown_combined}A).

Immigration concerns are either inconsequential to the respondents or they are extremely important. For example, 5.1\% more respondents indicated that immigration concerns were extremely important compared to bureaucracy concerns, while 20.8\% more respondents were not at all concerned by immigration issues compared to bureaucracy. Similar to early career scientists who accepted job offers, Postdocs applying for jobs tend to rate immigration issues with greater importance compared to PhD students. Broken down by race (Fig. \ref{Q29_breakdown_combined}D), white early career scientists tend to be the respondents not at all concerned with immigration concerns, while other racial groups found immigration concerns to be more important.

There was a small group of early career scientists (4.6\%) who wrote about other concerns they find important to their future in HEPA; 27.6\% of that group indicated their additional concern was extremely important. These concerns included compensation or salary concerns, finding a position in tandem with a romantic partner, and work environment. 

\begin{figure}[H]
    \includegraphics[scale=0.83]{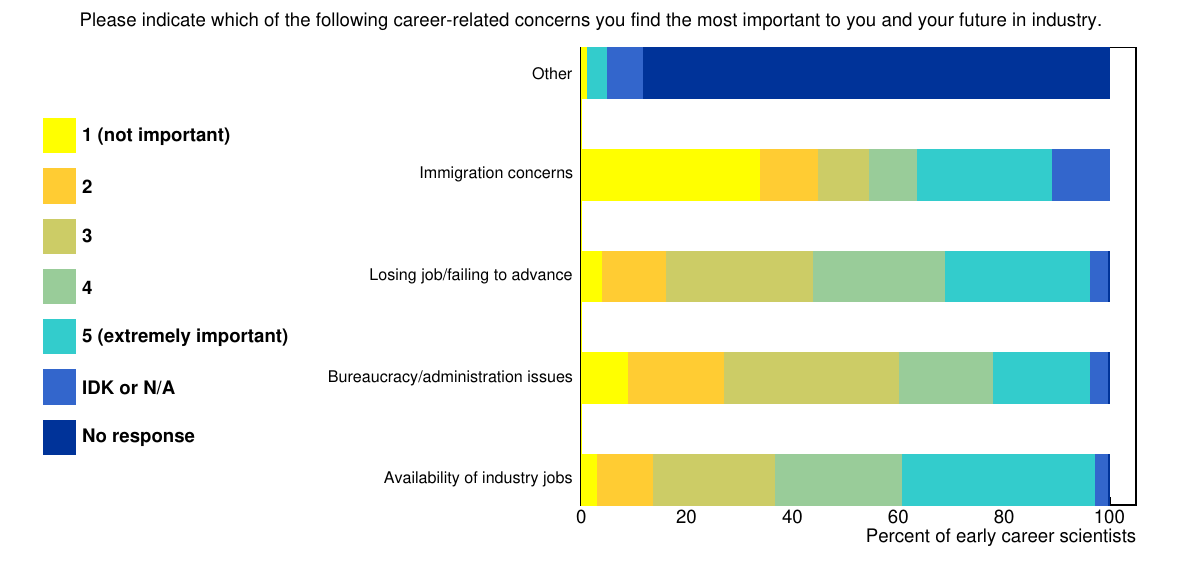}
    \caption{We asked early career scientists who were likely to apply to industry positions about their career-related concerns. A little over half of the early career scientists we surveyed saw this question.}
    \label{Q30_overall}
\end{figure}

For the 53.5\% of early career respondents who indicated that they were somewhat likely (or more) to apply to industry positions, we asked respondents to indicate which of the following career-related concerns they find the most important to their future in industry. Job availability is rated as the most important concern overall, with 36.7\% of respondents rating it as extremely important. Similar to the academia version of this question, immigration concerns either don't impact the respondent or they are extremely important. When broken down by Snowmass frontier, respondents in the Energy and Neutrino frontiers were the most concerned about immigration issues. Graduate students tend to be less concerned about immigration issues compared to Postdocs. Respondents who listed ``other'' concerns included work environment and job interest as important for their future in industry.

\begin{figure}[H]
    \includegraphics[scale=0.5]{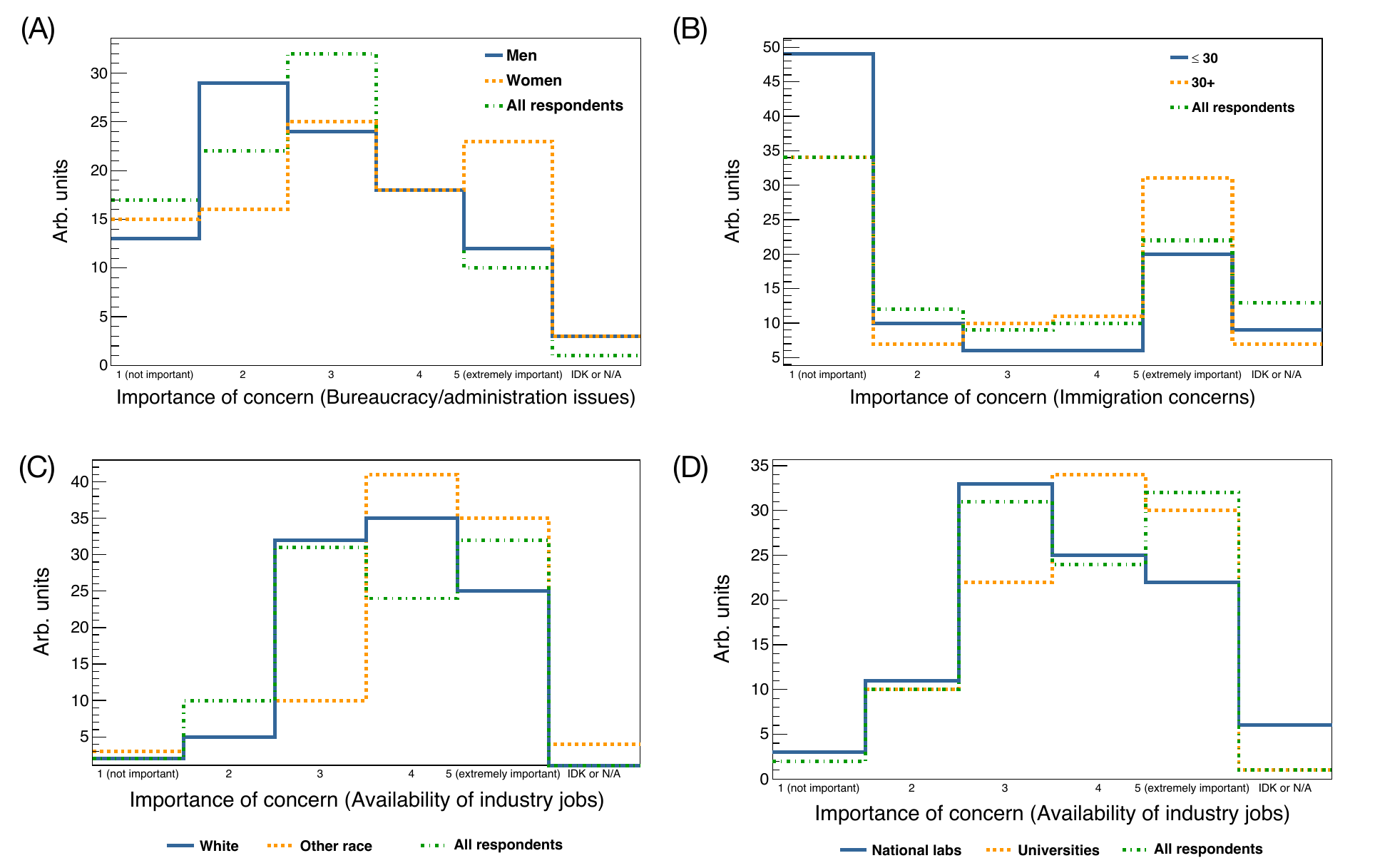}
    \caption{We asked early career scientists who were likely to apply to industry positions about their career-related concerns. Using demographic information, we broke down (A) bureaucracy issues by gender; (B) immigration concerns by age; and availability of industry jobs by (C) race and (D) primary workplace.}
    \label{Q30_breakdown_combined}
\end{figure}

Broken down by race (Figure \ref{Q30_breakdown_combined}C), white early career scientists are overall less concerned about job availability compared to early career scientists in other racial groups. Respondents at national labs are less concerned about job availability compared to other institutions (Figure \ref{Q30_breakdown_combined}D); in particular, early career scientists outside of universities and national labs are much more concerned about job availability. Over a quarter of respondents (27.5\%) rated failing to advance or losing your job as extremely important. Respondents were more mixed on the other listed concerns, although  women and NB-GQ-AG gender groups are more concerned about bureaucracy and administration issues compared to men (Fig. \ref{Q30_breakdown_combined}A).

\subsubsection{Faculty}

Faculty members -- including teaching, tenure-track, tenured, and retired faculty -- comprised 31.3\% of all survey respondents. Two-thirds of faculty respondents are tenured, with the remaining 22.7\% being tenured-track and 10\% being teaching or retired faculty. Around 62\% of our faculty respondents told us they are men -- they made up a little over one-third of all men who took our survey. Nearly 16\% of faculty respondents are women. The majority of faculty respondents (60\%) indicated they are White, making up 35\% of White respondents who took the survey. Faculty also comprised 23\% of Asian respondents and 30.6\% of Hispanic/Latino respondents. The majority of faculty respondents (68.8\%) reported that their primary workplace is a U.S. university; a further 10\% of faculty respondents indicated a university outside of the United States. Only about 5\% of faculty respondents reported a national lab (including those outside of the U.S.) as their primary workplace. The average age of the 65.3\% of tenure-track faculty who provided their ages was 37 years old.

When we separated tenure-track faculty from tenured faculty, some interesting differences emerged. Nearly two-thirds of tenured faculty indicated they are men, and an additional 14\% indicated they are women. Compare this to tenure-track faculty, where a little less than half are men (-18.6\%) and 26\% are women (+12.2\%). We saw something similar when it comes to race: 65.9\% of tenured faculty indicated they are white, and 13.3\% indicated they are another race. On the other hand, 44.4\% of tenure-track faculty indicated they are white (-21.5\%), and 30.6\% indicated they are another race (+17.3\%). It is also worth noting that 22\% of tenure-track and 16\% of tenured faculty did not respond to the survey's demographics questions.

We also compared tenure-track and tenured faculty's responses about their postdoc and job-hunting experiences. Based on the 98.1\% who provided a response, tenured faculty spent an average of 4.8 +/- 0.1 years as a postdoc, and 96.7\% reported spending an average of 17.6 +/- 1.8 months looking for a faculty position. Around 92\% of tenure-track faculty reported spending, on average, a longer amount of time as a postdoc and searching for a job, with respondents spending an average of 5.1 +/- 0.3 years as a postdoc and spending an average of 21.2 +/- 2.6 months looking for a faculty position. While comparing responses between how long faculty spent as Postdocs and how long they spent looking for a job, we uncovered two primary camps for tenure-track faculty: Those who spent 3 to 4 years as Postdocs, or those who spent 6 to 7 years as Postdocs. Those faculty who reported longer times as a postdoc also tended to report a longer job hunt, albeit by a margin of 1-3 months. This two-camp behavior did not appear in the tenured faculty responses. 

The vast majority of faculty (83.6\%) reported that they have always been in academia; only 11\% reported having experience outside of HEPA. Those who have outside experience are predominantly white men. When asked about whether they have considered leaving academia, over half of all faculty (56.6\%) reported that they have considered leaving. Broken down, 57.6\% of these respondents are men and 19.8\% are women; 61.6\% are white and 15.3\% are another race. Conversely, 38.8\% of faculty reported that they have not considered leaving academia. This group was 71.5\% men and 12.2\% women, and 63.4\% are white and 20.4\% are another race.

We asked teaching faculty whether they are continuing research; of the faculty who responded, half of them indicated that they are currently continuing research, while the other half are interested or might be interested in continuing research. The majority of teaching faculty are interested in pursuing a tenure-track position.

\begin{figure}[H]
    \includegraphics[scale=0.53]{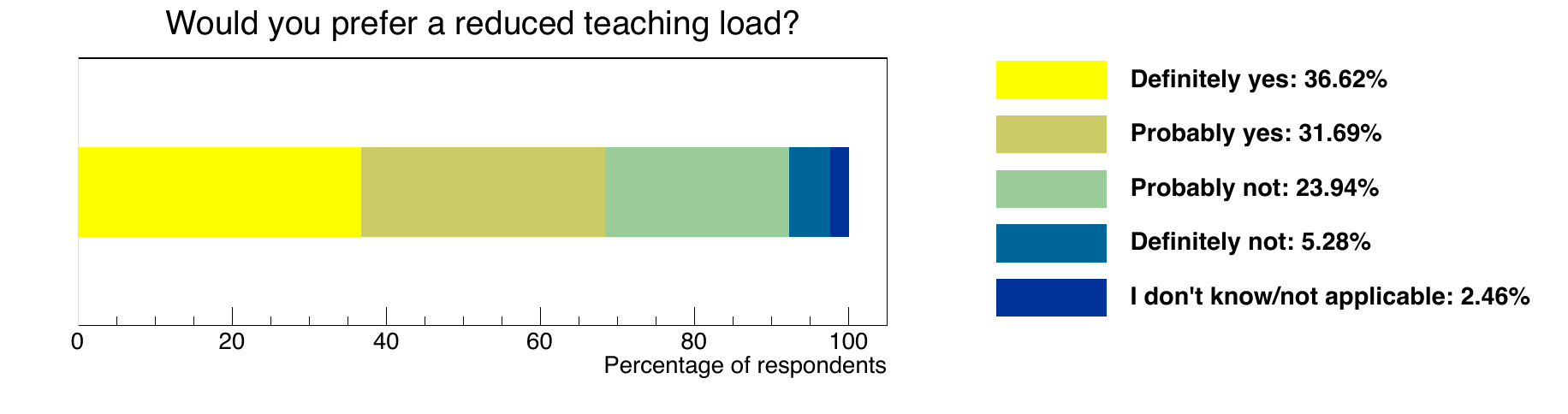}
    \caption{We asked teaching, tenure-track, and tenured faculty whether they would prefer a reduced teaching load. About 95\% of the faculty we surveyed answered this question.}
    \label{Q46_breakdownBar}
\end{figure}

The majority of teaching, tenure-track, and tenured faculty (62\%) would prefer a reduced teaching load on some level. This group of faculty is comprised of 59.3\% men and 20.6\% women, 63.4\% white faculty and 17.1\% faculty of another race. When we compare these demographics to the other side, the 27.9\% of faculty who would not prefer a reduced teaching load on some level, we see more men (+7\%) and less women (-9.8\%). There is also less white faculty (-6.8\%) and only slightly more faculty of another race (+1\%). We also asked retired faculty if they would have preferred a reduced teaching load. Only a limited fraction of the retired faculty surveyed responded to this question, but the majority would not have preferred a reduced teaching load on some level. Perhaps this is a sign that expectations and responsibilities have changed for faculty with regards to teaching.

\begin{figure}[H]
    \includegraphics[scale=0.57]{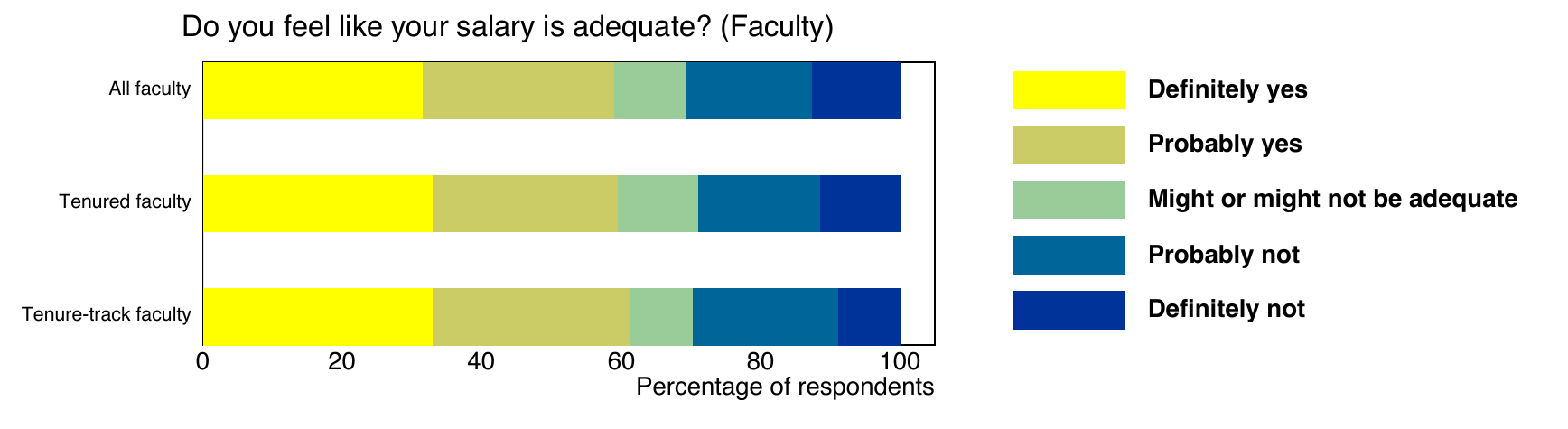}
    \caption{We asked faculty whether they feel their salary is adequate. Nearly all of the faculty we surveyed (94.3\%) answered this question; about 5\% more tenured faculty responded to this question compared to tenure-track faculty.}
    \label{Q48_combined_faculty}
\end{figure}

When asked about their salary, 57\% of tenure-track respondents reported their salary is probably or definitely adequate; around 28\% of tenure-track faculty think their salary probably or definitely isn't adequate. When we only consider the faculty who responded to this question, tenure-track faculty have a slightly larger population who think negatively of their salary compared to tenured faculty. Around 3\% more tenured faculty think their salary definitely isn't adequate. In a way, tenured faculty were more mixed on their opinions -- 3\% more tenured faculty fell into the middle ground compared to tenure-track respondents. However, we see very similar responses when we compared the two faculty groups: A majority of tenured faculty respondents (+1.3\%) think their salary probably or definitely is adequate. Compared to tenure-track faculty, a similar proportion of tenured faculty respondents (+0.4\%) reported that their salary is inadequate on some level.

\begin{figure}[H]
    \includegraphics[scale=0.84]{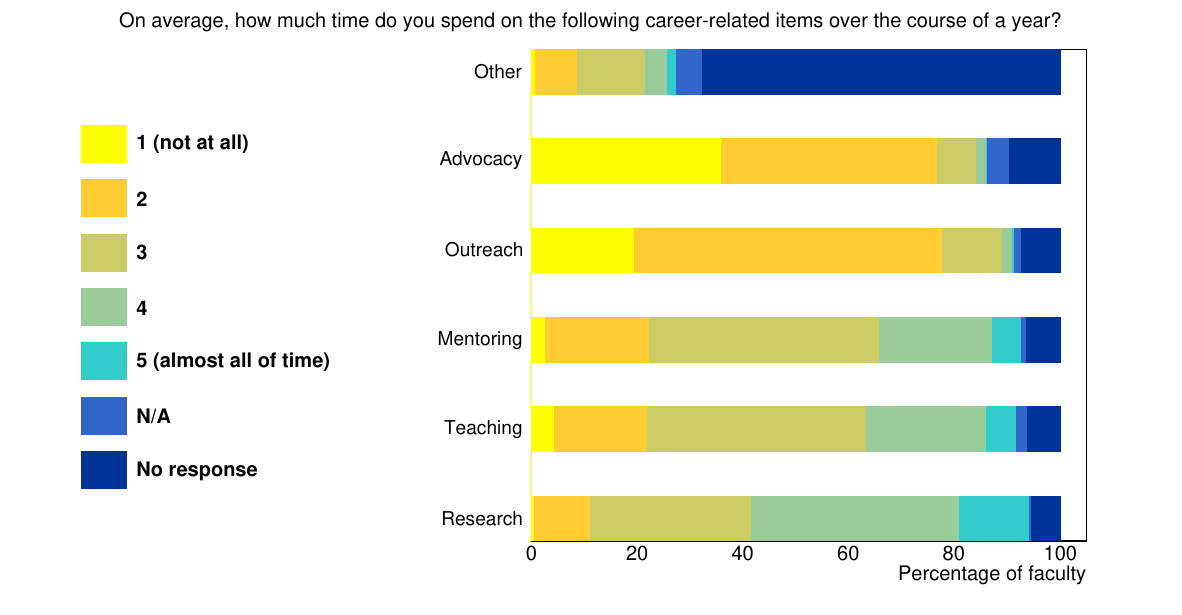}
    \caption{We asked all faculty about their yearly time commitments to various career-related items. Over 94\% of the faculty we surveyed responded to at least one part of this question.}
    \label{Q40_overall}
\end{figure}

We asked about how faculty spend their time on common career-related items over the course of a year with options ranging from 1 (``not at all'') to 5 (``almost all of my time''). A subset of faculty (18.9\%) reported spending almost all of their time on at least one item, and only 5\% of all faculty reported spending almost all of their time on two or more items. On average, faculty respondents reported committing the most amount of time to research, fairly equal amounts of time committed to teaching and mentoring and ``other'', and less time committed to outreach and advocacy. 

\begin{figure}[H]
    \includegraphics[scale=0.41]{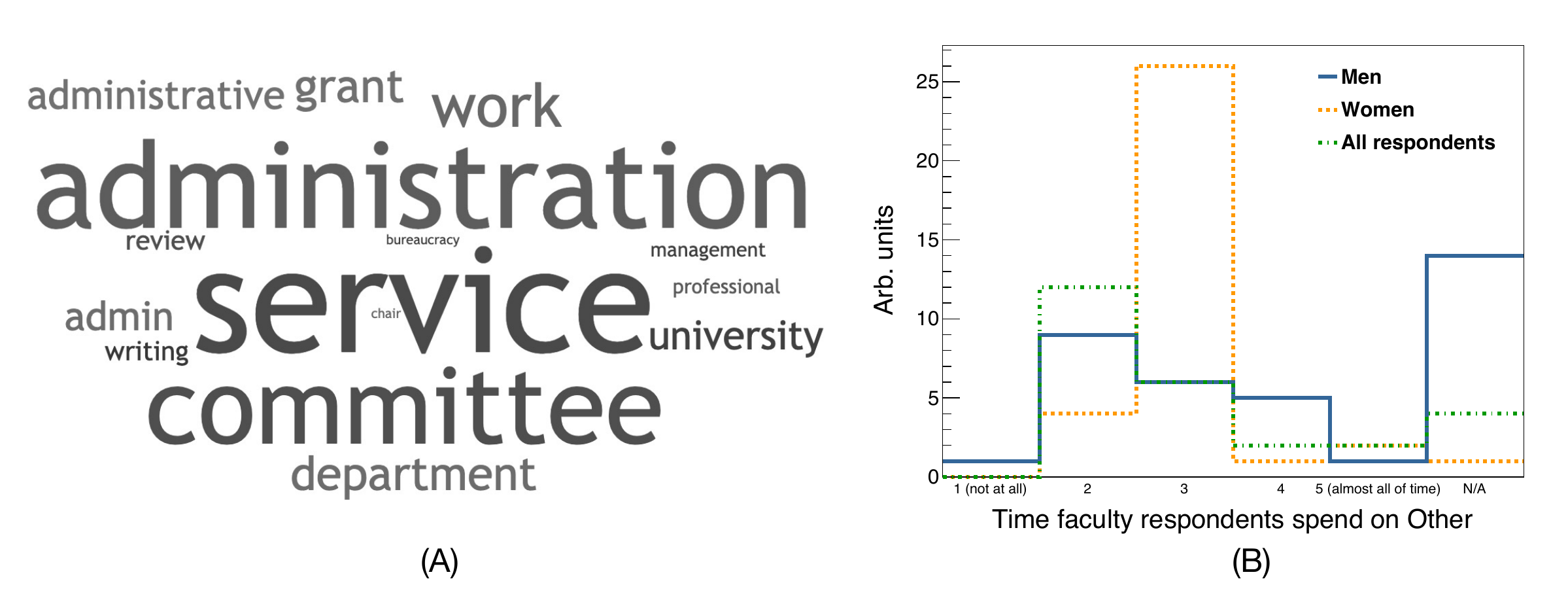}
    \caption{We asked all faculty about their yearly time commitments to self-reported career items. (A) A world cloud with the top 15 words based on faculty responses. (B) Responses for the self-reported category broken down by gender.}
    \label{Q40_Other_combined}
\end{figure}

We note here that only 27.1\% of all faculty reported some time commitment to the ``other'' category. A word cloud with the top 15 most commonly used words can be found in Figure \ref{Q40_Other_combined}A; text input revealed that this category was mostly used by faculty respondents who have service or administrative commitments. For the faculty who indicated they had those commitments, they reported spending approximately the same amount of time on their ``other'' commitments compared to teaching and mentoring. 

\begin{figure}[H]
    \includegraphics[scale=0.54]{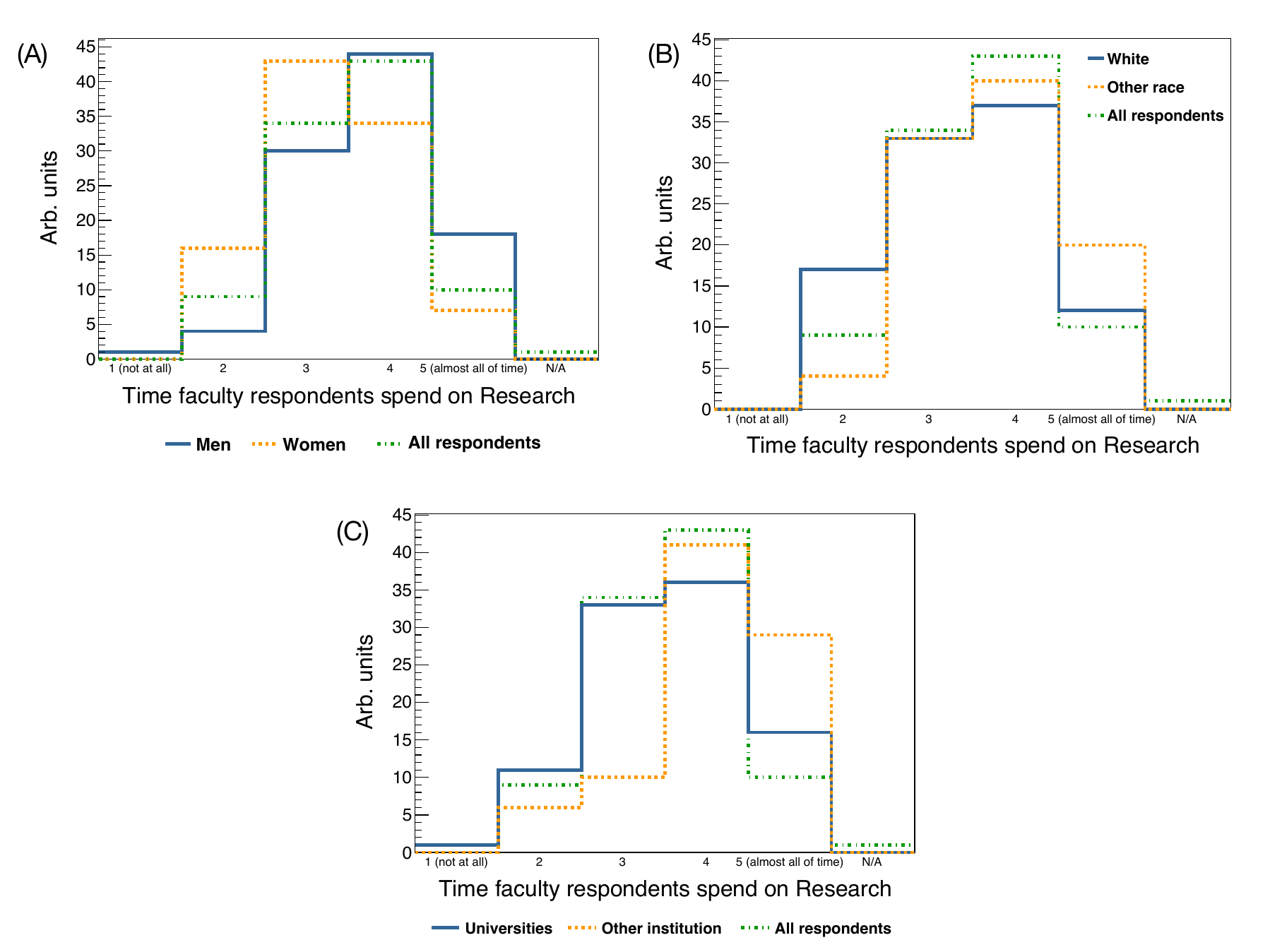}
    \caption{We asked all faculty about their time commitment to research over the course of a year. Using demographic information, we broke down these responses by (A) gender, (B) race, and (C) primary workplace.}
    \label{Q40_Research_combined}
\end{figure}

When we broke down this question using different demographic information, some interesting differences emerged. When broken down by gender, we found that male faculty report slightly more research time commitment compared to women (Fig. \ref{Q40_Research_combined}A). The small sample of NB-GQ-AG faculty reported spending the majority of their time on research (more than men and women), less time on teaching, and even less time on advocacy and ``other'' career-related items. All gender groups reported spending similar amounts of time on mentoring. When we focused on the 9.5\% of all faculty respondents who reported 3 or higher on Advocacy, nearly a third of this group were women. Finally, female faculty reported spending more time on ``other'' time commitments (i.e., administrative and service duties) compared to male faculty (Fig. \ref{Q40_Other_combined}B).

When broken down by race, white faculty respondents report slightly less research time commitment compared to other faculty in other racial groups (Fig. \ref{Q40_Research_combined}B). White faculty also reported more 3's for mentoring, while a greater proportion of faculty in other racial groups reported more 4's and 5's. When we focused on the 13.6\% of all faculty respondents who reported a 3 or higher on Outreach, a greater percentage of faculty in other racial groups reported 4's and 5's. Over half of white faculty respondents reported 2 on Advocacy time commitments; for the 9.5\% of faculty respondents who reported 3 or higher on Advocacy, more faculty in other racial groups fell into this population.

When broken down by primary workplace, we found that faculty at non-university primary workplaces report slightly more research time commitment compared to faculty at universities (Fig. \ref{Q40_Research_combined}C). We also found that the majority of faculty at universities reported a 3 for mentoring time commitments, while faculty at other institutions reported a larger proportion of 5's. 

 Finally, when broken down by age, we found that younger faculty report spending more time on teaching compared to older faculty.
 
\begin{figure}[H]
    \includegraphics[scale=0.54]{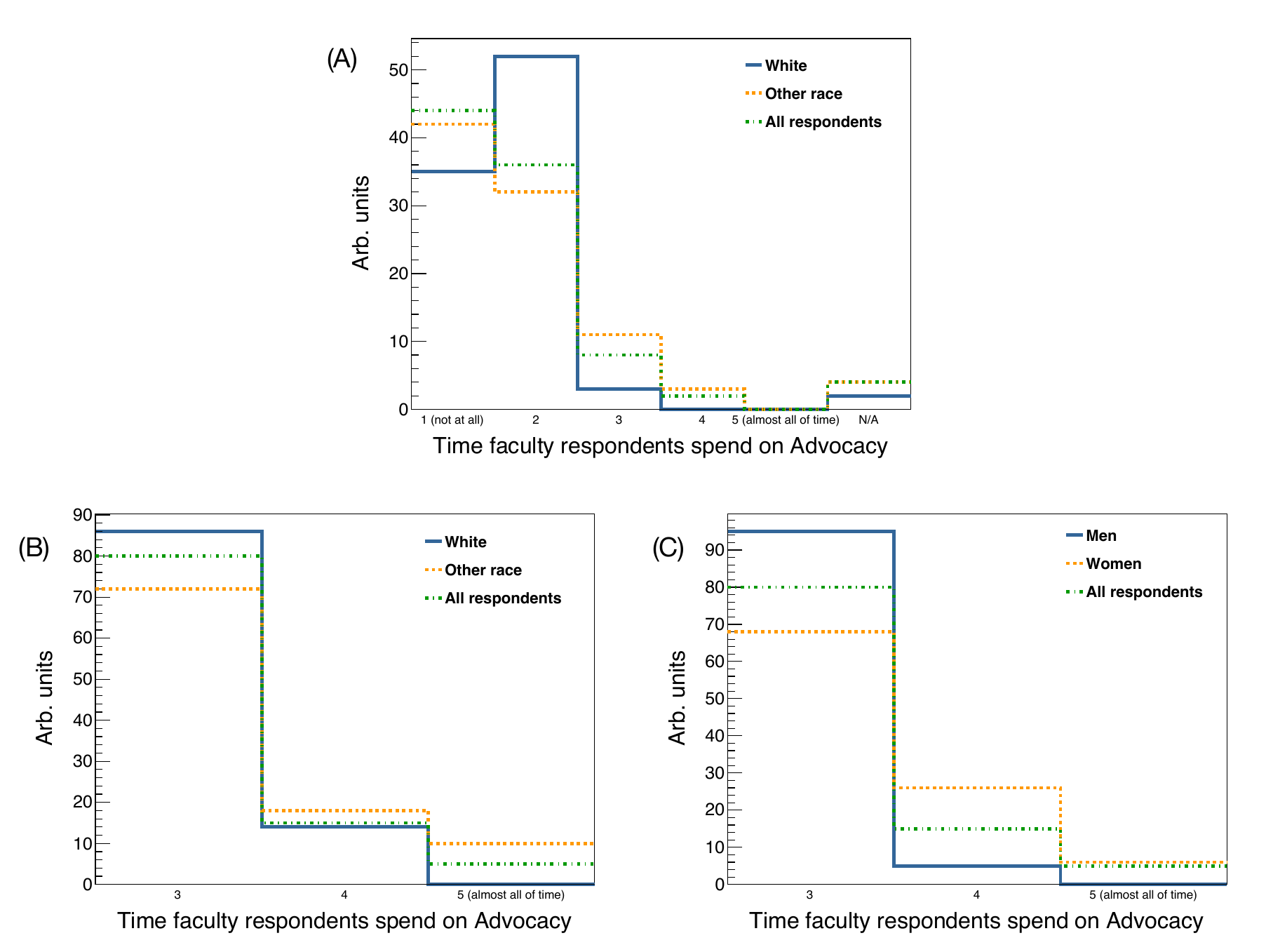}
    \caption{We asked all faculty about their time commitment to advocacy over the course of a year. Using demographic information, we broke down all of the responses by (A) race. We also looked more closely at the faculty who reported a significant time commitment to advocacy, we broke down those responses by (B) race and (C) gender.}
    \label{Q40_Advocacy_combined}
\end{figure}
 
\begin{figure}[H]
    \includegraphics[scale=0.54]{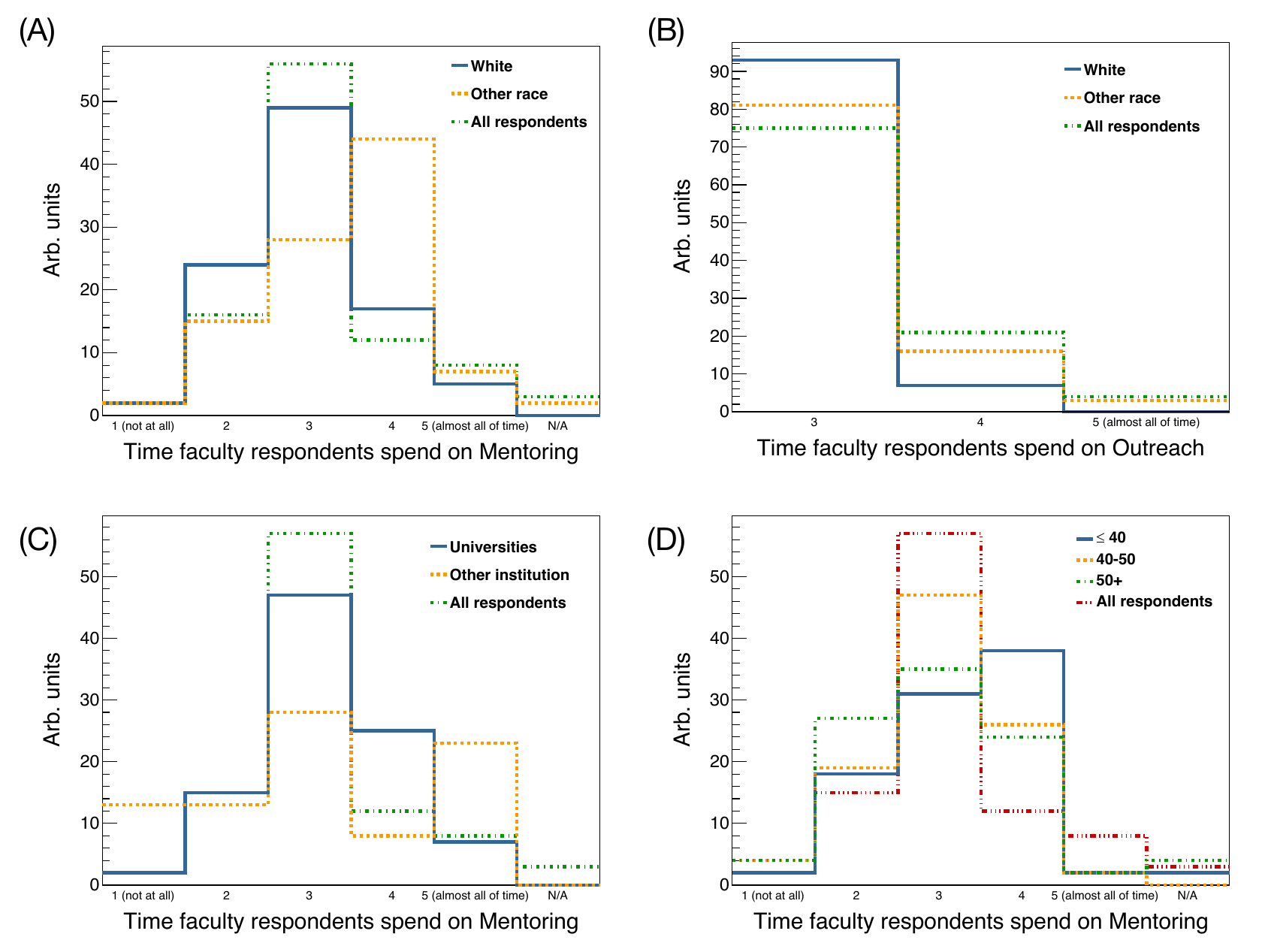}
    \caption{We asked all faculty about their time commitment to mentoring and outreach over the course of a year. Using demographic information, we broke down mentoring and outreach responses by race (A and B, respectively); we also broke down mentoring responses by (C) primary workplace and (D) age.}
    \label{Q40_Mentoring_Outreach_combined}
\end{figure}

When broken down by Snowmass frontier, we found that faculty working in the Community Engagement frontier reported spending more time on outreach and advocacy efforts while spending less time on research. Faculty in the Energy, Theory, and Instrumentation frontiers reported spending more time on research compared to faculty in other frontiers. Faculty in the Accelerator and Underground Facilities frontiers reported spending the least amount of time on advocacy efforts. The frontier that reported the most time spend on ``other'' career-related items (e.g., administrative duties) was the Neutrino frontier, followed by the Cosmic and Theory frontiers. 
 
We assumed that faculty who report a 4 or 5 on an item (e.g., research) spend at least 50\% of their time on this particular item. Armed with this assumption along with how many hours per week faculty reported they spend working, we calculated the minimum hours per week faculty commit to different career items. The outreach, advocacy, and ``other'' categories contained very few faculty who responded with a 4 or 5. Therefore, we will only report the calculation for the remaining categories: 
\begin{itemize}
    \item Forty-five percent of faculty commit at least 22.8 +/- 0.6 hours per week to research.
    \item Twenty-five percent of faculty commit at least 22.9 +/- 0.7 hours per week to teaching.
    \item Twenty-two percent of faculty commit at least 24.2 +/- 0.7 hours per week to mentoring.
\end{itemize}

\begin{figure}[H]
    \includegraphics[scale=0.83]{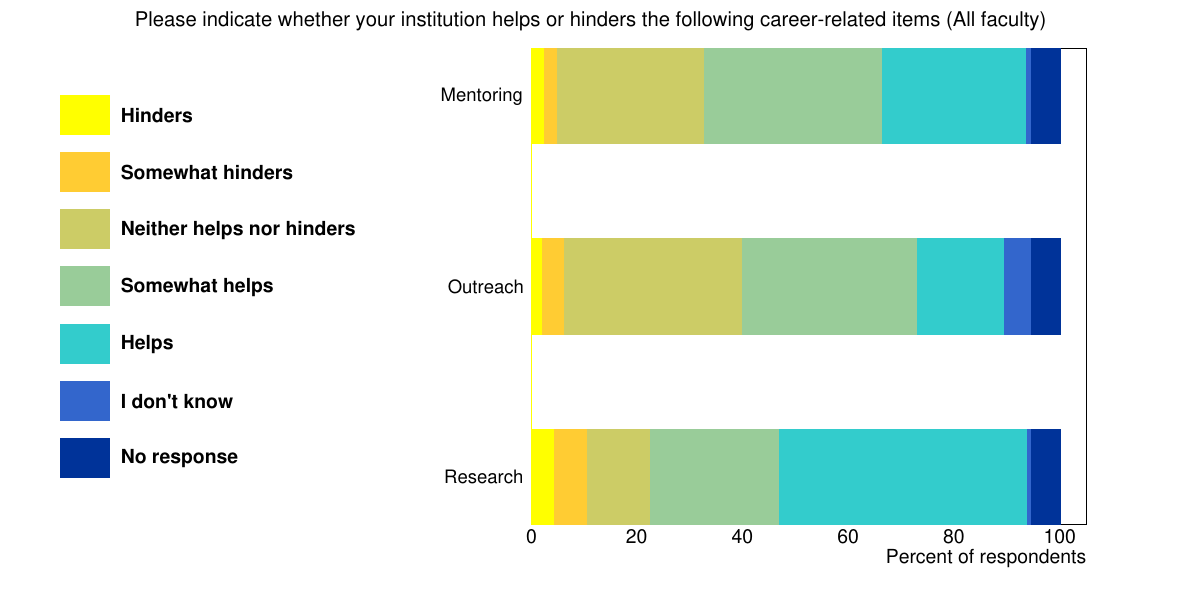}
    \caption{We asked all faculty about their institution's effect on research, outreach, and mentoring. Over 94\% of faculty answered all parts of this question.}
    \label{Q45}
\end{figure}

We asked faculty if they feel like their institution is helping or hindering them in their research, mentoring, and outreach. Almost half of faculty reported that their institution helps with research, much more than outreach (-32\%) and mentoring (-21\%). Only 10\% of faculty reported that their institution somewhat or outright hinders their research. The majority of faculty who reported their institution somewhat or outright hinders their research were white men at US universities, although a significant portion were women -- in fact, 14\% of female faculty reported their institution somewhat or outright hinders their research (compared to 10.2\% of male faculty with the same report). More faculty reported that their institution neither helps nor hinders mentoring (+15.8\%) and outreach (+21.8\%) compared to just 12\% of faculty who reported their institution neither helps nor hinders research. Moreover, outreach contained the most amount of faculty respondents who didn't know whether their institution helps or hinders (5.4\%).

When broken down by Snowmass frontiers, faculty in the Cosmic frontier had a larger proportion of its faculty reporting that their institutions help with outreach efforts. Meanwhile, faculty in the Energy frontier reported that their institutions were (on average) more helpful in their research efforts compared to other Snowmass frontiers. Faculty in the Computational frontier contained a larger proportion reporting their institution has some amount of hindering in their research and mentoring efforts, while a larger proportion indicated their institutions are somewhat helpful for outreach efforts. Finally, faculty in the Theory frontier were more mixed in their opinions about their institutions' involvement with their mentoring efforts, with a larger proportion of faculty falling into the ``somewhat helps'' and ``somewhat hinders'' categories. 

\begin{figure}[H]
    \includegraphics[scale=0.83]{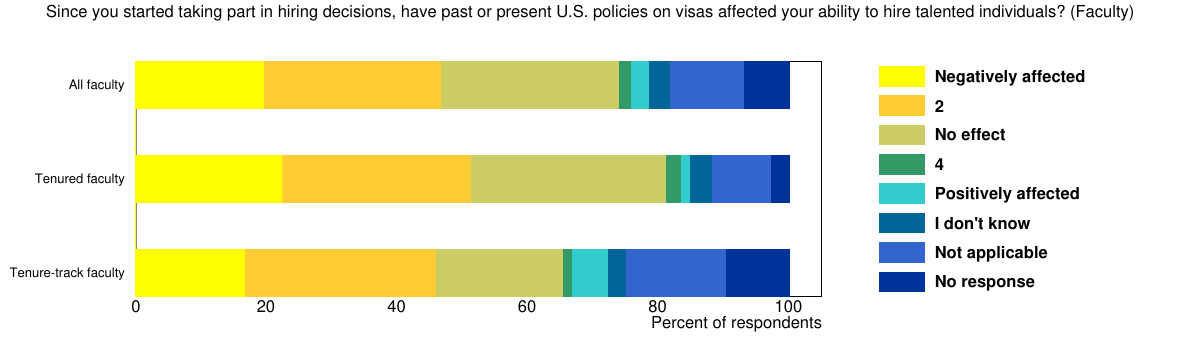}
    \caption{We asked faculty who take part in hiring whether U.S. policies on Visas have affected their ability to hire. A little over 93\% of all faculty surveyed responded to this question; broken down by seniority, nearly 7\% more tenured faculty answered this question compared to tenure-track faculty.}
    \label{Q59_faculty_combined}
\end{figure}

We asked faculty, specifically those who take part in hiring, if past or present U.S. policies on Visas have affected their ability to hire talented individuals. Of the faculty who answered this question, about half reported that past or present U.S. policies on Visas had some negative effect on their hiring ability. Comparing tenure-track and tenured faculty, 1.9\% more tenured faculty report negative effects. More tenure-track faculty reported a positive effect compared to tenured faculty (-3.8\%) and senior scientists (-3.4\%). It's worth noting that about 9\% more tenured faculty reported no effect, and nearly 8\% more tenure-track faculty answered with ``not applicable''. 

\begin{figure}[H]
    \includegraphics[scale=0.81]{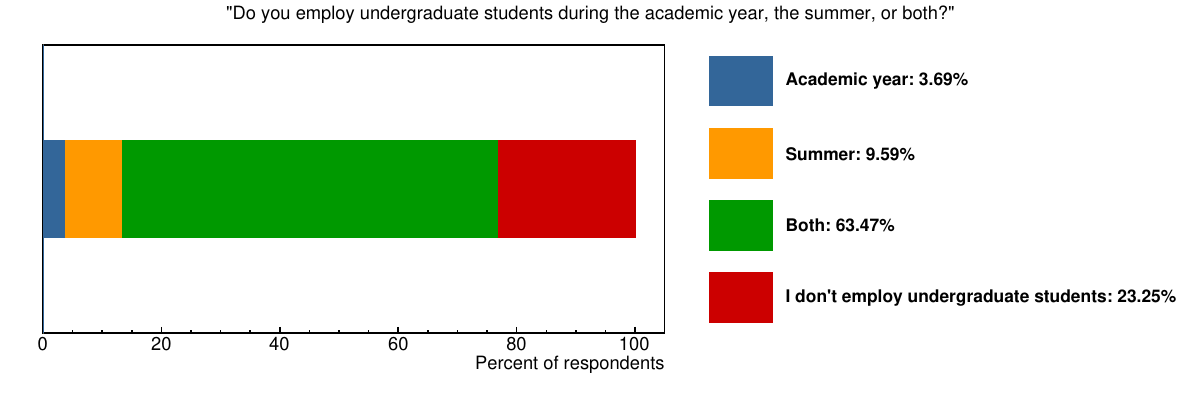}
    \caption{We asked tenure-track, tenured, and retired faculty about when and whether they employ undergraduate students.}
    \label{Q38_breakdownBar}
\end{figure}

We asked tenure-track, tenured, and retired faculty whether they employ undergraduate students and how those students are compensated. From the 89.7\% of faculty who responded to the former, 63.5\% of faculty reported that they employ undergraduate students during both the summer and academic year; 23.2\% of respondents do not employ undergraduate students. Looking further, we found that 23.2\% of the tenured faculty who responded to this question reported that they do not employ undergraduate students -- the equivalent number for tenure-track faculty is 15.6\%. Breaking down by gender, 26.7\% of male faculty and 8.5\% of female faculty who responded to this question reported that they do not employ undergraduate students. When we asked about compensation, 60.6\% of all faculty reported that they pay their undergraduate students. Furthermore, 42.7\% of all faculty reported offering course credit for their undergraduate students. The fraction of tenure-track and tenured faculty are very similar for both of these forms of compensation: comparatively, 1.6\% more tenure-track faculty reported paying their undergraduate students, while nearly 2\% more tenured faculty reported offering course credit. However, direct payment and course credit are not the only methods of compensation: 5.3\% of all faculty told us their various methods, including institutional fellowships and internships; work study programs; the NSF REU program \cite{NSF_REU}; or volunteer work. Finally, there were a nonzero number of faculty who reported that they do not compensate their undergraduate students: Specifically, 11.3\% of all faculty surveyed. This amounts to 9.3\% of all tenure-track faculty and 19.5\% of tenured faculty who responded to this question. 

\begin{figure}[H]
    \includegraphics[scale=0.56]{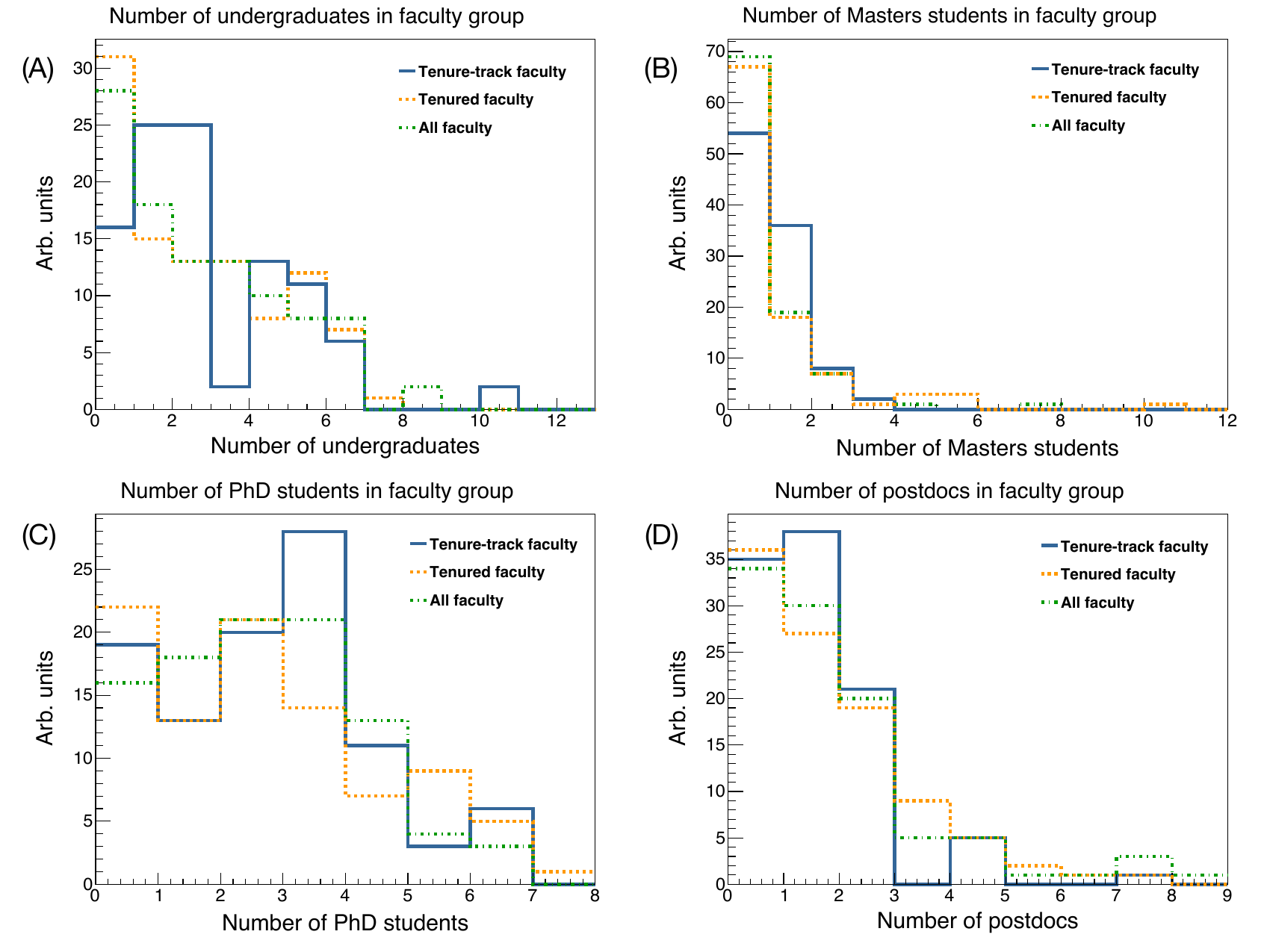}
    \caption{We asked faculty about the number of students and Postdocs in their research group. From the faculty we surveyed, between 90\% and 93\% responded about undergraduate students, PhD students, and Postdocs; conversely, 79.1\% of faculty answered about Master's students.}
    \label{Q36_combined}
\end{figure}

We asked faculty how many students and Postdocs they have in their groups. On average, faculty reported having 2 undergraduate students, 1 postdoc, and no Master's students in their research group. Tenured faculty reported slightly more Postdocs compared to tenure-track faculty, but the real difference arises in PhD students: On average, tenure-track faculty report close to 2 PhD students, while tenured faculty report closer to 3 PhD students. 

Furthermore, we asked tenure-track and tenured faculty what percent of students and Postdocs are or were employed after graduation. Out of the 48\% of faculty who responded about their Master's students, tenure-track faculty report 44.5$\pm$9.0 percent of their Master's students are employed, while tenured faculty report 75.9$\pm$3.8 percent. From those who answered the question, over half of tenure-track faculty (51.7\%) and 19.8\% of tenured faculty reported 0\% employment for their Master's students; broken down by gender, 0\% employment for Master's was reported by 22.8\% of male faculty and 39.3\% of female faculty who responded.

We received more responses from faculty (72.8\%) about the employment status of their PhD students. Tenure-track faculty reported 53.2$\pm$8.6 percent of their PhD students are employed, and tenured faculty reported 90.8$\pm$1.9 percent. Eleven percent of faculty who responded to the question reported that 0\% of their PhD students are employed. This sample size was small but contained more tenure-track faculty.

Finally, 70.9\% of faculty answered about the employment status of their Postdocs. Tenure-track faculty report 61.1$\pm$8.0 percent of their Postdocs are employed, while tenured faculty report 87.5$\pm$2.3 percent. Similar to PhD students, only a small fraction of faculty (13\%) reported 0\% of their Postdocs are employed -- and again, more tenure-track faculty reported 0\% employment for their Postdocs. 

\subsubsection{Scientists}
Scientists made up 18\% of all survey respondents. Both ``scientist'' and ``senior scientist'' were included as answer options in the survey; 46\% reported they were scientists, and 54\% reported they were senior scientists. One-fifth of all men who took the survey indicated they are scientists, making up 64.5\% of all the scientists we surveyed. Women comprised 14.8\% of scientists; alternatively, 15.5\% of all women surveyed indicated they are scientists. The majority of scientists (69.4\%) reported that their primary workplace is a national lab (including those outside of the U.S.), while 13.7\% indicated their primary workplace is a U.S. or international university. The majority of the scientists we surveyed (60.1\%) indicated they are White, and almost  15\% indicated they are part of another racial group. 

We asked all scientists how long they spent as Postdocs and how long they spent looking for a scientist position. Based on the average of 96\% of scientists who provided responses, scientists reported spending 4.3$\pm$0.3 years as a postdoc, and they reported spending 12.8$\pm$1.9 months looking for scientist position. Similarly for the roughly 94\% of senior scientists who provided responses, senior scientists reported that they spent 4.1$\pm$0.3 years as postdoc and 10.5$\pm$1.3 months looking for scientist position. 

\begin{figure}[H]
    \includegraphics[scale=0.57]{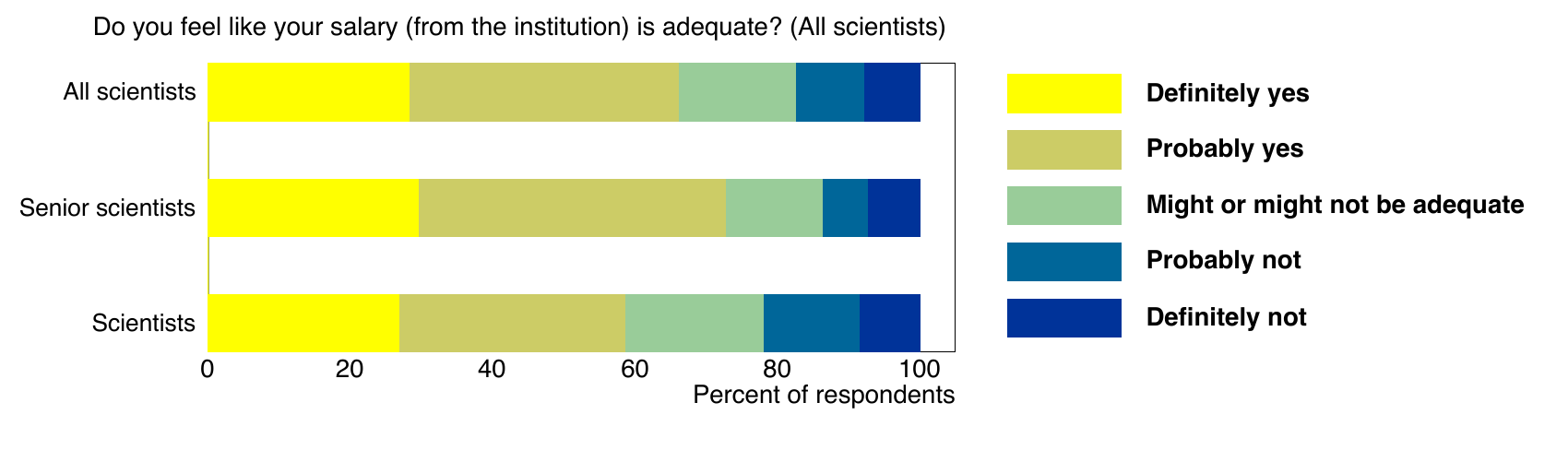}
    \caption{We asked all scientists if they feel like the salary from their institution is adequate. The majority of scientists (97.6\%) and senior scientists (96\%) answered this question.}
    \label{Q58_breakdownBar_updated}
\end{figure}

We asked all scientists whether they feel like their salary from their institution is adequate; the normalized results for all responses, along with split results for scientists and senior scientists, is shown in Figure \ref{Q58_breakdownBar_updated}. Of the scientists who answered the question, 66.1\% reported that their salary is probably or definitely adequate. This is higher than faculty who reported the same (+7.2\%). When we split scientists by seniority, the differences are striking: 58.5\% of scientists and 72.7\% of senior scientists reported their salary is adequate at some level. So scientists fall below both tenure-track (-2.7\%) and tenured faculty (-0.9\%), while senior scientists are much higher than both tenured faculty (+11.5\%) and tenure-track faculty (+13.3\%).

On the other hand, 17.5\% of scientists who answered the question reported that their salary probably or definitely isn't adequate, much lower than faculty (-13.3\%). When split by seniority, we again see a striking difference: 21.9\% of scientists find their salary is inadequate at some level versus 13.7\% of senior scientists. These are both much lower fractions compared to their faculty counterparts, although the difference for senior scientists is again more striking (-15.3\% and -16.3\% for tenured and tenure-track faculty, respectively). 

Broken down by gender, 73.1\% of female scientists reported that their salary is adequate on some level, 2.9\% higher than male scientists. We see the same slight difference on the opposite end: 15.8\% of male scientists reported that their salary is inadequate on some level, 4.3\% higher than female scientists. Broken down by Snowmass frontiers, scientists from the Computational and Accelerator frontiers contained the largest proportion of scientists indicating that their salary is inadequate on some level. 

\begin{figure}[H]
    \includegraphics[scale=0.83]{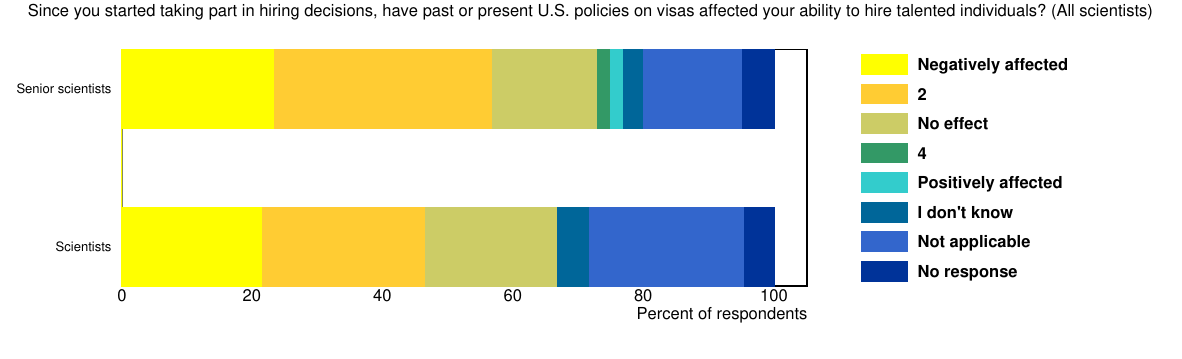}
    \caption{We asked scientists who take part in hiring whether U.S. policies on Visas have affected their ability to hire. Around 95\% of scientists and senior scientists responded.}
    \label{Q59_scientists_noResponse}
\end{figure}

We asked scientists, specifically those who take part in hiring, if past or present U.S. policies on Visas have affected their ability to hire talented individuals. The normalized results (split into scientists and senior scientists) is shown in Figure \ref{Q59_scientists_noResponse}. Nearly half of scientists (48.8\%) reported that past or present US policies on Visas had some negative effect on their hiring ability. More senior scientists (+10.8\%) reported a negative effect compared to scientists. We saw a similar situation with faculty, where more senior members reported negative effects. No scientists reported a positive effect on hiring, and only 4.3\% of senior scientists reported some level of positive impact on hiring. Twenty-five percent of scientists responded with ``not applicable'', 9\% higher than senior scientists. 

\begin{figure}[H]
    \includegraphics[scale=0.83]{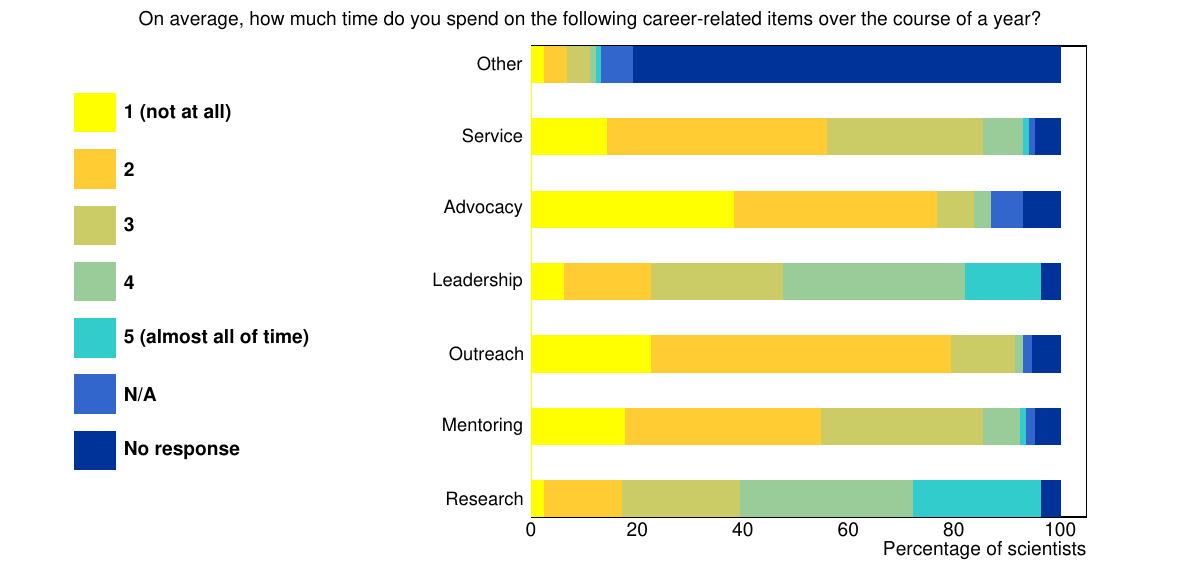}
    \caption{We asked all scientists about their yearly time commitments to various career-related items. Nearly 97\% of the scientists surveyed answered at least one part of this question.}
    \label{Q54_overall}
\end{figure}

We asked about how scientists spend their time on common career-related items over the course of a year -- Figure \ref{Q54_overall} shows the normalized responses to all parts of the question. Around 38\% of all scientists reported spending almost all of their time on at least one item, a larger proportion (+19.4\%) compared to faculty. Only 3.3\% of scientists surveyed reported spending almost all of their time on two or more items, a slightly smaller amount (-1.7\%) compared to the faculty we surveyed. Scientists reported spending the majority of their time on research and leadership: 56.8\% of all scientists reported a 4 or higher on research, and 48.6\% reported a 4 or higher on leadership. Moreover, 90.2\% of all scientists surveyed reported having some leadership responsibilities (meaning they indicated a 2 or higher for the leadership category). Similar to the faculty version of this question, we assumed that scientists who report a 4 or 5 on an item spend at least 50\% of their time on that particular item. Using this assumption along with how many hours per week scientists reported they spend working, we calculated the minimum hours per week that scientists commit to research and leadership: 
\begin{itemize}
    \item Forty-eight percent of scientists surveyed commit at least $21.2\pm7.0$ hours per week to research.
    \item Forty-three percent of scientist respondents commit at least $22.2\pm6.8$ hours per week to leadership. 
\end{itemize}

When broken down by Snowmass frontiers, we found that scientists in the Computational frontier reported more time than average to research. Scientists in the Cosmic frontier reported the most time spent on mentoring compared to all other Snowmass frontiers; they also reported more time than average to leadership responsibilities. Scientists in the Rare Processes frontier reported spending more time than average to research and advocacy. Compared to all other Snowmass frontiers, scientists in the Energy frontier reported the least amount of time to advocacy; similarly, a larger proportion of scientists in the Computational frontier put ``not applicable'' to advocacy efforts. Finally, scientists in the Neutrino frontier reported the most time spent on outreach among all Snowmass frontiers, and they reported more time than average to research, leadership, mentoring, and advocacy.

Scientists reported committing similar amounts of time to mentoring and service work: Thirty-eight percent of all scientists surveyed reported a 3 or higher in both the mentoring and service work categories. Broken down by gender, female scientists reported spending more time on service (+20.5\%) compared to male scientists. Scientists reported spending less time on mentoring compared to faculty: 70.3\% of faculty reported a ``3'' or higher time commitment on mentoring, nearly twice when compared to the scientists we surveyed (-31.5\%). Very few scientists reported committing time in the ``other'' category, but some career-related items reported in that category included teaching, project management, or other administrative tasks (e.g., paperwork or applying for grants).

\begin{figure}[H]
    \includegraphics[scale=0.46]{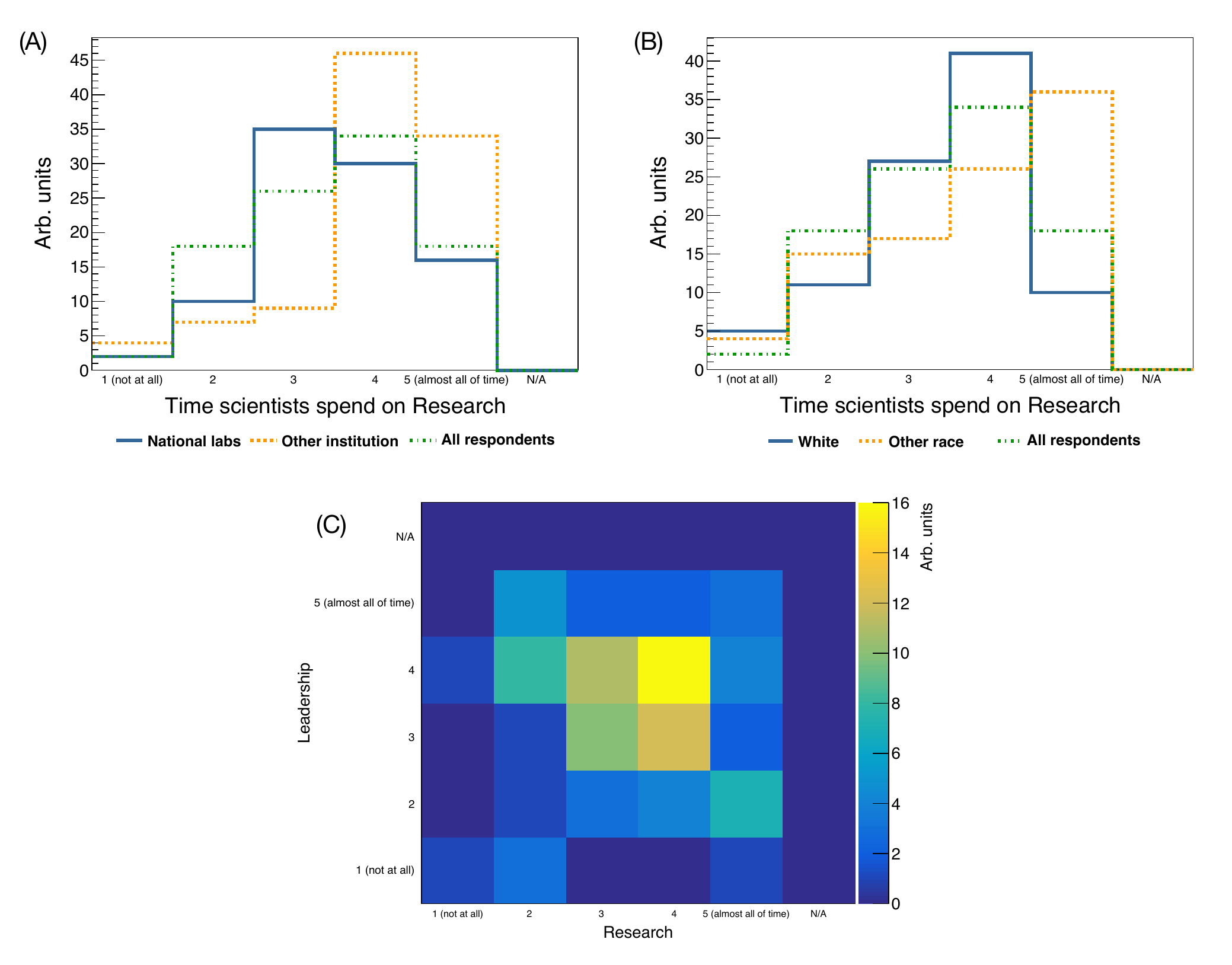}
    \caption{We asked all scientists about their yearly time commitment to research. We broke down the responses by (A) primary workplace and (B) race. (C) We also compared responses about time commitments to research and leadership.}
    \label{Q54_Research_combined}
\end{figure}

We considered scientists' responses about their research commitments more closely in Figure \ref{Q54_Research_combined}. Broken down by primary workplace (Fig. \ref{Q54_Research_combined}A), scientists outside of national labs reported more time commitment to research (+31.6\%) compared to national lab scientists. Broken down by race (Fig. \ref{Q54_Research_combined}B), 61.3\% of scientists in other racial groups reported a 4 or higher on research, indicating that they commit more time to research compared to White scientists (-8.6\%). Finally, we compared scientists' responses about research to their responses about leadership (Fig. \ref{Q54_Research_combined}C); fifty-three percent of scientists reported committing more time to leadership compared to research, 10.4\% more than scientists who reported the opposite. 

\begin{figure}[H]
    \includegraphics[scale=0.54]{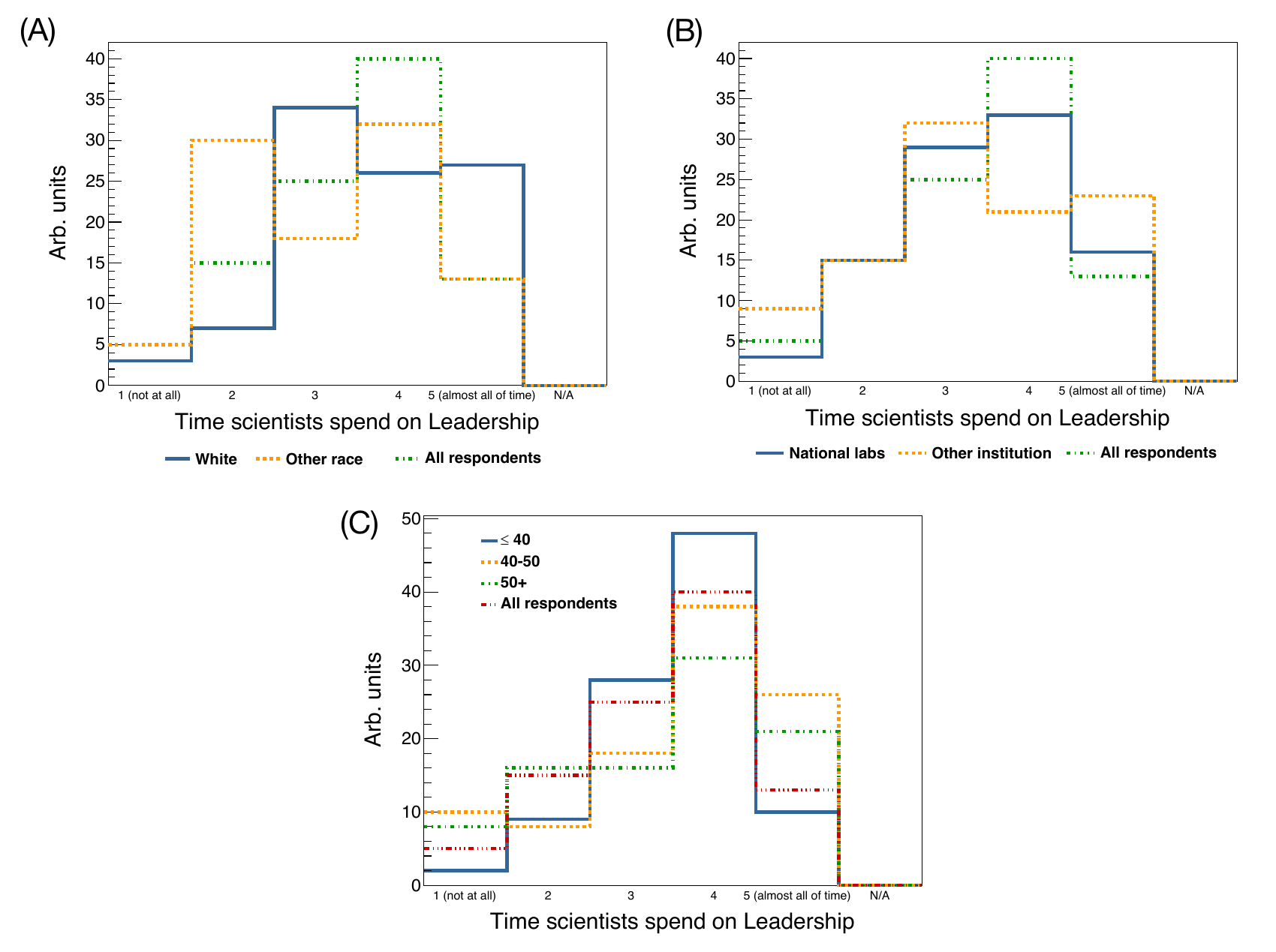}
    \caption{We asked all scientists about their yearly time commitment to leadership. We broke down the responses by (A) race, (B) primary workplace, and (C) age.}
    \label{Q54_Leadership_combined}
\end{figure}

We looked more closely at scientists' reported time commitment to leadership responsibilities. Broken down by race (Fig. \ref{Q54_Leadership_combined}A), 76.4\% of white scientists surveyed reported a 3 or higher in the leadership category, indicating that they commit more time to leadership compared to scientists part of other racial groups (-11.8\%). Broken down by primary workplace (Fig. \ref{Q54_Leadership_combined}B), 7.6\% more scientists outside of national lab institutions reported having little to no leadership time commitments when compared to national lab scientists. Finally, when we broke down the responses by age (Fig. \ref{Q54_Leadership_combined}C), we found that a quarter of scientists older than 50 years old reported little to no time commitment to leadership responsibilities; around 20\% of scientists younger than 50 years old made the same report. Looking at the other extreme, a little less than half of scientists younger than 40 years old reported a 4 or higher on leadership; 60\% of scientists between 40 and 50 years old made the same report. Comparing these proportions to the 43\% of scientists older than 50 years old who reported a 4 or higher on leadership, we saw that older scientists reported less overall time commitment to leadership, although scientists between 40 and 50 years old reported more overall time commitment compared to scientists younger than 40 years old.  

\begin{figure}[H]
    \includegraphics[scale=0.83]{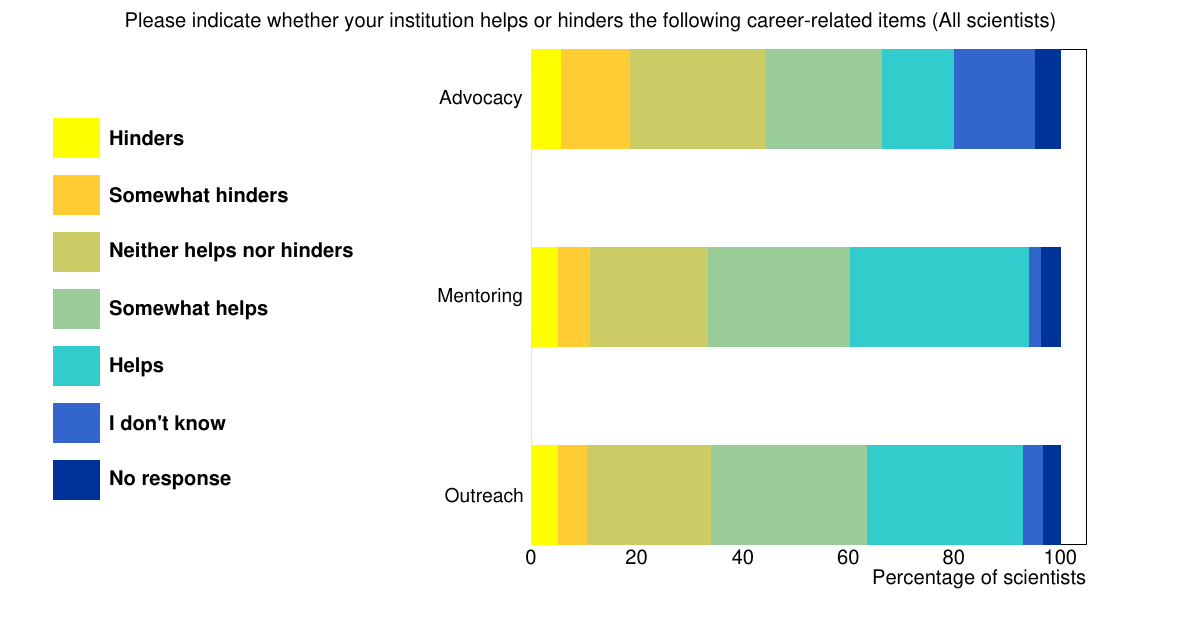}
    \caption{We asked all scientists about their institution's effect on outreach, mentoring, and advocacy. Around 96\% of scientists answered at least one part of this question.}
    \label{Q57}
\end{figure}

We asked scientists whether their institutions help or hinder with outreach, mentoring, and advocacy; the normalized responses from all scientists are shown in Figure \ref{Q57}. Outreach and mentoring received similar responses -- around 10\% of scientists surveyed reported that their institution hinders outreach and mentoring on some level. Compared to faculty who were asked the same questions, more scientists reported that their institution hinders outreach (+4.4\%) and mentoring (+6.2\%) on some level. Nearly nineteen percent of scientists surveyed reported that their institution hinders advocacy on some level. Compared to male scientists' responses, 24.3\% more female scientists reported that their institution hinders advocacy; broken down by race, 5\% more scientists of other racial groups reported this compared to white scientists. Finally, 15.3\% of scientists surveyed reported that they don't know whether their institution helps or hinders advocacy efforts. Compared to female scientists' responses, 8.4\% more male scientists indicated that they did not know. When broken down by Snowmass frontiers, more scientists in the Computational frontier reported that their institution hinders outreach efforts at some level; conversely, more scientists in the Cosmic, Accelerator, and Instrumentation frontiers reported that their institutions help with outreach efforts. A larger proportion of scientists in the Accelerator frontier skipped the portion of the question asking about advocacy; furthermore, a majority of the ``I don't know'' responses originated from scientists in the Accelerator and Computational frontiers. Compared to scientists in other Snowmass frontiers, a larger proportion of scientists in the Neutrino frontier reported that their institutions hinder advocacy efforts on some level.

We asked scientists if they have experience outside of HEPA, to which 90.7\% of scientists responded. The majority of scientists surveyed (63.4\%) reported that they have always been at a national lab, and 27.3\% reported that they have experience outside of HEPA. Broken down by gender, 27.1\% of male scientists reported having experience outside of HEPA, a larger proportion compared to the female scientists we surveyed (-8.6\%). 

We asked scientists whether they have thought about leaving the lab, to which 92.3\% of scientists surveyed provided a response. The majority of scientists (62.3\%) reported that they have thought about leaving the lab. When broken down by gender, more female scientists (+6.4\%) have thought about leaving compared to male scientists. When split by seniority, 71.4\% of scientists surveyed reported that they have thought about leaving, much more than senior scientists (-16.9\%). On the other hand, 30\% of scientists surveyed reported that they have not thought about leaving the lab. When we considered just the scientists who have not thought about leaving, 2.6\% more male scientists have not thought about leaving when compared to female scientists. When split by seniority, 36.4\% of senior scientists surveyed reported that they have not thought about leaving, much more than scientists surveyed (-13.8\%). 

\begin{figure}[H]
    \includegraphics[scale=0.66]{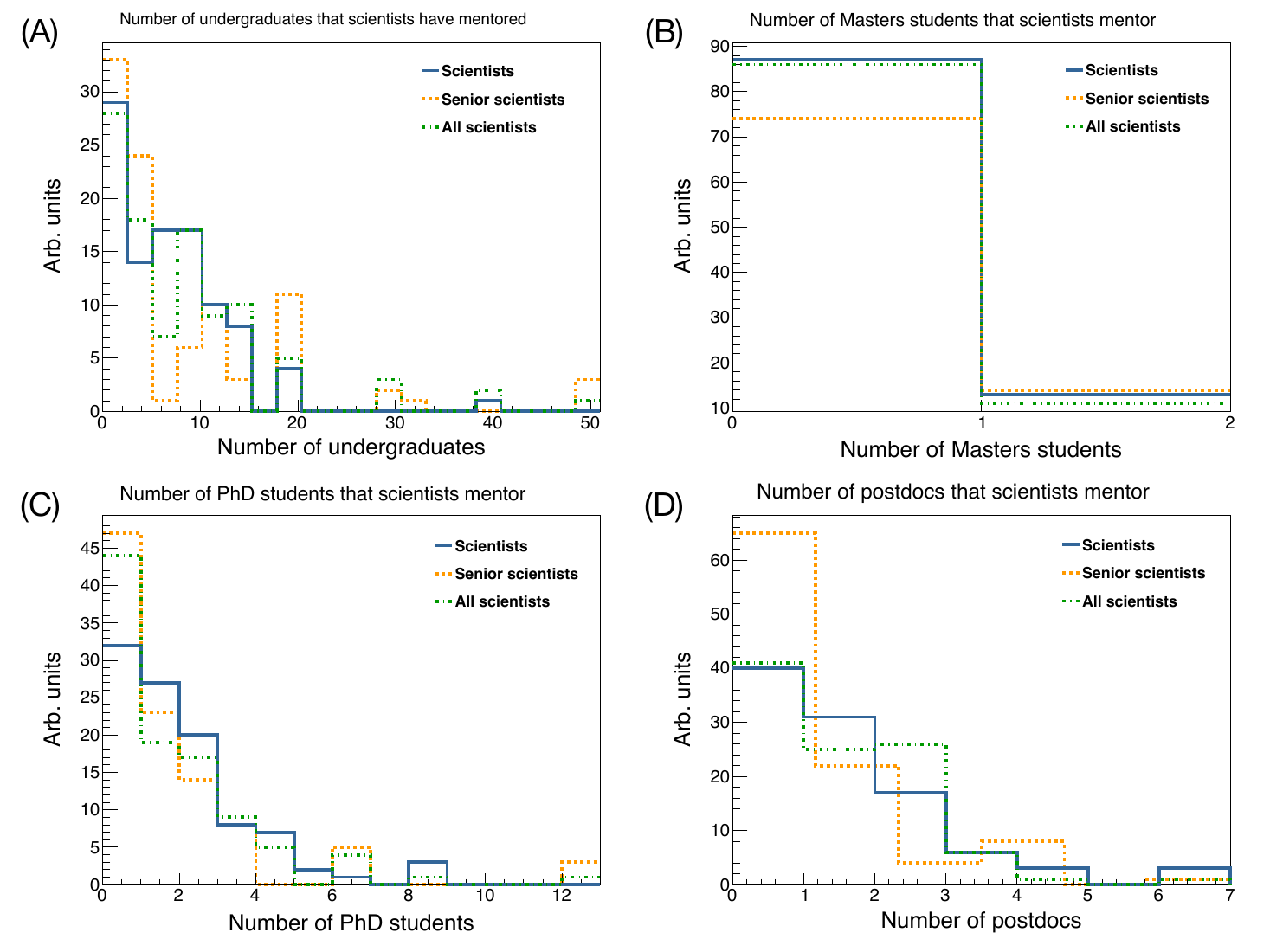}
    \caption{We asked scientists about (A) the total number of undergraduate students they have mentored, along with the number of (B) Master's students, (C) PhD students, and (D) Postdocs they currently mentor. We received responses about undergraduate students from 91.8\% of scientists; we received slightly less responses about Master's students (72.1\%), PhD students (84.2\%), and Postdocs (89.6\%).}
    \label{Q51_Q52_combined}
\end{figure}

We asked scientists how many graduate students and Postdocs they currently mentor, along with how many undergraduate students they have mentored. We converted the responses into normalized histograms found in Figure \ref{Q51_Q52_combined}. All scientists are currently mentoring an average of one PhD student, one postdoc, and no Master's students. When split by seniority, senior scientists reported mentoring more Master's students and Postdocs, while scientists reported mentoring more PhD students. The differences between scientists and senior scientists are much more apparent when it comes to undergraduate students: Not surprisingly, senior scientists reported they have mentored more with an average of 12.5$\pm$2.2 undergraduate students. Meanwhile, scientists reported they have mentored an average of 6.8$\pm$0.8 undergraduate students.

We also asked scientists about how many of the graduate students and Postdocs they mentor are employed. Of the 45.9\% of all scientists who responded about the Master's students they mentor, scientists reported 40.1$\pm$7.6 percent of the Master's students they mentor are employed, while senior scientists reported that 73.7$\pm$6.4 percent of the Master's students they mentor are employed. When we considered those who reported 0\% employment for the Master's students they mentor, we found that far more scientists (57.1\%) compared to senior scientists (16.7\%) made this report. Broken down by gender, 0\% employment for Master's students was reported by 37.5\% of the male scientists and 62.5\% of the female scientists who responded.

For PhD students and Postdocs, scientists and senior scientists reported similar levels of employment: For scientists, the average was around 65\% of the PhD students and Postdocs they mentor, while for senior scientists, that number was closer to 85\%. Nearly 27\% of scientists reported that 0\% of the PhD students or Postdocs they mentor are employed, much more than senior scientists (-19\%).

\subsubsection{Engineers and technicians}
\label{engineers_technicians_section}
Engineers and technicians comprised about 2\% of everyone surveyed. We received more responses from engineers than we did from technicians. A significant portion of the engineers and technicians did not provide demographic information. Nonetheless, engineers and technicians made up 1.4\% of all male respondents and 2.3\% of all female respondents; they also made up 2\% of the White survey respondents. A quarter of engineers and technicians reported their primary workplace as a university (either in or out of the U.S.), 30\% reported being at at national labs, and some reported being at government institutions. 

We asked engineers and technicians about the importance of the following career-related concerns in their job:
\begin{enumerate}
    \item Bureaucracy and administrative difficulties to conducting your work
    \item Losing your job/failing to advance
    \item Any other (specified) concerns 
\end{enumerate}
Overall, engineers and technicians rated the second concern as more important compared to the first. A small sample told us about specific concerns which they rated highly, e.g., funding. 

We also asked engineers and technicians whether they planned to switch jobs in the near future. The majority indicated that they probably or definitely will not switch jobs.

\subsubsection{Respondents not in academia}
As we mentioned at the beginning of this section, 6.9\% of survey respondents reported that they are not currently in academia. The majority of non-academics (71.4\%) are men, and 5.7\% are women; said another way, 8.7\% of male respondents and 2.3\% of female respondents reported that they are not currently in academia. Sixty-one percent of non-academics indicated they are white, and 14.3\% indicated they are in another racial group. Keeping in mind that we asked for respondents' primary workplaces prior to the COVID-19 pandemic, 32.9\% of non-academics reported their place workplace as a private industry institutions, 28.6\% reported an academic institution (i.e., university or national laboratory anywhere in the world), and 24.3\% reported some other type of institutions (e.g., government institution, private foundation). We received slightly more ``young'' non-academics based on the ages provided by 77.1\% of non-academics: 40\% told us they are 40 years old or younger, while 37.1\% are older than 40 years old. 

We first asked non-academics at what capacity they were most recently involved with or in HEPA, and the majority (98.6\%) responded. About 31.5\% of non-academics were graduate students (including Master's and PhD students) before leaving academia, 21.4\% were faculty or scientists, and 17.1\% were Postdocs. We also received some former engineers, technicians, and undergraduate students, along with some self-identified positions. For the 20\% of non-academics in the latter category, the majority were either retired or industry scientists, but this category also included people involved in science communication. Overall, about half of non-academics surveyed left HEPA as early career scientists, and the other half left in more senior positions.  

\begin{figure}[H]
    \includegraphics[scale=0.82]{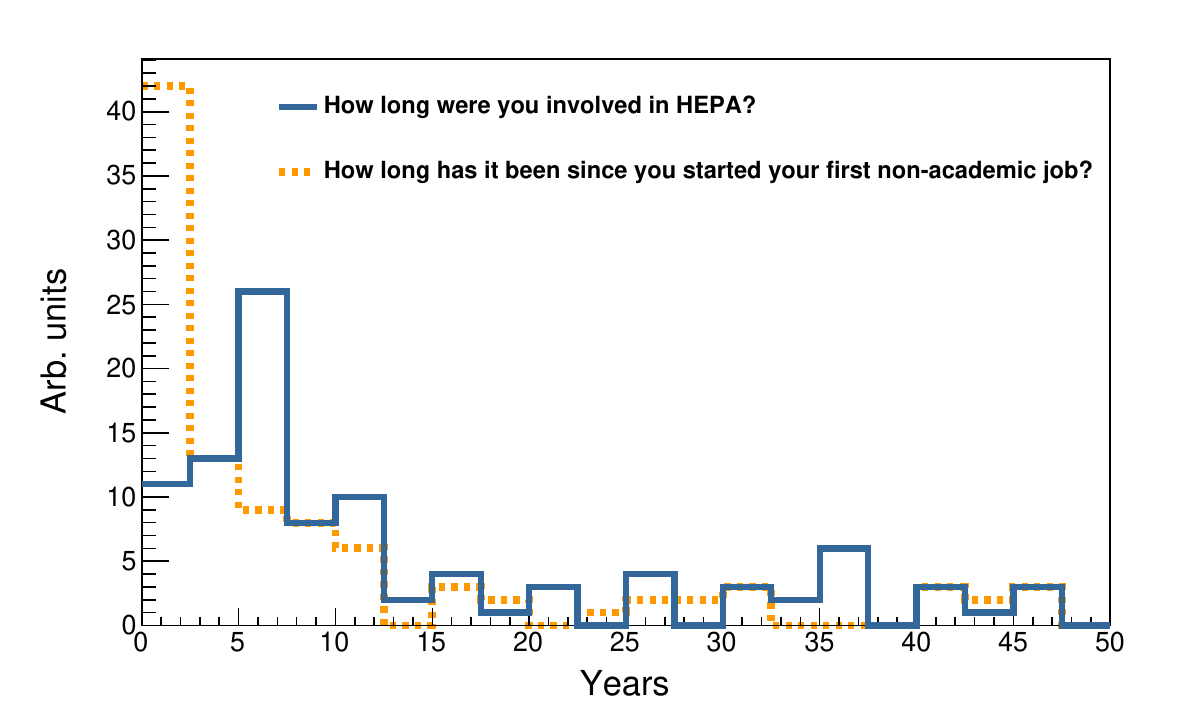}
    \caption{We asked non-academic respondents about how long they were involved in HEPA (in which 94.3\% of non-academics answered) and how long ago they started their first non-academic job (in which 88.6\% responded).}
    \label{Q67_overall}
\end{figure}

We asked non-academics about their length of time in HEPA and when they left. The normalized results are shown in Figure \ref{Q67_overall}. Based on the answers from 94.3\% of the group, non-academic respondents spent an average of $13.5\pm1.6$ years in HEPA. Slightly less people (88.6\%) responded to our second inquiry, but based on their answers, non-academic respondents began their first non-academic job $10.7\pm1.9$ years ago. This average might be inflated due to our retired non-academic respondents, since nearly half of non-academics started their first non-academic job within 5 years of taking the survey. Since we asked for respondents to tell us their primary workplace prior to the COVID-19 pandemic, we looked specifically at non-academic respondents who left academia within the last two years: These people were primarily at universities -- mostly outside of the U.S. -- or some other (academic or non-academic) institution. 

When asked about whether they attempted to find a job in HEPA, 95.7\% of non-academics were split fairly evenly between ``yes'' and ``no''. We found that non-academics were split evenly when we broke down this question by gender and racial groups. However, when split by age, we found a difference: 53.8\% of non-academic respondents over 40 years old told us that they attempted to find a job in HEPA, 7.4\% higher than non-academic respondents 40 years old or younger. Furthermore, when we looked at respondents who got their first non-academic jobs 5 or more years ago, 53.6\% told us that they attempted to find an academic job. For respondents who left within 5 years of taking the survey, closer to half told us they attempted to find an academic job. 

\begin{figure}[H]
    \includegraphics[scale=0.83]{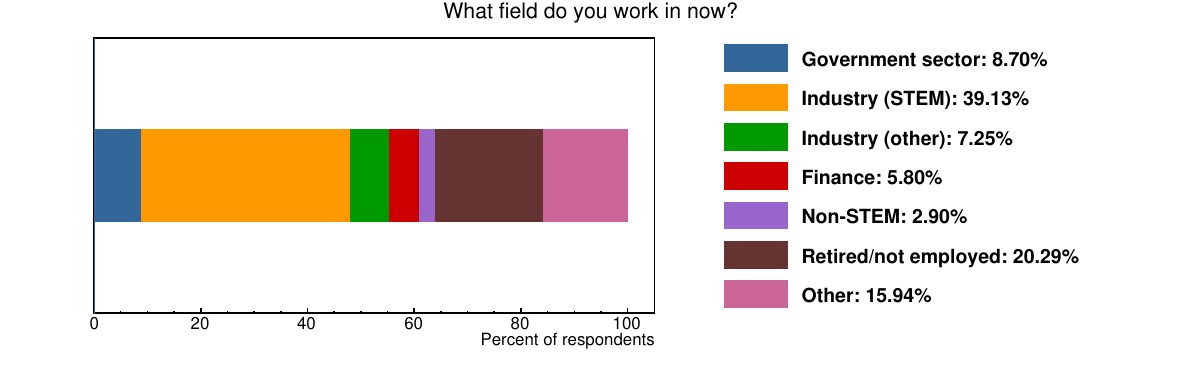}
    \caption{We asked non-academic respondents about what fields they currently work in. The majority (98.6\%) responded to this question.}
    \label{Q68_overall}
\end{figure}

We asked about what field non-academic respondents currently work in; the responses to this question are shown in Figure \ref{Q68_overall}. The majority of non-academics (45.7\%) are currently in industry; twenty percent are retired or not currently employed; and 15.7\% responded with ``other'' and self-identified their fields. Some of these respondents include semi-retired individuals, people working at non-profits, and people working in science communication. 

Ninety percent of non-academics told us about what resources they found useful in obtaining their job outside of HEPA; respondents could multi-select from a list of nine answer options, including text boxes to enter specific websites or other resources (see Appendix \ref{appendix_surveyQs} for more details). On average, respondents selected two resources. The most widely selected resources picked by non-academics were networking/LinkedIn (58.6\%) and coworkers (32.9\%). Other commonly picked resources were internships/training programs and self-identified resources including invited talks, word-of-mouth, conferences, and luck. LinkedIn was not the only online resource used by non-academics; others include Coursera, \url{angel.io}, Indeed, \url{leercode.com}, and \url{algoexpert.com}. When we looked for correlations between pairs of resources, we found that the top two selected resources (networking/LinkedIn and coworkers) were a highly correlated pair, and they each correlated highly with meetups/hackathons. On its own, networking/LinkedIn correlated highly with internships/training programs, while coworkers correlated highly with seminars/webinars, books, and career fairs. 

\begin{figure}[H]
    \includegraphics[scale=0.85]{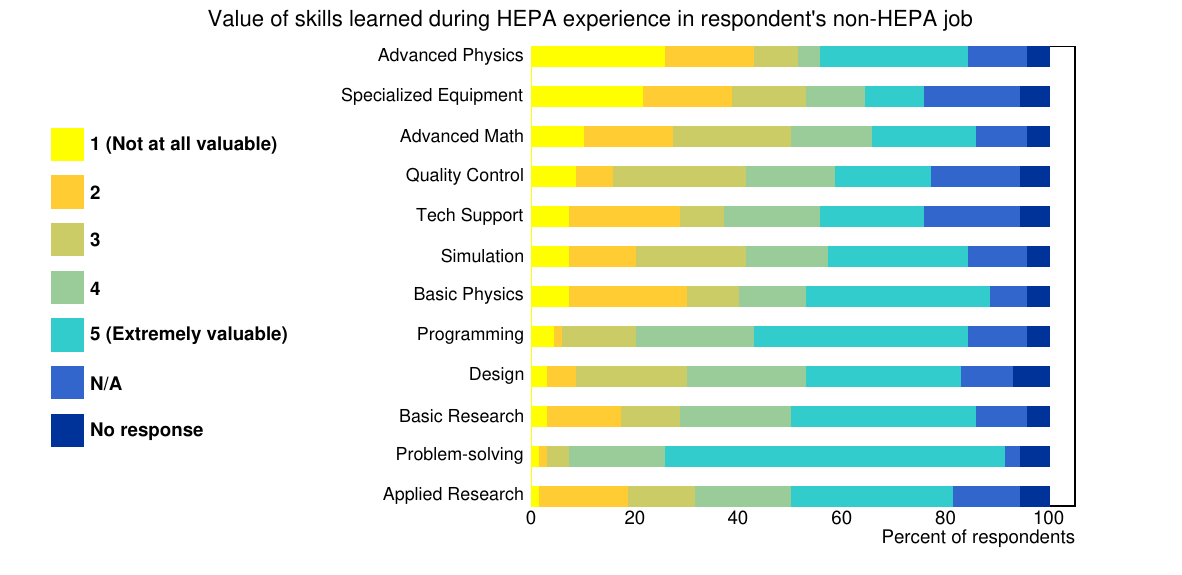}
    \caption{We asked non-academic respondents about the value of skills they learned in HEPA to their current job. Nearly 96\% of non-academics answered at least one part of this question.}
    \label{Q71_overall}
\end{figure}

We asked non-academics about how valuable the skills learned during their HEPA experience are to them in their current job; the normalized responses to this question are shown in Figure \ref{Q71_overall}. The most valuable skills were determined by those which received the most 4's and 5's by non-academic respondents, while the least valuable skills are those which received the most 1's and 2's. The most valuable skills include problem-solving (84.3\%), programming (64.3\%), basic research (57.1\%), and design (52.9\%). Meanwhile, skills such as basic physics, tech support, and advanced math received more mixed results, with 27-30\% indicating little to no value in their current job. To be fair, nearly half of non-academic respondents find value in basic physics, while closer to 37\% of respondents find value in the latter two skills. The least valuable skills include specialized equipment (38.6\%) and advanced physics (42.9\%). Moreover, a couple of skills received a significant number of ``not applicable'' responses: Both tech support and specialized equipment were not applicable to 18.6\% of non-academics, and quality control was not applicable to 17.1\% of non-academics. 

We compared our results to a list of the most commonly used skills by 2015/2016 Physics PhDs in physics, engineering, and computer science positions as found by the American Institute of Physics (AIP) Statistical Research Center \cite{AIP_Statistics_Plot}. The list was formed by looking at skills in which over 50\% of AIP survey respondents reported daily use. These skills include solving technical problems, programming, design and development, and advanced math -- a list that mirrors our findings. If we only look at AIP results for commonly used skills in physics and engineering positions (i.e., excluding computer science positions), the list expands to include basic physics principles, applied research, and specialized equipment. 

\subsection{Workplace Culture}

We asked respondents to provide their thoughts on the work culture of HEPA, in particular the time commitment, competitiveness and workplace location preferences. We also asked about support from colleagues and supervisors. Due to limited statistics, we limit this section to only PhD students, Postdocs, and faculty and scientists at all levels. 

We found that across all of these career stages, a majority of respondents reported working more than $40$ hours a week on average. PhD students were the least likely to work overtime ($59.0\%$), followed by Postdocs ($76.8\%$), lab scientists ($78.1\%$), senior lab scientists ($81.2\%$), tenure-track faculty ($84.5\%$), and tenured faculty ($90.1\%$). Some respondents reported working more than $60$ hours a week on average, ranging from $10.4\%$ (PhD students) to $19.0\%$ (tenure-track faculty). When asked whether they are able to maintain a healthy work-life balance, respondents in general disagreed. On a five-point likert scale ($1 =$ strongly agree), faculty and scientists reported the best balance ($3.14 \pm 0.08$ and $3.14 \pm 0.09$, respectively), followed by PhD students ($3.32 \pm 0.10$) and Postdocs ($3.50 \pm 0.09$). 

We asked respondents to grade on a five-point likert scale the level of competitiveness at their primary workplace ($1 =$ not competitive, $5 =$ very competitive) as well as whether that competitiveness was healthy or unhealthy ($1=$ very unhealthy, $5=$ very healthy). PhD students reported the lowest score for competitiveness (average = $2.66 \pm 0.10$). Postdocs and faculty reported higher competitiveness scores, with averages of $3.37 \pm 0.09$ and $3.39 \pm 0.07$, respectively. Scientists reported the highest competitiveness score: $3.72 \pm 0.08$. The four groups reported similar scores for how healthy the competition was, with PhD students reporting the healthiest competition ($3.65 \pm 0.10$), followed by faculty ($3.50 \pm 0.07$), scientists ($3.38 \pm 0.08$) and Postdocs ($3.37 \pm 0.09$).

PhD students reported the highest levels of satisfaction with support from colleagues, with faculty reporting the lowest. On a five-point likert scale, with $1$ = very unsatisfied, the averages were $3.85 \pm 0.10$ (PhD students), $3.66 \pm 0.08$ (Postdocs), $3.50 \pm 0.08$ (scientists), and $3.42 \pm 0.08$ (faculty).

We asked PhD students and Postdocs to rate their advisor's availability on a five-point likert scale, with an average of $3.92$ ($1=$ never/almost never available). We also asked about their satisfaction level with their advisor's support ($1=$ very unsatisfied) and found an average of $3.81$. Advisors were asked to rate their availability for the students when they need a meeting. We found that advisors rate their availability higher ($4.36$) than students did. Advisors rated their belief of the sufficiency of their support at $3.87$ ($1=$ very insufficient), a similar score to the students' satisfaction level.

The COVID-19 pandemic changed working arrangements, and we inquired about current and preferred teleworking arrangements. This survey was conducted over several weeks, and we asked respondents to tell us their current teleworking arrangement. We found that $69.2\%$ of respondents spent a majority of their time working from home.  A large group, $44.5\%$ prefer to work $1-2$ days from home, while $26.7\%$ prefer their workplace, and $9.25\%$ would prefer to work only from home. 

\subsection{Diversity / Racism}

\begin{figure}[htbp]
  \centering
  	\includegraphics[width=150mm]{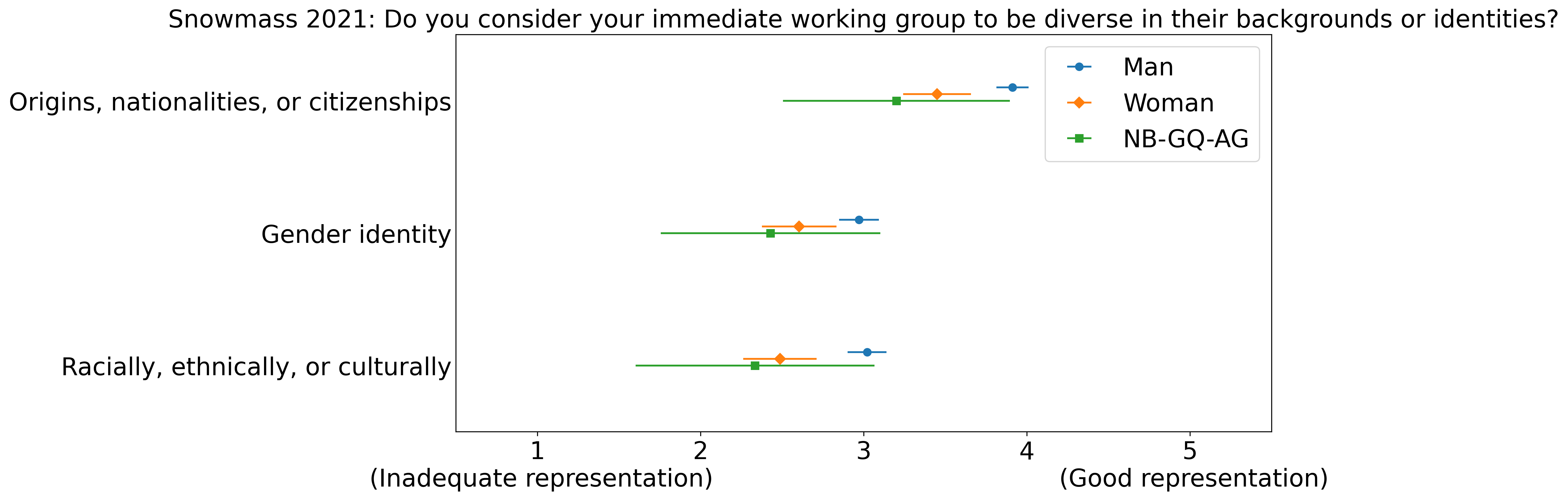}
  \caption{We asked respondents how diverse they believe their immediate working group is based on three sets of criteria.}
  \label{fig:divgroup}  
\end{figure}

We asked the respondents whether they consider their immediate working group to be diverse regarding different aspects: (1) race/ethnicity, (2) gender and (3) nationalities/citizenships. \autoref{fig:divgroup} shows the responses split by different gender groups; on average, women and non-binary/gender-queer/a-gender (NB-GQ-AG) seem to have lower responses values in their groups than men.         

All groups considered that their working groups have a better representation regarding origins, nationalities or citizenships. However, women and NB-GQ-AG considered that their groups do not have a good representation regarding gender and race/ethnicity or culture. In contrast, men had a more neutral opinion regarding these categories.       

\begin{figure}[htbp]
\centering
\subfloat[Experience of racism outside the work environment.\label{fig:rac1a}]{\includegraphics[width=0.5\textwidth]{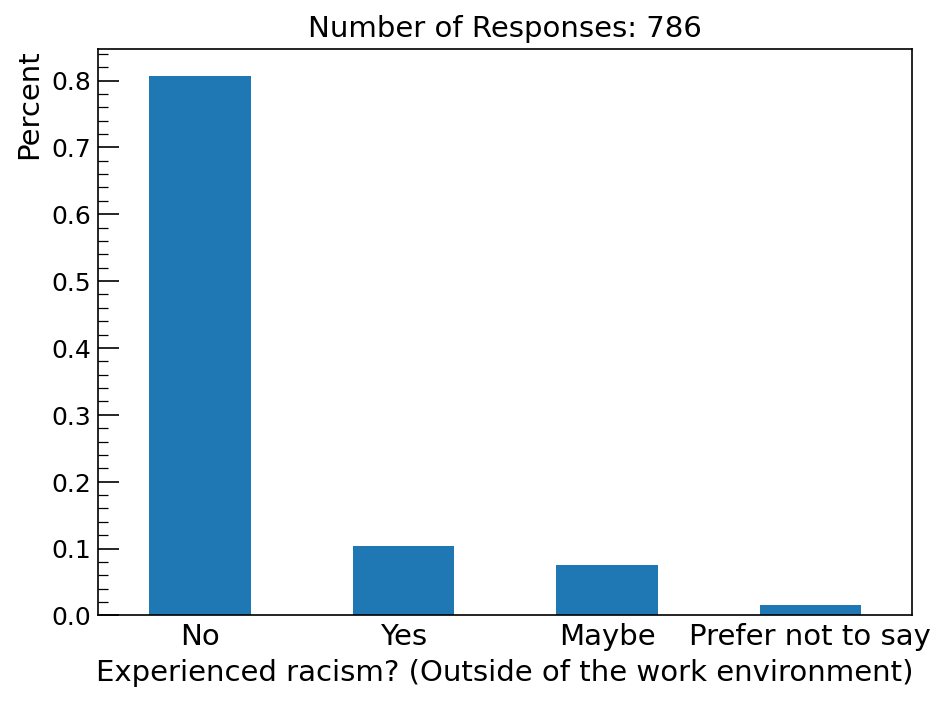}}\hfill
\subfloat[Witness of racism outside the work environment.\label{fig:rac1b}] {\includegraphics[width=0.5\textwidth]{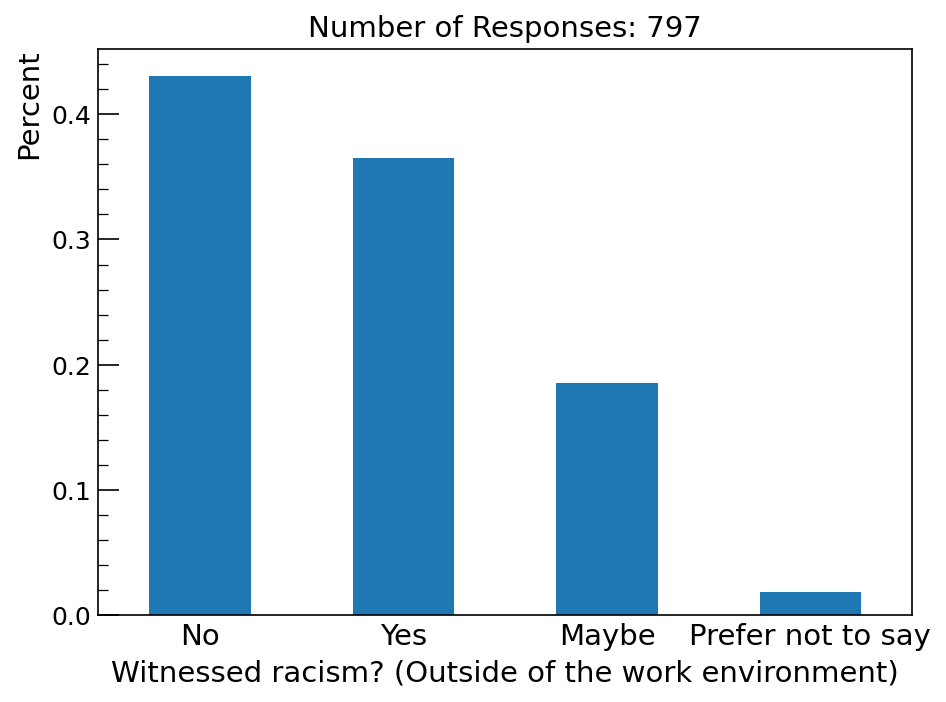}}\hfill
\subfloat[Experience of racism in the work environment.\label{fig:rac1c}] {\includegraphics[width=0.5\textwidth]{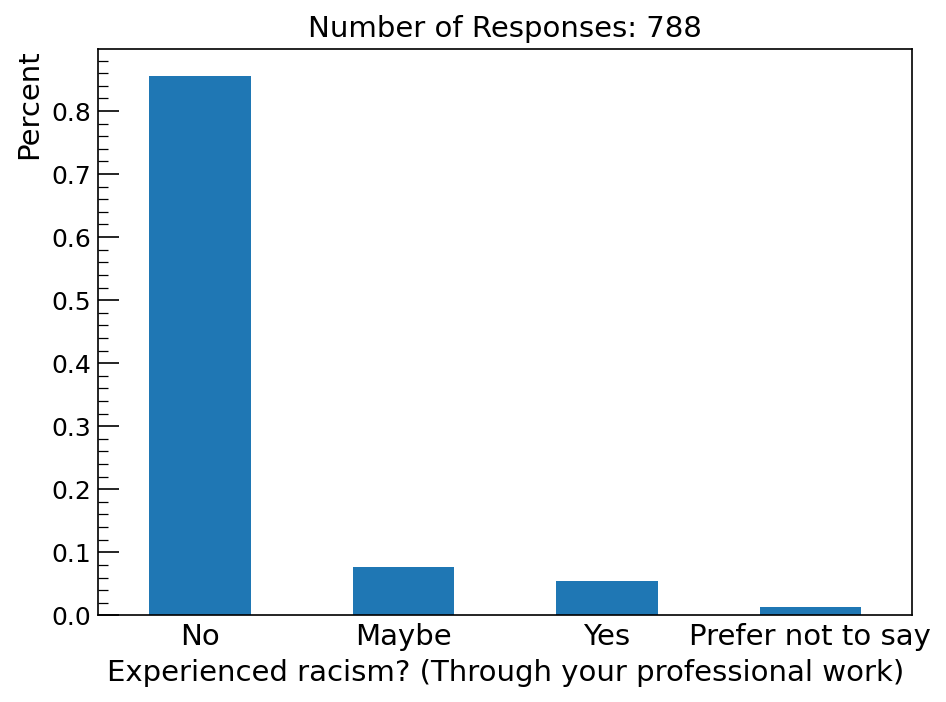}}\hfill
\subfloat[Witness of racism in the work environment.\label{fig:rac1d}]{\includegraphics[width=0.5\textwidth]{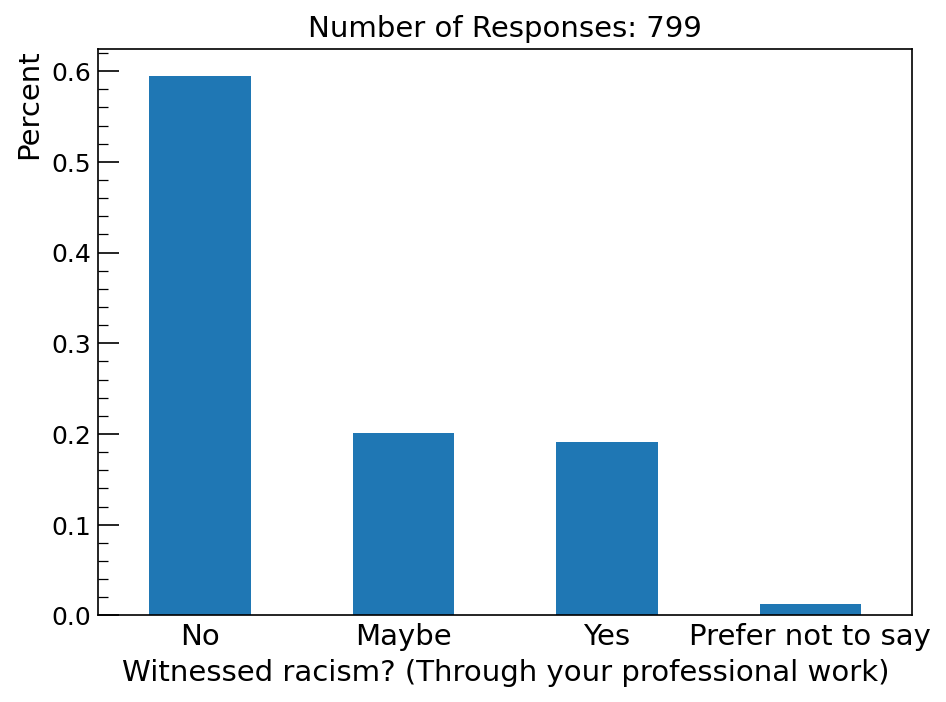}}
\caption{Respondents reports of whether they experienced or witnessed racism inside or outside the work environment. All questions were optional, and thus we see a different number of responses for each question.} \label{fig:rac}
\end{figure}

We asked respondents if they have experienced or witnessed racism either outside or inside their work environment. Most respondents said they did not experience racism either outside ($\sim$ 81\%) or inside ($\sim$ 86\%) their work environment. However, this percentage dropped to $\sim$ 43\% and $\sim$ 60\% for witnessed racism outside and inside their work environment, respectively (see \autoref{fig:rac}).  

We looked more closely at respondents who reported experiencing or witnessing racism -- including those who might have experienced or witnessed racism -- both inside and outside of the work environment:
\renewcommand{\arraystretch}{1.15}
\begin{table}[H]
\centering
\begin{tabular}{c|c|c|c}
Experienced or witnessed? & In or out of work environment? &  ``Yes'' & ``Maybe'' \\ \hline

\multirow{2}{*}{Experienced} & Out & 10\% & 8\% \\
  & In  & 5\%  & 8\% \\ 
  
  &&&\\

\multirow{2}{*}{Witnessed} & Out & 36\% & 19\% \\
  & In  & 19\%  & 20\% \\
  
\end{tabular}
\end{table}

We can see that respondents are more likely to witness racism rather than experience it themselves, occurring more frequently outside their work environment.

\subsection{Caregiving Responsibilities}
\label{sec:caregiving}

We asked all respondents whether they have caregiving responsibilities, including caring for children, seniors, or a person with a disability or other medical condition. Of the 77.8\% of respondents who answered this question, 28.3\% indicated that they have some form of caregiving responsibilities, and 68.2\% reported no caregiving responsibilities. The majority of caregivers (61.4\%) are based at a U.S. or international university, while 31.4\% of caregivers are based at a national laboratory (including labs outside of the U.S.). Broken down by gender, 29.3\% of all male respondents and 25.6\% of all female respondents reported that they are caregivers. Broken down by race, 69.5\% of caregivers are white while 24.3\% are apart of another racial group. Broken down by career stage, 51.6\% of caregivers are faculty, 25.6\% are scientists, and 12.1\% are early career scientists (i.e., PhD student or postdoc). Said another way, 36.3\% of all faculty respondents, 33.9\% of all scientist respondents, and 7\% of all early career respondents reported that they are caregivers. The majority of caregivers (86.1\%) were willing to answer additional questions about their caregiving responsibilities. 

\begin{figure}[H]
    \includegraphics[scale=0.48]{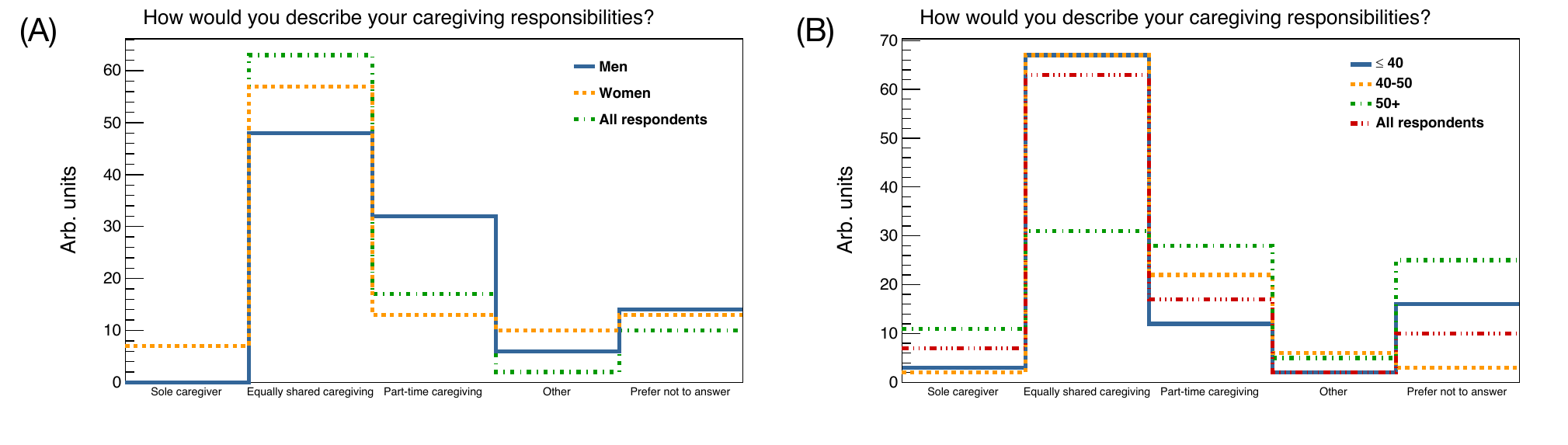}
    \caption{We asked caregivers to describe their caregiving responsibilities; the majority of caregivers (85.7\%) responded. Using demographic information, we broke down this answers by (A) gender and (B) age.}
    \label{Q153_combined}
\end{figure}

We asked caregivers how they describe their responsibilities: sole, equally shared, part-time, or some other (self-input) description. The majority of caregivers surveyed (56.5\%) described their caregiving responsibilities as equally shared, and 19.3\% described their responsibilities as part-time caregiving. Many of the caregivers in the ``other'' category described themselves as a primary caregiver. Broken down by gender (Fig. \ref{Q153_combined}A), more female caregivers reported they are sole or primary (``other'') caregivers compared to male caregivers. More male caregivers described their responsibilities as part-time caregiving. Broken down by age (Fig. \ref{Q153_combined}B), more caregivers older than 40 reported sole or part-time caregiving responsibilities compared to caregivers 40 or younger.

We asked caregivers about who they care for in which respondents could multi-select from the following options: children (under 18), seniors, and a person or people with a disability or medical condition. We received responses from 85.2\% of all caregivers. A small group of caregivers (6.7\%) reported that they care for a person with a disability or medical condition. Another small group of caregivers (6.7\%) reported that they have multiple groups of people to care for, usually involving children and an additional group. There were also a nonzero number of caregivers who reported caring for all three groups. The majority of caregivers with multiple responsibilities were tenured faculty or senior scientists, and the same proportion of male and female caregivers reported having caregiving responsibilities for multiple groups. 

The vast majority of caregivers (70\%) reported that they care for children under 18. Approximately the same proportion of male and female caregivers reported caring for children. Broken down by race, 78.1\% of white caregivers reported caring for children, much more than caregivers in other racial groups (-26.2\%). Caregivers with children are mostly based at universities (62.2\%), although one-third of caregivers with children reported being based at a national lab. Looking more closely at career stage, 56.4\% of caregivers with children are faculty on some level; the majority are tenured. The next largest group of caregivers (24.4\%) are scientists or senior scientists, and about 9\% of caregivers with children are early career scientists. The majority of caregivers who are tenure-track faculty, tenured faculty, scientists, and postdocs reported caring for children.

A smaller group of caregivers (16.6\%) reported that they care for seniors. Broken down by gender, only slightly more female caregivers (+1.7\%) reported caring for seniors compared to male caregivers. Broken down by race, 25.9\% of caregivers in other racial groups reported caring for seniors, significantly more than white caregivers (-11.7\%). Respondents who care for seniors are mostly based at universities (51\%), but some are at national labs (35.1\%). When we considered career stages, we found that 37.8\% of respondents who care for seniors are faculty (with the majority being tenured faculty), 24.3\% are scientists, and 21\% are early career scientists. 

We asked caregivers with children about their children's age groups and their current living situation: 
\begin{itemize}
    \item Thirty-five percent of caregivers have preschool-aged children. More female caregivers (+17.5\%) reported having preschool-aged children compared to male caregivers. More caregivers working at universities and male caregivers reported not living with their preschool-aged children.
    \item Thirty-nine percent of caregivers have 6-13 year old children. More male caregivers (+5.5\%) reported having 6-13 year old children compared to female caregivers. More caregivers older than 40, caregivers in other racial groups, and female caregivers reported part-time living or not living with their 6-13 year old children.
    \item Twenty-one percent of caregivers have 14-18 year old children. More male caregivers (+12.5\%) reported having 14-18 year old children compared to female caregivers. More white caregivers, caregivers working at national labs, and caregivers older than 40 reported living part-time with their 14-18 year old children. 
\end{itemize}

\begin{figure}[H]
    \includegraphics[scale=0.64]{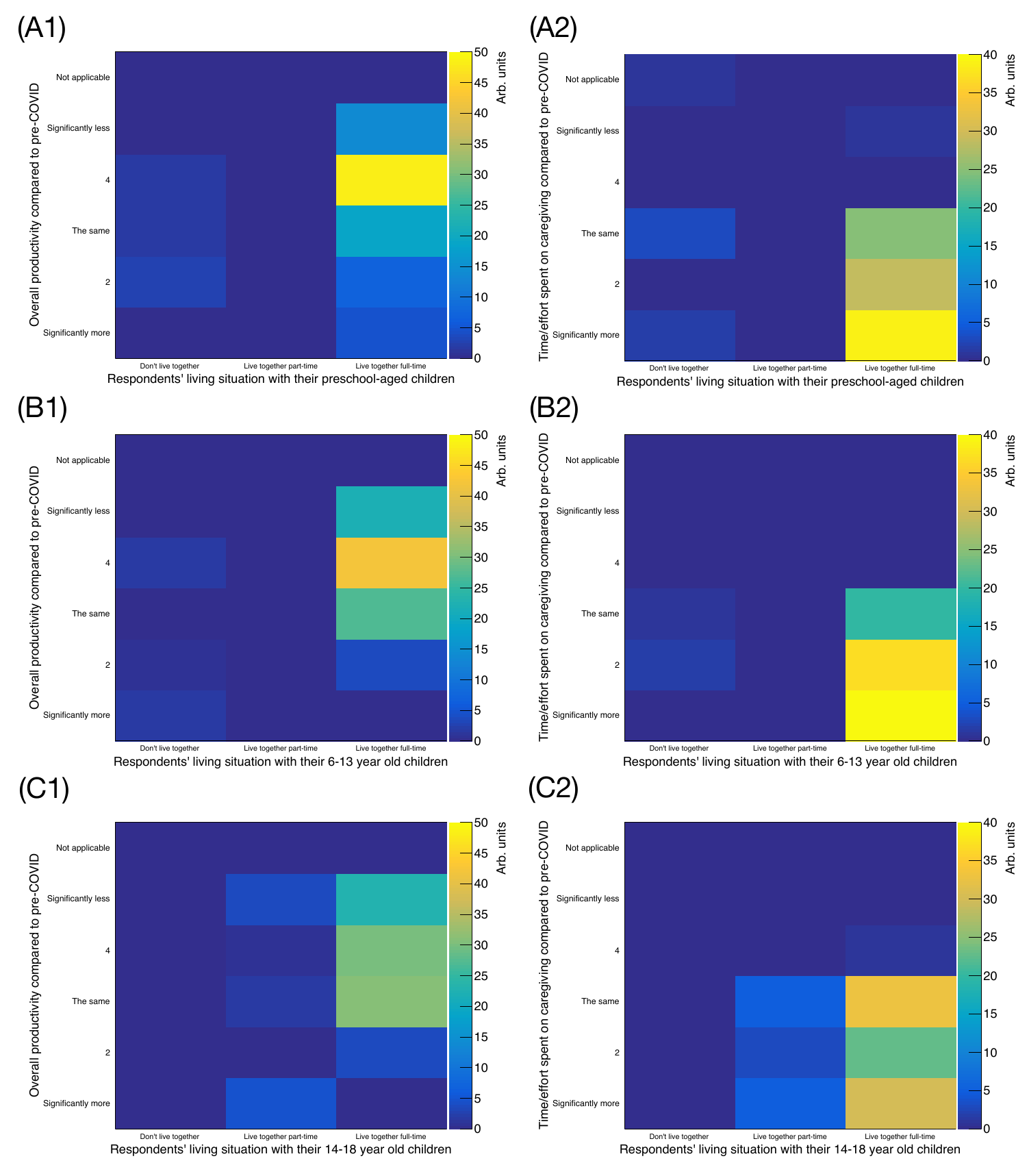}
    \caption{We combined reported living situations for caregivers with (A) preschool aged children, (B) 6-13 year old children, and (C) 14-18 year old children with caregivers' responses to other questions.  \textbf{Left}: Overall productivity compared to pre-COVID. \textbf{Right}: Time and effort spent on caregiving compared to pre-COVID.}
    \label{Q155_combined}
\end{figure}

We combined responses about children's age groups and living situations with caregivers' responses concerning their workload. We were specifically interested in two questions: Overall productivity compared to pre-COVID productivity levels, and ``At the time of this survey, how much time and effort do you spend on caregiving responsibilities compared to pre-COVID time and effort?". The normalized results are shown in Figure \ref{Q155_combined}.

The majority of all caregivers with children reported less overall productivity compared to pre-COVID (Figure \ref{Q155_combined}, left column). By splitting the responses based on the reported age groups of caregivers' children, a trend appeared: For the two older age groups, more caregivers with children in those groups report the same overall productivity compared to pre-COVID compared to caregivers with children under 6 years old. However, more overall caregivers with children in those older age groups reported significantly less productivity compared to pre-COVID (Figures \ref{Q155_combined}B1 and C1). The overall fraction of caregivers in all three groups is about the same: 61.5\% of caregivers with preschool-aged children reported less overall productivity compared to pre-COVID, while the fraction is around 59\% for caregivers with children older than 6 years of age.

The majority of caregivers with children younger than 14 reported spending more time and effort on caregiving compared to pre-COVID (Figures \ref{Q155_combined}A2 and B2). Caregivers with children in school seem to be affected more: 76.7\% of caregivers with 6-13 year old children reported more time and effort spent on caregiving compared to pre-COVID, 8.8\% more than caregivers with preschool-aged children. Caregivers with teenaged children were more mixed in their experiences, but the majority (53.2\%) reported more time and effort on caregiving during the pandemic (Fig. \ref{Q155_combined}C2). 

\begin{figure}[H]
    \includegraphics[scale=0.46]{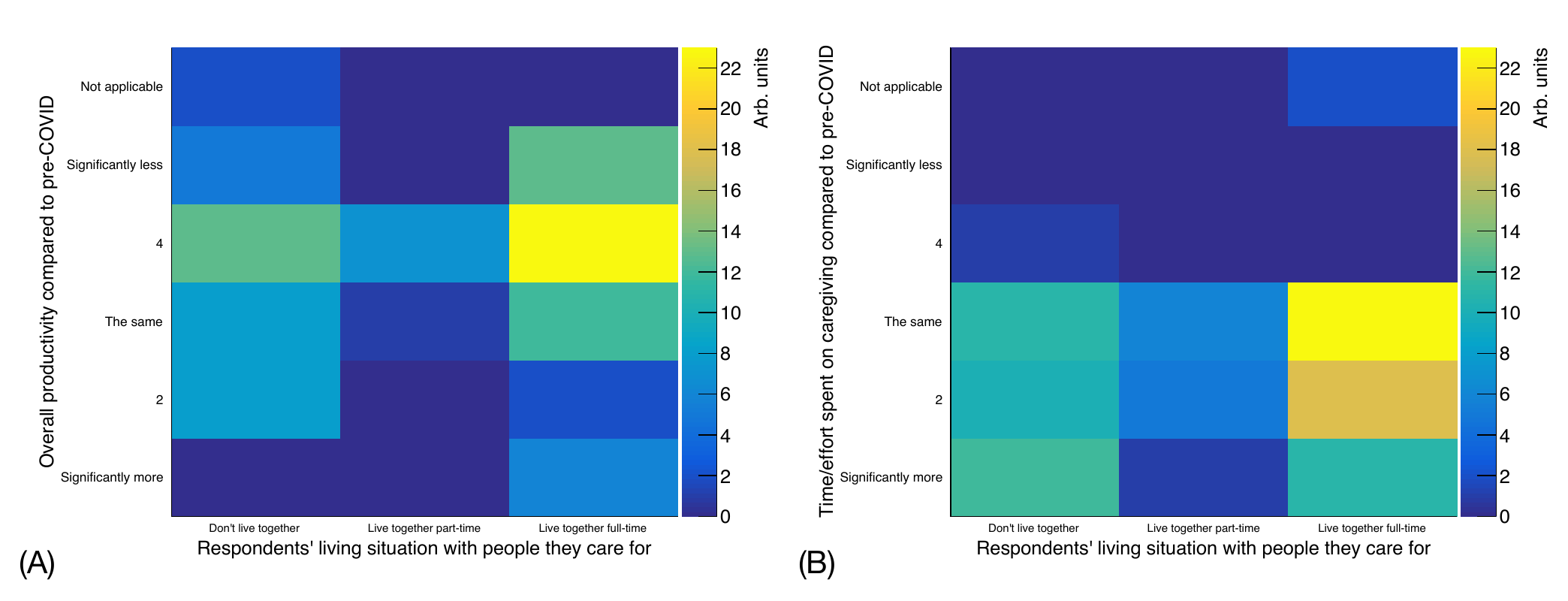}
    \caption{We combined reported living situations for caregivers of seniors or people with a disability or medical condition to other questions. \textbf{Left}: Overall productivity compared to pre-COVID. The majority of these caregivers (95.7\%) provided responses about their living situation and overall productivity. \textbf{Right}: Time and effort spent on caregiving compared to pre-COVID. A smaller majority of these caregivers (89.1\%) provided responses about their living situation and amount of time and effort.}
    \label{Q156_combined}
\end{figure}

We also asked about the living situation for caregivers of seniors or people with a disability or medical condition; the majority of these caregivers (56.5\%) live full-time with the person they care for. We performed the same type of correlation with COVID workload questions, and the normalized correlated results are shown in Figure \ref{Q156_combined}. The majority of caregivers of seniors or people with a disability or medical condition (56.5\%) reported more time and effort spent on caregiving compared to pre-COVID. Half of these caregivers reported less overall productivity compared to pre-COVID. 
\begin{figure}[H]
    \includegraphics[scale=0.83]{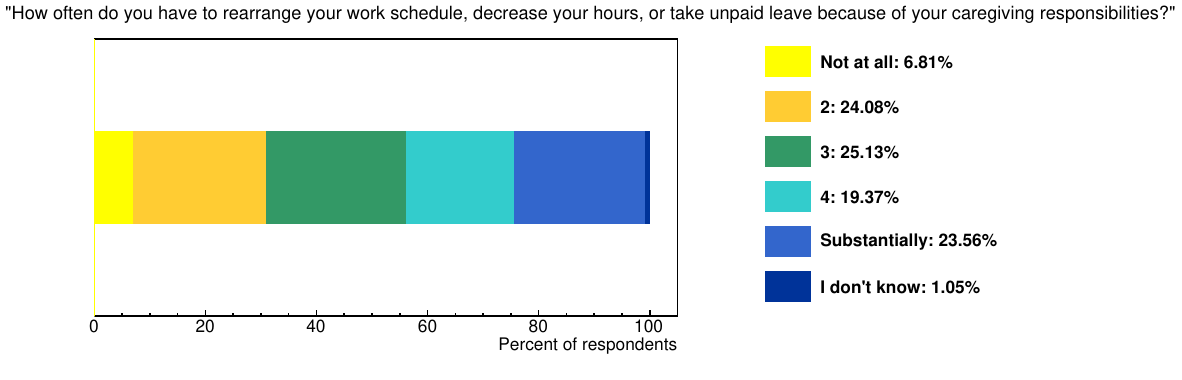}
    \caption{We asked caregivers about how their caregiving responsibilities negatively impact their work. The majority of caregivers surveyed (85.7\%) responded to this question.}
    \label{Q157_overall}
\end{figure}

We asked caregivers about how often they have to rearrange their work schedule, decrease their hours, or take unpaid leave due to their caregiving responsibilities. Looking at the overall results (Fig. \ref{Q157_overall}),  the experiences were mixed across the board. One group of caregivers (26.5\%) reported less or none of these effects due to their caregiving responsibilities; a larger group (36.8\%) reported more or substantial effects (i.e., answered with a 4 or ``substantial''). 

\begin{figure}[H]
    \includegraphics[scale=0.5]{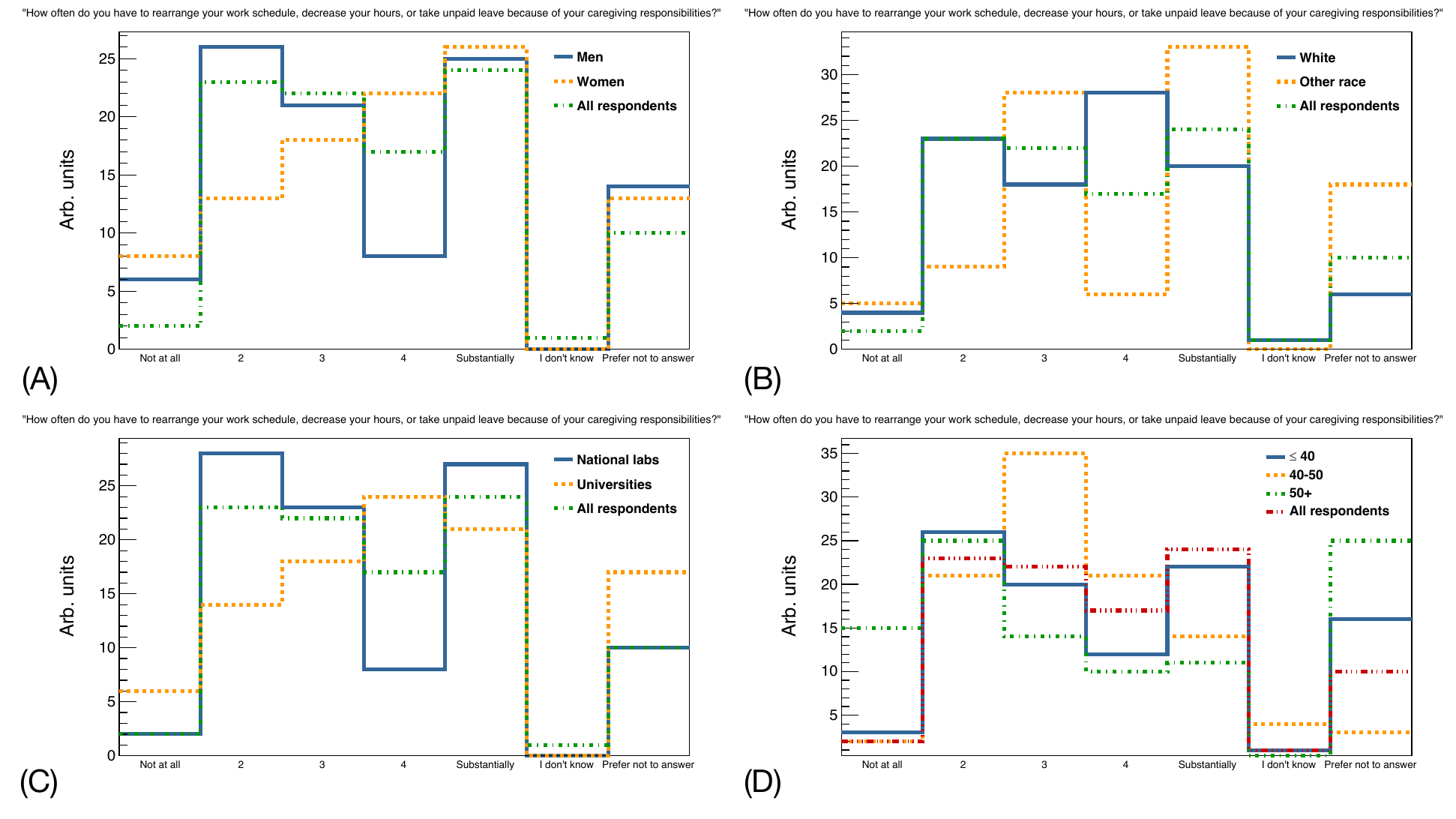}
    \caption{We asked caregivers about how their caregiving responsibilities negatively impact their work. We broke down the responses by (A) gender, (B) race, (C) primary workplace, and (D) age.}
    \label{Q157_combined}
\end{figure}

Broken down by gender (Figure \ref{Q157_combined}A), 48.8\% of all female caregivers indicated more or substantial effects, significantly more than male caregivers (-13.9\%). Broken down by primary workplace (Figure \ref{Q157_combined}C), 43.1\% of caregivers at universities indicated more or substantial effects, significantly more (+18.8\%) than caregivers at national labs reported. Focusing on the faculty and scientist responses, around 44\% of faculty and 26\% of scientists reported more or substantial effects. Broken down by age (Figure \ref{Q157_combined}D), more caregivers older than 50 years old reported little to no ill effects compared to caregivers younger than 50 years old.

\begin{figure}[H]
    \includegraphics[scale=0.6]{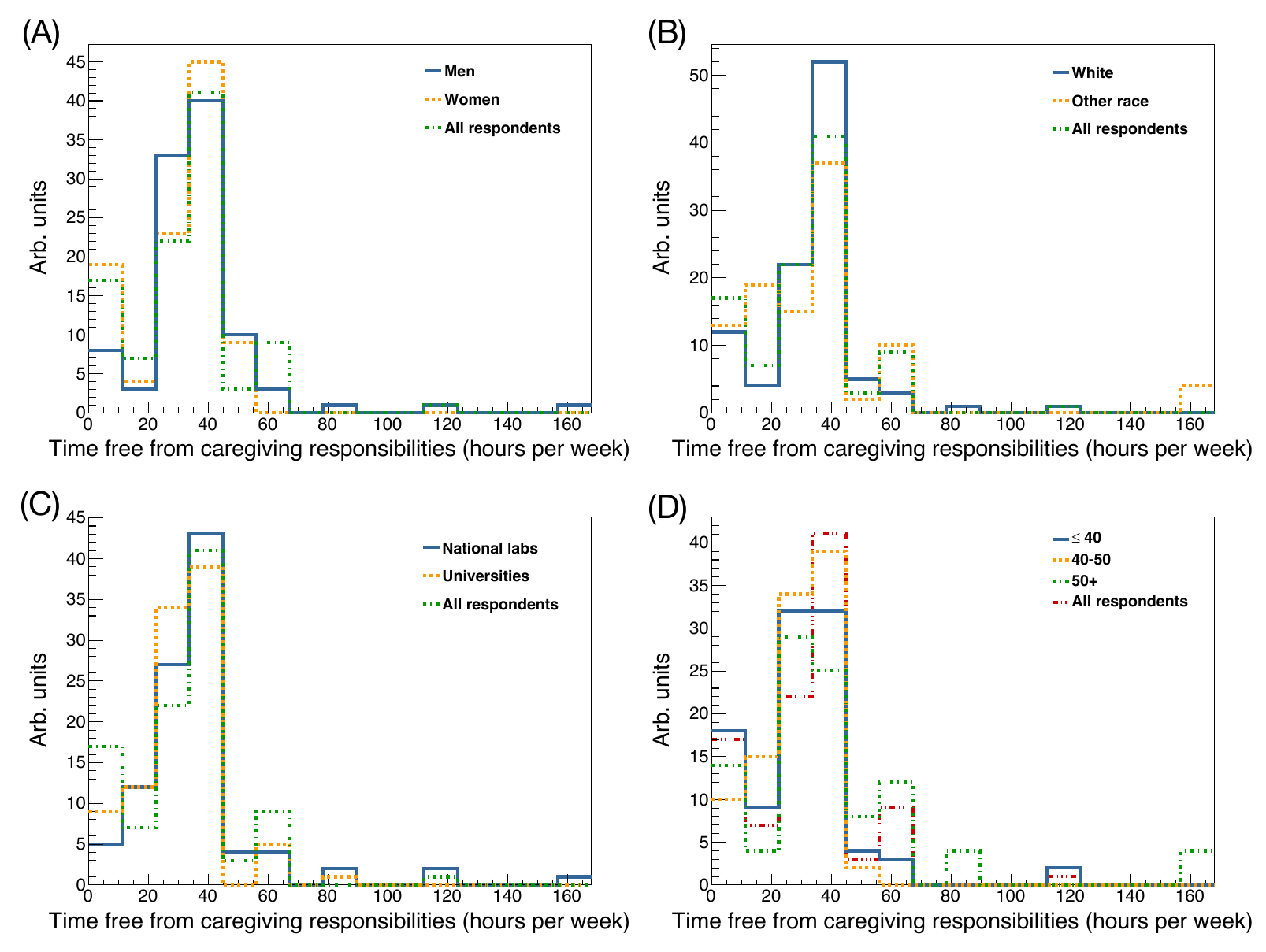}
    \caption{We asked caregivers to tell us approximately how many hours per week they are free from all caregiving responsibilities. The x-axis maximum was set to the total number of hours in a week. We broke down the responses by (A) gender, (B) race, (C) primary workplace, and (D) age.}
    \label{Q158_combined}
\end{figure}

We asked caregivers to tell us approximately how many hours per week they are free from all caregiving responsibilities, e.g., their children are at school or with a babysitter. We received responses from 71.7\% of all caregivers; on average, caregivers reported being free of all caregiving responsibilities for 33.7$\pm$1.6 hours per week. Broken down by gender, we received responses from 73.5\% of all male caregivers and 74.4\% of all female caregivers. We found that these groups reported being free of caregiving responsibilities for approximately the same amount of time per week, with male caregivers reporting an average of 34.2$\pm$1.9 hours per week, and female caregivers reporting an average of 31.3$\pm$2.3 hours per week (Fig. \ref{Q158_combined}A). Broken down by race, we received responses from 78.1\% of White caregivers and 61.1\% of caregivers in other racial groups. White caregivers reported an average of 32.6$\pm$1.6 hours per week free from all caregiving responsibilities; caregivers in other racial groups reported an average of 36.8$\pm$ 4.9 hours per week (Figure \ref{Q158_combined}B). 

Broken down by primary workplace, we received responses from 63.5\% of caregivers at universities and 71.4\% of caregivers at national labs (Fig. \ref{Q158_combined}C). Caregivers at national labs indicated more time free from caregiving responsibilities compared to their university counterparts; on average, caregivers at national labs reported 7.9$\pm$4.0 more free hours per week. Caregivers at other institutions reported less free hours per week compared to caregivers at either national labs or universities. Broken down by age, we received responses from 74.7\% of caregivers 40 years old or younger and 74.2\% of caregivers older than 40 years old (Fig. \ref{Q158_combined}D). Caregivers younger than 40 years old reported an average of 32.7$\pm$ 2.2 hours per week free from caregiving responsibilities, and caregivers older than 40 years old reported an average of 35.1$\pm$2.2 hours per week.

\begin{figure}[H]
    \includegraphics[scale=0.49]{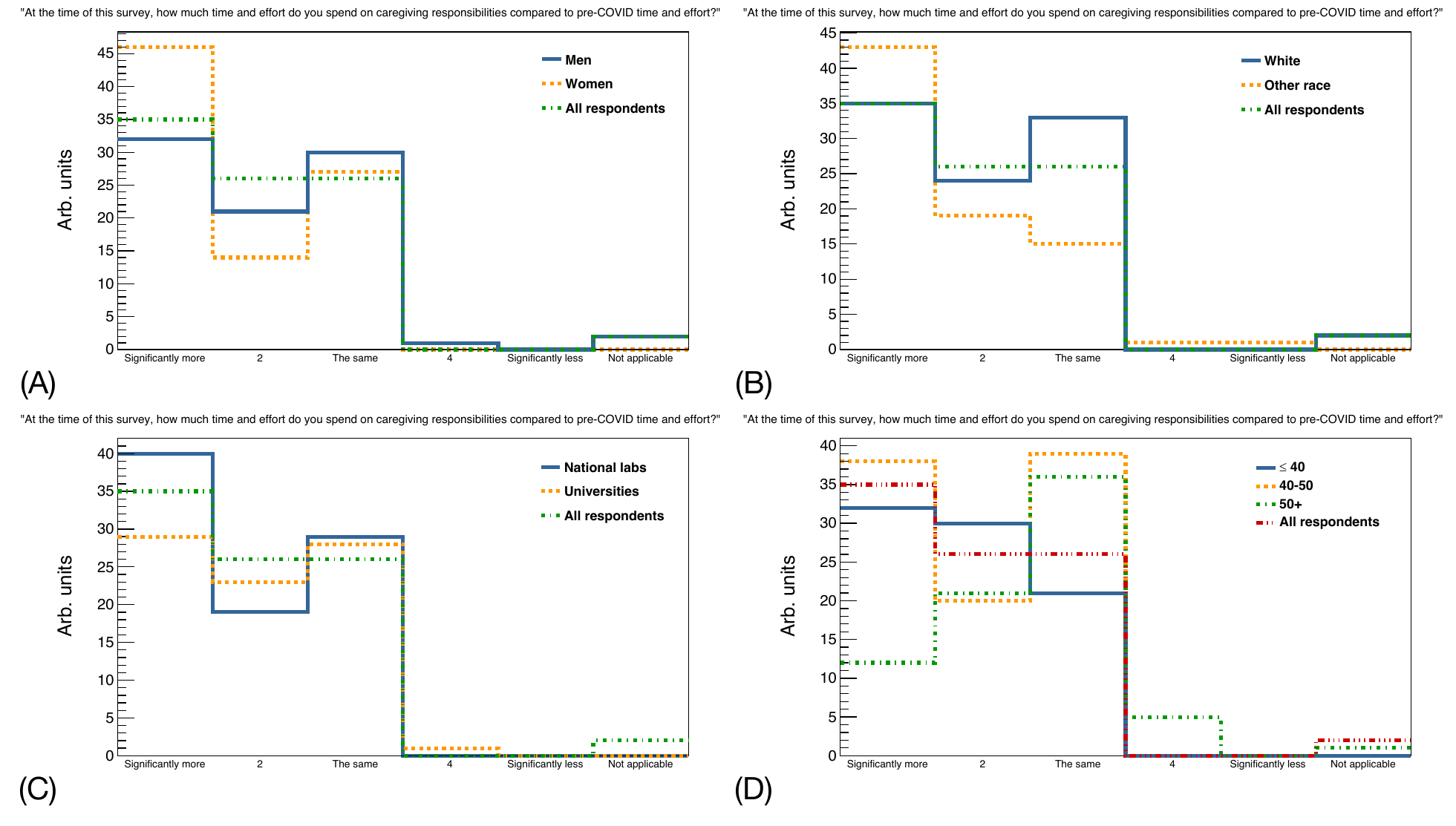}
    \caption{We asked caregivers to compare their time and effort on caregiving responsibilities to their pre-COVID time and effort. The majority of caregivers (84.8\%) responded to this question.}
    \label{Q159_combined}
\end{figure}

We asked caregivers how much time and effort they are spending on caregiving responsibilities compared to pre-COVID time and effort; we first thought about the responses to this question while we were learning about caregivers with children (see Figures \ref{Q155_combined} and \ref{Q156_combined}). The vast majority of caregivers reported the same (27.8\%) or more time and effort (54.7\%) on caregiving compared to pre-COVID; only 1\% of caregivers reported less time and effort. Broken down by gender (Fig. \ref{Q159_combined}A), 60.5\% of female caregivers reported more effort on caregiving, more than male caregivers (-5.1\%). Furthermore, we found that more caregivers in other racial groups (Fig. \ref{Q159_combined}B), more caregivers working at national labs (Fig. \ref{Q159_combined}C), and more caregivers younger than 50 years old (Fig. \ref{Q159_combined}D) reported spending more time and effort on caregiving compared to pre-COVID time and effort. 

\begin{figure}[H]
    \includegraphics[scale=0.83]{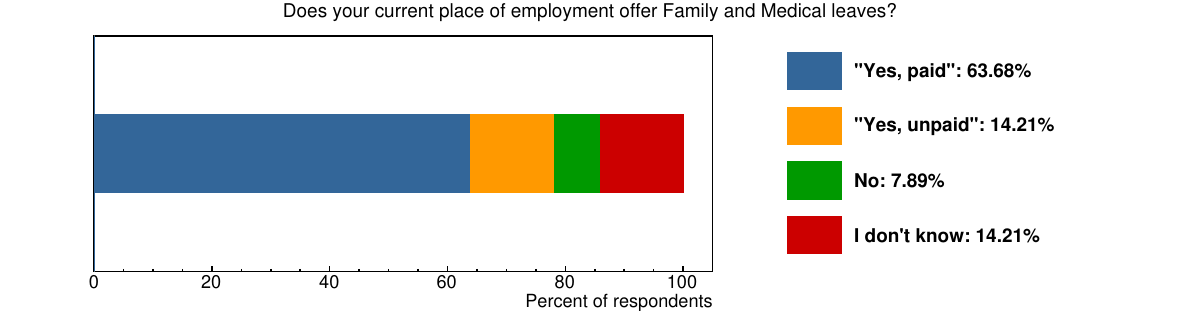}
    \caption{We asked caregivers whether their employers offer Family and Medical leaves. The majority of caregivers surveyed (85.2\%) provided an answer. }
    \label{Q160_overall}
\end{figure}

We asked caregivers whether their current place of employment offers Family and Medical leaves. The majority of caregivers (66.4\%) reported that their place of employment offers some form of leaves, either paid (54.3\%) or unpaid (12.1\%). A smaller group of caregivers (18.8\%) reported that either their place of employment doesn't offer Family and Medical leaves or that they don't know. We looked closely at the respondents who didn't know to see if any patterns emerged; the majority are faculty (51.9\%) with the next highest population being early career scientists (29.6\%). Broken down by gender, 14.5\% of male caregivers reported that they don't know if their place of employment offers Family and Medical leaves, significantly higher than female caregivers (-12.5\%). Similarly, 14.2\% of white caregivers reported that they didn't know, significantly higher compared to caregivers in other racial groups (-8.7\%). Finally, 16.1\% of caregivers at universities reported that they didn't know, significantly higher compared to caregivers at national labs (-10.4\%).

\begin{figure}[H]
    \includegraphics[scale=0.82]{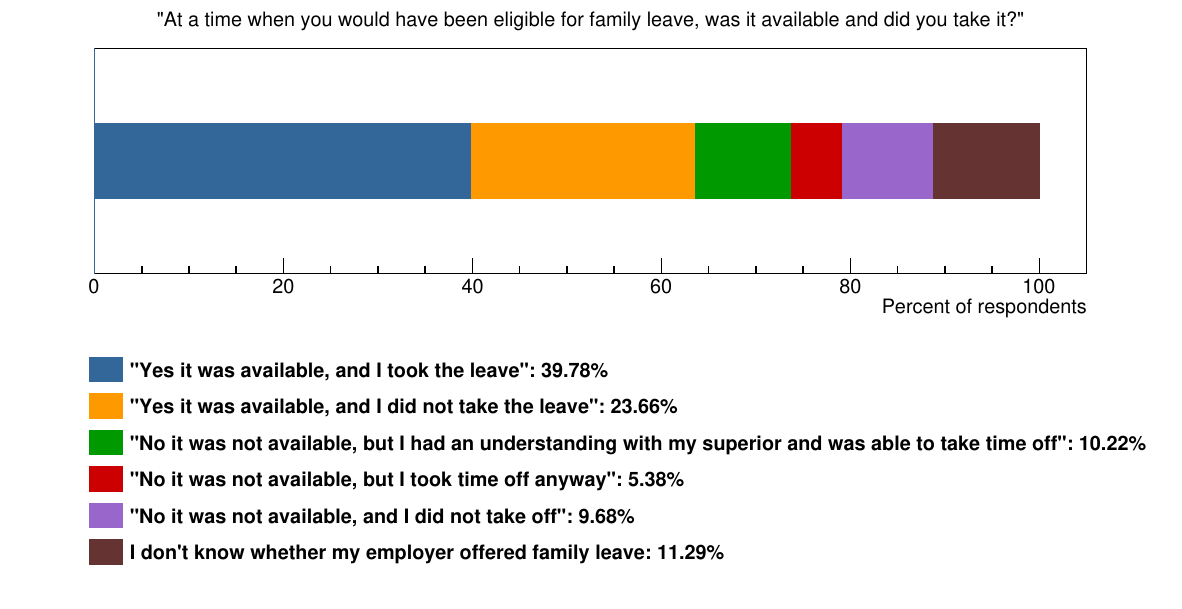}
    \caption{We asked caregivers about their experiences with Family and Medical leave. The majority of caregivers (83.4\%) responded to this question.}
    \label{Q161_overall}
\end{figure}

We also asked caregivers whether they accepted Family and Medical leave at a time when they would have been eligible, if it was available at all. The majority of caregivers (52.9\%) reported that family leave was available at a time when they would have been eligible; 21.1\% of caregivers reported that family leave was not available. We looked more closely at the caregivers who reported that they didn't know whether their employer offered the leave (9.4\%): One-third of the people who didn't know were faculty, and one-third were early career scientists. Broken down by gender, 11.4\% of male caregivers surveyed reported that they didn't know, significantly more than female caregivers (-6.7\%).

\begin{figure}[H]
    \includegraphics[scale=0.82]{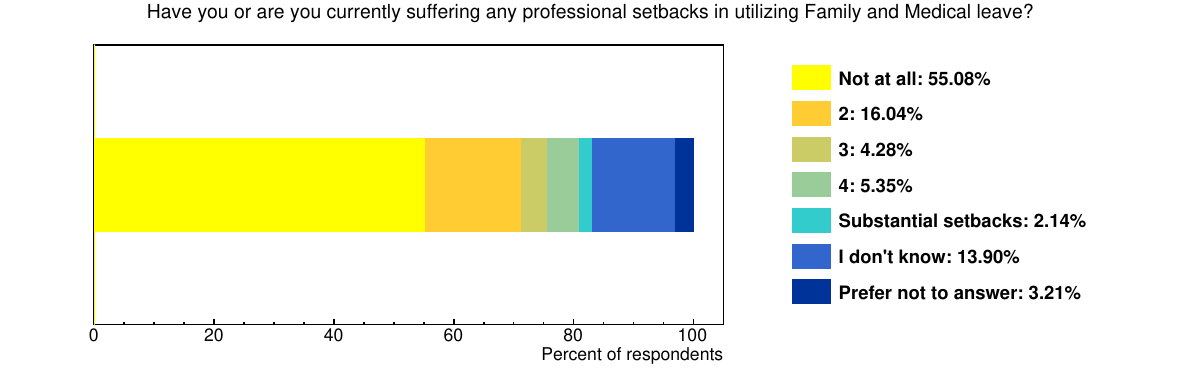}
    \caption{We asked caregivers about their experiences with Family and Mexical leave. The majority of caregivers (83.9\%) responded to this question. }
    \label{Q162_overall}
\end{figure}

Our final question on the topic of Family and Medical leave was whether caregivers have suffered or are currently suffering any professional setbacks in utilizing Family and Medical leave. Many of the caregivers we surveyed (46.2\%) reported not suffering any professional setbacks in utilizing Family and Medical leave. Another group of caregivers (23.3\%) reported some professional setbacks, with 2\% of caregivers reporting substantial setbacks. Again we looked at the caregivers who reported that they didn't know (11.7\% of caregivers): This group comprised of 53.8\% faculty and 26.9\% early career scientists. 

\begin{figure}[H]
    \includegraphics[scale=0.49]{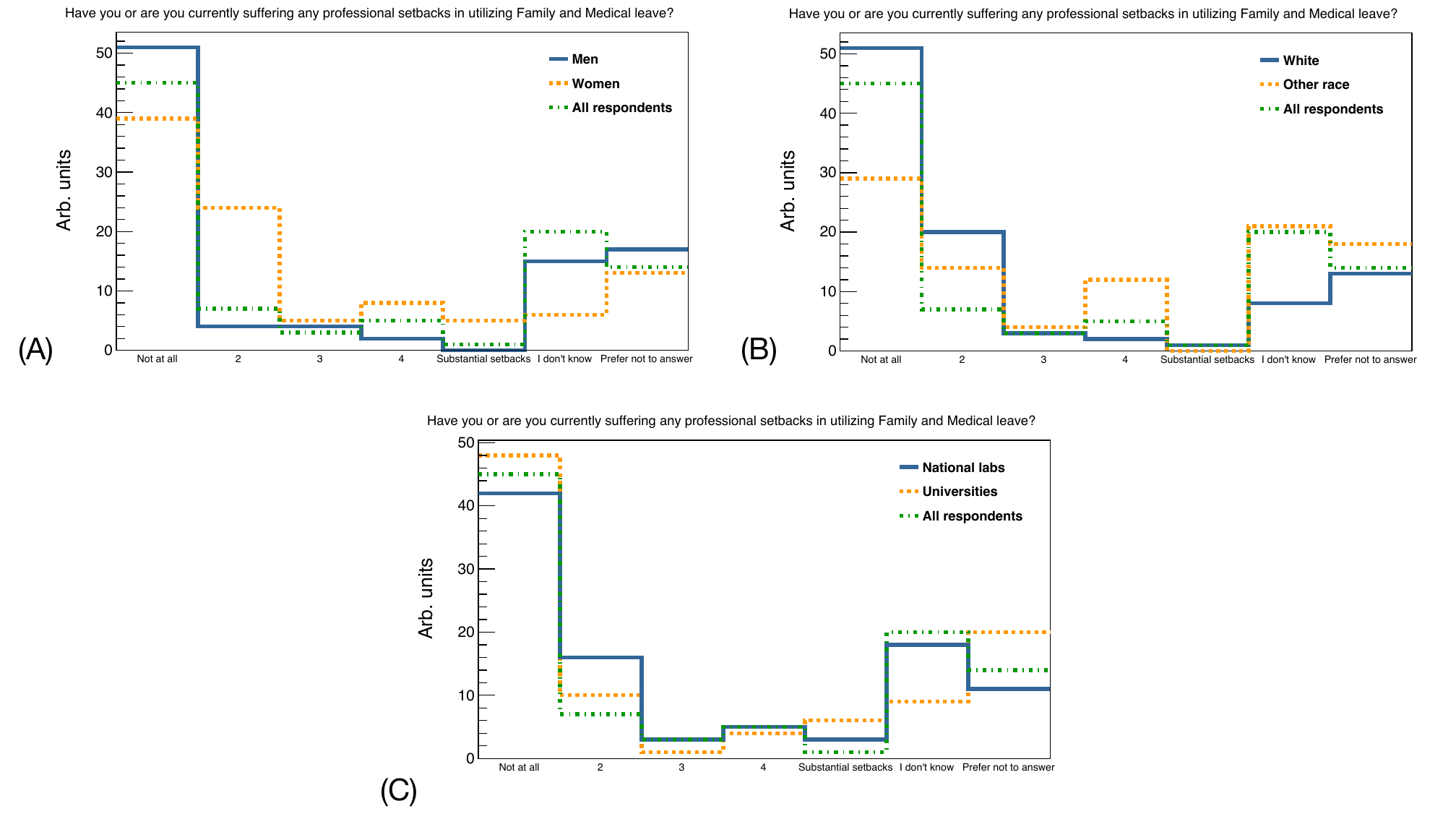}
    \caption{We asked caregivers about their experiences with Family and Mexical leave. We broke down the responses by (A) gender, (B) race, and (C) primary workplace. }
    \label{Q162_combined}
\end{figure}

Broken down by gender (Figure \ref{Q162_combined}A), 41.9\% of female caregivers reported some professional setbacks (meaning they answered with a 2 or higher), over twice as much compared to male caregivers (-22.6\%). Compared to female caregivers, 2.7\% more male caregivers reported that they don't know whether they have suffered professional setbacks. Broken down by race (Figure \ref{Q162_combined}B), 29.6\% of caregivers in other racial groups reported some professional setbacks, 7\% more than white caregivers. Broken down by primary workplace (Figure \ref{Q162_combined}C), 27.1\% of caregivers at national labs reported some professional setbacks, more than caregivers at universities (-6.7\%). Compared to caregivers at national labs, 3.1\% more caregivers at universities reported that they don't know whether they suffered professional setbacks. 

\begin{figure}[H]
    \includegraphics[scale=0.82]{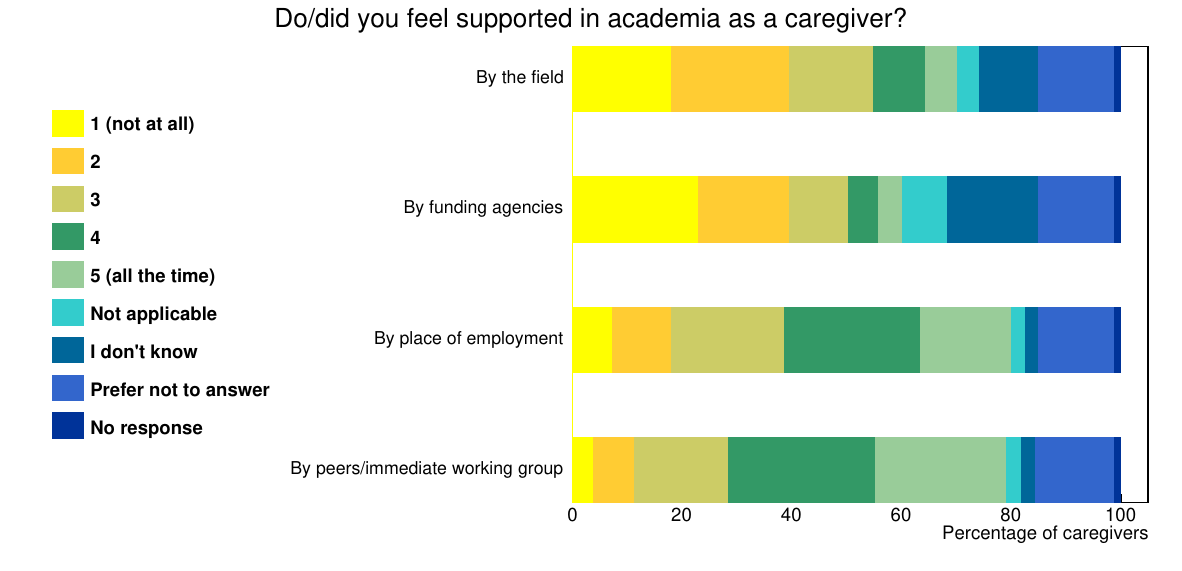}
    \caption{We asked caregivers whether they feel or felt supported in academia as a caregiver. The majority of caregivers (84.8\%) answered all parts to this question, but the breakdown bars include caregivers who did not answer additional questions.}
    \label{Q163_overall}
\end{figure}

Finally, we asked whether caregivers feel (or felt) supported in academia as a caregiver in four different areas: by their peers or immediate working group, by their place of employment, by funding agencies, and by the field as a whole. By and far caregivers feel the most support from their peers or immediate working group, where a little over half of caregivers answered with a 4 or 5. Similarly, 41.3\% of caregivers reported that they felt supported by their place of employment; on the other hand, 6.7\% more caregivers reported feeling little to no support from their place of employment compared to their peers or immediate working group. For funding agencies and the overall field, the responses indicated the opposite: 39.5\% of caregivers reported that they feel little to no support from both funding agencies and the field, although 5\% more caregivers reported feeling no support whatsoever from funding agencies. Our question about funding agencies also elicited the most ``I don't know'' responses from caregivers: 18.1\% of all male caregivers (+4.1\% compared to female caregivers), 21.4\% of caregivers at national labs (+6.8\% compared to caregivers at universities), and 20\% of white caregivers (+10.7\% compared to caregivers in other racial groups) reported that they don't know whether they feel or felt support from funding agencies. 

\begin{figure}[H]
    \includegraphics[scale=0.44]{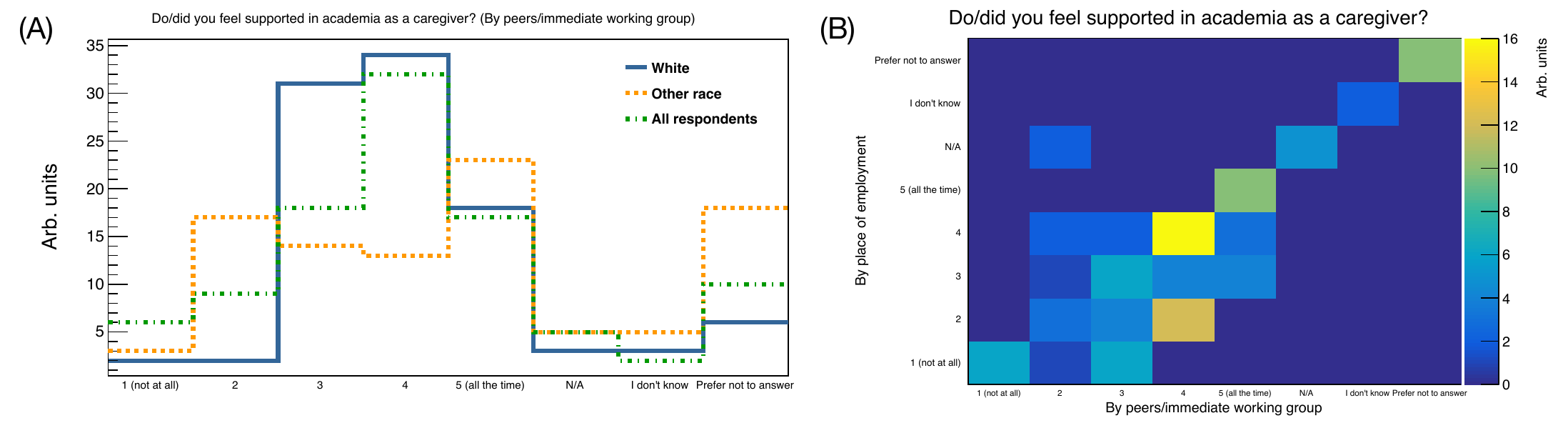}
    \caption{We asked caregivers whether they feel or felt supported by their peers or immediate working group as a caregiver. We broke down the responses by (A) race, and we compared these responses to caregivers' responses about their place of employment (B).}
    \label{Q163_Peers_combined}
\end{figure}

Looking more closely at caregiver support from peers and immediate working groups, we broke down the responses based using racial demographic information (Figure \ref{Q163_Peers_combined}A). Compared to white caregivers, about 3\% more caregivers in other racial groups reported little to no support from their peers or immediate working group. We also compared responses about peers to responses about place of employment (Figure \ref{Q163_Peers_combined}B). Most caregivers reported feeling the same level of support between the two categories; however, there is a group of caregivers who reported feeling support from their peers but also reported less support from their place of employment. In fact, it is more likely that a caregiver would rate the level of support from their place of employment lower than the support from their peers compared to the other way around.

\begin{figure}[H]
    \includegraphics[scale=0.5]{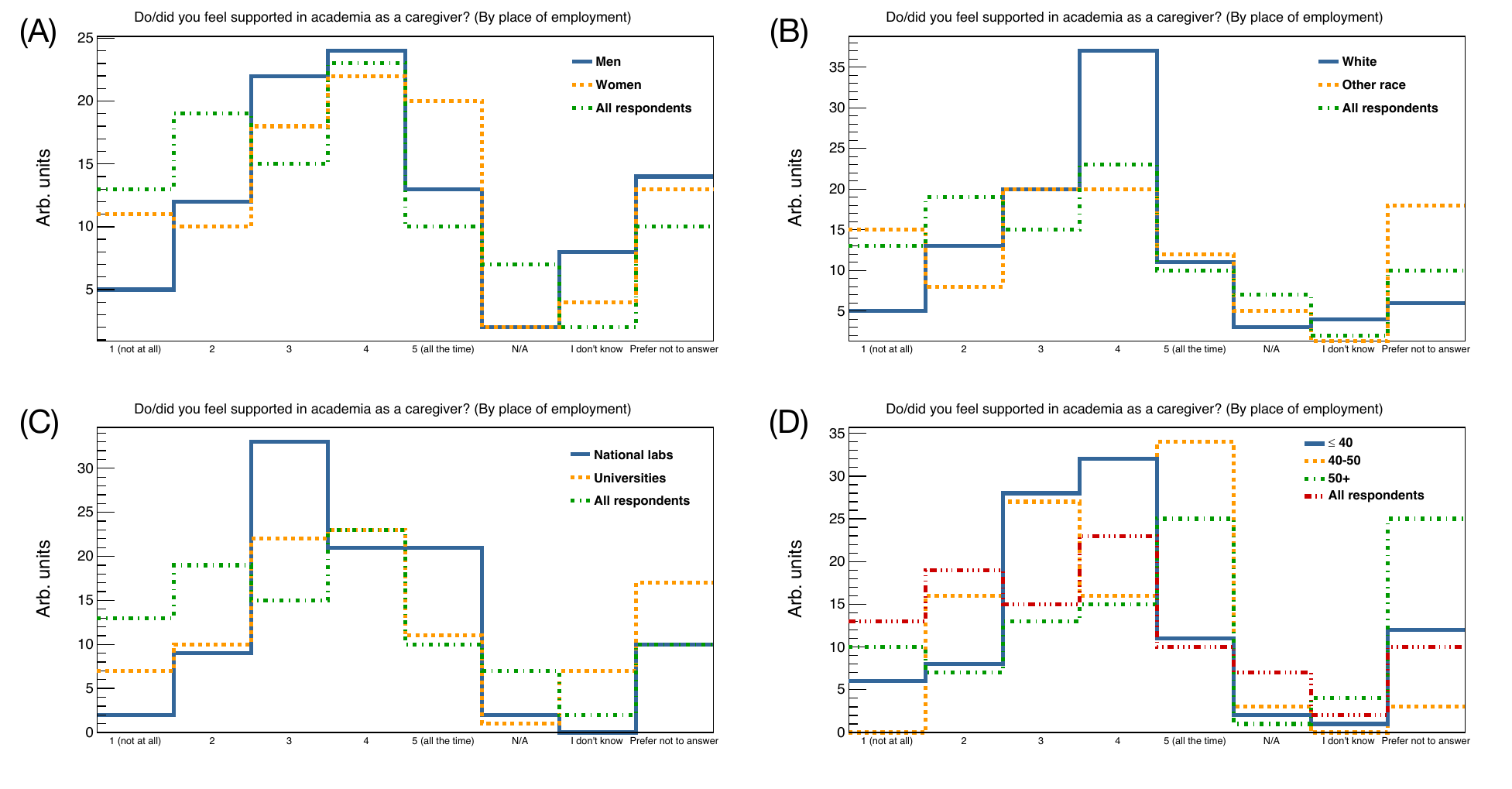}
    \caption{We asked caregivers whether they feel or felt supported by place of employment as a caregiver. We broke down the responses by (A) gender, (B) race, (C) primary workplace, and (D) age.}
    \label{Q163_Employment_combined}
\end{figure}

Considering the responses about level of support from places of employment, we noticed interesting trends while breaking down by different demographic information. Broken down by gender (Fig. \ref{Q163_Employment_combined}A), 25.6\% of female caregivers reported feeling little to no support from their place of employment, significantly higher than male caregivers (-9.3\%). Broken down by race (Figure \ref{Q163_Employment_combined}B), 44.5\% of white caregivers reported feeling supported by their place of employment, 5.6\% higher compared to caregivers of other racial groups. That is not to say that more caregivers in other racial groups reported less or no support from their place of employment, as the fractions are similar compared to white caregivers (17.4\% white caregivers versus 16.7\% caregivers in other racial groups). Broken down by primary workplace (Figure \ref{Q163_Employment_combined}C), caregivers at national labs reported feeling supported by their place of employment compared to caregivers at universities. Finally, broken down by age (Figure \ref{Q163_Employment_combined}D), 5.5\% more caregivers older than 40 years old reported feeling little to no support from their place of employment compared to caregivers younger than 40. Moreover, 1.5\% more caregivers older than 40 reported feeling support all the time from their place of employment. Caregivers younger than 40 tended to respond somewhere in the middle. 

\begin{figure}[H]
    \includegraphics[scale=0.48]{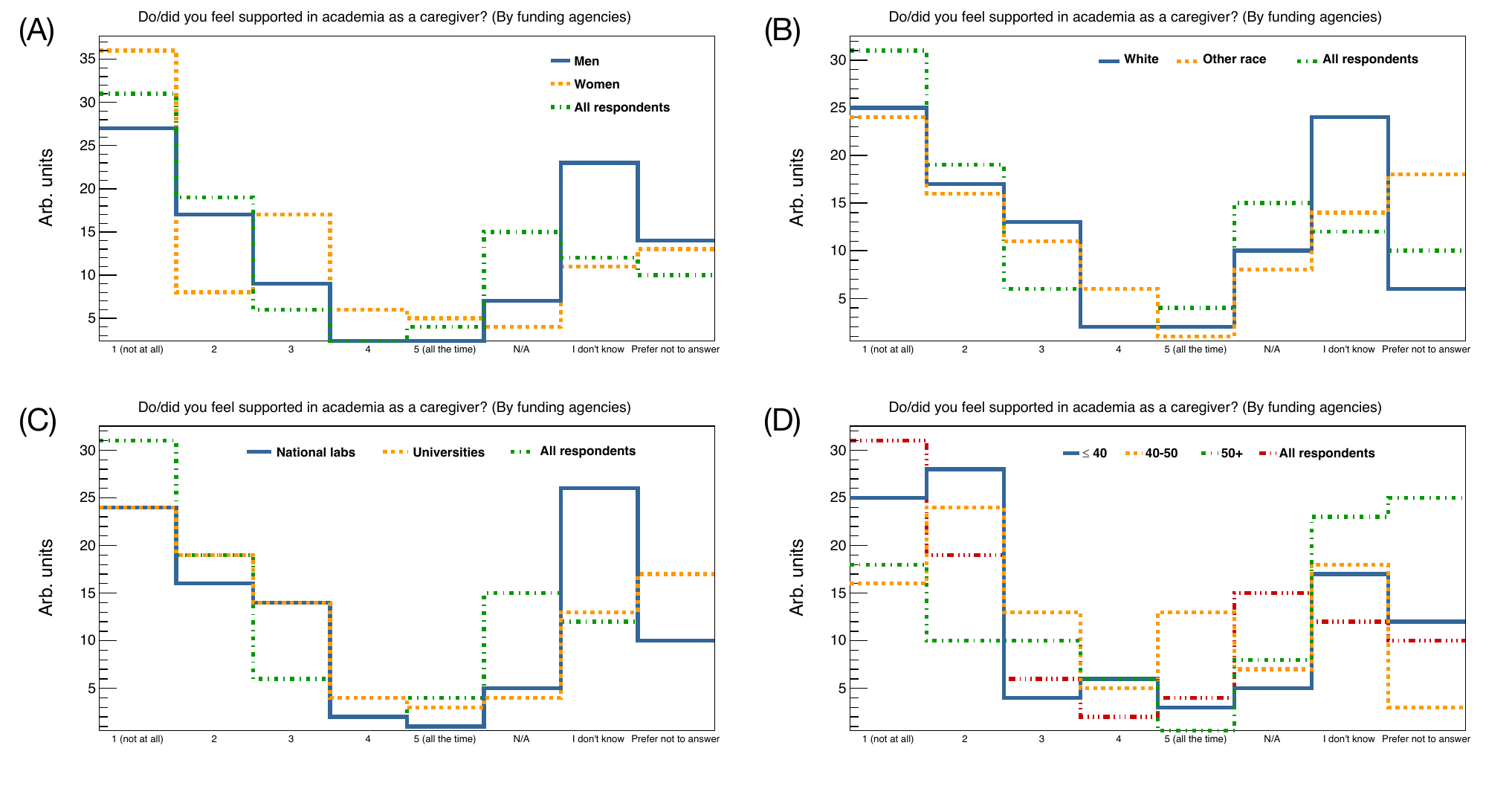}
    \caption{We asked caregivers whether they feel or felt supported by funding agencies as a caregiver. We broke down the responses by (A) gender, (B) race, (C) primary workplace, and (D) age.}
    \label{Q163_FundingAgencies_combined}
\end{figure}

Considering responses about funding agencies, we first noticed the differences in the ``I don't know'' categories for each of the breakdown plots: More male caregivers, white caregivers, caregivers at national labs, and caregivers older than 40 responded with ``I don't know''. Broken down by gender (Fig. \ref{Q163_FundingAgencies_combined}A), 32.6\% of female caregivers say they do not feel any support at all from funding agencies, significantly higher (+12.5\%) compared to male caregivers. Broken down by age (Figure \ref{Q163_FundingAgencies_combined}D), about 8.7\% more caregivers younger than 50 reported little to no support from funding agencies compared to caregivers older than 50. 

\begin{figure}[H]
    \includegraphics[scale=0.46]{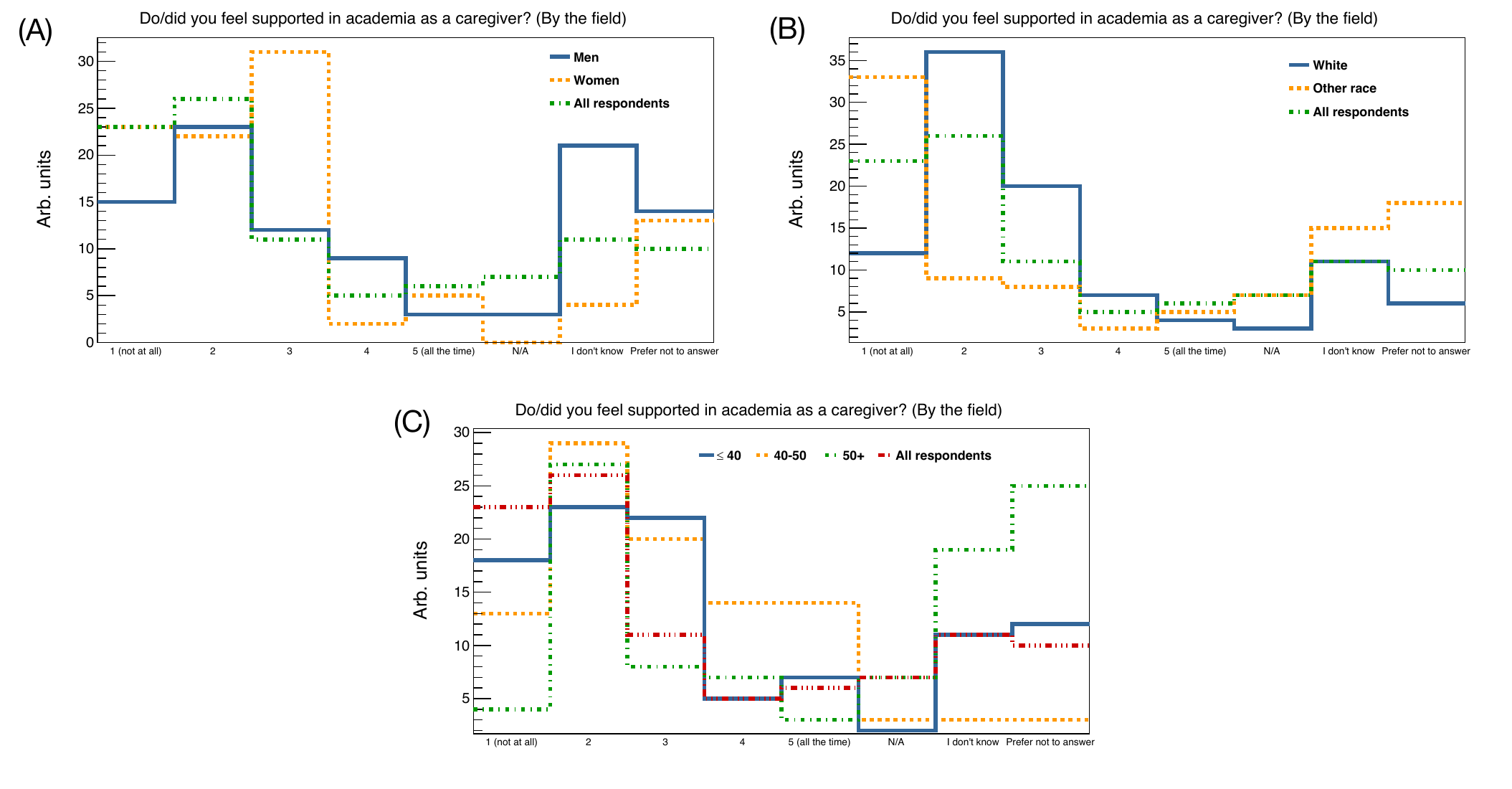}
    \caption{We asked caregivers whether they feel or felt supported by the field as a caregiver. We broke down the responses by (A) gender, (B) race, and (C) age.}
    \label{Q163_Field_combined}
\end{figure}

Finally, we looked at the responses about level of support from the field. Broken down by gender (Figure \ref{Q163_Field_combined}A), over half of female caregivers reported feeling little to no support from the field, 14.5\% more than male caregivers. On the other hand, around one-third of male and female caregivers respectively reported a 3 or higher. Broken down by race (Figure \ref{Q163_Field_combined}B), 25.9\% of caregivers in other racial groups reported feeling no support from the field, much more (+11.7\%) compared to white caregivers. Broken down by age (Figure \ref{Q163_Field_combined}C), around 19\% of caregivers younger than 50 reported feeling no support from the field, a higher fraction compared to the 10.5\% of caregivers older than 50 who reported feeling the same lack of support.

\begin{figure}[H]
    \includegraphics[scale=0.66]{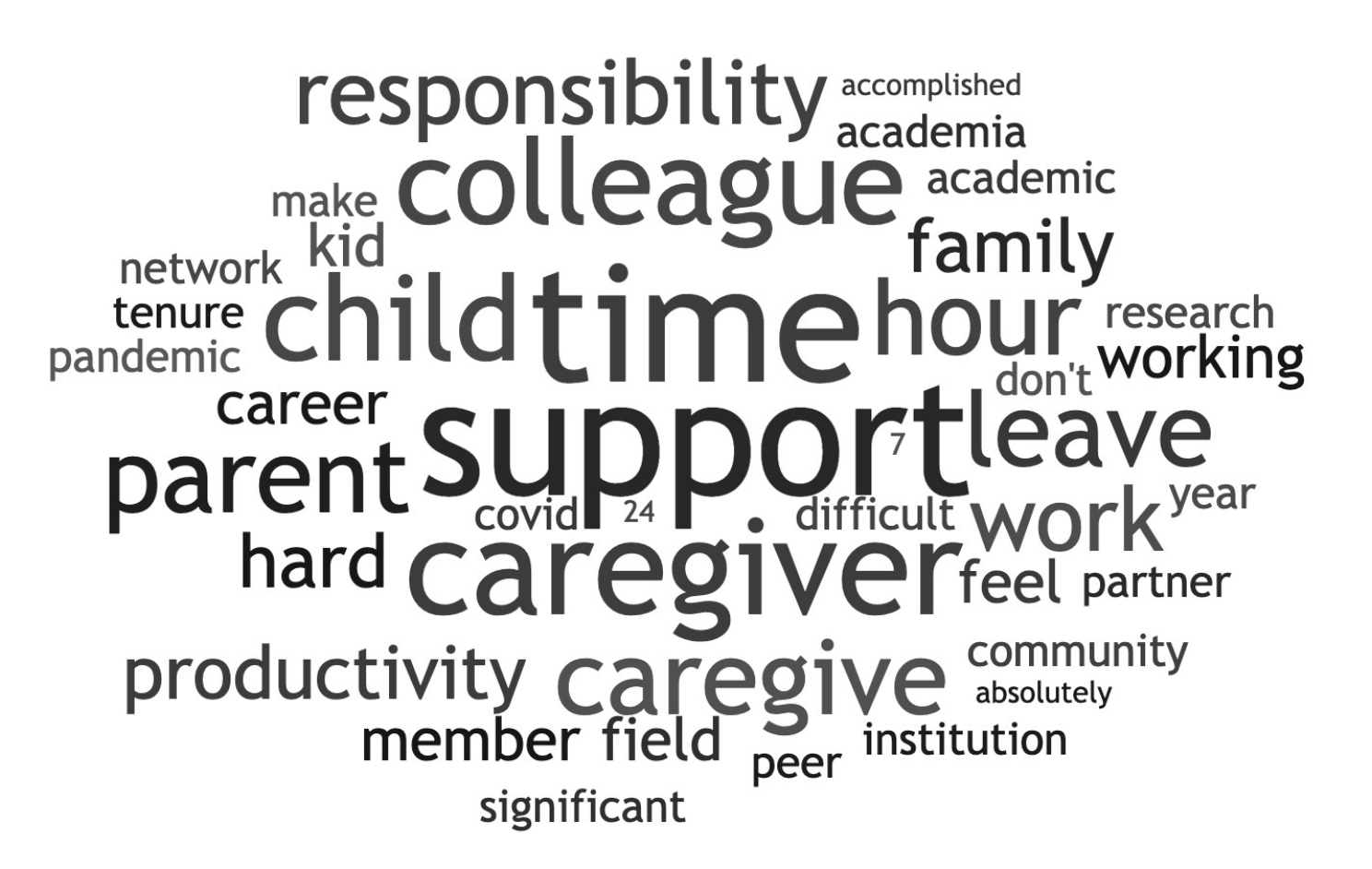}
    \caption{We asked caregivers to share their experiences in academia as a caregiver. The top 40 commonly used words are presented in this word cloud, where ``I'd'' was removed from the top 40 list. }
    \label{wordcloud_caregivers_top40}
\end{figure}

Finally, we asked caregivers to write about their experiences as a caregiver in academia. The 40 most commonly used words are presented as a word cloud in Figure \ref{wordcloud_caregivers_top40}. The top word from caregivers was ``support'', and caregivers used it to express variety of sentiments. Some caregivers wrote about the support they receive from their partner, colleagues, or institution, with an emphasis on how lucky the caregiver is to receive such support. Other caregivers wrote about the lack of support from their colleagues, funding agencies, or HEPA as a whole. Several caregivers noted that their colleagues rarely understand or empathize, or they wrote about how their colleagues expect the same level of productivity despite the respondent's caregiving responsibilities. Colleagues might ``get it'', but they ultimately view caregiving as an inconvenience, and caregivers wrote about feeling excluded from career opportunities due to their caregiving responsibilities. 

Many caregivers also wrote about the difficulties of caregiving during the COVID-19 pandemic. They wrote about the challenges of staying involved and participating in research, and they expressed dissatisfaction that they cannot achieve the same level of productivity compared to pre-COVID. Several caregivers also mentioned their disdain for the culture of constant availability since they are not available 24/7. Caregivers in temporary positions wrote about the level of pressure they face especially in terms of funding or advancing in their career. Caregivers wrote about the difficulties of balancing their work and life, and they wrote that people without caregiving responsibilities have a better chance at career advancement.

\textbf{}\subsection{The COVID-19 Pandemic}

In this subsection, we'll focus on several questions that were asked directly about the impacts of the COVID-19 pandemic and institutional responses to it.

The COVID-19 pandemic had a significant impact, both on the lives of people in the field of HEPA, and on the Snowmass process in particular, including a delay of the summer study to 2022. Given the broad impact, it is difficult to separate answers to a number of questions from the context of the pandemic. Several questions on the survey, however, were directly aimed at better understanding its effects, and this subsection is devoted to a discussion of these questions. Of course, the results presented below are not intended as a scientific representation of the impacts of COVID-19, only to give a snapshot. 
Note also that the survey was live during a period of rapid changes in the U.S., with many institutions starting to resume in-person activities  between the Spring 2021 and Fall 2021 terms, between which the survey took place, and also partly overlapping with the surge of cases associated with the Delta variant. The situation was also rapidly changing at different times in different places, so the results should be considered with all of these factors in mind.

\begin{figure}[H]
    \centering
    \includegraphics[width=0.95\linewidth]{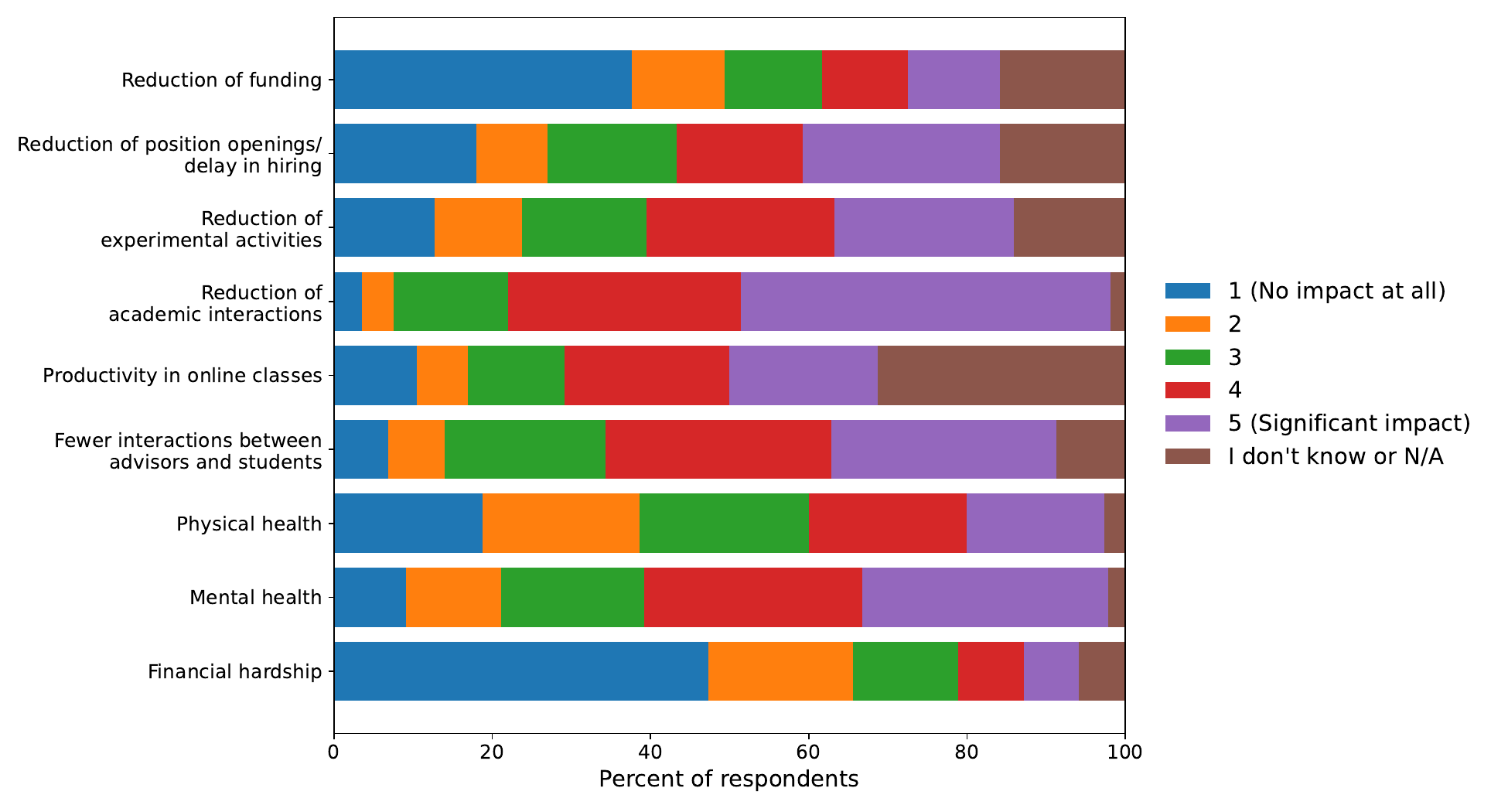}
    \caption{We asked all respondents to rate how much different factors were impacted due to COVID-19.}
    \label{fig:covid_1}
\end{figure}

We asked respondents to rate how much a number of factors in their work and life (such as funding or academic interactions) were impacted due to COVID-19.  The results are shown in \autoref{fig:covid_1}. Amongst the nine impacts that we considered, the three largest impacts were the reduction of academic interactions, impacts on mental health, and fewer interactions  between advisors and students.
The smallest impacts were seen in financial hardship, a reduction of funding, and in physical health.

\begin{figure}
    \centering
    \includegraphics[width=0.85\linewidth]{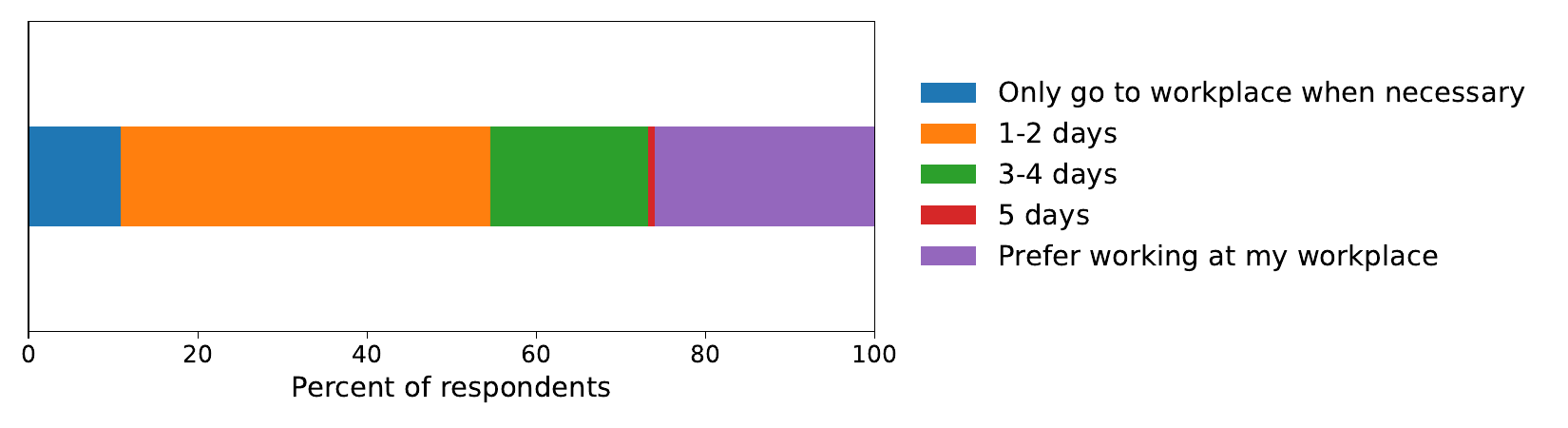}
    \caption{We asked all respondents if, given the flexibility to continue to do so, how many days of the work week they were likely to work from home.}
    \label{fig:covid_2}
\end{figure}

Relatedly, we also asked respondents how frequently they would choose to work from home, if given the flexibility to do so, as many members of the community had to adapt to such conditions during the pandemic. The results are shown in \autoref{fig:covid_2}.
Responses varied significantly. We found $74\%$ of respondents preferred to work from home at least one day of the week, but $70\%$ of respondents preferred to spend a majority ($\geq 3$ days) of the work-week at their workplace.

\begin{figure}[H]
    \centering
    \includegraphics[width=0.95\linewidth]{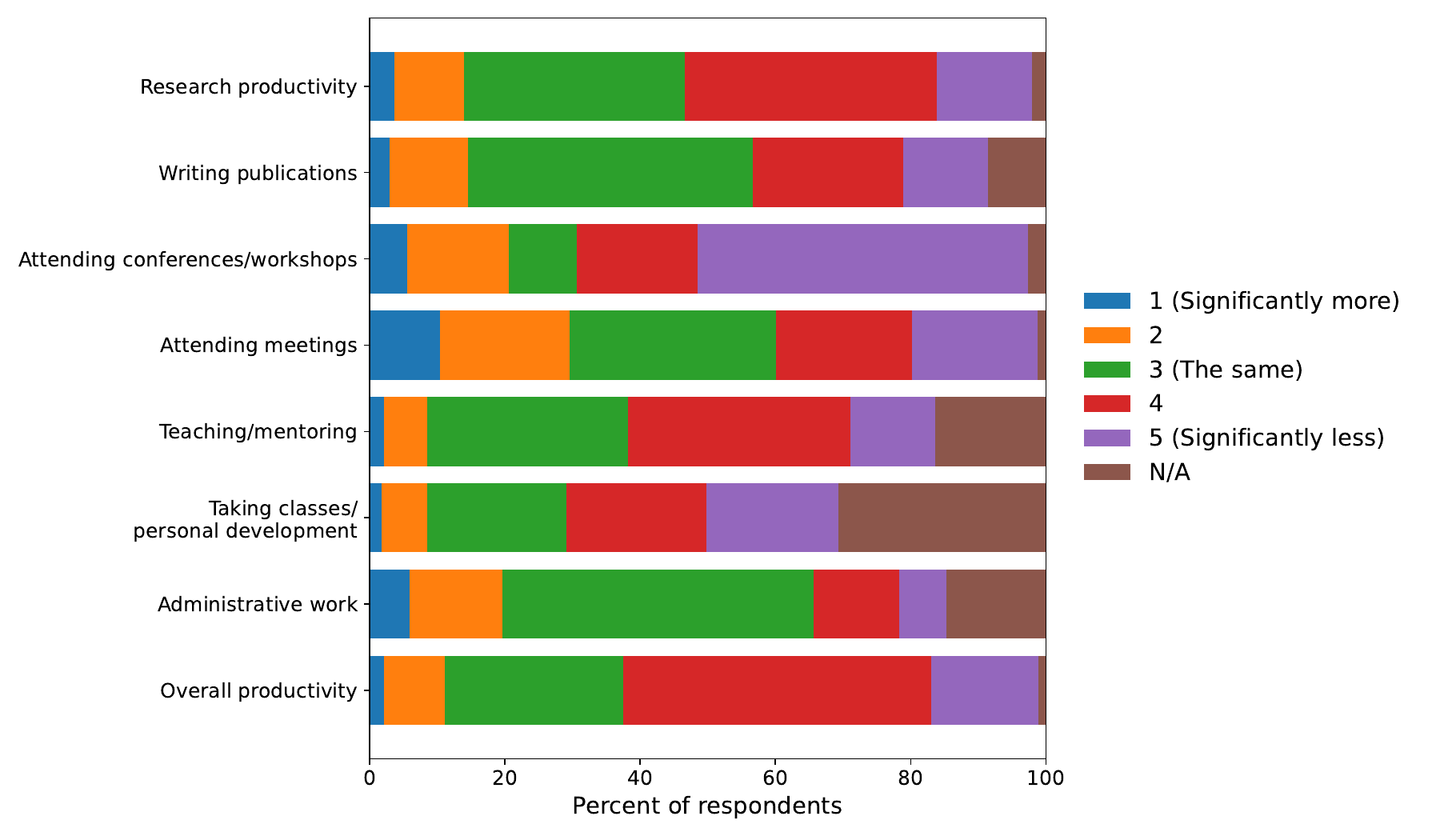}
    \caption{We asked respondents to rate their productivity on various activities compared to pre-covid levels.}
    \label{fig:covid_3}
\end{figure}

To understand the impacts on respondents' work productivity in more detail, we asked them to rate their productivity on different activities compared to pre-COVID levels. The activities considered were researching, writing publications, attending conferences / workshops, attending meetings, teaching and mentoring, taking classes or personal development, administrative work, and their overall productivity. The results are shown in \autoref{fig:covid_3}. 

Easily the biggest impact was seen in attending conferences / workshops, with nearly half of respondents saying they were ``significantly less'' productive in this way. Despite the prevalence of virtual conferences and workshops, only $20\%$ of respondents said they were ``more'' or ``significantly more'' productive.
Notable impacts were also seen on ``Research productivity'', ``Teaching and mentoring'', and on ``Overall productivity'', each of which saw a plurality of respondents reporting less or significantly less productivity.

We also broke the results for overall productivity down for respondents identifying as men versus women, and for junior versus senior researchers.
Overall, the impacts on overall productivity for men and women and for junior and senior faculty were roughly the same, though a slightly higher percentage of women reported ``Significantly Less'' overall productivity.

\begin{figure}[H]
    \centering
    \includegraphics[width=0.95\linewidth]{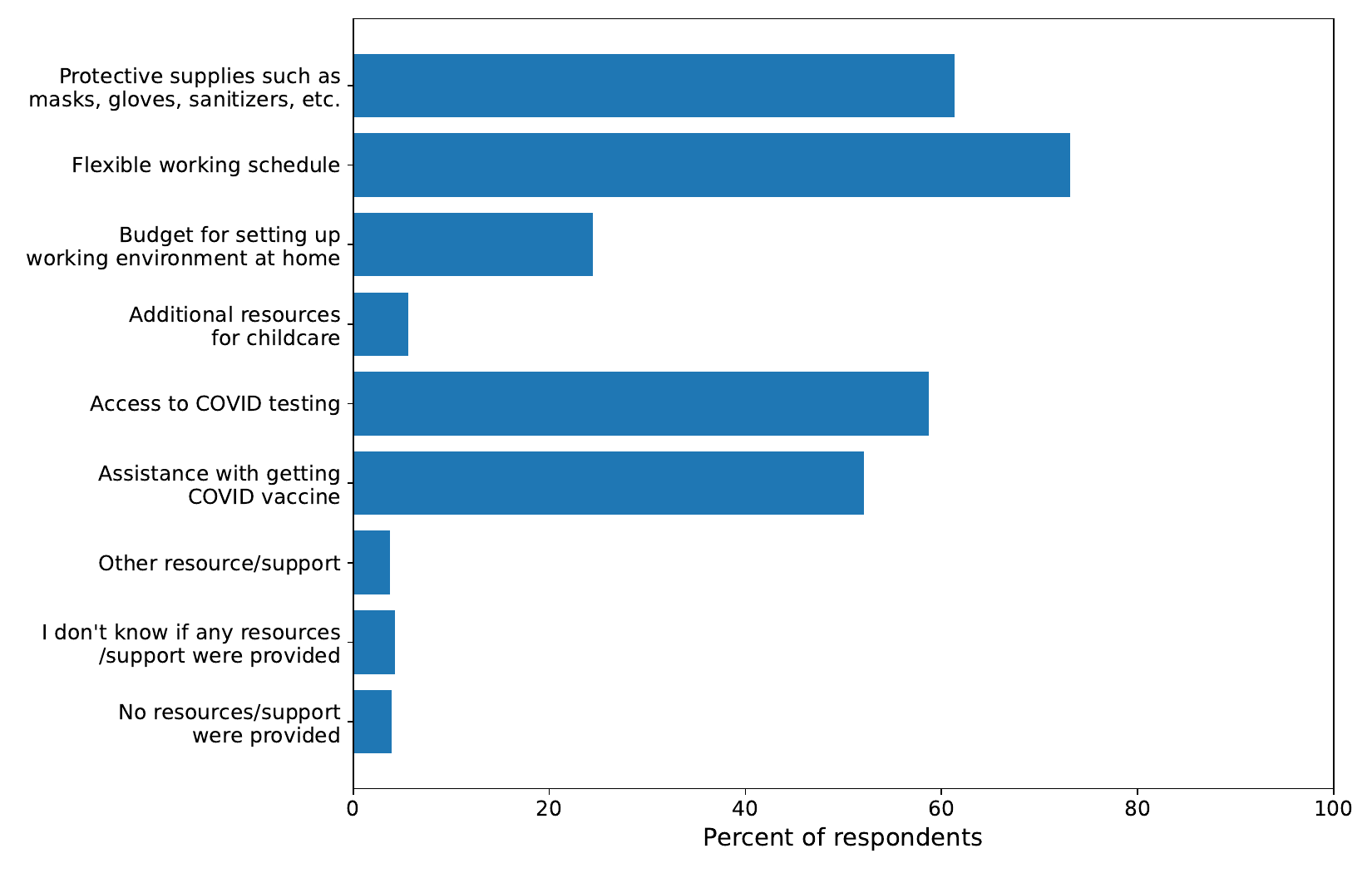}
    \caption{We asked respondents what resources and support (to their knowledge) were provided by their workplaces to help continue working during the COVID-19 pandemic. Responses are normalized to the total number of respondents to this question.}
    \label{fig:covid_4}
\end{figure}

Many institutions took concrete action to help researchers continue working through the pandemic. We asked respondents to report which of a list of resources and support were provided by their workplace. The results are shown in \autoref{fig:covid_4}. 
The most common steps reported were supplies of protective equipment, a flexible working schedule, and access to covid testing or support for getting the vaccine. Far fewer respondents said that their institutions provided any additional budget to help set up a working environment at home. Other means of support, such as additional resources for childcare, were not widely reported.

\begin{figure}[H]
    \centering
    \includegraphics[width=0.85\linewidth]{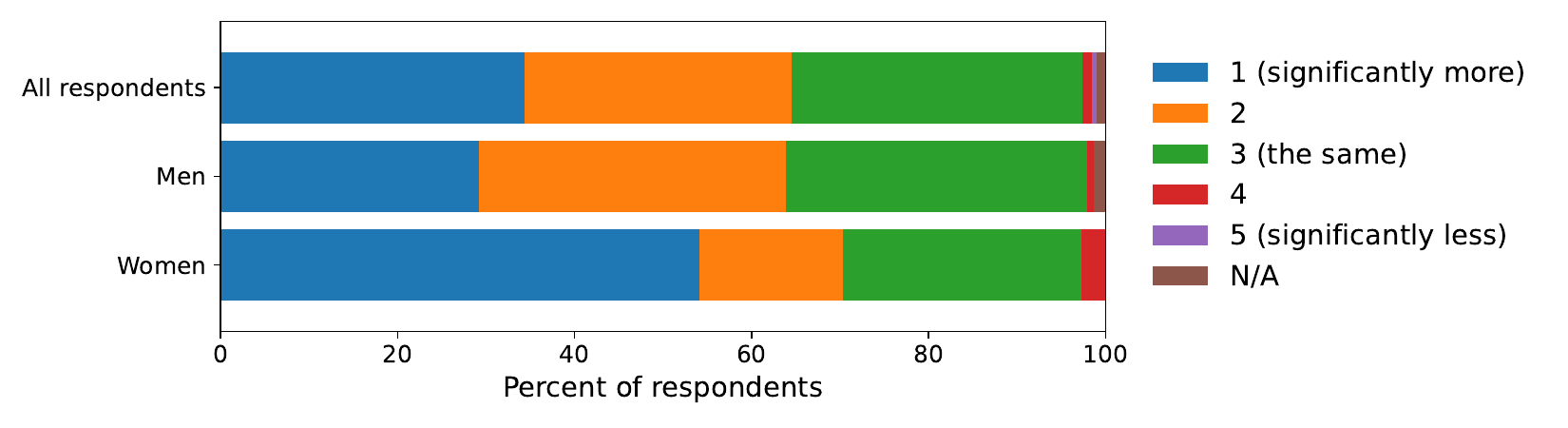}
    \caption{We asked respondents with caregiving responsibilities how much time and effort they spend on caregiving responsibilities compared to pre-COVID time and effort.}
    \label{fig:covid_5}
\end{figure}

Finally, along with the other questions asked of respondents who had caregiving responsibilities disucssed in Section~\ref{sec:caregiving}, we asked caregivers to rate the amount of time and effort they spent on caregiving responsibilities compared to pre-COVID levels. The results are shown in \autoref{fig:covid_5}, where we also break down the responses by those identifying as Men or Women. 
We see that a significant majority of all caregivers reported spending more time and effort (i.e., 1 or 2 on a 5-point scale) on caregiving compared to pre-COVID levels.
While the number of men and women reporting ``more"  time and effort was roughly the same, a larger percentage of women reported spending ``significantly more" time and effort on caregiving than men, potentially reflecting an imbalance on the impacts of the pandemic on different groups within the field.

\section{Discussions}
\label{sec:discussions}

\subsection{The state of early career scientists in HEPA}
The early career scientists who responded to the 2021 survey were a more diverse group compared to other career stages: about half of all female respondents were early career scientists, and the Asian, Hispanic/Latino, Black, and Middle Eastern respondents were largely composed of early career scientists. Hopefully future Snowmass processes will show these early career scientists ascending into permanent senior positions in HEPA, a hypothesis that could be verified in future Snowmass surveys. 

When 45.4\% of PhD students and 39.6\% of Postdocs report feeling their salary probably or definitely is inadequate, that indicates there might be a systemic issue with adequately compensating early career scientists for their contributions to HEPA. It's common for PhD students to be paid less for their contributions (e.g., due to the ``trainee'' or student aspect of the position), and there is little to no compromise for stipend amounts. Lack of compromise is also the case for Postdocs at some institutions, especially for institutions with high-demand positions or with strict salary requirements. Some Postdocs can negotiate their salaries, but those skills are not within the scope of the average U.S. graduate school curriculum. The 2021 survey results in this context grow more alarming; specific systemic changes should occur in order to improve early career scientists' compensation concerns. 

Salary amounts should (at minimum) cover the cost of living for the specific institution's location, and they should increase every year to cover the cost of inflation. For universities, increasing stipends for graduate students might be handled outside of the physics department, so faculty and scientists can be advocates and consistently place pressure on the administrators who have control over determining graduate school stipends and Postdoc salaries. Funding agencies should also take cost of living and inflation into account while awarding grants to faculty and scientists, so that research groups have adequate funds to compensate their students and Postdocs. Furthermore, graduate programs should offer training (e.g., in a class, seminar, or through a broader institutional effort) on how to negotiate salaries, and they should advertise or require the training at some point during the graduate curriculum.  
    
The early career scientists with recent job offers are largely staying in HEPA. Overall, the vast majority of early career scientists who accepted or plan to accept a position applied for (and accepted) Postdoc positions. Even if PhD students ultimately accepted a non-academic position, they likely also applied for academic positions. While PhD students collectively applied for academic positions, Postdocs were more varied in their job search, where about two-thirds reported applying for academic and non-academic jobs alike. Compared to graduate students, more Postdocs indicated they were switching to a different job sector. While this is not necessarily a concerning or surprising result, institutions should take action and adequately train early career scientists to move into a variety of job sectors, especially industry positions. According to a report from the Organisation for Economic Co-operation and Development (OECD), institutions should track career outcomes from their PhD students so that the U.S. part of HEPA can better understand job opportunities on a local level along with what challenges new PhD holders face while transitioning into the next stage of their career. Institutions should also provide opportunities (e.g. seminars or internships) to Postdocs so they can gain professional development skills \cite{oecd_2021}.

The survey also revealed that Postdocs also are affected more by visa restrictions compared to graduate students, and some Postdocs applying for jobs reported that immigration issues were an important concern for them. Our results indicate that immigration issues disproportionately affect early career scientists in other racial groups, while White early career scientists remain largely unaffected. Current U.S. visa policies are largely inadequate to support Postdocs' transitions into non-academic job sectors \cite{geng_2022}. Immigration concerns should always be taken into account while training Postdocs on how to navigate various job markets, and institutions should support their Postdocs' struggles with U.S. visas and immigration policies by advocating for updated policies and a streamlined application process.

The early career scientists who are applying for jobs are largely interested in remaining in HEPA. Overall, early career scientists expressed much more interest in applying for academic positions versus industry positions. The results indicate that universities might not adequately advertise or train their students on how to apply for industry jobs, as we found that early career scientists located outside of universities were more likely to apply for STEM industry positions. When asked to comment on why they chose their top sector, many early career scientists who rated academic positions highly wrote about their love for teaching and research. There was also a group of early career scientists who wrote about their struggles to remain in HEPA (e.g., not enough permanent positions, the expectation to complete multiple postdocs, etc.), and others who expressed dissatisfaction with HEPA and a desire to switch to a different sector. The early career scientists who took the 2021 survey are aware about the lack of permanent positions and highly competitive nature of HEPA. While the majority continue to apply for academic positions, some early career scientists are disappointed, dissatisfied, and moving out of HEPA. Furthermore, early career scientists who are applying for jobs are deeply concerned about job availability regardless of the sector. On the whole, early career scientists indicated much more concern about academic job availability compared to the industry job sector.  

Overall, early career scientists are applying for non-academic jobs (even if they ultimately remain in HEPA); there exists a subset of scientists who are dissatisfied with HEPA and wish to switch job sectors; and early career scientists expressed a deep concern for academic job availability. Outside survey data found that early career scientists (specifically Postdocs) are passionate about their research, but they also acknowledge the challenges of low pay, work-life balance, and a competitive job market \cite{afonja_salmon_quailey_lambert_2021}; similar results have also been observed in other surveys conducted during the pandemic \cite{woolston_2020}. Institutions should train their early career scientists to explore a variety of job sectors comfortably and confidently. PhD students and postdocs rely on the support of their advisors or supervisors, not just their work colleagues. Principal investigators should also offer support to their early career scientists, whether that be advocating for institutions to offer mental-health services; exhibiting flexibility and patience toward their students and Postdocs; or advocating for increased salaries or benefits for their students and Postdocs. 

\subsection{The state of faculty in HEPA}
The survey revealed that tenure-track faculty were more diverse in racial and ethnic backgrounds compared to tenured faculty. Similar to early career scientists, we hope that future Snowmass processes reflect this level of diversity in more senior career stages, and this could be checked in a future Snowmass survey. Another interesting difference between tenure-track and tenured respondents arose when asked about their salaries: More tenure-track faculty think their salary probably or definitely is inadequate.

While not a significant difference, the average tenure-track faculty who took our survey noted more time as a Postdoc and more time looking for faculty positions compared to their averaged tenured faculty counterpart. This indicates that scientists searching for permanent positions in HEPA may spend longer amounts of time as an early career scientist. Furthermore, tenure-track faculty respondents set themselves apart from their tenured colleagues by forming two distinct groups: Those who spent 3-4 years as Postdocs, and others who spent 6-7 years as Postdocs. This seems to indicate that some tenure-track faculty completed two or more postdocs before securing their faculty position, a result that is not too surprising. Since tenured faculty respondents roughly reported 3-5 years as a Postdoc with some outlying responses, we find the difference troubling. These results might hint of changes within HEPA; expectations for faculty have changed over the past three decades in places such as collaborations and departments. Furthermore, the job market is more competitive now compared to the recent past. Another hint to HEPA's changing culture is that the majority of teaching, tenure-track, and tenured faculty would prefer a reduced teaching load on some level, while the limited number of retired faculty respondents reported that they would not have preferred a reduced teaching load. Perhaps teaching expectations have changed over the past three decades, or perhaps other responsibilities have taken priority over teaching. 

One concerning result from the survey was that about one-tenth of tenure-track faculty and one-fifth of tenured faculty reported not compensating their undergraduate students for their research responsibilities. This result is coupled with reports from some undergraduate student respondents that they are not compensated for their research responsibilities. Similar to an unpaid internship, uncompensated research disproportionately affects students from lower income backgrounds or those who lack financial support for their education expenses \cite{rothschild_rothschild_2020}. As one method to increase inclusiveness within HEPA, undergraduate students should be compensated for their research responsibilities, whether it is through financial compensation, course credit, outside fellowships, work study programs, or other means.  This may need to be handled on a department-by-department basis, but funding agencies should also allocate adequate funding toward compensating undergraduate students in faculty research groups. Funding agencies should also continue to fund fellowships and programs (e.g., the NSF REU program \cite{NSF_REU}) which support and nourish undergraduate student research opportunities.

One other thing about faculty when it comes to their relationship with students -- we found it interesting that advisors rated their availability higher than what students reported, although the sufficiency of support was similarly rated between students and advisors.

This is not the only concerning result related to compensation. Compared to faculty respondents, tenure-track faculty reported far less of their PhD students and Postdocs as employed after graduation. More tenure-track faculty also reported 0\% of their PhD students and Postdocs as employed. This could be related to early career scientists' concern over job availability, especially if both groups of scientists are applying for academic positions. Furthermore, we discovered a similar report between scientists versus senior scientists for the PhD students and Postdocs they mentor. Because there is a difference between the responses of tenure-track and tenured faculty members, this once again indicates that the HEPA's culture and job market have evolved over the past three decades. 

The survey results indicate that institutions are not systemically hindering research, outreach, or mentoring for their faculty members. On the other hand, faculty reported that their institutions helps with research much more than outreach or mentoring. Furthermore, more faculty reported that their institution neither helps nor hinders mentoring and outreach compared to research. These results indicate that institutions are not actively helping with faculty outreach or mentoring efforts. Female faculty might also be disproportionately affected by their institutions compared to male faculty, as our results showed more female faculty reporting that their institution hinders their research on some level.

The vast majority of faculty work more than the stereotypical 40-hour work week, and this comes as no surprise based on our findings: A little less than half of the faculty respondents reported committing around 23 hours per week to research at the very least, and a quarter of faculty reported committing at least 23 hours a week to teaching. Overall, the average faculty respondent commits the most amount of time to research, fairly equal amounts of time committed to teaching and mentoring, and far less time to outreach and advocacy -- a hierarchy that seems reasonable but might be unbalanced in terms of actual time committed. When we compared White faculty with faculty in other racial groups, we found that the latter group reported committing more time to research, outreach, and advocacy. Over a quarter of all faculty reported committing time to other career items -- mostly service and administrative work -- and female faculty reported spending more time on these responsibilities compared to male faculty. The faculty members commit similar amounts of time to their administrative duties and their teaching/mentoring, an unbalanced set of commitments. 

Not surprisingly, the survey shows that more tenured faculty take place in hiring compared to tenure-track faculty, where around half report that past or present U.S. policies on Visas had some negative effect on their hiring ability. Since current policies are currently inadequate \cite{geng_2022}, senior members of HEPA should advocate for stronger U.S. visa policies to encourage international collaborations and support more streamlined and inclusive hiring processes.

Compared to other career stages, faculty reported the highest amount of competitiveness at their workplaces and the lowest satisfaction level with the support from their colleagues. While the majority of faculty reported feeling like their salary probably or definitely is adequate, about one-fifth of faculty feel the opposite. A significant portion of the faculty who responded to the 2021 survey face enormous time commitments and responsibilities at work and home while lacking support from their colleagues and the field overall. Funding agencies don't have to play a passive role in light of these results; for example, they could shift funds into block grants in which universities directly pay salaries to their faculty members \cite{woolston_2021}. This could reduce some of the competition faced by faculty who compete for grants, and it also reduces the workload and headache for faculty with administrative or grant/funding time commitments. 

\subsection{The state of scientists in HEPA}
Comparing responses from scientists and senior scientists proved to be a similar exercise compared to their faculty counterparts while also highlighting distinct differences between faculty and scientists on the whole. Scientists and senior scientists reported spending similar amounts of time as Postdocs, and although the difference is not significant, scientists did report spending a couple more months looking for a job compared to senior scientists. Similar to tenure-track faculty, the results from scientists might indicate the high competition currently being experienced in the HEPA job market.

When we compared responses about salaries, scientists ranked the lowest (below senior scientists, tenure-track faculty, and tenured faculty) in terms of salary adequacy, and they also ranked higher than senior scientists in terms of salary inadequacy. The results continue to indicate that scientists in more junior positions are less satisfied with their salary, and the same sentiment is not shared with scientists in more senior positions. Conversely, more senior scientists reported that their salary is adequate at some level compared to the other three groups, and far less scientists and senior scientists reported that their salary is inadequate on some level compared to faculty. This difference could stem from the difference in pay between national labs and universities, and it indicates that national labs might be offering a more reasonable salary compared to universities.

When it comes to comparing time commitments, nearly two-fifths of scientists reported spending almost all of their time on at least one career item, one-fifth more than faculty who made the same report. With forty-three percent of scientists committing at least 22 hours per week to leadership and ninety percent of scientists reporting some leadership responsibilities, the results indicate that national labs provide ample opportunity for leadership opportunities. We also found that scientists older than 50 years old commit less time to leadership compared to those younger than 50, a sensible and reasonable result. More scientists reported committing more time to leadership than research, but this isn't too concerning if these leadership responsibilities bear research-rich fruit. We found differences between the amount of time scientists commit to career-related items across gender and racial lines -- particularly when it came to research, service, and leadership. Female scientists disproportionately reported spending more time on service work compared to male scientists; White scientists reported committing more time to leadership; and scientists in other racial groups reported committing more time to research.

With the other half of their 40-hour work week committed to research, perhaps it's not too surprising that scientists commit far less time to items outside of research and leadership. For example, scientists reported spending less time on mentoring compared to faculty reports, a result that is unsurprising, although dedicated funding for national lab-based student research opportunities could also help increase the mentoring opportunities for scientists. Scientists commit much less time to advocacy than all other career items over the course of a year. Finally, scientists on the whole reported that past or present U.S. policies on Visas had some negative effect on their hiring ability. Considering we saw the same report from faculty members, this indicates that past and present U.S. policies on Visas have affected hiring at both universities and national labs in similar ways. 

Another hint of institutional differences arose when we asked about whether institutions help or hinder outreach and mentoring. Compared to faculty, more scientists reported that their institution hinders outreach and mentoring on some level. Nearly a quarter more female scientists reported that their institution hinders advocacy compared to reports from male scientists, and more male scientists indicated that they didn't know whether their institution helps or hinders advocacy efforts. This indicates that advocacy efforts might not be prioritized by national labs and might even be hindered in some cases, and this hindrance is disproportionately noticed by female scientists.  

\subsection{What we can learn from respondents who left HEPA}
The 2021 survey received a balanced mix of respondents with a variety of age ranges, with a group who recently left HEPA and another group who left decades ago. This allowed us to compare their responses to look for changes in HEPA's job market and culture. The results indicated that younger and recently departed non-academic respondents seem less interested in an academic future: Significantly more non-academic respondents over 40 years old reported that they attempted to find a job in HEPA, and less respondents who left HEPA recently reported looking for an academic job. Networking is important within HEPA and, according to the survey results, also important for those who are planning to leave HEPA. These results are concerning for early career scientists in HEPA who are planning to apply or currently applying for jobs, especially with the reports from early career scientists in HEPA about their dissatisfaction with HEPA and desires to switch sectors. Surveys conducted outside of the Snowmass process show that Postdocs are concerned about their career prospects and lack of support from their supervisor during the pandemic~\cite{nature_news_2020}, and these sentiments could, at worst, lead to a generational abandonment of HEPA as a whole. Institutions need to train early career scientists for a variety of job paths and support professional development throughout graduate programs and job contracts. Collaborations between industry partners and HEPA could offer opportunities for early career scientists, exposing them to both jobs and mentors in industry. 

Finally, we want to comment on the difference between our survey results and the results from AIP \cite{AIP_Statistics_Plot}. Several of the same skills -- for instance, problem-solving, programming, and design -- were reported as valuable by non-academic respondents and also appeared as commonly used skills for new physics PhDs. However, the ``specialized equipment'' skill notably appeared on the AIP list but was not reported as valuable by the non-academics we surveyed. This could be due to a difference between HEPA and other areas of physics academia. 

\subsection{The state of caregivers in HEPA}
The world is full of caregivers, and HEPA is no exception. While the average 2021 survey respondent is not a caregiver, a significant portion of the survey respondents (including early career scientists) had some form of caregiving responsibilities. The majority of caregivers reported caring for children, although some respondents reported caring for seniors, a person with a disability or medical condition, or some combination of those three groups. Disproportionately more female caregivers reported they are sole or primary caregivers compared to male caregivers, while more male caregivers described their responsibilities as part-time caregiving. Perhaps it's not surprising that female caregivers reported slightly less free hours per week on average, although the difference between male and female caregivers is not significant enough to make any strong statements. More female caregivers also indicated substantial career effects because of their caregiving responsibilities, including rearrangement of their work schedule, decrease of hours, or unpaid leave. These results seem to hint that female caregivers are not as widely supported within HEPA, and some of them face extra struggles like sole caregiving at the same time.

When asked about the level of support they receive in academia as a caregiver, the scientists who participated in this survey feel the most support from their peers or immediate working group, and they feel the least amount of support from funding agencies and the overall field. More male caregivers, caregivers at national labs, and white caregivers reported that they don't know whether they feel or felt support from funding agencies. For the primary workplace result, it indicates that caregivers at universities and other institutions where caregivers deal more directly with funding agencies face an extra set of pressure and competition with no additional support to match.

While comparing caregivers' reports about their support from peers versus their place of employment, we found that the responses formed two main groups: caregivers who reported feeling the same levels of support between their peers and their place of employment, and caregivers who reported less support from their place of employment. Recall that faculty also reported the lowest satisfaction level with the support from their colleagues when compared to other career stages -- a troubling result when it seems like some caregivers might need to rely on their peers at times when their place of employment does not provide adequate support. Additionally, the first group of caregivers contains those who reported feeling no support from either their peers or their place of employment, highlighting that institutions can continue to improve how they support scientists with caregiving responsibilities.  

The HEPA field overall is providing inadequate support to many of its caregivers, and according to the 2021 survey results, the lack of support disproportionately affects caregivers across different gender, racial, and age lines. Disproportionately more female caregivers reported feeling little to no support from their place of employment, funding agencies, and the field. More caregivers in other racial groups and more caregivers younger than 50 years old also reported feeling lack of support from funding agencies and the field. Keep in mind that the majority of all caregivers reported little to no support by the field; when broken down by gender, the same proportion of male and female caregivers respectively reported feeling some support as a caregiver by HEPA, and more male caregivers reported that they don't know whether they feel supported by the field. Clearly the field can extend more support to all its scientists with caregiving responsibilities.  

Caregivers in HEPA were profoundly impacted by the COVID-19 pandemic, and they continued to feel these impacts at the time they took the survey. The majority of all caregivers with children reported less overall productivity compared to pre-COVID levels; they also reported significantly more time and effort spent on caregiving responsibilities during the pandemic. More female caregivers, caregivers in other racial groups, and caregivers younger than 50 years old reported spending more time and effort on caregiving compared to pre-COVID time and effort. For caregivers of seniors or people with a disability or medical condition, half of these caregivers reported less overall productivity compared to pre-COVID levels, and over half reported more time and effort spent on caregiving during the COVID-19 pandemic. Caregivers also expressed similar sentiments when asked to comment about their experiences; they also expressed an inability to return to pre-COVID productivity levels despite their best attempts. Finally, caregivers did not widely report additional resources for childcare provided by their institutions during the COVID-19 pandemic.

When comparing caregivers with children in different age groups, the results revealed that more caregivers with younger children reported severe effects due to the COVID-19 pandemic, and caregivers with younger children (notably, young children in school) reported more time and effort spent on caregiving during the COVID-19 pandemic. The results revealed that more caregivers with older children reported the same overall productivity compared to pre-COVID levels; at the same time, more caregivers with children 14 years old or older -- or younger than 6 -- reported the same amount of time and effort spent on caregiving. Overall, more caregivers with children younger than 14 years old reported spending more time and effort on caregiving compared to pre-COVID time and effort. More female caregivers reported having preschool-aged children compared to male caregivers, while more male caregivers reported having children older than 5 years old. All of these results indicate that female caregivers in HEPA are disproportionately affected by the COVID-19 pandemic, impacting overall productivity levels and requiring more time and effort spent on their caregiving responsibilities. 

Several interesting differences arose when we split caregivers' responses by different primary workplaces, notably universities and national labs. More caregivers at universities disproportionately indicated substantial career effects because of their caregiving responsibilities. When broken down by career stage, faculty caregivers reported more or substantial effects due to their caregiving responsibilities compared to scientist caregivers. The COVID-19 pandemic also affected caregivers differently depending on their primary workplaces: Caregivers working at national labs reported spending more time and effort on caregiving compared to pre-COVID time and effort. These results indicate that universities might not be providing adequate support for scientists with caregiving responsibilities.

Faculty and scientists (which made up a large portion of the respondents with caregiving responsibilities) reported working long hours, with significant time commitments toward a variety of career-related items. The average caregiver reported being free of all caregiving responsibilities for about 34 hours per week (far fewer hours than what the majority of faculty and scientists spend working per week), and caregivers at national labs reported several more free hours compared to caregivers at other institutions. Around a quarter of faculty and one-fifth of scientists reported that their salaries are probably or definitely inadequate -- so there's room for institutions to encourage reasonable work hours and increase salary, two effective methods of supporting those with caregiving responsibilities \cite{collins_2021}. 

The average respondent with caregiving responsibilities has access to some form of Family and Medical leaves from their employer, and they are not not suffering any professional setbacks in the event they utilized Family and Medical leave. While the caregiver might have access to Family and Medical leaves, the leaves might be unpaid; some caregivers also reported that their institution does not offer Family and Medical leaves. More caregivers at universities and white caregivers reported that they don't know if their place of employment offers Family and Medical leaves. Disproportionately more male caregivers reported that they didn't know when it came to a few caregiving topics: More male caregivers didn't know if their place of employment offers Family and Medical leaves, whether their employer offered leaves during a time when they could have taken leaves, or whether they suffered any professional setbacks due to their taking of leaves. Caregivers also reported some professional setbacks while utilizing Family and Medical leaves, and this was disproportionately reported by female caregivers, caregivers in other racial groups, and caregivers at national labs. Institutions can offer paid Family and Medical leaves as one means of supporting the scientists with caregiving responsibilities \cite{collins_2021}, and institutions can actively encourage and support the scientists who need to utilize the leave.

With the COVID-19 pandemic predominantly affecting access to childcare and productivity levels, caregivers in HEPA require more support than ever from all angles: Work colleagues, department or institution administrators, and funding agencies. While specific action can be taken by the U.S. government concerning caregivers in all U.S. job sectors \cite{collins_2021}, institutions do not need to wait to implement such changes. Institutions should offer adequate salaries and paid Family and Medical leaves for all employees. Institutions can also support caregivers by subsidizing or offering childcare to their employees. Hiring and promotion committees should fairly evaluate caregivers' drop in productivity in the context of current events. Scientists with caregiving responsibilities make valuable contributions and bring an irreplaceable perspective to HEPA; they should not be left behind by the field. 

The responses and comments from caregivers were deeply honest, personal, and sometimes heartbreaking. We should not ignore or dismiss the real and valid responsibilities that scientists in HEPA face alongside their research commitments.

\section{Conclusions}
\label{sec:conclusions}

The 2021 Snowmass process offered a truly unique experience for its scientists, a process set apart by the COVID-19 pandemic and its padded lifetime. With this context in mind, the 2021 Snowmass Community Survey served to be a comprehensive overview of the people working in and around HEPA instead of a study of present experimental and theoretical ideas and their funding priorities. The SEC Survey Initiative sought to understand human experiences, and those who filled out the survey were able to inform the SEC Survey Initiative about the experiences of different career stages, genders, and caregivers, among other groups. Over the course of several months, the initiative constructed and finalized the questions; the 2021 survey was officially live for nearly two months, collecting nearly 1500 interactions and over 1000 responses over the course of the survey's run-time. The results from the 2021 survey and our conclusions can hopefully be used to inform the Snowmass community about the aspects of HEPA which work and the aspects that require improvement. 

Because the Snowmass process heavily deals with funding prioritization and future directions of the HEPA field, the 2021 survey results can hopefully offer some hints and recommendations for these priorities and directions. The 2021 survey results indicated less of a desire to remain in HEPA, both from the early career scientists currently in HEPA and the non-academics who recently departed HEPA for industry jobs. A majority of faculty and scientists also reported that they have thought about leaving HEPA or their institution. Even if scientists strive to remain in HEPA, they are predominantly concerned about job availability and already feel the competition surrounding them in their current workplace. Non-academics reported that networking (LinkedIn or otherwise) was the most useful skill in obtaining their first non-academic job, a skill that might not be adequately developed for early career scientists who ultimately leave HEPA. Institutions should provide early career scientists opportunities and training for professional development skills (e.g., negotiating salaries), so that they can effectively prepare for variety of job sectors beyond HEPA. Funding agencies can also create opportunities for early career scientists to collaborate with industry partners, offering a bridge between HEPA and industry while also exposing early career scientists to a larger diversity of job sectors and mentors. 

While various funding opportunities for outreach efforts already exist and are being utilized by those who responded to the survey, institutions might not be advertising and supporting their senior faculty and scientists in accessing those opportunities. Knowing that faculty and scientists currently juggle several key responsibilities that carry demanding time commitments, perhaps it's unsurprising that they commit far less time to outreach compared to their research or mentoring responsibilities. However, faculty and scientists reported far less help (and sometimes hindrance) by their institutions when it comes to outreach efforts. These results suggest there's room for improvement on an institutional level when it comes to helping faculty and scientists in all endeavours of their career, especially outreach. Respondents told us about the funding opportunities they already pursue in areas of their career like outreach and funded undergraduate student research; these could be starting places for others, or inspiration for future opportunities offered by funding agencies.

The survey revealed that more faculty and scientists spend no time on advocacy compared to all other career items asked about. Now more than ever there is much to advocate for; faculty and scientists should increase their advocacy efforts in support of the early career scientists they interact with, mentor, and advise. For example, many Postdocs expressed concerns about immigration issues in several areas throughout the survey, and senior members of HEPA should advocate for updated U.S. visa policies which support and encourage international collaboration. Faculty and scientists reported past and present U.S. visa laws have a similar effect on hiring, so senior members at universities and national labs alike have a reason increase their advocacy efforts. Finally, current advocacy opportunities like traveling to Washington, D.C. already have strong interest from the community according to the survey results, and those opportunities should be explored and promoted by faculty and scientists.

Of course, the COVID-19 pandemic had an inseparable effect on all of the results of the 2021 survey. In addition to the direct effects on the Snowmass process (both in the timeline and the nature of the work), the pandemic fundamentally altered the every day lives and work of all the members of the HEPA field. Our results found that these effects---some temporary, others permanent and career-altering---had varying impacts across different groups and career stages, while the levels of support offered by institutions were similarly varied. Caregivers in particular were heavily impacted, and it may be a few years before the true extent of the impacts of the COVID-19 pandemic can be properly assessed. In particular, we found varied responses on how frequently researchers would prefer to work from home moving forward, which could indicate a shift in how the workplace is utilized for members of the field in the future.

In examining human experiences in the 2021 survey, we found so much honesty and openness within the Snowmass community. People were honest about how much the COVID-19 pandemic has affected their career and personal life; they were honest about the experiences they faced involving harassment and racism; and they were honest about where they think the field is heading over the next decade. The survey uncovered unique perspectives and experiences for different career stages, primary workplaces, caregiving responsibilities, and across gender and racial lines. Yet the survey also revealed many similarities among the members of the Snowmass community, including concerns about salary, HEPA's environmental impact, and directions in which they would like the field to progress. Finally, many 2021 survey respondents reported feeling a lack of support in various places -- between peers and work colleagues, from an advisor, at the institutional level, or by the field as a whole. We each face a unique set of challenges for however long we work in HEPA, and on an individual level, we can extend more sympathy and understanding to our immediate working group, students, and other collaborators. Individual cultural changes are not enough -- systemic changes should also be made to provide effective support for everyone, especially for early career scientists and those with caregiving responsibilities. By exploring the varied experiences of scientists in and out of HEPA, the 2021 survey exposed many concerns and disparities that should be addressed through meaningful action and cultural changes. At the time of the next Snowmass process, the next survey can check for improvements in these areas, an optimistic indication that the field is committed to assisting all of its scientists.

\section*{Acknowledgments}

We are grateful to all the members of the Snowmass Community who took the time to respond to the survey. In various early stages of the survey's development, many gave us their thoughts on how we might better develop and improve the 2021 survey before distribution to the broader Snowmass and general particle physics community. We would like to thank some of them here directly: The Snowmass Early Career Diversity, Equity, and Inclusion Initiative, Felix King, Elizabeth Worcester, Diana Mendez, and Jason Stock.

JLB was supported partially supported by the Visiting Scholars Award Program of the Universities Research Association. Any opinions, findings, and conclusions or recommendations expressed in this material are those of the authors and do not necessarily reflect the views of the Universities Research Association, Inc. JLB would also like to thank The University of Tennessee at Knoxville Department of Physics and Astronomy, as well as the Zuckerman Institute, for partial support of this work at various later stages. EC was supported for this work by the U.S. Department of Energy Office of High Energy Physics. MESP is funded by the Deutsche Forschungsgemeinschaft (DFG, German Research Foundation) under Germany's Excellence Strategy -- EXC 2121 ``Quantum Universe'' -- 390833306. SH is supported in part by the DOE Grant DE-SC0013607 and in part by the Alfred P. Sloan Foundation Grant No. G-2019-12504.

\newpage
\appendix
\newpage

\section{Survey Questions}
\label{appendix_surveyQs}

Italicized notes indicate when display logic was used. Bracketed notes provide more information about the question. Questions added by the SEC CEF groups have been marked.

\begin{multicols}{2}

\begin{itemize}
    \item Are you currently in academia (i.e, teaching or conducting research within a university or national lab)? 
    \begin{itemize}
        \itemsep0em
        \item Yes
        \item No
        \item Prefer not to respond
    \end{itemize}
    \item \textit{(In academia)} What is your current position? Choose the one that is closest to your position:
    \begin{itemize}
        \itemsep0em
        \item Undergraduate student
        \item Master's student/candidate
        \item PhD student/candidate 
        \item PostDoc 
        \item Engineer  
        \item Technician  
        \item Teaching faculty  
        \item Tenure-track faculty  
        \item Tenured faculty 
        \item Retired faculty  
        \item Scientist  
        \item Senior scientist  
        \item Other (please specify)
        \item Prefer not to answer
    \end{itemize}
    \item How long have you been in your current position? Please indicate the approximate number of months or years.
    \item Are you pursuing an eventual permanent position in the U.S.?
    \begin{itemize}
        \itemsep0em
        \item Already in a permanent position
        \item Yes
        \item Maybe
        \item I haven't thought about it
        \item No
    \end{itemize}
    \item \textit{(Not pursuing eventual permanent position in U.S.)} Feel free to specify your desired place of future work/residence. [Answer options were a list of 196 locations.]
    \item Where were you first exposed to the subject of High Energy Physics \& Astrophysics (HEPA)?\footnotemark{}
    \begin{itemize}
        \itemsep0em
        \item High school or earlier
        \item Undergraduate institution
        \item Graduate school 
        \item Internship/Work experience 
        \item Informal Education Settings and Experiences (such as NOVA shows or museums) 
        \item National Laboratory Visits or Tours 
        \item Online (YouTube, podcasts, etc.)
        \item Other (please specify)
    \end{itemize}
    \item Was that exposure to HEPA a significant influence in your career choice?\footnotemark[\value{footnote}] 
    \begin{itemize}
        \itemsep0em
        \item Yes, it was the primary reason that I chose my career in academia.
        \item Yes, it was the primary reason that I chose my career in industry/non-academia.
        \item It somewhat (not actively) shaped my decision. 
        \item No, my career choice was not significantly influenced by this first exposure to the subject.
    \end{itemize}
    \item From the education you received in the course of your undergraduate and graduate studies, please indicate which career paths you were trained for.\footnotemark[\value{footnote}] 
    \begin{itemize}
        \itemsep0em
        \item Academic careers in HEPA
        \item Academic careers NOT in HEPA
        \item Industry or other non-academic paths
        \item Other (please specify)
    \end{itemize}
    \item \textit{(Undergraduate student)} Please indicate your plans for the future:
    \begin{itemize}
        \itemsep0em
        \item I'm planning to apply to graduate school
        \item I have accepted or plan to accept a graduate school offer 
        \item I'm planning to apply for jobs 
        \item I have accepted or plan to accept a job offer 
        \item I'm planning on taking a break 
        \item I don't know  
        \item Other (please specify)
    \end{itemize}
    \item \textit{(Undergraduate student, applying to jobs or have accepted a job offer)} Which job sector do you plan to enter or have accepted a position in?
    \begin{itemize}
        \itemsep0em
        \item Government sector
        \item Industry (STEM)  
        \item Industry (other)  
        \item Finance  
        \item Non-STEM  
        \item Business/entrepreneurship  
        \item Other (please specify)
    \end{itemize}
    \item \textit{(Undergraduate student, planning to apply to graduate school or accepted a graduate school offer)} What do you want to study in graduate school?
    \begin{itemize}
        \itemsep0em
        \item Physics
        \item Math 
        \item Engineering 
        \item Computer Science 
        \item Astrophysics/Astronomy  
        \item Other (please specify)
    \end{itemize}
    \item \textit{(Undergraduate student)} Do you currently participate in research?
    \begin{itemize}
        \itemsep0em
        \item Yes
        \item No
    \end{itemize}
    \item \textit{(Undergraduate student)} Have you ever traveled for a conference?
    \begin{itemize}
        \itemsep0em
        \item Yes
        \item No 
    \end{itemize}
    \item \textit{(Undergraduate student, participates in research)} When do you participate in research?
    \begin{itemize}
        \itemsep0em
        \item During the academic year
        \item During the summer
        \item All year
    \end{itemize}
    \item \textit{(Undergraduate student, participates in research)} How many hours a week do you typically commit to research? Please enter an integer number.
    \item \textit{(Undergraduate student, participates in research)} How are you compensated for your research? Select all that apply:
    \begin{itemize}
        \itemsep0em
        \item I am paid
        \item I receive course credit  
        \item I am not compensated 
        \item Other (please specify)
    \end{itemize}
    \item \textit{(Undergraduate student, participates in research)} Do you enjoy the research you participate in?
    \begin{itemize}
        \itemsep0em
        \item Definitely enjoy 
        \item Somewhat enjoy  
        \item Neither like nor dislike  
        \item Somewhat dislike  
        \item Definitely dislike
    \end{itemize}
    \item \textit{(Undergraduate student, participates in research)} How did you find out or get into your research group? Select all that apply:
    \begin{itemize}
        \itemsep0em
        \item Professor recruited me  
        \item I contacted a professor  
        \item Department or university-led effort  
        \item Other (please specify)
    \end{itemize}
    \item \textit{(Master's student, PhD student, or PostDoc)} Please indicate your current career plans:
    \begin{itemize}
        \itemsep0em
        \item I plan to accept or have accepted a job offer (in academia, industry, etc.) within the last year 
        \item I'm currently applying for or have applied for jobs (in academia, industry, etc.) within the last year  
        \item I haven’t applied for jobs recently but plan to (in academia, industry, etc.)
        \item \textit{(Master's student)} I'm planning on getting a PhD in Physics
        \item \textit{(Master's student)} I'm planning on continuing graduate or professional study in another field
        \item I'm planning on taking a break
        \item I don't know
        \item Other (please specify)
    \end{itemize}
    \item \textit{(Master's student, PhD student, or PostDoc who is planning to apply or currently applying for jobs)} How prepared or informed do you feel to pursue jobs outside HEPA?\footnote{Contributed by the SEC CEF groups. }
    \begin{itemize}
        \itemsep0em
        \item Not at all prepared or informed (1) 
        \item 2 
        \item 3 
        \item 4  
        \item Very well prepared or informed (5)
        \item I don't know/not applicable
    \end{itemize}
    \item \textit{(Master's student, PhD student, or PostDoc who is planning to accept or accepted a job offer)} What sector did you accept or plan to accept an offer in?
    \begin{itemize}
        \itemsep0em
        \item \textit{(PhD student or PostDoc)} Post-doctoral position (including national labs)
        \item \textit{(PhD student or PostDoc)} Non tenure-track research faculty at university
        \item \textit{(PhD student or PostDoc)} Non tenure-track research faculty at national lab
        \item \textit{(PhD student or PostDoc)} Tenure-track research faculty
        \item Teaching faculty
        \item Government sector  
        \item Industry (STEM) 
        \item Industry (other)  
        \item Finance  
        \item Business/Entrepreneurship  
        \item Not for employment  
        \item Non-professorial or non-tenure research-based position  
        \item Other (please specify)
    \end{itemize}
    \item \textit{(Master's student, PhD student, or PostDoc who is planning to accept or accepted a job offer)} Why did you choose this sector? Please select all that apply.
    \begin{itemize}
        \itemsep0em
        \item Compensation 
        \item Personal or family-related reasons 
        \item Intellectually challenging/found it interesting 
        \item Job availability  
        \item Freedom in research topics  
        \item Possibility for career advancement  
        \item Opportunity to work with a specific person  
        \item Switch to different sector  
        \item Necessary step to get future position  
        \item Visa restrictions limited my options  
        \item More inclusive working environment  
        \item Other reason (please specify)
    \end{itemize}
    \item \textit{(Master's student, PhD student, or PostDoc who is planning to accept or accepted a job offer)} What other sectors did you apply to (select all that apply):
    \begin{itemize}
        \itemsep0em
        \item \textit{(PhD student or PostDoc)} Post-doctoral position (including national labs)
        \item \textit{(PhD student or PostDoc)} Research faculty at university
        \item \textit{(PhD student or PostDoc)} Research faculty at national lab
        \item \textit{(PhD student or PostDoc)} Tenure-track research faculty
        \item Teaching faculty
        \item Government sector 
        \item Industry (STEM)  
        \item Industry (other)  
        \item Finance  
        \item Business/Entrepreneurship  
        \item Not for employment  
        \item I didn't apply to other sectors  
        \item Other (please specify)
    \end{itemize}
    \item \textit{(Master's student, PhD student, or PostDoc who is planning to apply or currently applying for jobs)} Please indicate how likely you are to apply for jobs in the following sectors:
    \begin{itemize}
        \itemsep0em
        \item Research academia at university
        \item Research academia at national lab
        \item Teaching academia
        \item Government sector
        \item Industry (STEM)
        \item Industry (other)
        \item Finance
        \item Business/Entrepreneurship
        \item Other (please specify)
    \end{itemize}
    \item \textit{(Master's student, PhD student, or PostDoc who is planning to apply or currently applying for jobs)} Please comment on why you picked your top sector.
    \item \textit{(Master's student, PhD student, or PostDoc who is planning to apply or currently applying for jobs, and who rated academia sectors as likely to apply)} Please indicate which of the following career-related concerns you find the most important to you and your future in HEPA.
    \begin{itemize}
        \itemsep0em
        \item Availability of university-based jobs
        \item Availability of laboratory-based jobs
        \item Funding for large-scale, long lead time experiments in the future
        \item Funding for small-scale, short lead time experiments in the future
        \item Proximity to your home institution of future planned experiments
        \item Bureaucracy and administrative difficulties to conducting research
        \item Immigration concerns
        \item Other (please specify) 
    \end{itemize}
    \item \textit{(Master's student, PhD student, or PostDoc who is planning to apply or currently applying for jobs, and who rated industry sectors as likely to apply)} Please indicate which of the following career-related concerns you find the most important to you and your future in industry.
    \begin{itemize}
        \itemsep0em
        \item Availability of industry-based jobs
        \item Bureaucracy and administrative difficulties to conducting your work
        \item Losing your job/failing to advance
        \item Immigration concerns
        \item Other (please specify)
    \end{itemize}
    \item \textit{(Undergraduate student, Master's student, PhD student, or PostDoc)} How satisfied are you with the amount of career mentorship you receive?\footnotemark{}
    \begin{itemize}
        \itemsep0em
        \item Very unsatisfied  (1) 
        \item 2 
        \item 3  
        \item 4  
        \item Very satisfied (5)  
        \item I don't know/not applicable
    \end{itemize}
    \item \textit{(Undergraduate student, Master's student, PhD student, or PostDoc)} What type of career mentorship is available to you? Select all that apply [Answer options were ``From your supervisor'' and ``From other mentors'']:\footnotemark[\value{footnote}] 
    \begin{itemize}
        \itemsep0em
        \item Research-dominated HEPA career paths (academia, national laboratories)
        \item Teaching-dominated HEPA career paths
        \item Private or government sector career paths
        \item Other (please specify)
    \end{itemize}
    \item \textit{(Undergraduate student, Master's student, PhD student, or PostDoc)} Would you like to participate in projects that connect HEPA and Industry partners?\footnotemark[\value{footnote}]
    \begin{itemize}
        \itemsep0em
        \item Yes
        \item Maybe
        \item No
        \item I don't know
    \end{itemize}
    \item \textit{(Undergraduate student, Master's student, PhD student, or PostDoc who answered the previous question with ``Yes'')} Why, and in which cases would you be interested? Choose all that apply:\footnotemark[\value{footnote}]
    \begin{itemize}
        \itemsep0em
        \item This would be beneficial to my career  
        \item Only if it does not require a significant time commitment  
        \item Only if the costs of the project are covered by HEPA collaborations and/or Industry 
        \item Other (please specify)
    \end{itemize}
    \footnotetext{Contributed by the SEC CEF groups.}
    \item \textit{(Teaching faculty, tenure-track faculty, tenured faculty, retired faculty, scientist, or senior scientist)} Indicate the approximate amount of time for the following items by entering an integer number:
    \begin{itemize}
        \itemsep0em
        \item Number of \textbf{years} you spent as a postdoc
        \item Number of \textbf{months} you spent looking for a faculty/scientist position
    \end{itemize}
    \item \textit{(Teaching faculty, tenure-track faculty, tenured faculty, or retired faculty)} Please indicate the number of students and PostDocs you currently have in your research group:
    \begin{itemize}
        \itemsep0em
        \item Undergraduate students
        \item Master's student/candidates
        \item PhD students/candidates
        \item PostDocs
    \end{itemize}
    \item \textit{(Tenure-track faculty or tenured faculty)} Approximately what percentage of your students and PostDocs were/are employed after graduation?
    \begin{itemize}
        \itemsep0em
        \item Master's student/candidates
        \item PhD students/candidates
        \item PostDocs
    \end{itemize}
    \item \textit{(Tenure-track faculty, tenured faculty, or retired faculty)} Do you employ undergraduate students during the academic year, the summer, or both?
    \begin{itemize}
        \itemsep0em
        \item Academic year 
        \item Summer 
        \item Both 
        \item I don't employ undergraduate students  
        \item I can't employ to hire undergraduate students
    \end{itemize}
    \item \textit{(Tenure-track faculty, tenured faculty, or retired faculty who employ undergraduate students)} How do you compensate your undergraduate student researchers? Please select all that apply:
    \begin{itemize}
        \itemsep0em
        \item Pay them 
        \item Course credit  
        \item They are not compensated  
        \item Other (please specify)
    \end{itemize}
    \item \textit{(Teaching faculty, tenure-track faculty, tenured faculty, or retired faculty)} On average, how much time do you spend on the following career-related items over the course of a year?
    \begin{itemize}
        \itemsep0em
        \item Research
        \item Teaching
        \item Mentoring students
        \item Outreach
        \item Advocacy
        \item Other work-related items (please specify)
    \end{itemize}
    \item \textit{(Teaching faculty, tenure-track faculty, tenured faculty, or retired faculty)} Have you always been in academia, or do you have experience outside of the field?
    \begin{itemize}
        \item I have always been in academia 
        \item I have experience outside of the field (please indicate the number of years)
    \end{itemize}
    \item \textit{(Teaching faculty, tenure-track faculty, tenured faculty, or retired faculty)} Have you ever considered leaving academia?
    \begin{itemize}
        \itemsep0em
        \item Yes
        \item No
    \end{itemize}
    \item  \textit{(Teaching faculty)} Are you interested in (or have you been) continuing research in HEPA?
    \begin{itemize}
        \itemsep0em
        \item  Yes, continuing research 
        \item Yes, interested in continuing research  
        \item Maybe  
        \item No  
        \item I don't know
    \end{itemize}
    \item \textit{(Teaching faculty)} If given the opportunity, would you (or are you actively trying to) switch to a tenure track faculty position?
    \begin{itemize}
        \itemsep0em
        \item Yes
        \item Maybe
        \item No
        \item I don't know
    \end{itemize}
    \item \textit{(Teaching faculty, tenure-track faculty, tenured faculty, or retired faculty)} Please indicate whether your institution helps or hinders the following career-related items:
    \begin{itemize}
        \itemsep0em
        \item Research
        \item Outreach
        \item Mentoring
    \end{itemize}
    \item \textit{(Teaching faculty, tenure-track faculty, or tenured faculty)} Would you prefer a reduced teaching load?
    \begin{itemize}
        \item Definitely yes  
        \item Probably yes 
        \item Probably not  
        \item Definitely not  
        \item I don't know/not applicable
    \end{itemize}
    \item \textit{(Retired faculty)} Would you have preferred a reduced teaching load?
    \begin{itemize}
        \itemsep0em
        \item Definitely yes  
        \item Probably yes  
        \item Probably not  
        \item Definitely not  
        \item I don't know/not applicable
    \end{itemize}
    \item \textit{(PhD student, PostDoc, teaching faculty, tenure-track faculty, tenured faculty, or retired faculty)} Do you feel like your salary is adequate? 
    \begin{itemize}
        \itemsep0em
        \item Definitely yes 
        \item Probably yes 
        \item Might or might not be adequate  
        \item Probably not  
        \item Definitely not 
    \end{itemize}
    \item \textit{(Master's student, PhD student, PostDoc, teaching faculty, tenure-track faculty, or tenured faculty)} Referring to the last round of interviews you've been through: on average, how “connected” did you feel to the local research groups?\footnotemark{}
    \begin{itemize}
        \itemsep0em
        \item Not at all connected (1)
        \item 2  
        \item 3  
        \item 4  
        \item Very well connected (5)
        \item I don't know/not applicable
    \end{itemize}
    \item \textit{(Master's student, PhD student, PostDoc, teaching faculty, tenure-track faculty, or tenured faculty who answered ``Not at all connected (1)'' on the previous question)} What issues made you feel less connected? Select all that apply:\footnotemark[\value{footnote}]
    \begin{itemize}
        \item Differences in cultures/backgrounds 
        \item Language barrier  
        \item Due to a specific person/interaction 
        \item Discrimination to race/gender/identity  
        \item I don't know  
        \item Other (please specify)
    \end{itemize}
    \footnotetext{Contributed by the SEC CEF groups.}
    \item \textit{(Scientist or senior scientist)} How many students (Master's, PhD, PostDocs) do you currently mentor for their research?
    \begin{itemize}
        \itemsep0em
        \item Master's student/candidates
        \item PhD students/candidates
        \item PostDocs
    \end{itemize}
    \item \textit{(Scientist or senior scientist)} How many undergraduate students have you mentored?
    \item \textit{(Scientist or senior scientist)} Approximately what percentage of the students and PostDocs you mentor are employed?
    \begin{itemize}
        \itemsep0em
        \item Master's student/candidates
        \item PhD students/candidates
        \item PostDocs
    \end{itemize}
    \item \textit{(Scientist or senior scientist)} On average, how much time do you spend on the following career-related items over the course of a year?
    \begin{itemize}
        \itemsep0em
        \item Research
        \item Mentoring students
        \item Outreach 
        \item Leadership/coordination roles
        \item Advocacy
        \item Professional service roles (committees, Snowmass, workshop organizing, etc.)
        \item Other work-related items (please specify)
    \end{itemize}
    \item \textit{(Scientist or senior scientist)} Have you always been at a national lab, or do you have experience outside of the field?
    \begin{itemize}
        \itemsep0em
        \item I have always been at a national lab
        \item I have experience outside of the field
    \end{itemize}
    \item \textit{(Scientist or senior scientist)} While at this job, have you ever thought about leaving the lab?
    \begin{itemize}
        \itemsep0em
        \item Yes
        \item No
    \end{itemize}
    \item \textit{(Scientist or senior scientist)} Please indicate whether your institution helps or hinders the following career-related items:
    \begin{itemize}
        \itemsep0em
        \item Outreach
        \item Mentoring
        \item Advocacy
    \end{itemize}
    \item \textit{(Scientist or senior scientist)} Do you feel like your salary (from the institution) is adequate?
    \begin{itemize}
        \itemsep0em
        \item Definitely yes
        \item Probably yes  
        \item Might or might not be adequate  
        \item Probably not  
        \item Definitely not
    \end{itemize}
    \item \textit{(Teaching faculty, tenure-track faculty, tenured faculty, retired faculty, scientist, or senior scientist)} Since you started taking part in hiring decisions, have past or present U.S. policies on visas affected your ability to hire talented individuals?
    \begin{itemize}
        \itemsep0em
        \item Negatively affected (1)
        \item 2 
        \item No effect (3)
        \item 4 
        \item Positively affected (5)
        \item I don't know  
        \item N/A
    \end{itemize}
    \item \textit{(Engineer or technician)} Please indicate which of the following career-related concerns you find the most important to you and your job: 
    \begin{itemize}
        \itemsep0em
        \item Bureaucracy and administrative difficulties to conducting your work
        \item Losing your job/failing to advance
        \item Other (please specify)
    \end{itemize}
    \item \textit{(Teaching faculty, tenure-track faculty, tenured faculty, retired faculty, scientist, or senior scientist)} Does your institute/lab have the infrastructure to collaborate for long-term research with faculty members and students at a non-research/undergraduate-only university? Select all that apply:\footnote{Contributed by the SEC CEF groups.}
    \begin{itemize}
        \itemsep0em
        \item Yes, I have experience with this and find it beneficial to both universities. 
        \item Yes, but only for faculty members whose institution is near mine.  
        \item Yes, but only for faculty members who have already obtained funding for themselves and any students.  
        \item Yes, but I don’t know how it could work and would need guidance.  
        \item No  
        \item Not applicable  
        \item Other (please specify)
    \end{itemize}
    \item \textit{(Teaching faculty, tenure-track faculty, tenured faculty, retired faculty, scientist, or senior scientist)} How important do you feel is it for HEPA supervisors to prepare trainees or students for career paths outside HEPA?\footnotemark{}
    \begin{itemize}
        \itemsep0em
        \item Very unimportant (1)
        \item 2 
        \item 3 
        \item 4 
        \item Very important (5)
        \item I don't know
    \end{itemize}
    \item \textit{(Teaching faculty, tenure-track faculty, tenured faculty, retired faculty, scientist, or senior scientist)} How much of your mentorship is focused on preparing the student/trainee to pursue career paths outside HEPA?\footnotemark[\value{footnote}]
    \begin{itemize}
        \itemsep0em
        \item None (1)
        \item 2 
        \item 3 
        \item 4 
        \item A significant fraction (5)
    \end{itemize}
    \item \textit{(Teaching faculty, tenure-track faculty, tenured faculty, retired faculty, scientist, or senior scientist)} Would you support your student’s or trainee's participation in a HEPA-Industry partnership and mobility program? Select all that apply:\footnotemark[\value{footnote}]
    \begin{itemize}
        \itemsep0em
        \item Yes, I would support my student’s or trainee's participation as part of their HEP training
        \item Yes, I would support my student’s or trainee's participation, but only if the program has separate funding for labor or travel costs  
        \item Yes, I would support my student’s or trainee's participation, but only for short amounts of time 
        \item No 
        \item I don't know 
        \item Other (please specify)
    \end{itemize}
    \footnotetext{Contributed by the SEC CEF groups.}
    \item \textit{(Engineer or technician)} Do you plan to switch jobs in the near future?
    \begin{itemize}
        \itemsep0em
        \item Definitely yes 
        \item Probably yes  
        \item Might or might not  
        \item Probably not  
        \item Definitely not 
        \item I don't know
    \end{itemize}
    \item \textit{(Not in academia)} At what capacity were you most recently involved with in HEPA?
    \begin{itemize}
        \itemsep0em
        \item Undergraduate student 
        \item Master's student/candidate  
        \item PhD student/candidate  
        \item PostDoc  
        \item Engineer  
        \item Technician 
        \item Teaching faculty  
        \item Tenure-track faculty  
        \item Tenured faculty  
        \item Scientist or senior scientist (national lab or university)  
        \item Other (please specify)
    \end{itemize}
    \item \textit{(Not in academia)} Please indicate the number of years for the following items (if either have lasted less than a year, enter 0):
    \begin{itemize}
        \itemsep0em
        \item How long you were involved in HEPA
        \item How long since you've started your first non-academic job
    \end{itemize}
    \item \textit{(Not in academia)} What field do you work in now?
    \begin{itemize}
        \itemsep0em
        \item Government sector 
        \item Industry (STEM) 
        \item Industry (other) 
        \item Finance  
        \item Non-STEM  
        \item Business/Entrepreneurship  
        \item Retired/not employed  
        \item Other (please specify)
    \end{itemize}
    \item \textit{(Not in academia)} Before going into your current field of work, did you attempt to find a job in academia?
    \begin{itemize}
        \itemsep0em
        \item Yes
        \item No
    \end{itemize}
    \item \textit{(Not in academia)} Please indicate resources that you found useful in obtaining your first non-academia job (select all that apply): \label{test}
    \begin{itemize}
        \item Networking/LinkedIn  
        \item Other online resources (specify a website here)
        \item Coworkers  
        \item Seminars/webinars  
        \item Career fairs  
        \item Meet-ups/hackathons  
        \item Internships/training programs  
        \item Books  
        \item Other resource (please specify)
    \end{itemize}
    \item \textit{(Not in academia)} Please indicate how valuable the skills learned during your HEPA experience are to you in your current job:
    \begin{itemize}
        \itemsep0em
        \item Solving technical problems
        \item Programming
        \item Design and development 
        \item Basic physics principles
        \item Applied research
        \item Advanced physics principles
        \item Basic research
        \item Advanced math
        \item Using specialized equipment 
        \item Simulation and modeling
        \item Quality control
        \item Tech support
    \end{itemize}
    \item \textit{(Not in academia)} Would you be interested in returning to HEPA for any reason? Choose all that apply:\footnote{Contributed by the SEC CEF groups.}
    \begin{itemize}
        \itemsep0em
        \item Yes, I would return full-time if the compensation was competitive with the private sector 
        \item Yes, if research-only positions were available (e.g., research scientist)  
        \item Yes, if I am able to continue my work with my current employer (i.e., work part-time in HEPA)  
        \item Yes, I would be interested in contributing to HEPA research while continuing to work in industry if access were available 
        \item No, I am no longer interested in a HEPA academic career 
        \item I don't know 
        \item Other (please specify)
    \end{itemize}
    \item Please indicate your interest and involvement in the Snowmass Frontiers (or most recently involved in if you are no longer active in HEPA). Please leave blank if you're not interested in the frontier. [Answer options were ``Working on'' and ``Personally interested in, but not working on'']
    \begin{itemize}
        \itemsep0em
        \item Accelerator Frontier
        \item Cosmic Frontier
        \item Community Engagement Frontier
        \item Computational Frontier
        \item Energy Frontier
        \item Instrumentation Frontier
        \item Neutrino Frontier
        \item Rare Processes and Precision Measurements Frontier
        \item Theory Frontier
        \item Underground Facilities Frontier
    \end{itemize}
    \item \textit{(Currently in academia)} Each year representatives of the HEPA community travel to Washington, D.C. to communicate with the United States Congress the value of our research and to lobby for continued funding. Please indicate your knowledge and participation in this HEPA advocacy:
    \begin{itemize}
        \itemsep0em
        \item To what extent do you know about this HEPA advocacy?
        \item To what extent have you participated in this HEPA advocacy?
        \item Are you interested in participating in this HEPA advocacy?
    \end{itemize}
    \item \textit{(Indicated involvement in at least one Snowmass Frontier)} Please indicate your interest and involvement in various topical groups within the Snowmass Frontiers you previously selected (leave blank if you're not interested). [Answer options were ``Working on'' and ``Personally interested in, but not working on'']
    \begin{itemize}
        \itemsep0em
        \item \textit{Indicated involvement in the Accelerator Frontier:}
        \begin{itemize}
            \itemsep0em
            \item AF01: Beam Physics and Accelerator Education
            \item AF02: Accelerators for Neutrinos
            \item AF03: Accelerators for EW/Higgs
            \item AF04: Multi-TeV Colliders
            \item AF05: Accelerators for PBC and Rare Processes
            \item AF06: Advanced Accelerator Concepts
            \item AF07: Accelerator Technology R\&D
        \end{itemize}
        \item \textit{Indicated involvement in the Cosmic Frontier:}
        \begin{itemize}
            \itemsep0em
            \item CF01: Dark Matter: Particle-like
            \item CF02: Dark Matter: Wave-like
            \item CF03: Dark Matter: Cosmic Probes
            \item CF04: Dark Energy and Cosmic Acceleration: The Modern Universe
            \item CF05: Dark Energy and Cosmic Acceleration: Cosmic Dawn and Before
            \item CF06: Dark Energy and Cosmic Acceleration: Complementarity of Probes and New Facilities
            \item CF07: Cosmic Probes of Fundamental Physics
        \end{itemize}
        \item \textit{Indicated involvement in the Community Engagement Frontier:}
        \begin{itemize}
            \itemsep0em
            \item CommF01: Applications \& Industry
            \item CommF02: Career Pipeline \& Development
            \item CommF03: Diversity \& Inclusion
            \item CommF04: Physics Education
            \item CommF05: Public Education \& Outreach
            \item CommF06: Public Policy and Government Engagement
        \end{itemize}
        \item \textit{Indicated involvement in the Computational Frontier:}
        \begin{itemize}
            \itemsep0em
            \item CompF01: Experimental Algorithm Parallelization
            \item CompF02: Theoretical Calculations and Simulation
            \item CompF03: Machine Learning
            \item CompF04: Storage and processing resource access (Facility and Infrastructure R\&D)
            \item CompF05: End user analysis
            \item CompF06: Quantum computing
            \item CompF07: Reinterpretation and long-term preservation of data and code
        \end{itemize}
        \item \textit{Indicated involvement in the Energy Frontier:}
        \begin{itemize}
            \itemsep0em
            \item EF01: EW Physics: Higgs Boson properties and couplings
            \item EF02: EW Physics: Higgs Boson as a portal to new physics
            \item EF03: EW Physics: Heavy flavor and top quark physics
            \item EF04: EW Physics: EW Precision Physics and constraining new physics
            \item EF05: QCD and strong interactions: Precision QCD
            \item EF06: QCD and strong interactions: Hadronic structure and forward QCD
            \item EF07: QCD and strong interactions: Heavy Ions
            \item EF08: BSM: Model specific explorations
            \item EF09: BSM: More general explorations
            \item EF10: BSM: Dark Matter at colliders
        \end{itemize}
        \item \textit{Indicated involvement in the Instrumentation Frontier:}
        \begin{itemize}
            \itemsep0em
            \item IF01: Quantum Sensors
            \item IF02: Photon Detectors
            \item IF03: Solid State Detectors and Tracking
            \item IF04: Trigger and DAQ
            \item IF05: Micro Pattern Gas Detectors (MPGDs)
            \item IF06: Calorimetry
            \item IF07: Electronics/ASICs
            \item IF08: Noble Elements
            \item IF09: Cross Cutting and Systems Integration
        \end{itemize}
        \item \textit{Indicated involvement in the Neutrino Frontier:}
        \begin{itemize}
            \itemsep0em
            \item NF01: Neutrino Oscillations
            \item NF02: Sterile Neutrinos
            \item NF03: BSM
            \item NF04: Neutrinos from natural sources
            \item NF05: Neutrino properties
            \item NF06: Neutrino Interaction Cross Sections
            \item NF07: Applications
            \item NF08: Theory of Neutrino Physics
            \item NF09: Artificial Neutrino Sources
            \item NF10: Neutrino Detectors
        \end{itemize}
        \item \textit{Indicated involvement in the Rare Processes and Precision Measurements Frontier:}
        \begin{itemize}
            \itemsep0em
            \item RPF01: Weak decays of b and c quarks
            \item RPF02: Weak decays of strange and light quarks
            \item RPF03: Fundamental Physics in Small Experiments
            \item RPF04: Baryon and Lepton Number Violating Processes
            \item RPF05: Charged Lepton Flavor Violation (electrons, muons and taus)
            \item RPF06: Dark Sector Studies at High Intensities
            \item RPF07: Hadron Spectroscopy
        \end{itemize}
        \item \textit{Indicated involvement in the Theory Frontier:}
        \begin{itemize}
            \itemsep0em
            \item TF01: String theory, quantum gravity, black holes
            \item TF02: Effective field theory techniques
            \item TF03: CFT and formal QFT
            \item TF04: Scattering amplitudes
            \item TF05: Lattice gauge theory
            \item TF06: Theory techniques for precision physics
            \item TF07: Collider phenomenology
            \item TF08: BSM model building
            \item TF09: Astro-particle physics \& cosmology
            \item TF10: Quantum Information Science
            \item TF11: Theory of neutrino physics
        \end{itemize}
        \item \textit{Indicated involvement in the Underground Facilities Frontier:}
        \begin{itemize}
            \itemsep0em
            \item UF01: Underground Facilities for Neutrinos
            \item UF02: Underground Facilities for Cosmic Frontier
            \item UF03: Underground Detectors
            \item UF04: Supporting Capabilities
            \item UF05: Synergistic Research
            \item UF06: An Integrated Strategy for Underground Facilities and Infrastructure
        \end{itemize}
    \end{itemize}
    \item \textit{(In academia)} Please indicate how informed you feel about future scientific directions within the frontiers:
    \begin{itemize}
        \itemsep0em
        \item Accelerator
        \item Cosmic
        \item Community Engagement 
        \item Computational
        \item Energy
        \item Instrumentation
        \item Neutrino Physics
        \item Rare Processes and Precision Measurements
        \item Theory
        \item Underground Facilities
    \end{itemize}
    \item \textit{(In academia)} What sources do you rely on to form your opinions on the future of HEPA? Select all that apply:
    \begin{itemize}
        \itemsep0em
        \item Wikipedia/encyclopedias
        \item Reading scientific papers  
        \item Talking to colleagues/advisor  
        \item Attending workshops/conferences/ talks/seminars  
        \item Joining working groups/performing studies  
        \item Newspapers/magazines (including electronic equivalents) 
        \item Social Media (including blogs, forums, or discussion boards)  
        \item Podcasts, radio, YouTube, documentaries, or other media format 
        \item Other (please specify)
    \end{itemize}
    \item \textit{(In academia)} The next set of questions consider your broad impressions of the direction of the field of High Energy Physics \& Astrophysics (HEPA). For each question, we’ll ask where you think the field is \textbf{currently going} and where you think the field \textbf{should go}. [All questions included ``I don't know'' as an answer option.]
    \begin{itemize}
        \itemsep0em
        \item Small versus large collaborations or experiments [Answer options ranged from ``Smaller'' to ``Balanced'' to ``Larger'']
        \item Focused versus broad experimental programs/facilities [Answer options ranged from ``More focused'' to ``Balanced'' to ``Broader'']
        \item New experimental directions or continuing established programs [Answer options ranged from ``New experimental directions'' to ``Balanced'' to ``Continuing established programs'']
        \item New theoretical ideas or established topics [Answer options ranged from ``More new ideas'' to ``Balanced'' to ``Established topics (no new ideas)'']
        \item Ease versus difficulty in hierarchy ascension across universities, labs, and/or collaborations [Answer options ranged from ``Easier'' to ``No change'' to ``Harder'']
    \end{itemize}
    \item \textit{(In academia)} Do you think long timescales of experimental programs in HEPA are concerning for the field?
    \begin{itemize}
        \itemsep0em
        \item Yes
        \item Maybe
        \item No
        \item I don't know
    \end{itemize}
    \item Which of the following data/software/analysis code do you think should be made open source alongside published results? Select all that apply:
    \begin{itemize}
        \itemsep0em
        \item Raw experimental data for all results 
        \item Raw experimental data only for important / controversial results 
        \item Minimally processed (ready for analysis) data  
        \item Data/results as it appears in publications  
        \item Publication-specific analysis code and simulations  
        \item Fully corrected and reconstructed data/Legacy samples  
        \item Other (please specify)
        \item I don't know/not applicable
    \end{itemize}
    \item \textit{(In academia)} What aspects of research do you think are underfunded across the field? Select all that apply:
    \begin{itemize}
        \itemsep0em
        \item Public data releases and associated storage
        \item Development and maintenance of open source software 
        \item Free access to (potentially costly) software for early-career researchers 
        \item Opportunities for early-career researchers to attend workshops, schools, conferences, collaboration meetings, etc. 
        \item Membership opportunities in collaborations for scientists with limited funding 
        \item Undergraduate research experiences 
        \item Other (please specify)
    \end{itemize}
    \item \textit{(In academia)} Please indicate your knowledge and concern about carbon emissions and environmental impact in HEPA research: [Answer options ranged from ``Not at all (1)'' to ``A great deal (5)'']
    \begin{itemize}
        \itemsep0em
        \item To what extent do you know about HEPA's environmental impact?
        \item How concerned are you about HEPA's environmental impact?
        \item How important is it for you that the environmental impact is taken into account when making decisions on future HEPA projects?
    \end{itemize}
    \item \textit{(In academia)} Taking into account the current Snowmass process so far, please indicate your level of agreement with the following sentences. Feel free to answer regardless of your level of engagement in the process. [Answer options ranged from ``Strongly disagree'' to ``Strongly agree'', including ``I don't know'']
    \begin{itemize}
        \itemsep0em
        \item I feel that my ideas were heard, well represented, and taken seriously
        \item There is no space for me to express my divergent ideas
        \item The process is inclusive towards the whole HEP community
        \item I do not understand the current process and/or its final products
        \item The current Snowmass process structure allows more effective community consensus than previous editions
    \end{itemize}
    \item \textit{(In academia)} Please use this space to add any comments, criticism, or suggestions related to the current Snowmass process. Please be mindful that your comments may be shared with Snowmass process organizers.
    \item The next three questions refer to your workplace prior to COVID-19.
    \item Please select the type of your primary workplace (not necessarily your employer):
    \begin{itemize}
        \itemsep0em
        \item U.S. national lab
        \item National lab outside of the U.S.  
        \item U.S. university  
        \item University outside of the U.S. 
        \item Private industry 
        \item Private foundation  
        \item Government institution  
        \item Other type of institution (feel free to specify)
    \end{itemize}
    \item Please indicate the level of competitiveness at your workplace and whether you consider that level to be healthy:
    \begin{itemize}
        \item Competitive: Answer options were 1-5 from ``Not competitive'' to ``Very competitive''
        \item Health: Answer options were 1-5 from ``Very unhealthy'' to ``Very healthy'' 
    \end{itemize}
    \item \textit{(Undergraduate student, Master's student, PhD student, or PostDoc)} How would you rate the overall availability of your advisor when you need a meeting?
    \begin{itemize}
        \itemsep0em
        \item Never/almost never available (1)
        \item 2
        \item 3
        \item 4 
        \item Always/almost always available (5)
    \end{itemize}
    \item \textit{(Undergraduate student, Master's student, PhD student, or PostDoc)} Are you satisfied with the support from your advisor?
    \begin{itemize}
        \itemsep0em
        \item Very unsatisfied (1)
        \item 2 
        \item 3
        \item 4 
        \item Very satisfied (5)
    \end{itemize}
    \item \textit{(Teaching faculty, tenure-track faculty, tenured faculty, retired faculty, scientist, or senior scientist)} How would you rate your availability for your students when they need a meeting?
    \begin{itemize}
        \itemsep0em
        \item Never/almost never available (1)
        \item 2  
        \item 3  
        \item 4 
        \item Always/almost always available (5)
        \item N/A 
    \end{itemize}
    \item \textit{(Teaching faculty, tenure-track faculty, tenured faculty, retired faculty, scientist, or senior scientist)} How sufficient do you feel your support is to your students?
    \begin{itemize}
        \itemsep0em
        \item Very insufficient (1)
        \item 2
        \item 3 
        \item 4  
        \item Very sufficient (5)
        \item N/A
    \end{itemize}
    \item Were you satisfied with the support from your colleagues?
    \begin{itemize}
        \itemsep0em
        \item Very unsatisfied (1)
        \item 2 
        \item 3
        \item 4 
        \item Very satisfied (5)
    \end{itemize}
    \item \textit{(In academia)} Please rate how much the following factors due to COVID-19 have impacted or will impact you: [Answer options were 1-5 ranging from ``No impact at all'' to ``Significant impact'', including an option for ``I don't know/not applicable'']
    \begin{itemize}
        \item Reduction of funding
        \item Reduction of position openings/delay in hiring
        \item Reduction of experimental activities
        \item Reduction of academic interactions
        \item Less productivity of online classes 
        \item Less interactions between advisors and students
        \item Physical health 
        \item Mental health 
        \item Financial hardship 
    \end{itemize}
    \item How many hours do you spend working (studying, teaching, research, etc.) in a typical week?
    \begin{itemize}
        \itemsep0em
        \item $<20$ hours
        \item 20-30 hours  
        \item 30-40 hours  
        \item 40-50 hours  
        \item 50-60 hours  
        \item 60-80 hours  
        \item $>80$ hours
    \end{itemize}
    \item What fraction of your work is currently being carried out via teleworking?
    \begin{itemize}
        \itemsep0em
        \item All
        \item A majority  
        \item About half  
        \item A minority  
        \item Almost none
    \end{itemize}
    \item The COVID-19 pandemic forced us to adapt to working from home. If given the flexibility to continue to do so, how many days of the work week are you likely to work from home?
    \begin{itemize}
        \itemsep0em
        \item I prefer working at my workplace 
        \item 1-2 days 
        \item 3-4 days 
        \item 5 days 
        \item I'll only go to the workplace when necessary
    \end{itemize}
    \item Please indicate your level of agreement with the following statement: I am able to maintain a healthy work-life balance.
    \begin{itemize}
        \itemsep0em
        \item Strongly agree (1)
        \item 2  
        \item 3  
        \item 4  
        \item Strongly disagree (5)
        \item I don't know  
        \item Prefer not to answer
    \end{itemize}
    \item \textit{(In academia)} Currently, how would you rate your productivity on the following activities compared to the pre-COVID level? [Answer options were 1-5 ranging from ``Significantly more'' to ``The same'' to ``Significantly less'', including an N/A option]
    \begin{itemize}
        \itemsep0em
        \item Researching
        \item Writing publications
        \item Attending conferences/workshops 
        \item Attending meetings 
        \item Teaching/mentoring 
        \item Taking classes/personal development 
        \item Administrative work 
        \item Overall productivity 
    \end{itemize}
    \item \textit{(In academia)} In the last 12 months, how many entirely virtual conferences/workshops have you attended... [Answer options were ``0'', ``1-2'', ``3-5'', ``6-10'', or ``10+'']
    \begin{itemize}
        \itemsep0em
        \item ...as a speaker/presenter?
        \item ...without a formal presenting role? 
    \end{itemize}
    \item \textit{(In academia)} How does the number of conferences/workshops you attended in the last 12 months compare to a typical 12 month period (before the COVID-19 Pandemic)? [Answer options were ``I usually present at (attend) more conferences/workshops'', ``I have presented (attended) at roughly the same number of conferences/workshops'', ``I usually present at (attend) fewer conferences/workshops'', and ``N/A''] 
    \begin{itemize}
        \itemsep0em
        \item As a speaker/presenter 
        \item Without a formal presenting role
    \end{itemize}
    \item \textit{(In academia)} Select all factors that have played a role in the number of conferences/workshops you have attended in the last 12 months: 
    \begin{itemize}
        \itemsep0em
        \item I have not attended conferences  
        \item More conferences available with no registration fee 
        \item Easier to attend conferences due to lack of travel 
        \item Spent more time looking for conferences to replace in-person contact 
        \item More flexible working schedule to make time for conferences 
        \item Lack of time due to additional time constraints related to the pandemic (childcare, etc)
        \item Lack of interest in attending conferences during the pandemic 
        \item Lack of accessibility of virtual conferences  
        \item Conferences I would like to attend are at inconvenient times in my time zone 
        \item Conferences I would like to attend are canceled due to COVID-19 
        \item At a stage where I can participate in conferences and gain more out of them  
        \item Other (please specify)
    \end{itemize}
    \item \textit{(In academia)} Which of the following tools have you found helpful in having a more meaningful virtual conference experience? Please select all that apply:
    \begin{itemize} 
        \itemsep0em
        \item Dedicated Slack/Mattermost/etc. workspace or channel for discussion at a particular conference or session 
        \item Dedicated Zoom (or equivalent) breakout sessions
        \item Video recording of the session so I can watch/rewatch later
        \item Attending multiple parallel sessions at the same time using multiple devices 
        \item Other (please specify)
    \end{itemize}
    \item \textit{(In academia)} What factors will influence the number of in-person conferences you attend after travel restrictions are lifted? Please select all that apply:
    \begin{itemize}
        \itemsep0em
        \item More availability of online conferences
        \item Carbon footprint/environmental impacts 
        \item Availability of funding for in-person conferences 
        \item Building new collaborations/relationships
        \item Reaching herd immunity
        \item Other (please specify)
    \end{itemize}
    \item To your knowledge, which of the following resources/support were provided by your workplace to help continue working during the COVID-19 pandemic? Select all that apply:
    \begin{itemize}
        \itemsep0em
        \item Protective supplies such as masks, gloves, sanitizers, etc.
        \item Flexible working schedule  
        \item Budget for setting up working environment at home  
        \item Additional resources for childcare  
        \item Access to COVID testing  
        \item Assistance with getting COVID vaccine  
        \item Other (please specify)
        \item I don't know if any resources/support were provided  
        \item No resources/support were provided
    \end{itemize}
    \item Do you think harassment (e.g., sexual harassment, racial harassment, power abuse, moral harassment, verbal abuse, disability-based harassment, physical harassment, or gender/sexual orientation-based harassment) is a problem in HEPA?
    \begin{itemize}
        \itemsep0em
        \item Yes
        \item No
        \item I don't know
    \end{itemize}
    \item Trigger warning: Do you feel comfortable answering specific questions about work-related harassment? 

    Disclaimer: You will be asked about sexual harassment or assault experiences you may have had in the past, albeit in a general way. These questions may be triggering, and stimulation of these memories may make you uncomfortable. If you are worried that thinking about or sharing your experiences about sexual harassment or assault will be too mentally harmful, please answer No. 
    \begin{itemize}
        \itemsep0em
        \item Yes
        \item No
    \end{itemize}
    \item \textit{(Willing to answer harassment questions)} Have you experienced or witnessed any type of harassment at work or in work-related relationships? (Examples: sexual harassment, racial harassment, power abuse, moral harassment, verbal abuse, disability-based harassment, physical harassment, or gender/sexual orientation-based harassment)
    \begin{itemize}
        \itemsep0em
        \item Yes, I have experienced
        \item Yes, I have witnessed
        \item No
    \end{itemize}
    \item \textit{(Reported experiencing harassment)} What type of harassment did you \textbf{experience}? Mark all that apply:
    \begin{itemize}
        \itemsep0em
        \item \textbf{Sexual Harassment} (Including but not limited to sharing sexual photos/pornography; sexual comments, jokes, or questions; inappropriate sexual touching or gestures; catcalling; unsolicited explicit pictures; quid pro quo; stalking)
        \item \textbf{Racial Harassment} (Including but not limited to racial slurs, insults, or jokes; degrading comments; disgust; intolerance of differences; intolerance toward religious traditions and customs; racial profiling; stalking) 
        \item \textbf{Power Abuse} (Including but not limited to excessive demands that are impossible to meet; demeaning demands far below the employee's capability; intrusion into the employee's personal life; quid pro quo) 
        \item \textbf{Moral Harassment} (Including but not limited to isolating or denying the victim's presence; belitting or trivializing the victim's thoughts; discrediting or spreading rumors about the victim; opposing or challenging everything the victim says; bullying) 
        \item \textbf{Verbal Abuse} (Including but not limited to threatening, yelling, or cursing at a victim in public or in private)  (5)
        \item \textbf{Disability-Based Harassment} (Harassment directed towards individuals that have a disability, are acquainted with a disable person, or use disability services) 
        \item \textbf{Physical Harassment} (Including but not limited to direct threats of intent to inflict harm; physical attacks such as hitting, shoving, or kicking; threatening behavior such as shaking fists angrily; destroying property to intimidate) 
        \item \textbf{Gender or Sexual Orientation-Based Harassment} (Hostile, offensive, or intimidating behavior because of your gender or sexual orientation)
    \end{itemize}
    \item \textit{(Reported experiencing harassment)} How many times did you \textbf{experience} harassment?
    \begin{itemize}
        \itemsep0em
        \item Once 
        \item More than once a year 
        \item More than once a month
        \item More than once a week 
        \item Daily
    \end{itemize}
    \item \textit{(Reported experiencing harassment)} Where did you \textbf{experience} harassment? Mark all that apply.
    \begin{itemize}
        \itemsep0em
        \item Personal work space (office/cubicle) 
        \item Perpetrator work space 
        \item University/laboratory campus, but not in an office 
        \item University classroom 
        \item During work at a laboratory 
        \item During a conference
        \item Outside of work, but by a work-related person 
        \item Online 
        \item Other (please specify)
    \end{itemize}
    \item \textit{(Reported experiencing harassment)} Where was the perpetrator(s) in their career when you \textbf{experienced} harassment? Mark all that apply:
    \begin{itemize}
        \itemsep0em
        \item Undergraduate student
        \item Master's student/candidate
        \item PhD student/candidate 
        \item PostDoc 
        \item Engineer 
        \item Technician 
        \item My advisor
        \item Teaching faculty 
        \item Tenure-track faculty 
        \item Tenured faculty 
        \item Retired faculty 
        \item Scientist 
        \item Senior Scientist  
        \item Other (please specify)
    \end{itemize}
    \item \textit{(Reported experiencing harassment)} Did you report the incidents you \textbf{experienced}?
    \begin{itemize}
        \itemsep0em
        \item Yes
        \item No
    \end{itemize}
    \item \textit{(Reported witnessing harassment)} What type of harassment did you \textbf{witness}? Mark all that apply: 
    \begin{itemize}
        \itemsep0em
        \item \textbf{Sexual Harassment} (Including but not limited to sharing sexual photos/pornography; sexual comments, jokes, or questions; inappropriate sexual touching or gestures; catcalling; unsolicited explicit pictures; quid pro quo; stalking)
        \item \textbf{Racial Harassment} (Including but not limited to racial slurs, insults, or jokes; degrading comments; disgust; intolerance of differences; intolerance toward religious traditions and customs; racial profiling; stalking) 
        \item \textbf{Power Abuse} (Including but not limited to excessive demands that are impossible to meet; demeaning demands far below the employee's capability; intrusion into the employee's personal life; quid pro quo) 
        \item \textbf{Moral Harassment} (Including but not limited to isolating or denying the victim's presence; belitting or trivializing the victim's thoughts; discrediting or spreading rumors about the victim; opposing or challenging everything the victim says; bullying) 
        \item \textbf{Verbal Abuse} (Including but not limited to threatening, yelling, or cursing at a victim in public or in private)  (5)
        \item \textbf{Disability-Based Harassment} (Harassment directed towards individuals that have a disability, are acquainted with a disable person, or use disability services) 
        \item \textbf{Physical Harassment} (Including but not limited to direct threats of intent to inflict harm; physical attacks such as hitting, shoving, or kicking; threatening behavior such as shaking fists angrily; destroying property to intimidate) 
        \item \textbf{Gender or Sexual Orientation-Based Harassment} (Hostile, offensive, or intimidating behavior because of your gender or sexual orientation)
    \end{itemize}
    \item \textit{(Reported witnessing harassment)} How many times did you \textbf{witness} harassment?
    \begin{itemize}
        \itemsep0em
        \item Once 
        \item More than once a year 
        \item More than once a month
        \item More than once a week 
        \item Daily
    \end{itemize}
    \item \textit{(Reported witnessing harassment)} Where did you \textbf{witness} harassment? Mark all that apply.
    \begin{itemize}
        \itemsep0em
        \item Personal work space (office/cubicle) 
        \item Perpetrator work space 
        \item University/laboratory campus, but not in an office 
        \item University classroom 
        \item During work at a laboratory 
        \item During a conference
        \item Outside of work, but by a work-related person 
        \item Online 
        \item Other (please specify)
    \end{itemize}
    \item \textit{(Reported witnessing harassment)} Where was the perpetrator(s) in their career when you \textbf{witnessed} harassment? Mark all that apply:
    \begin{itemize}
        \itemsep0em
        \item Undergraduate student
        \item Master's student/candidate
        \item PhD student/candidate 
        \item PostDoc 
        \item Engineer 
        \item Technician 
        \item My advisor
        \item Teaching faculty 
        \item Tenure-track faculty 
        \item Tenured faculty 
        \item Retired faculty 
        \item Scientist 
        \item Senior Scientist  
        \item Other (please specify)
    \end{itemize}
    \item \textit{(Reported witnessing harassment)} Did you report the incidents you \textbf{witnessed}?
    \begin{itemize}
        \itemsep0em
        \item Yes
        \item No
    \end{itemize}
    \item Have you experienced or witnessed racial discrimination or racist acts of any kind within interactions \textbf{outside of the work environment} in the location you currently reside? [Answer options were ``Yes'', ``Maybe'', ``No'', and ``Prefer not to say'']
    \begin{itemize}
        \itemsep0em
        \item Experienced
        \item Witnessed
    \end{itemize}
    \item Have you experienced or witnessed racial discrimination or racist acts of any kind \textbf{through your professional work} (including conferences, etc)? [Answer options were ``Yes'', ``Maybe'', ``No'', and ``Prefer not to say'']
    \begin{itemize}
        \itemsep0em
        \item Experienced
        \item Witnessed
    \end{itemize}
    \item \textit{(In academia)} Do you consider your immediate working group (i.e., members of the HEPA community you regularly work with on scientific projects) to be diverse in their backgrounds or identities? [Answer options were ``No representation'', 1-5 ranging from ``Inadequate representation'' to ``Good representation'', and ``I don't know'']
    \begin{itemize}
        \itemsep0em
        \item Racially, ethnically, or culturally
        \item Gender identity
        \item Origins, nationalities, or citizenships
    \end{itemize}
    \item What factors do you feel are significant in limiting racial, ethnic, and cultural diversity within HEPA? [Answer options were ``Not significant'', ``Significant'', and ``I don't know'']
    \begin{itemize}
        \itemsep0em
        \item Lack of diversity in candidates for positions at any level 
        \item Lack of networking/outreach with members of underserved communities 
        \item Lack of financial support (e.g., scholarships) available to members of underserved communities
        \item Lack of leadership and or teaching expertise within underserved communities
        \item Lack of blinded application processes
        \item Lack of openly focused hiring practices prioritizing employment of various underserved communities
        \item Professional priority differences across various racial or ethnic groups
        \item Closeted or subconscious discrimination or racial bias against racial, ethnic, or cultural groups
        \item Open or overt discrimination or racial bias against various racial, ethnic, or cultural groups
        \item General indifference to racial, ethnic, or cultural issues
    \end{itemize}
    \item Please enter any other factors you feel are significant in limiting diversity within your department or field (these can include factors derived from norms, biases, or past/present laws)
    \item What steps do you feel should be taken to ensure equitable hiring practices in HEPA? Choose all that apply:\footnote{Contributed by the SEC CEF groups.}
    \begin{itemize}
        \itemsep0em
        \item Current practices are adequate
        \item Identity-hidden applications and/or letters of reference (candidate’s name/institute hidden) 
        \item Mentorship and professional’s development programs 
        \item Culturally sensitive admins/Human Resources departments 
        \item Affirmative action principles 
        \item Better bias training for hiring committees 
        \item Enacting policies that support better work-life balance 
        \item I don't know 
        \item Other (please specify)
    \end{itemize}
    \item What is your age?
    \begin{itemize}
        \item Your age (please enter an integer number):
        \item Prefer not to answer
    \end{itemize}
    \item What is your race or ethnicity? Select all that apply to you:
    \begin{itemize}
        \itemsep0em
        \item American Indian or Alaska Native (e.g., Navajo Nation, Blackfeet Tribe, Native Village of Barrow Inupiat Traditional Government, Nome Eskimo Community)
        \item Asian (e.g., Chinese, Filipino, Asian Indian, Vietnamese, Korean, Japanese) 
        \item Black or African American (e.g., Jamaican, Haitian, Nigerian, Ethiopian, Somalian) 
        \item Hispanic or Latino (e.g., Mexican or Mexican American, Puerto Rican, Cuban, Salvadoran, Dominican, Colombian, Brazilian) 
        \item Middle Eastern or North African (e.g., Lebanese, Iranian, Egyptian, Syrian, Moroccan, Algerian)
        \item Native Hawaiian or Other Pacific Islander (e.g., Native Hawaiian, Samoan, Chamorro, Tongan, Fijian, Marshallese)
        \item White (e.g., German, Irish, English, Italian, Polish, French)
        \item Some other race, ethnicity, or origin (please specify)
        \item Prefer not to answer
    \end{itemize}
    \item How do you currently describe your gender? Select all that apply:
    \begin{itemize}
        \itemsep0em
        \item Man
        \item Woman 
        \item Nonbinary, agender, or genderqueer
        \item Self identity (please specify)
        \item Prefer not to answer
    \end{itemize}
    \item Do you identify as Transgender?
    \begin{itemize}
        \itemsep0em
        \item Yes
        \item No
        \item Prefer not to answer
    \end{itemize}
    \item Which of the following best describe your sexual orientation? Select all that apply:
    \begin{itemize}
        \itemsep0em
        \item Asexual 
        \item Bisexual or pansexual 
        \item Gay or lesbian 
        \item Straight or heterosexual 
        \item Self identify (please specify)
        \item Prefer not to answer
    \end{itemize}
    \item Location of citizenship (if you are a citizen of multiple locations, you may select up to 3 by using the 'ctrl' or 'cmd' key)
    \begin{itemize}
        \itemsep0em
        \item My country/region is not listed
        \item Prefer not to answer
        \item A list of 196 locations 
    \end{itemize}
    \item Please select the appropriate location for the following items: [Answer options were the same as the previous question]
    \begin{itemize}
        \itemsep0em
        \item Location of your current employer (e.g., lab or university)
        \item Location you currently live at
    \end{itemize}
    \item \textit{(Selected ``My country/region is not listed'' at some point in the last two questions)} If your country/region was not listed for any of the previous questions, please indicate here for all the appropriate questions: 
    \begin{itemize}
        \itemsep0em
        \item Location of citizenship
        \item Location of current employer
        \item Location of current residence
    \end{itemize}
    \item \textit{(Not a citizen of the U.S. but lives or works in the U.S.)} Are you a Green card holder or permanent resident of the United States?
    \begin{itemize}
        \itemsep0em
        \item Yes
        \item No
    \end{itemize}
    \item \textit{(Not a Green card holder or permanent resident of the U.S.)} What US visa do you hold? [Answer options were a list of 129 different U.S. visa types]
    \item \textit{(Not a citizen of the U.S. but lives or works in the U.S.)} What are your concerns related to U.S. policies on visa issuance, maintaining status, renewal, work permit, etc.? Check all that apply.
    \begin{itemize}
        \itemsep0em
        \item Unpredictability of the visa application process 
        \item Lack of clarity on reason for visa denial
        \item Lack of accessible options for amending an application (e.g., no options for amendment, high amendment fees) 
        \item Lack of consular support as a researcher/scientist 
        \item Inadequate support from the university’s international student/scholar office
        \item Undefined/long wait time for cases with administrative processing 
        \item High and non-refundable visa application/renewal fees 
        \item Managing your current student/employment-based visa status (e.g., F1, J1, H1B) 
        \item Navigating the Optional Practical Training (OPT) process
        \item Navigating the Curricular Practical Training (CPT) process 
        \item Managing the non-immigrant visa status of your family or significant other/partner
        \item Travel restrictions between the U.S. and your home country
        \item Impact on future employment-based visa (e.g. H1B) 
        \item Concerns about your personal safety 
        \item Other (please specify) 
        \item Not applicable
    \end{itemize}
    \item \textit{(Not a citizen of the U.S. but lives or works in the U.S.)} Do the U.S. policies on visa affect your career decisions in the short term (e.g., 2 years from now)?
    \begin{itemize}
        \itemsep0em
        \item No. I don’t see a problem for my current or future visa status in the U.S. 
        \item No. I don’t plan/want to stay in the U.S. 
        \item Maybe. I am not sure about my next position. 
        \item Yes. I plan to move out of the U.S. due to immigration issues/policies on visas.
        \item I'm planing to move out of the U.S. for other reasons.  
        \item I don't know. 
        \item Other (please specify)
        \item Not applicable
    \end{itemize}
    \item What is your current marital status? Select all that apply:
    \begin{itemize}
        \itemsep0em
        \item Married 
        \item Marriage-like or long-term relationship 
        \item Widowed 
        \item Separated/divorced 
        \item Single 
        \item Prefer not to answer
    \end{itemize}
    \item \textit{(Married or in marriage-like relationship)} Do you live with your partner(s)?
    \begin{itemize}
        \itemsep0em
        \item Yes
        \item No
        \item Other (please specify)
        \item Prefer not to answer
    \end{itemize}
    \item \textit{(Married or in marriage-like relationship)} Does your partner(s) work in academia?
    \begin{itemize}
        \itemsep0em
        \item Yes
        \item No
        \item Prefer not to answer
    \end{itemize}
    \item Do you have caregiving responsibilities? These include parenting, caring for a senior, or caring for a person with a disability or medical condition.
    \begin{itemize}
        \itemsep0em
        \item Yes
        \item No
        \item Prefer not to answer
    \end{itemize}
    \item What was (or is) your personal experience in affording higher education (college or university)? Select all that apply, and also indicate whether you obtained (or are obtaining) any of your degrees outside of the U.S. [Answer options were ``Bachelor's'', ``Master's'', and ``PhD'']
    \begin{itemize}
        \itemsep0em
        \item Obtained outside U.S.
        \item Family support 
        \item Personal income (not from your family) 
        \item Merit/need-based scholarships 
        \item Stipends from your school for work performed (e.g. teaching/research assistantship). 
        \item Loans
        \item Military service or equivalent (e.g. the G. I. Bill)
        \item Free public education
        \item Other 
        \item Not applicable
    \end{itemize}
    \item What is the highest level of degree (or international equivalent) obtained by your immediate family members, excluding yourself?
    \begin{itemize}
        \itemsep0em
        \item US high school diploma or General Educational Development (GED)
        \item Associate degree
        \item Bachelor's degree
        \item Master's degree
        \item Doctorate degree
        \item I don't know
    \end{itemize}
    \item While you were a student, did you ever consider \textbf{withdrawing from} college or graduate study for any reason? Select all that apply:
    \begin{itemize}
        \itemsep0em
        \item Financial difficulties 
        \item Mental health difficulties 
        \item General health difficulties 
        \item Loss of interest in your studies 
        \item Lack of guidance or advice from teachers or mentors 
        \item Family difficulties
        \item Romantic difficulties 
        \item Child-rearing or caretaking difficulties 
        \item Desire for better benefits (e.g., family leave, health care, child care, retirement, etc.) 
        \item Toxic environment due to harassment, racism, etc. 
        \item Other difficulties (please specify)
        \item No, I did not consider leaving college or graduate study. 
        \item Not applicable
    \end{itemize}
    \item \textit{(Has caregiving responsibilities)} Are you willing to answer some questions about your caregiving responsibilities? 
    \begin{itemize}
        \itemsep0em
        \item Yes
        \item No
    \end{itemize}
    \item \textit{(Willing to answer additional caregiving questions)} How would you describe your caregiving responsibilities?
    \begin{itemize}
        \itemsep0em
        \item Sole caregiver
        \item Equally shared caregiving 
        \item Part-time caregiving 
        \item Other (please specify)
        \item Prefer not to answer
    \end{itemize}
    \item \textit{(Willing to answer additional caregiving questions)} Who do you care for? Select all that apply:
    \begin{itemize}
        \itemsep0em
        \item Children (under 18) 
        \item Senior 
        \item Person with disability or medical condition
    \end{itemize}
    \item \textit{(Cares for children)} What age group do you children belong to, and do they live with you? Mark all that apply. [Answer options were ``They do not live with me'', ``They live with me part-time'', and ``They live with me full-time''.]
    \begin{itemize}
        \itemsep0em
        \item Preschool (birth to 5 years old)
        \item Elementary (6 to 13 years old)
        \item Adolescent (14 to 18 years old) 
    \end{itemize}
    \item \textit{(Cares for senior or person with a disability or medical condition)} Do you live with the person you care for?
    \begin{itemize}
        \itemsep0em
        \item I do not live with them.
        \item I live with them part-time.
        \item I live with them full-time.
    \end{itemize}
    \item \textit{(Willing to answer additional caregiving questions)} How often do you have to rearrange your work schedule, decrease your hours, or take unpaid leave because of your caregiving responsibilities?
    \begin{itemize}
        \itemsep0em
        \item Not at all (1)
        \item 2 
        \item 3 
        \item 4
        \item Substantially (5)
        \item I don't know 
        \item Prefer not to answer
    \end{itemize}
    \item \textit{(Willing to answer additional caregiving questions)} Approximately how many hours per week are you free of all caregiving responsibility (e.g., kids at school, daycare, with a babysitter, etc)? Please enter an integer number.
    \item \textit{(Willing to answer additional caregiving questions)} At the time of this survey, how much time and effort do you spend on caregiving responsibilities compared to pre-COVID time and effort?
    \begin{itemize}
        \itemsep0em
        \item Significantly more (1)
        \item 2 
        \item The same (3)
        \item 4 
        \item Significantly less (5)
        \item N/A
    \end{itemize}
    \item \textit{(Willing to answer additional caregiving questions)} Does your current place of employment offer Family and Medical leaves? (Family leaves include maternity/paternity leaves for newborn/adopted child, care for immediate family member with serious medical condition, or medical leave when employee is unable to work due to serious illness.)
    \begin{itemize}
        \itemsep0em
        \item Yes, it's paid 
        \item Yes, it's unpaid 
        \item No 
        \item I don't know
    \end{itemize}
    \item \textit{(Willing to answer additional caregiving questions)} At a time when you would have been eligible for family leave, was it available and did you take it?
    \begin{itemize}
        \itemsep0em
        \item Yes it was available, and I took the leave
        \item Yes it was available, and I did not take the leave 
        \item No it was not available, but I had an understanding with my superior and was able to take time off 
        \item No it was not available, but I took time off anyway 
        \item No it was not available, and I did not take off 
        \item I don't know whether my employer offered family leave
    \end{itemize}
    \item \textit{(Willing to answer additional caregiving questions)} Have you or are you currently suffering any professional setbacks in utilizing Family and Medical leave?
    \begin{itemize}
        \itemsep0em
        \item Not at all (1)
        \item 2 
        \item 3 
        \item 4 
        \item Substantial setbacks (5)
        \item I don't know
        \item Prefer not to answer
    \end{itemize}
    \item \textit{(Willing to answer additional caregiving questions)}  Do/did you feel supported in academia as a caregiver? [Answer choices were 1-5 from ``Not at all'' to ``All the time'', ``N/A'', ``I don't know'', and ``Prefer not to answer'']
    \begin{itemize}
        \item By your peers/immediate working group
        \item By your place of employment 
        \item By funding agencies 
        \item By the field 
    \end{itemize}
    \item \textit{(Willing to answer additional caregiving questions)} Please feel free to write about your experience as a caregiver in academia.
    \item Thank you for taking the Snowmass Community Survey! We truly appreciate your participation, and we wish you a productive Snowmass Meeting. Please click the arrow in the bottom-right corner of this page to officially record your response.

    If you care to give us any comments on any section(s) or question(s), please feel free to enter your thoughts in the text box below (500 character limit).
    
\end{itemize}

\end{multicols}
\newpage

\section{Additional Survey Results}

\begin{figure}[H]
    \includegraphics[scale=0.8]{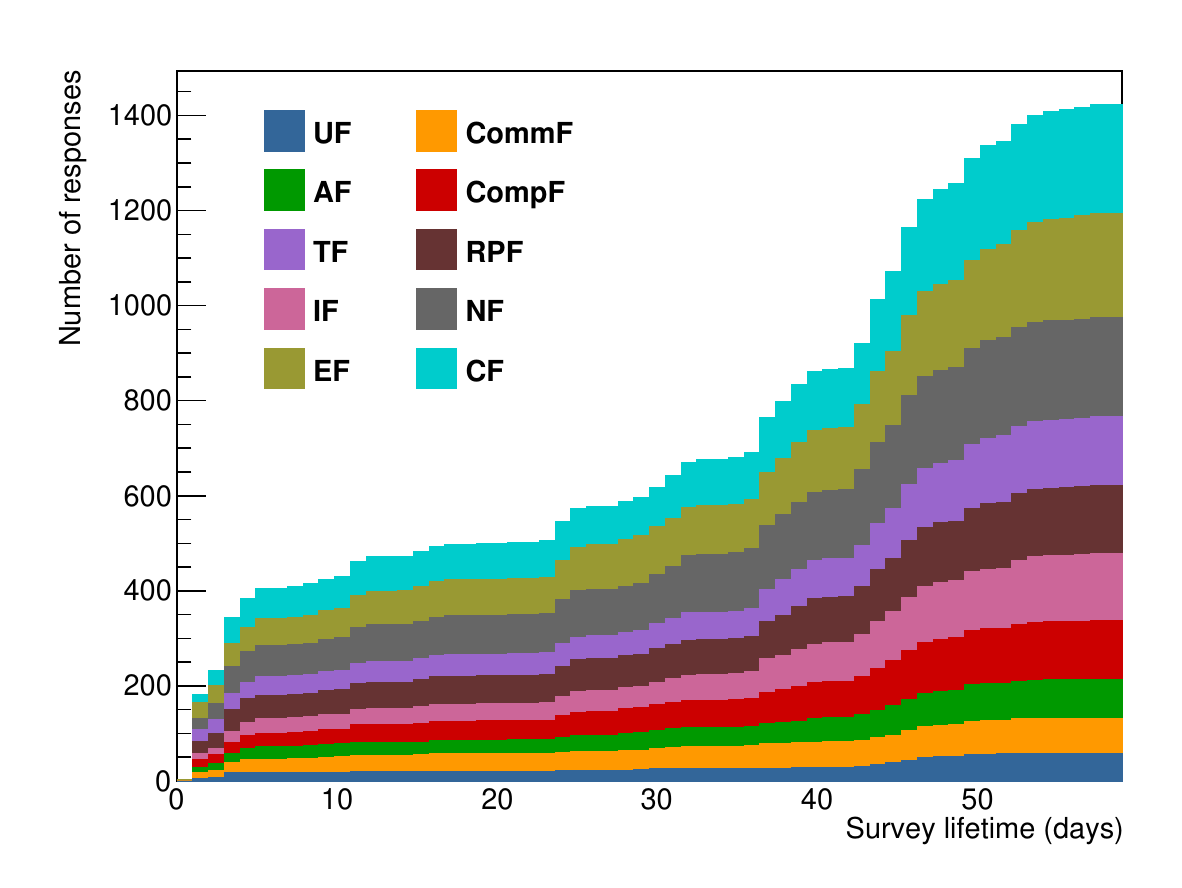}
    \caption{Responses over survey lifetime broken into different Snowmass frontiers. A little over two-fifths of respondents selected more than one frontier in which they work, which is why the y-axis extends beyond the total number of respondents.}
    \label{metadata_interactions_time_frontiers}
\end{figure}

\begin{figure}[H]
    \includegraphics[scale=0.8]{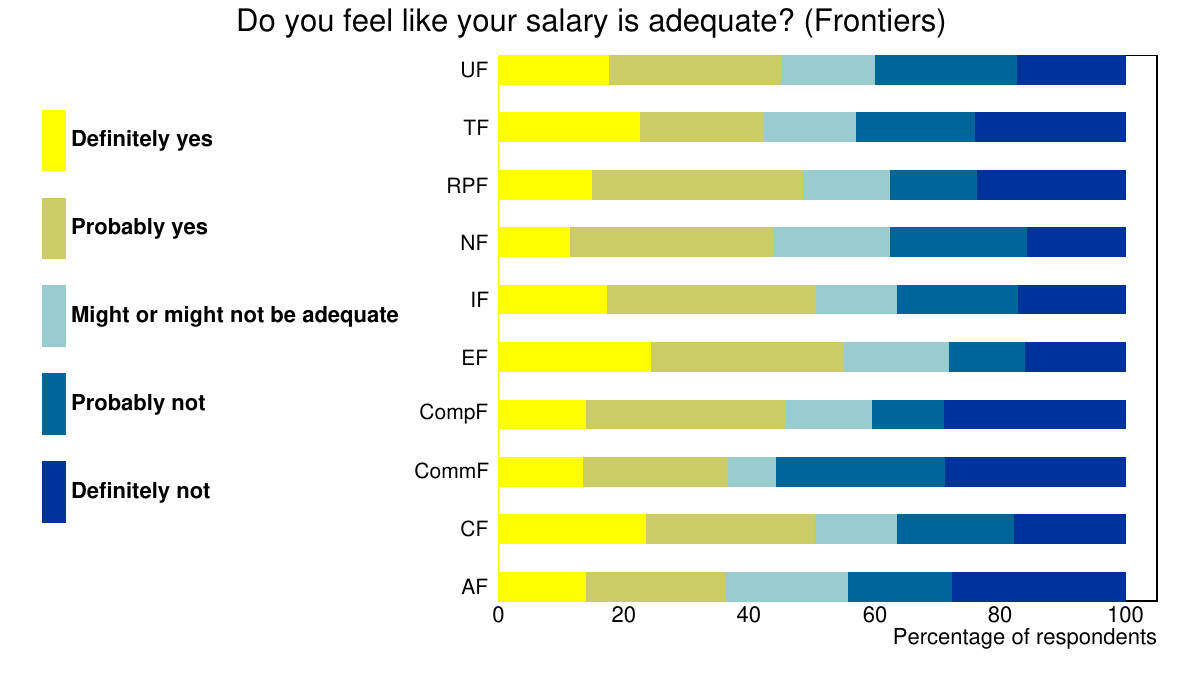}
    \caption{We asked PhD students, Postdocs, and faculty whether they feel like their salary is adequate. We split the responses by the Snowmass frontier(s) the respondents work in.}
\end{figure}

\begin{figure}[H]
    \includegraphics[scale=0.47]{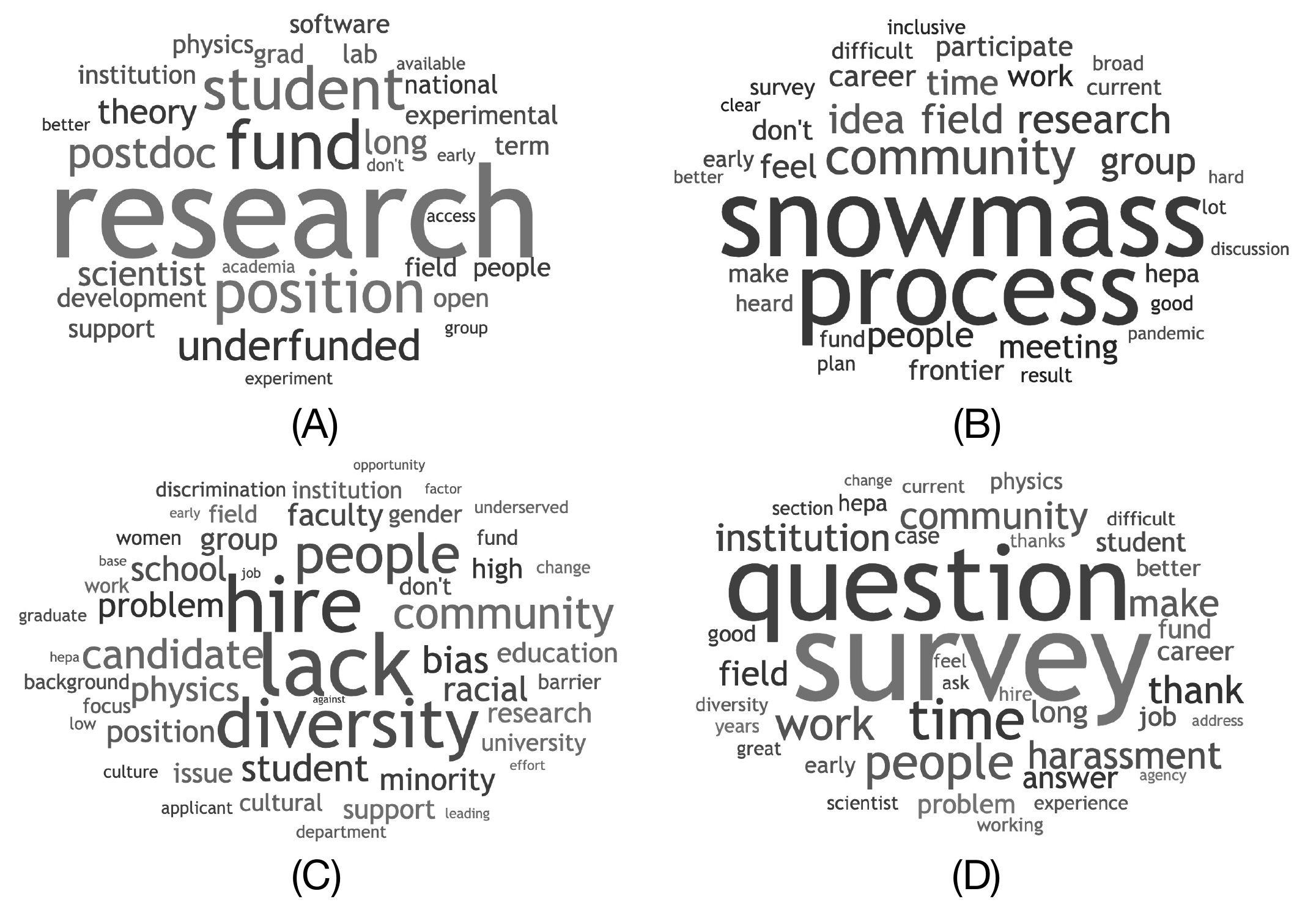}
    \caption{Word clouds for questions where respondents could provide additional comments. (A) The top 30 commonly used words by respondents in academia when asked about what aspects of research are underfunded by the field. Around 6\% of survey respondents provided a response to this world cloud. (B) The top 35 commonly used words from survey respondents in academia who provided comments, criticism, or suggestions related to the current Snowmass process. Nearly 11\% of survey respondents provided a comment for this question. (C) The top 50 commonly used words from respondents who were asked to comment on any other factors they felt are significant in limiting diversity within their department or field. Fourteen percent of all survey respondents provided a comment. (D) The top 40 commonly used words from respondents with additional comments about the 2021 Snowmass Community Survey. Around 9\% of survey respondents provided a comment.}
\end{figure}
\newpage

\section{Thoughts for Future Surveyors}

In this appendix, the authors hope to provide some additional context behind the 2021 survey's construction. We also hope to provide some starting points and potential pitfalls for future Snowmass surveyors so that they might have a better grasp of the Snowmass Community Survey's undertaking. Perhaps in a decade from now, our recommendations will be outdated; let us hope for some compromise in between.

Starting as early as possible in developing the survey proved a valuable decision, as we committed several months to merely discussing the survey's construction. The Survey Initiative developed a Gantt chart which helped set a schedule to meet deadlines and accomplish goals. We also utilized a shared Google Drive, Slack messaging, and GitHub to securely store minutes, presentations, data, ideas, drafts, and analysis. By working on Slack, some tasks could be dealt with more quickly and easily -- but this only worked because everyone in the SEC Survey Initiative used Slack on a regular basis.

We spent a couple of weeks discussing how long the 2021 survey should be, and several more weeks trimming down the survey. To trim down the survey, we discussed whether the questions were useful, data could be obtained with outside of the 2021 survey, and whether two or more questions could be combined. Using Qualtrics features as rough estimation tools, we aimed to have the survey take at maximum one hour of the respondent's time (i.e., in the event where the respondent writes detailed comments in several text boxes, display logic opens up extra questions in several sections, etc.). As the survey progressed, we saw a decrease in response rate of questions visible to all respondents: 99\% of academic respondents responded to the question about their current career stage, while 78\% answered the question about caregiving responsibilities in the demographics section. The drop-off could be due to the length of the 2021 survey, and this possibility had been discussed during the construction phase. We ultimately decided to make the majority of the survey optional due to the number of questions included, and we ultimately saw a sacrifice in the response rate due to that decision. We also received several constructive and a few bewildered comments in the 2021 survey's final question about the survey's length and the time it took to complete. In any event, the response rate for the 2021 survey was tremendously and graciously successful. 

Using display logic allowed the survey to show sets of questions to different groups of people (e.g., in different career stages), or show additional answers for certain groups of people. Dealing with display logic could be a challenging and sometimes irritating experience (hopefully this comment will be in the ``outdated'' category). However, display logic was extraordinarily useful in trimming down the run length of the survey, although display logic also made it difficult to gauge the time to complete the survey.

One of the most effective processes the team undertook was to distribute a draft of the 2021 survey to outside reviewers for feedback. The team opted for a group which had a diverse set of backgrounds and experiences; in this way, many possible display logic routes would be tested. Some people pointed out bugs or grammatical errors, while others provided suggestions for alternate syntax or additional answer options. The team also had another preliminary estimation of how long the survey took to complete. This experience prompted rich discussions during the final weeks of the survey's construction, and many of the reviewers' suggestions were incorporated into the finalized version of the 2021 survey.

While constructing the 2021 survey, the team discussed at length about whether the survey required Institutional Review Board (IRB) approval. The team reached out to several IRBs at our various institutions and researched the requirements at length. We began researching IRB approval because in the event that the survey required IRB approval, the overall process could take months to complete. Ultimately we determined that the 2021 survey did not require IRB approval, and this information was included in the survey's preface. 

An exercise that would have been useful for us to complete is to develop an advertising plan. There is a ``lull'' in the beginning of the response rate shown in Figure \ref{overall_num_respondents_dates}; admittedly, we developed a preliminary advertisement plan but did not have concrete plans for the survey's full lifetime. The team eventually developed an effective plan on the fly, but there remained room for improvement. For example, we could have developed specific strategies to advertise to different career stages such as undergraduate students, engineers, technicians, and non-academics. We also could have pursued other forums or media outlets such as APS to advertise the 2021 survey. Some of these efforts can take weeks or months. Finally, the social media copy of the 2021 survey did not receive heavy traffic, but perhaps would have received more with a developed advertising plan.

The survey's livetime is an opportunity to develop analysis tools. Survey software such as Qualtrics offers tools to generate mock data. While it might not be realistic, the mock data provides an opportunity to test code and other analytical tools without opening the box early.


\newpage
\bibliographystyle{style}
\bibliography{bibliography}

\end{document}